\documentclass[fleqn,usenatbib]{mnras}

\usepackage{newtxtext,newtxmath}

\usepackage[T1]{fontenc}

\DeclareRobustCommand{\VAN}[3]{#2}
\let\VANthebibliography\thebibliography
\def\thebibliography{\DeclareRobustCommand{\VAN}[3]{##3}\VANthebibliography}

\usepackage{xspace}
\usepackage{graphicx}
\usepackage{natbib}
\usepackage{booktabs}
\usepackage{array}
\usepackage{pdflscape}
\usepackage{orcidlink}
\usepackage{longtable}
\usepackage{caption}

\renewcommand{\arraystretch}{\bodystretch}

\newcommand{\panelgap}{\hspace{2.5em}}

\makeatletter
\newcommand{\resetpageheight}{%
  \global\@colht\textheight
  \global\@colroom\textheight
  \global\vsize\textheight}
\makeatother

\newcommand{\bodystretch}{1.15}
\newcommand{\litstretch}{1.32}
\newcommand{\medianstretch}{1.68}
\newcommand{\bimodalstretch}{1.82}

\newcommand\tess{\textit{TESS}\xspace}
\newcommand\gaia{\textit{Gaia}\xspace}
\newcommand\exofast{\texttt{EXOFASTv2}\xspace}
\newcommand\mj{M$_{\rm J}$\xspace}
\newcommand\msun{M$_\odot$\xspace}

\newcommand\rj{R$_{\rm J}$\xspace}

\newcommand{\bjdtdb}{\ensuremath{\rm {BJD_{TDB}}}}

\newcommand{\rsun}{R$_\odot$\xspace}
\newcommand{\lsun}{L$_\odot$\xspace}
\newcommand{\teff}{$T_{\rm eff}$\xspace}

\newcommand{\totHJs}{669\xspace} 
\newcommand{\nearetrievaldate}{2026 Sep 18\xspace}

\graphicspath{{./}{figures/}}

\title[MEEP III: 29 Giant Planets]{Migration and Evolution of giant ExoPlanets (MEEP). III. Twenty-Nine Giant Planets from the \tess Mission}

\author[Schulte et al.]{
\parbox{\textwidth}{
Jack Schulte$^{1\orcidlink{0000-0002-7382-0160}}$\thanks{E-mail: jschulte@msu.edu},
Joseph~E.~Rodriguez$^{1\orcidlink{0000-0001-8812-0565}}$,
David~W.~Latham$^{2\orcidlink{0000-0001-9911-7388}}$,
Allyson Bieryla$^{2\orcidlink{0000-0001-6637-5401}}$,
Karen A.~Collins$^{2\orcidlink{0000-0001-6588-9574}}$,
Samuel~N.~Quinn$^{2\orcidlink{0000-0002-8964-8377}}$,
Xian-Yu Wang$^{3\orcidlink{0000-0002-0376-6365}}$,
Rafael Brahm$^{4\orcidlink{0000-0002-9158-7315}}$,
Abdullah Al-Haji$^{5\orcidlink{0009-0005-6680-8373}}$,
Robert Aloisi$^{6\orcidlink{0000-0003-2822-616X}}$,
Caitlyn Altermatt$^{7,8\orcidlink{0009-0002-4848-7631}}$,
Nikola Antonov$^{9,10\orcidlink{0009-0002-1484-7029}}$,
David Baker$^{11\orcidlink{0000-0002-2970-0532}}$,
Pavel Balanutsa$^{12\orcidlink{0009-0007-2413-3851}}$,
Khalid Barkaoui$^{13,14,15\orcidlink{0000-0003-1464-9276}}$,
\"Ozg\"ur Ba\c{s}t\"urk$^{16,17\orcidlink{0000-0002-4746-0181}}$,
Slawomir Bednarz$^{18\orcidlink{0009-0000-1171-3370}}$,
Paul Benni$^{19\orcidlink{0000-0001-6981-8722}}$,
Krzysztof Bernacki$^{18\orcidlink{0000-0003-4647-7114}}$,
Mario Billiani$^{20\orcidlink{0000-0002-3278-9590}}$,
Ventsislav Bodakov$^{21}$,
Pau Bosch-Cabot$^{22\orcidlink{0000-0002-1514-5558}}$,
Valerio Bozza$^{23,24\orcidlink{0000-0003-4590-0136}}$,
Ahmet Cem Kutluay$^{25\orcidlink{0009-0007-0502-7359}}$,
David R. Ciardi$^{26\orcidlink{0000-0002-5741-3047}}$,
Catherine A.~Clark$^{26\orcidlink{0000-0002-2361-5812}}$,
Jerome de Leon$^{27\orcidlink{0000-0002-6424-3410}}$,
Jason~D.~Eastman$^{2\orcidlink{0000-0003-3773-5142}}$,
\.{I}pek Aleyna Ert\"{u}rk$^{16,17\orcidlink{0009-0000-8026-2104}}$,
Thomas M. Esposito$^{28,29,30\orcidlink{0000-0002-0792-3719}}$,
Phil Evans$^{31\orcidlink{0000-0002-5674-2404}}$,
Miranda Felsmann$^{32\orcidlink{0009-0002-3987-5798}}$,
Raquel For\'es-Toribio$^{33,34\orcidlink{0000-0002-6482-2180}}$,
Isaac Fournier$^{1}$,
Akihiko Fukui$^{27,35\orcidlink{0000-0002-4909-5763}}$,
Tianjun Gan$^{35,36\orcidlink{0000-0002-4503-9705}}$,
Cristilyn Gardner-Watkins$^{2\orcidlink{0000-0001-8621-6731}}$,
Purvi Garg$^{1\orcidlink{0009-0009-1914-6054}}$,
Eric Girardin$^{37\orcidlink{0000-0002-5443-3640}}$,
Tateki Goto$^{20\orcidlink{0000-0002-9540-6112}}$,
Ferran Grau Horta$^{38\orcidlink{0000-0001-9927-7269}}$,
Thomas Henning$^{39\orcidlink{0000-0002-1493-300X}}$,
Eric G.~Hintz$^{40\orcidlink{0000-0002-9867-7938}}$,
Keith Horne$^{41\orcidlink{0000-0003-1728-0304}}$,
Kai Ikuta$^{42\orcidlink{0000-0002-5978-057X}}$,
Jon M.~Jenkins$^{43\orcidlink{0000-0002-4715-9460}}$,
Andr\'es Jord\'an$^{4,44\orcidlink{0000-0002-5389-3944}}$,
Jenson Keen$^{20\orcidlink{0009-0003-7663-7112}}$,
Adam Kraus$^{45\orcidlink{0000-0001-9811-568X}}$,
Valentina Kuzmina$^{46\orcidlink{0009-0002-8921-3581}}$,
Artem Kuznetsov$^{12\orcidlink{0000-0002-6010-4114}}$,
Adam Lark$^{47\orcidlink{0009-0009-2881-7112}}$,
Rodrigo Leiva$^{48\orcidlink{0000-0002-6477-1360}}$,
Vladimir Lipunov$^{12\orcidlink{0000-0001-9133-7175}}$,
Michael B.~Lund$^{26\orcidlink{0000-0003-2527-1598}}$,
Franck Marchis$^{28,29\orcidlink{0000-0001-7016-7277}}$,
Joshua Margherone$^{1\orcidlink{0009-0002-9567-7518}}$,
Robert Massey$^{49\orcidlink{0000-0001-8879-7138}}$,
Kim McLeod$^{50\orcidlink{0000-0001-9504-1486}}$,
Edward Michaels$^{51}$,
Mayuko Mori$^{52,53\orcidlink{0000-0003-1368-6593}}$,
J. A. Mu\~noz$^{54,55\orcidlink{0000-0001-9833-2959}}$,
Felipe Murgas$^{35,36\orcidlink{0000-0001-9087-1245}}$,
Norio Narita$^{27,52,35\orcidlink{0000-0001-8511-2981}}$,
Takaya Okada$^{20\orcidlink{0009-0001-3799-3814}}$,
Shane Painter$^{1,56\orcidlink{0009-0003-5765-7978}}$,
Enric Palle$^{35,36\orcidlink{0000-0003-0987-1593}}$,
Ivan Panchenko$^{12}$,
Hannu Parviainen$^{35,36\orcidlink{0000-0001-5519-1391}}$,
Anastasiia Pechko$^{57,20\orcidlink{0009-0000-0893-2329}}$,
Adam Popowicz$^{18\orcidlink{0000-0003-3184-5228}}$,
Don Radford$^{58\orcidlink{0000-0002-3940-2360}}$,
Fabian Rodriguez Frustaglia$^{59}$,
Sylvain Roux$^{60,20}$,
Subhroja Roy$^{1,61\orcidlink{0009-0005-5402-8900}}$,
Matthew Ryno$^{62,20\orcidlink{0000-0001-8337-0020}}$,
G\"{u}lben \c{S}anc{\i}$^{16,17\orcidlink{0009-0009-5777-3374}}$,
Richard P. Schwarz$^{2\orcidlink{0000-0001-8227-1020}}$,
Victor Senik$^{12\orcidlink{0009-0002-1848-6136}}$,
Lauren A. Sgro$^{28\orcidlink{0000-0001-6629-5399}}$,
Avi Shporer$^{63\orcidlink{0000-0002-1836-3120}}$,
Gregor Srdoc$^{64}$,
Denise Stephens$^{40\orcidlink{0000-0003-4658-7567}}$,
Chris Stockdale$^{65\orcidlink{0000-0003-2163-1437}}$,
Ashley Stone$^{1,66,67\orcidlink{0009-0005-8110-6062}}$,
Ivan Strakhov$^{68\orcidlink{0000-0003-0647-6133}}$,
Marcelo Tala Pinto$^{69\orcidlink{0009-0004-8891-4057}}$,
Thiam-Guan Tan$^{70\orcidlink{0000-0001-5603-6895}}$,
Suzanne Taylor$^{71\orcidlink{0009-0004-7966-2812}}$,
Neil Thomas$^{72\orcidlink{0000-0002-2146-3894}}$,
Nataly Tiurina$^{12\orcidlink{0000-0002-7457-1928}}$,
Julian C.~van Eyken$^{26\orcidlink{0000-0003-2192-5371}}$,
Songhu Wang$^{3\orcidlink{0000-0002-7846-6981}}$,
Noriharu Watanabe$^{73\orcidlink{0000-0002-7522-8195}}$,
Levi R. Webb$^{1\orcidlink{0009-0008-9894-6641}}$,
Francis P. Wilkin$^{74\orcidlink{0000-0003-2127-8952}}$,
Joe Williams$^{40\orcidlink{0009-0006-0339-7551}}$,
Sel\c{c}uk Yal\c{c}{\i}nkaya$^{16,17,14\orcidlink{0000-0002-5224-247X}}$,
Mesut Y{\i}lmaz$^{16,17\orcidlink{0000-0002-3276-0704}}$,
and Carl Ziegler$^{75\orcidlink{0000-0002-0619-7639}}$
}
\\
Affiliations are listed at the end of the paper
}

\date{Accepted XXX. Received YYY; in original form ZZZ}

\pubyear{\the\year{}}

\begin{document}
\label{firstpage}
\pagerange{\pageref{firstpage}--\pageref{lastpage}}
\maketitle

\begin{abstract}
We present the discovery and characterization of 29 hot and warm Jupiter systems transiting bright ($G < 12.8$) FGK stars using \tess data and ground-based photometry and spectroscopy. Through the use of high angular resolution imaging, we discovered three bound stellar companions, named TOI-3988\,B, TOI-6171\,B, and TOI-7266\,B. The planets that we confirmed span an orbital period range of $1.28 - 16.9$ days, a mass range of $0.35 - 8.7$ \mj, and an orbital eccentricity range of $0 - 0.6$, and add to the growing self-consistent sample of hot and warm Jupiter systems detected using \tess and analyzed using the Markov-Chain Monte Carlo code \exofast. By performing two-sample Kolmogorov-Smirnov tests on a set of host star and planetary parameters, we find that this growing self-consistent sample of giant planets has a different distribution of host star masses, surface gravities, and metallicities, compared to the HJ and WJ systems in the literature, likely due to the additional constraint on the host star density from the planetary transits included in our global fits. We find that seven of the 29 newly confirmed planets have significant orbital eccentricity, and that one of these, TOI-3365\,b, is in a system which likely hosts a massive outer companion. The continued confirmation of HJ and WJ systems and the follow-up RV monitoring on these systems to identify outer companions will pave the way for important investigations of giant planet migration.
\end{abstract}





\section{Introduction}

Hot Jupiters (HJs) are planets unlike any in the Solar System. They orbit their host stars in fewer than 10 days, much shorter than even Mercury's orbit around the Sun, while having the mass and composition of a giant planet such as Jupiter. While rare (occurring around $\sim 1 \%$ of Sun-like stars; \citealt{Wright:2012}), HJs have a large impact on the systems they inhabit. The vast majority of HJs have no nearby planetary companions \citep{Huang:2016, Sha:2026}, a likely consequence of the violent migration that many of these planets undergo. Warm Jupiters (WJs), which occupy the orbital period space between $\sim10-200$ days, are more likely to have nearby companions, indicating that their evolutionary history may be less dramatic \citep{Huang:2016}. For these reasons, understanding how HJs form and evolve, and what conditions are key to shaping the HJ population, are integral to understanding the architectures of exoplanetary systems and the survivability of Earth-like planets.

Models of planetary system evolution, informed by the Solar System, suggest that giant planets readily form beyond the snow line, where water ice can condense to rapidly form planetary cores that are sufficient for runaway gas accretion \citep{Kennedy:2008}. Contrasting hypotheses have been offered that argue that hot Jupiters may form in situ \citep{Batygin:2016}, but the aggregate community opinion thus far has been that in situ HJ formation is unlikely to explain the majority of the HJ population \citep{Dawson:2018}. Instead, it is argued that most HJs likely form beyond the snow line, before migrating inward to short orbital periods, either through the protoplanetary disk \citep{Goldreich:1980, Lin:1986}, or via dynamical interactions with other planets or stars \citep{Kozai:1962, Lidov:1962, Rasio:1996, Naoz:2016}. Inward migration through a gas-rich disk is possible under certain disk conditions \citep{Bitsch:2015} and can lead to planetary systems in which nearby planets are unperturbed \citep[e.g.,]{Becker:2015, Canas:2019, Wu:2023}. Alternatively, planets that undergo high-eccentricity migration (HEM) do so violently, leading to the ejection or engulfment of nearby planets and the eventual isolation of the HJ \citep{Huang:2016, Sha:2026}. HJs which migrate via HEM can have their migration triggered by a wide variety of mechanisms, such as planet-planet scattering \citep{Rasio:1996}, von Zeipel-Kozai-Lidov \citep[vZKL;][]{vonZeipel:1910, Kozai:1962, Lidov:1962} oscillations, and stellar flyby events \citep{winter_stellar_2020, rodet_correlation_2021}. The common property of these different flavors of HEM is that the HJ's orbit is excited to a high eccentricity which may, if the periastron distance is close enough to the host star, circularize due to tidal dissipation. While it is unlikely to find HJs as they are actively migrating as a consequence of the relatively short timescale \citep{Dawson:2018}, statistical studies of orbital properties such as the HJ's orbital period, eccentricity, and the spin-orbit alignment angle, along with discoveries of additional companions to HJs, make it possible to place constraints on the relative frequency of each giant planet migration pathway. Such constraints have far-reaching implications on the survivability, and consequently, the occurrence rate of terrestrial planets.

In the 30 years since the discovery of 51\,Pegasi\,b \citep{Mayor:1995}, \totHJs HJs have been discovered by a variety of instruments (NASA Exoplanet Archive\footnote{\url{https://exoplanetarchive.ipac.caltech.edu/index.html}}, accessed \nearetrievaldate). While this sample is large, especially considering the low occurrence rate of HJs, it suffers from several pitfalls that make it difficult to perform statistical analyses. First, these discoveries span a long time period, over which time techniques have evolved and instrument sensitivities have improved. Furthermore, the sample is incomplete and governed by a number of selection biases which are difficult to quantify. Finally, the software used to perform each analysis varies, and decisions made in the fitting process vary from author to author, introducing a number of additional biases that make the resulting sample of HJs not self-consistent. To address this, \cite{Schulte:2024} introduced the Migration and Evolution of giant ExoPlanets (MEEP) survey, with the aim of using data from the Transiting Exoplanet Survey Satellite \citep[\tess;][]{Ricker:2015} to discover and characterize all remaining HJs transiting FGK host stars brighter than $G = 12.5$. The MEEP survey joined other collaborating efforts \citep{Rodriguez:2019, Rodriguez:2021, Rodriguez:2023, Ikwut-Ukwa:2022, Yee:2022, Yee:2023a, Yee:2025, RodriguezMartinez:2025} to build a complete sample of HJs with well-quantified biases, so that this sample may be used to address a large number of science questions regarding the formation and evolution of giant planets. This paper is the third in the MEEP series, following \cite{Schulte:2025}, which provided the characterization of five giant planets which populated sparse regions of the mass-radius distribution of HJs.

In this paper, the third in the MEEP series, we announce the discovery and characterization of 29 transiting hot and warm Jupiter systems detected by \tess. In \S\ref{sec:observations}, we summarize the photometric, spectroscopic, and astrometric observations used to confirm each planetary system. In \S\ref{sec:exofast}, we present the global fits used to derive parameters for each planet and host star. In \S\ref{sec:discussion}, we investigate the properties of the growing self-consistent sample of giant planets and compare it to the literature population. Finally, in \S\ref{sec:summary}, we summarize our results.

\section{Observations}\label{sec:observations}

To confirm the planets in this article, we collected an array of ground-based and space-based observations of each target. In this section, we discuss the \tess photometry used to identify each transit signal and the ground-based time-series photometry collected to confirm that each transit is on target and shows no signs of chromaticity. Then, we review the ground-based spectroscopy and radial velocity measurements used to determine the mass of each planet and confirm that our target sources are not spectroscopic binaries. We also review the high-resolution imaging techniques used to identify any nearby contaminating sources of light. Finally, we discuss the archival ground-based and space-based photometry from the \gaia mission \citep{GaiaDR3}, the Two Micron All-Sky Survey (2MASS; \citealt{Cutri:2003, Skrutskie:2006}), and the Wide-field Infrared Survey Explorer (WISE; \citealt{Wright:2010, Cutri:2012}) that were used to construct the spectral energy distributions (SEDs) of each star and determine their distances.

\subsection{\tess Photometry}\label{subsec:TESS}

All 29 of the systems in this article were first identified as giant planet candidates by the Transiting Exoplanet Survey Satellite (\tess) mission. The \tess spacecraft observes large $24^\circ \times 96^\circ$ regions ($\sim 5\%$ of the entire sky) continuously for 27-day intervals, using a vertical stack of four $4096 \times 4096$ pixel cameras. This setup allows \tess to observe many stars at once, enabling automated pipelines to discover more than 7900 planet candidates since it launched in April of 2018. 

Several different pipelines were used to identify the transit-like signals in the lightcurves of our targets. TOI-3365 was discovered by the \tess Science Processing Operations Center (SPOC) pipeline \citep{Jenkins:2016, Caldwell:2020, tess2min, tessffi}. TOI-3041, TOI-3972, TOI-4088, and TOI-4140 were first established as ``Community \tess Objects of Interest'' (CTOIs) by \cite{Olmschenk:2021} before later being secondarily detected by the Massachusetts Institute of Technology (MIT)'s Quick-Look Pipeline (QLP; \citealt{Huang:2020a, Huang:2020b, Kunimoto:2021, tessqlp}). Specifically, they were secondarily detected by the independent vetting pipeline called the QLP faint-star search \citep{Kunimoto:2022}. All remaining systems in this work were discovered by the QLP faint-star search. Following their discovery and establishment as \tess Objects of Interest (TOIs), the lightcurves for each system were reduced by the QLP and SPOC pipelines. When both pipelines have produced data products for a given \tess sector, we use the SPOC data instead of the QLP data because of its more robust handling of \tess systematics. We make use of all available \tess data up to Sector\,99 for each of our targets. All \tess data products used in our fits are listed in Table~\ref{tab:tess}. In this table, SPOC data products from the full-frame images are labeled as ``TESS-SPOC,'' whereas SPOC products generated from shorter-cadence data from Target Pixel Files (TPFs) are labeled simply as ``SPOC.''

\begin{table*}
\centering
\caption{Summary of \tess Observations}
\label{tab:tess}
\footnotesize
\setlength{\tabcolsep}{4pt}
\begin{tabular}[t]{cccc}
\toprule
TOI & Sector & Cadence (s) & Source \\
\midrule
3041 & 10, 11 & 1800 & TESS-SPOC \\
--- & 36, 37, 38 & 600 & TESS-SPOC \\
--- & 63, 65, 90 & 120 & SPOC \\
3365 & 09 & 1800 & TESS-SPOC \\
--- & 35, 36 & 600 & TESS-SPOC \\
--- & 62, 63, 89, 90, 99 & 120 & SPOC \\
3601 & 17 & 1800 & QLP \\
--- & 57, 84 & 120 & SPOC \\
3788 & 20 & 1800 & TESS-SPOC \\
--- & 60, 73 & 120 & SPOC \\
3972 & 17, 18, 24 & 1800 & TESS-SPOC \\
--- & 58, 78, 85 & 120 & SPOC \\
3988 & 17, 18, 24 & 1800 & QLP \\
--- & 58, 78, 85 & 120 & SPOC \\
3998 & 17, 18, 24 & 1800 & QLP \\
--- & 57, 58, 84, 85 & 120 & SPOC \\
4009 & 18, 24 & 1800 & QLP \\
--- & 25 & 1800 & TESS-SPOC \\
--- & 52, 58, 59, 78, 79, 85, 86 & 120 & SPOC \\
4079 & 18, 24, 25 & 1800 & QLP \\
--- & 58, 78, 85 & 120 & SPOC \\
4088 & 19, 20, 26 & 1800 & TESS-SPOC \\
--- & 40, 47, 59, 60, 73, 79 & 120 & SPOC \\
4140 & 19, 20, 26 & 1800 & TESS-SPOC \\
--- & 40, 53, 59, 60, 73, 79 & 120 & SPOC \\
4144 & 19, 20, 26 & 1800 & TESS-SPOC \\
--- & 40, 52, 53, 59, 60, 73, 79, 86 & 120 & SPOC \\
5236 & 14, 15 & 1800 & QLP \\
--- & 41, 55 & 600 & QLP \\
--- & 56, 75, 76 & 120 & SPOC \\
5479 & 22 & 1800 & TESS-SPOC \\
--- & 45, 46, 49 & 600 & TESS-SPOC \\
--- & 72 & 120 & SPOC \\
\bottomrule
\end{tabular}
\panelgap
\begin{tabular}[t]{cccc}
\toprule
TOI & Sector & Cadence (s) & Source \\
\midrule
5925 & 55 & 600 & TESS-SPOC \\
--- & 82 & 120 & SPOC \\
6148 & 56 & 200 & TESS-SPOC \\
--- & 82, 83 & 120 & SPOC \\
6166 & 15, 16 & 1800 & QLP \\
--- & 56 & 200 & QLP \\
--- & 76 & 120 & SPOC \\
6171 & 56 & 200 & TESS-SPOC \\
--- & 83 & 120 & SPOC \\
6184 & 56 & 200 & TESS-SPOC \\
--- & 83 & 120 & SPOC \\
6191 & 16 & 1800 & TESS-SPOC \\
--- & 56, 57 & 200 & QLP \\
--- & 76, 77, 83, 84 & 120 & SPOC \\
6208 & 16, 17 & 1800 & QLP \\
--- & 56, 57 & 200 & QLP \\
--- & 76, 77, 83, 84 & 120 & SPOC \\
6334 & 18 & 1800 & TESS-SPOC \\
--- & 58 & 200 & QLP \\
--- & 85 & 120 & SPOC \\
6417 & 34 & 600 & TESS-SPOC \\
--- & 61 & 200 & QLP \\
--- & 88 & 120 & SPOC \\
6443 & 07 & 1800 & QLP \\
--- & 34 & 600 & QLP \\
--- & 61 & 200 & QLP \\
--- & 88 & 120 & SPOC \\
7219 & 54 & 600 & TESS-SPOC \\
--- & 81, 92 & 200 & QLP \\
7266 & 55 & 600 & QLP \\
--- & 82 & 200 & QLP \\
7404 & 17 & 1800 & QLP \\
--- & 57, 84 & 200 & QLP \\
7425 & 18, 24 & 1800 & QLP \\
--- & 25 & 1800 & TESS-SPOC \\
--- & 58, 85 & 200 & QLP \\
7574 & 87 & 200 & QLP \\
\bottomrule
\end{tabular}
\end{table*}

Each of the lightcurves shown in Table~\ref{tab:tess} was downloaded using a custom pipeline\footnote{\url{https://github.com/jackschulte/exofast_tools/tree/main/scripts/lightcurve_pipeline}} which uses the \texttt{lightkurve} package to download lightcurve data from the Mikulski Archive for Space Telescopes (MAST\footnote{\url{https://archive.stsci.edu/}}). In the case of our SPOC lightcurves, we chose to download the Presearch Data Conditioning Simple Aperture Photometry (PDCSAP) flux, which has been processed to remove systematic trends and account for crowding. Then, each lightcurve was flattened using the Python-based spline-fitting tool \texttt{keplersplinev2}\footnote{\url{https://github.com/avanderburg/keplersplinev2}} \citep{Vanderburg:2014}. Specifically, we masked out the transits and used the \texttt{choosekeplersplinev2} function, which finds the spline break point spacing that minimizes the Bayesian Information Criterion (BIC). Then, we divided the downloaded lightcurve by the best-fit spline to remove out-of-transit stellar variability. Finally, we removed unnecessary baseline to reduce the computational cost of our fits, ensuring that at least one full transit duration is kept on each side of every transit.

The \tess photometry we use enables precise size and ephemeris constraints for all of the systems in this work. The large field-of-view of the \tess cameras enables the efficient sky-coverage, allowing the construction of a large, self-consistent sample of HJ systems. However, because of the large pixel scale of the \tess cameras (21 arcsec $\mathrm{pixel}^{-1}$), there are often several unresolved stars in each pixel, each contributing to the downloaded lightcurves. To account for this, we obtained higher-resolution observations from the ground to confirm that the detected transit signal is truly an exoplanet orbiting the target star and that the radius of each planet is not being systematically underestimated.

\subsection{Ground-based Follow-up Photometry}\label{subsec:followup}

To complement the \tess photometry, we obtained additional time-series photometry from the ground of each target system in this work. This ground-based follow-up photometry complements the \tess photometry in a few ways. First, the angular resolution of these observations is much better than is possible with the \tess cameras, allowing the confirmation that the transit signal is on target, rather than from a nearby star which is blended with the target star in a single \tess pixel. In fact, the worst angular resolution of the ground-based observations, 2.92 arcsec $\mathrm{pixel}^{-1}$, is nearly a factor of 10 better than the angular resolution of \tess. Second, the follow-up photometry was obtained in a variety of filters which differ from the \tess filter. Multiple observations of the same star in different filters enables the determination of whether or not each system shows evidence of chromaticity, or a significant change in transit depth as a function of wavelength. Planetary transits are expected to have much smaller variations in transit depth across different filters, unlike a common false positive scenario, eclipsing binaries \citep{Tingley:2004}. Evidence of chromaticity has been used to reclassify planet candidates as hierarchical eclipsing binaries \citep[e.g.,][]{Lillo-Box:2024}.

We obtained a total of 134 ground-based time-series observations of our targets from 54 different telescopes as part of the \tess Follow-up Observing program (TFOP; \citealt{Collins:2018}). This global network of professional astronomers and citizen scientists produces high-quality follow-up photometry for nearly every TOI, enabling the efficient confirmation of hundreds of planets. 36 of the follow-up lightcurves that we used in our global fits were collected by the 0.35\,m, 0.4\,m, 1.0\,m, and 2.0\,m telescopes of the Las Cumbres Observatory Global Telescope \citep[LCOGT;][]{Brown:2013} network nodes. All of our follow-up time-series observations are listed with additional details, such as their angular resolutions, aperture sizes, and the full-width-half-maximum (FWHM) of each target star's point spread function (PSF) in Table~\ref{tab:followup}. All of these observations are also available for download on ExoFOP-TESS\footnote{https://exofop.ipac.caltech.edu/tess/}.

\onecolumn
\resetpageheight
\scriptsize
\setlength{\tabcolsep}{2.5pt}
\begin{longtable}{@{}lllllllll@{\hspace{4pt}}l@{}}
\caption{Summary of Follow-up Observations} \label{tab:followup} \\
\toprule
TIC ID & TOI & Telescope & Tel. Size (m) & Date & Camera & Filter & Pix. Scale ($\arcsec$/pix) & PSF FWHM ($\arcsec$) & Aper. Rad. ($\arcsec$) \\
\midrule
\endfirsthead
\caption[]{Summary of Follow-up Observations} \\
\toprule
TIC ID & TOI & Telescope & Tel. Size (m) & Date & Camera & Filter & Pix. Scale ($\arcsec$/pix) & PSF FWHM ($\arcsec$) & Aper. Rad. ($\arcsec$) \\
\midrule
\endhead
\midrule
\multicolumn{10}{r}{Continued on next page} \\
\midrule
\endfoot
\bottomrule
\endlastfoot
371864043 & 3041 & El Sauce CDK20 & 0.51 & 2022 Jan 28 & STT-1603 & $R_c$ & 1.08 & 2.9 & 5 \\
 &  & El Sauce CDK20 & 0.51 & 2022 Feb 06 & STT-1603 & $R_c$ & 1.08 & 2.84 & 5 \\
 &  & Brierfield & 0.36 & 2023 Jan 09 & Moravian 16803 & $B$ & 0.735 & 3.68 & 10.6 \\
34297761 & 3365 & PEST & 0.305 & 2022 Feb 04 & QHY183M & $r'$ & 0.71 & 3.21 & 8 \\
 &  & El Sauce CDK20 & 0.51 & 2023 Jan 21 & Moravian C3-26000 & $B$ & 0.449 & 2.2 & 11 \\
191202679 & 3601 & KeplerCam & 1.2 & 2021 Oct 27 & KeplerCam & $i'$ & 0.672 & 3.11 & 8 \\
312811620 & 3788 & KeplerCam & 1.2 & 2021 Nov 19 & KeplerCam & $i'$ & 0.672 & 1.87 & 6 \\
 &  & WCO & 0.36 & 2023 Jan 23 & SBIG XLT 6303E & $g'$ & 0.66 & 2.1 & 8 \\
 &  & BBO & 0.4 & 2024 Feb 22 & STXL 1002M & $B$ & 1.087 & 4.46 & 11 \\
 &  & Celestron C11 & 0.28 & 2025 Jan 21 & Moravian Instruments G2-1600 & Clear & 1 & --- & 7 \\
 &  & Celestron C11 & 0.28 & 2025 Mar 21 & Moravian Instruments G2-1600 & Clear & 1 & --- & 10 \\
284206913 & 3972 & Hamilton & 0.31 & 2021 Sep 25 & ZWO ASI 1600 & A \#23 & 0.709 & 3.46 & 10 \\
 &  & CALOU & 0.4 & 2022 Jan 18 & FLI PL1001 & $R$ & 1.11 & 2.93 & 5 \\
 &  & SUTO1 & 0.3 & 2022 Mar 01 & ATIK11000 & $B$ & 0.754 & 4.2 & 10 \\
 &  & KeplerCam & 1.2 & 2022 Nov 09 & KeplerCam & $i'$ & 0.672 & 2.13 & 7 \\
 &  & TCS-MuSCAT2 & 1.52 & 2023 Sep 30 & MuSCAT2 & $g'$ & 0.44 & --- & --- \\
 &  & TCS-MuSCAT2 & 1.52 & 2023 Sep 30 & MuSCAT2 & $i'$ & 0.44 & --- & --- \\
 &  & TCS-MuSCAT2 & 1.52 & 2023 Sep 30 & MuSCAT2 & $r'$ & 0.44 & --- & --- \\
 &  & TCS-MuSCAT2 & 1.52 & 2023 Sep 30 & MuSCAT2 & $z_s$ & 0.44 & --- & --- \\
 &  & TRAPPIST-North & 0.6 & 2024 Sep 11 & Andor ikon-L & $z'$ & 0.6 & 2.94 & 8 \\
452938965 & 3988 & WCO & 0.36 & 2021 Oct 26 & SBIG STXL-6303E & $r'$ & 0.659 & 1.85 & 6 \\
 &  & OAAlbanya & 0.4 & 2022 Jul 31 & Moravian G4-9000 & $I_c$ & 1.44 & 10.1 & 9 \\
 &  & Whitin CDK700 & 0.7 & 2022 Nov 20 & FLI & $g'$ & 0.68 & 2.6 & 7 \\
 &  & Whitin CDK700 & 0.7 & 2022 Nov 20 & FLI & $r'$ & 0.68 & 2.6 & 7 \\
 &  & TUG T100 & 1 & 2022 Dec 19 & SI 1100 Cryo-cooler CCD & $g'$ & 0.62 & 2.5 & 12 \\
 &  & TUG T100 & 1 & 2022 Dec 19 & SI 1100 Cryo-cooler CCD & $i'$ & 0.62 & 2.5 & 12 \\
 &  & CMO & 0.4 & 2024 Dec 21 & SBIG STXL & $I_c$ & 0.53 & --- & 11 \\
320281287 & 3998 & OAUV-T50 & 0.5 & 2021 Jul 23 & FLI ProLine PL16801 & $R$ & 0.54 & 2.33 & 8 \\
 &  & Wendelstein & 0.43 & 2021 Jul 23 & SBIG STX-16803 & $r'$ & 0.635 & 2.8 & 9 \\
 &  & TCS-MuSCAT2 & 1.52 & 2021 Sep 28 & MuSCAT2 & $g'$ & 0.44 & --- & --- \\
 &  & TCS-MuSCAT2 & 1.52 & 2021 Sep 28 & MuSCAT2 & $i'$ & 0.44 & --- & --- \\
 &  & TCS-MuSCAT2 & 1.52 & 2021 Sep 28 & MuSCAT2 & $r'$ & 0.44 & --- & --- \\
 &  & TCS-MuSCAT2 & 1.52 & 2021 Sep 28 & MuSCAT2 & $z_s$ & 0.44 & --- & --- \\
 &  & Wendelstein & 0.43 & 2021 Oct 20 & SBIG STX-16803 & $r'$ & 0.635 & 3.6 & 8 \\
 &  & SUTO-Otivar & 0.3 & 2022 Jul 19 & ASI1600MM & $R$ & 0.685 & 4.63 & 9 \\
 &  & Wendelstein & 0.43 & 2022 Aug 28 & SBIG STX-16803 & $r'$ & 0.635 & 3.5 & 9 \\
 &  & LX850 & 0.304 & 2022 Sep 08 & QHY 268m & $R$ & 0.302 & 2.4 & 15 \\
 &  & CALOU & 0.4 & 2023 Nov 03 & FLI PL1001 & $r'$ & 1.11 & 5.32 & 13.3 \\
 &  & Salerno University Observatory & 0.6 & 2025 Dec 06 & FLI PL-230 & $R$ & 0.61 & 3.56 & 11 \\
368734712 & 4009 & OAA & 0.4 & 2023 Nov 18 & Moravian G4-9000 & $I_c$ & 1.44 & 7.32 & 7 \\
 &  & GdP & 0.4 & 2024 Aug 08 & FLI4710 & $g'$ & 0.73 & 3.44 & 8 \\
 &  & Adams Observatory & 0.61 & 2025 Jan 16 & FLI PL16803 & $I_c$ & 0.38 & 2.6 & 13 \\
444558604 & 4079 & OAUV-T50 & 0.5 & 2021 Jul 15 & FLI ProLine PL16801 & $R$ & 0.54 & 2.46 & 9 \\
 &  & TCS-MuSCAT2 & 1.52 & 2021 Jul 17 & MuSCAT2 & $g'$ & 0.44 & --- & --- \\
 &  & TCS-MuSCAT2 & 1.52 & 2021 Jul 17 & MuSCAT2 & $i'$ & 0.44 & --- & --- \\
 &  & TCS-MuSCAT2 & 1.52 & 2021 Jul 17 & MuSCAT2 & $r'$ & 0.44 & --- & --- \\
 &  & TCS-MuSCAT2 & 1.52 & 2021 Jul 17 & MuSCAT2 & $z_s$ & 0.44 & --- & --- \\
 &  & Wendelstein & 0.43 & 2022 Jun 18 & SBIG STX-16803 & $i'$ & 0.635 & 3.2 & 10 \\
 &  & TRAPPIST-North & 0.6 & 2022 Dec 12 & Andor ikon-L & $z'$ & 0.6 & 2 & 4 \\
 &  & SUTO-Otivar & 0.3 & 2025 Nov 03 & ASI1600MM & $B$ & 0.642 & 3.98 & 6 \\
302381397 & 4088 & OAUV-TURIA2 & 0.3 & 2021 Oct 04 & QHY600 & $R$ & 1.36 & 5.61 & 7 \\
 &  & Celestron EdgeHD1400 f\/7.7 & 0.36 & 2021 Nov 20 & SBIG ST8XME & $r'$ & 0.69 & 3 & 8 \\
 &  & LCO-Teid & 0.4 & 2022 Mar 03 & STX-6303 & $g'$ & 0.57 & 3.1 & 12 \\
 &  & TUG T100 & 0.31 & 2023 Apr 30 & SI 1100 Cryo-cooler CCD & $g'$ & 0.31 & 1.5 & 15 \\
 &  & TUG T100 & 0.31 & 2023 Apr 30 & SI 1100 Cryo-cooler CCD & $z'$ & 0.31 & 1.4 & 12 \\
 &  & RFAC & 0.31 & 2023 Jul 26 & QHY268M & $R$ & 0.29 & 4.7 & 26 \\
 &  & Unistellar eVscope2 & 0.11 & 2024 Mar 09 & Sony IMX347 CMOS & Clear & 1.33 & 9.5 & 14.5 \\
 &  & Unistellar eVscope2 & 0.11 & 2024 Sep 05 & Sony IMX347 CMOS & Clear & 1.33 & 6.7 & 12.6 \\
 &  & Unistellar eVscope1 & 0.11 & 2025 Apr 11 & Sony IMX224 CMOS & Clear & 1.72 & 6.5 & 13.6 \\
 &  & Unistellar eVscope1 & 0.11 & 2025 May 11 & Sony IMX224 CMOS & Clear & 1.72 & 5.3 & 8.6 \\
 &  & Unistellar eVscope1 & 0.11 & 2025 May 11 & Sony IMX224 CMOS & Clear & 1.72 & 4.7 & 8.6 \\
400103802 & 4140 & KeplerCam & 1.2 & 2021 Oct 23 & KeplerCam & $i'$ & 0.672 & 1.85 & 5 \\
 &  & SUTO2 & 0.3 & 2022 Jan 28 & ZWO ASI1600MM & $B$ & 0.685 & 3.9 & 8 \\
 &  & Unistellar eVscope1 & 0.11 & 2024 Mar 08 & Sony IMX224 CMOS & Clear & 1.72 & 5.5 & 9.5 \\
313194972 & 4144 & Newtonian reflector & 0.25 & 2021 Oct 03 & ASI294MM PRO & $R$ & 0.8 & 3.72 & 8 \\
 &  & OAUV-TURIA2 & 0.3 & 2024 Feb 12 & QHY600 & $B$ & 0.68 & 3.66 & 10 \\
 &  & LCO-Teid & 1 & 2025 Dec 02 & SINISTRO & $z_s$ & 0.389 & 1.62 & 11 \\
193765661 & 5236 & LCO-HAL & 0.35 & 2026 Aug 28 & QHY600 & $g'$ & 0.73 & 2.32 & 3 \\
 &  & LCO-HAL & 0.35 & 2026 Aug 28 & QHY600 & $g'$ & 0.73 & 2.32 & 3 \\
 &  & LCO-HAL-M3 & 2 & 2026 Aug 28 & MuSCAT3 & $g'$ & 0.27 & 1.2 & 4 \\
 &  & LCO-HAL-M3 & 2 & 2026 Aug 28 & MuSCAT3 & $r'$ & 0.27 & 1.2 & 4 \\
 &  & LCO-HAL-M3 & 2 & 2026 Aug 28 & MuSCAT3 & $i'$ & 0.27 & 1.2 & 4 \\  &  & LCO-HAL-M3 & 2 & 2026 Aug 28 & MuSCAT3 & $z'$ & 0.27 & 1.2 & 4 \\
67444896 & 5479 & TRAPPIST-North & 0.6 & 2024 Jan 31 & Andor ikon-L & $z'$ & 0.6 & 2.18 & 7 \\
 &  & LCO-Teid & 1 & 2025 Apr 14 & SINISTRO & $g'$ & 0.389 & 1.95 & 14 \\
 &  & LCO-Teid & 1 & 2025 Apr 14 & SINISTRO & $i'$ & 0.389 & 1.95 & 14 \\
345193111 & 5925 & El Sauce CDK20 & 0.51 & 2023 Jun 02 & Moravian C3-26000 & $R_c$ & 0.449 & 2.8 & 13 \\
 &  & PEST & 0.305 & 2024 Jun 30 & QHY183M & $r'$ & 0.7 & 3.78 & 9 \\
 &  & Acton-Sky-Portal & 0.36 & 2024 Aug 25 & SBIG A4710 & $g'$ & 1 & 3 & 5 \\
 &  & LCO-CTIO & 0.35 & 2024 Aug 25 & QHY600 & $g'$ & 0.745 & 2.4 & 6 \\
371573539 & 6148 & LCO-Teid & 1 & 2024 Nov 04 & SINISTRO & $i'$ & 0.389 & 1.14 & 6 \\
 &  & LCO-HAL & 0.35 & 2025 Jun 08 & QHY600 & $g'$ & 0.73 & 2.13 & 7 \\
190822775 & 6166 & LCO-McD & 0.35 & 2023 Jul 24 & QHY600 & $g'$ & 0.73 & 2.43 & 10 \\
 &  & Acton-Sky-Portal & 0.36 & 2023 Jul 24 & SBIG A4710 & $r'$ & 1 & 4 & 6 \\
 &  & MSU & 0.6 & 2023 Aug 10 & Apogee Alta U47 & $V$ & 0.55 & 4.59 & 8.25 \\
 &  & LCO-HAL & 0.4 & 2023 Sep 07 & SBIG STX6303 & $i'$ & 0.57 & 1.93 & 9.5 \\
 &  & Unistellar eVscope1 & 0.11 & 2024 Aug 16 & Sony IMX224 CMOS & Clear & 1.72 & 5.5 & 10.8 \\
 &  & Unistellar eVscope1 & 0.11 & 2025 Jul 08 & Sony IMX224 CMOS & Clear & 1.72 & 5.8 & 9.6 \\
67478724 & 6171 & LCO-HAL & 0.35 & 2023 Sep 24 & QHY600 & $i'$ & 0.73 & 2.84 & 12 \\
 &  & LCO-McD & 0.35 & 2023 Sep 28 & SCICAM QHY600 & $i'$ & 0.73 & 2.89 & 8 \\
 &  & TCS-MuSCAT2 & 1.52 & 2024 Oct 06 & MuSCAT2 & $g'$ & 0.44 & --- & --- \\
 &  & TCS-MuSCAT2 & 1.52 & 2024 Oct 06 & MuSCAT2 & $i'$ & 0.44 & --- & --- \\
 &  & TCS-MuSCAT2 & 1.52 & 2024 Oct 06 & MuSCAT2 & $r'$ & 0.44 & --- & --- \\
 &  & TCS-MuSCAT2 & 1.52 & 2024 Oct 06 & MuSCAT2 & $z_s$ & 0.44 & --- & --- \\
197743152 & 6191 & LCO-HAL & 0.35 & 2023 Jun 26 & SCICAM QHY600 & $g'$ & 0.73 & 4.74 & 10 \\
 &  & Villa39 & 0.35 & 2023 Jul 15 & SX16803 & $I$ & 0.94 & 4.09 & 6 \\
 &  & LCO-HAL & 0.35 & 2023 Oct 25 & SCICAM QHY600 & $g'$ & 0.73 & 2.55 & 8 \\
 &  & TCS-MuSCAT2 & 1.52 & 2023 Nov 01 & MuSCAT2 & $i'$ & 0.44 & --- & --- \\
 &  & TCS-MuSCAT2 & 1.52 & 2023 Nov 01 & MuSCAT2 & $r'$ & 0.44 & --- & --- \\
 &  & TCS-MuSCAT2 & 1.52 & 2023 Nov 01 & MuSCAT2 & $z'$ & 0.44 & --- & --- \\
257207557 & 6184 & El Sauce CDK20 & 0.51 & 2023 Jul 04 & Moravian C3-26000 & $R_c$ & 0.449 & 4.2 & 15 \\
 &  & Brierfield & 0.36 & 2023 Oct 13 & Moravian 16803 & $B$ & 0.735 & 3.11 & 9 \\
 &  & LCO-SSO & 0.35 & 2023 Oct 13 & SCICAM QHY600 & $g'$ & 0.73 & 3.02 & 9 \\
 &  & MSU & 0.6 & 2024 Aug 11 & QHY600 & $g'$ & 0.159 & 2.11 & 2.07 \\
 &  & El Sauce CDK20 & 0.51 & 2024 Aug 12 & Moravian C3-26000 & $I_c$ & 0.449 & 2.3 & 12 \\
326475995 & 6208 & LCO-McD & 0.35 & 2023 Jul 18 & QHY600 & $g'$ & 0.73 & 3.05 & 10 \\
 &  & LCO-HAL & 0.35 & 2024 Sep 29 & QHY600 & $i'$ & 0.73 & 2.59 & 8 \\
63718617 & 6334 & LCO-McD & 1 & 2023 Oct 19 & SINISTRO & $i'$ & 0.389 & 2.75 & 12 \\
 &  & LCO-Teid & 1 & 2024 Aug 11 & SINISTRO & $i'$ & 0.389 & 1.6 & 8 \\
95191643 & 6417 & LCO-McD & 0.35 & 2024 Feb 01 & SCICAM QHY600 & $i'$ & 0.73 & 2.61 & 7 \\
 &  & LCO-SAAO & 0.4 & 2024 Feb 07 & SBIG & $g'$ & 0.57 & 2.38 & 8 \\
125563127 & 6443 & PEST & 0.305 & 2024 Dec 05 & QHY183M & $r'$ & 0.7 & 6.46 & 11 \\
 &  & LCO-SSO & 1 & 2025 Jan 04 & SINISTRO & $z_s$ & 0.389 & 1.53 & 4 \\
 &  & SOAR & 4.1 & 2026 Jan 08 & Goodman HTS & $i'$ & 0.15 & 2.94 & 6 \\
242674266 & 7219 & LCO-McD & 0.35 & 2025 May 01 & QHY600 & $i'$ & 0.73 & 2.69 & 10 \\
 &  & El Sauce CDK20 & 0.51 & 2025 May 04 & Moravian C3-26000 & $B$ & 0.449 & 2.8 & 13 \\
125695940 & 7266 & LCO-HAL & 0.35 & 2025 Jul 02 & QHY600 & $g'$ & 0.73 & 2.46 & 9 \\
 &  & LCO-HAL & 0.35 & 2025 Jul 02 & QHY600 & $i'$ & 0.73 & 2.46 & 9 \\
 &  & LCO-CTIO & 0.35 & 2025 Aug 14 & QHY600 & $g'$ & 0.73 & 2.76 & 8 \\
 &  & LCO-CTIO & 0.35 & 2025 Aug 14 & QHY600 & $i'$ & 0.73 & 2.76 & 8 \\
352409708 & 7404 & WCO & 0.36 & 2025 Oct 11 & QHY248M & $r'$ & 0.55 & --- & 10 \\
285592400 & 7425 & SUTO-Otivar & 0.3 & 2025 Jul 08 & ASI1600MM & $R$ & 0.636 & 3.27 & 8 \\
 &  & Unistellar eVscope1 & 0.11 & 2025 Aug 13 & Sony IMX224 CMOS & Clear & 1.72 & 7 & 9.5 \\
 &  & LCO-HAL & 0.4 & 2025 Sep 17 & QHY600 & $g'$ & 0.57 & 1.64 & 6 \\
 &  & LCO-HAL & 0.4 & 2025 Sep 17 & QHY600 & $i'$ & 0.57 & 1.73 & 7 \\
 &  & AUKR & 0.8 & 2025 Oct 02 & ALTA U47 & $g'$ & 0.72 & --- & 7 \\
 &  & AUKR & 0.8 & 2025 Oct 02 & ALTA U47 & $i'$ & 0.72 & --- & 6 \\
 &  & Maidanak NT & 0.6 & 2025 Oct 11 & Andor iKon-L 936 & $R$ & 0.77 & 2 & 6 \\
 &  & BYU-OPO & 0.6 & 2025 Oct 13 & SBIG STX-16803 3 & $B$ & 0.93 & 4.01 & 4 \\
 &  & BYU-OPO & 0.6 & 2025 Oct 13 & SBIG STX-16803 3 & $I$ & 0.93 & 3.78 & 5 \\
 &  & Stellar Society Rozhen SCA260 & 0.26 & 2025 Nov 12 & Player One ZEUS 455M Pro & $V$ & 0.6 & 3.3 & 10 \\
159084486 & 7574 & LCO-McD & 0.35 & 2025 Dec 19 & QHY 600 & $g'$ & 0.744 & 3.22 & 9 \\
 &  & LCO-McD & 0.35 & 2025 Dec 19 & QHY 600 & $i'$ & 0.744 & 3.22 & 9 \\
 &  & WCO & 0.36 & 2025 Dec 19 & QHY268M & $r'$ & 0.55 & 3.5 & 15 \\
 &  & C97 & 0.61 & 2025 Dec 23 & SBIG-16803 & $R_c$ & 0.77 & 2.46 & 5 \\
\end{longtable}

\twocolumn
\normalsize
\setlength{\tabcolsep}{4pt}

The vast majority of the ground-based time-series observations were reduced using the aperture photometry tool \texttt{AstroImageJ} \citep[AIJ;][]{Collins:2017}. A procedural guide on how these reductions were done is shown in \S 2.2 of \cite{Schulte:2024}. In addition to those that were reduced using \texttt{AstroImageJ}, several of our observations were reduced using custom pipelines. 

TOI-3972, TOI-3998, TOI-4079, TOI-6148, TOI-6171, and TOI-6191 were all observed by the MuSCAT2 instrument \citep{Narita:2019} on the 1.5~m Telescopio Carlos S\'anchez at Teide Observatory in Tenerife, Spain. MuSCAT2 is a multi-band imager that observes in the the \textit{Sloan} $g'$, $r'$, $i'$, and $z'$ filters simultaneously, enabling an independent measurement of chromaticity. The dedicated MuSCAT2 pipeline \citep{Parviainen:2019} was used to calibrate the science frames and perform aperture photometry. TOI-3365, TOI-5925, and TOI-6443 were observed by the Perth Exoplanet Survey Telescope (PEST), a 0.3~m telescope based in Australia. To reduce these data, a custom pipeline based on \texttt{C-Munipack}\footnote{\url{http://c-munipack.sourceforge.net}} was used to calibrate the frames and perform differential photometry. TOI-3788 and TOI-4144 were observed by the MASTER-Kislovodsk \citep{Lipunov2010,Lipunov22} wide-field observations twin 0.4~m telescope. These two observations were collected and reduced using the MASTER robotic planner \citep{Lipunov19,Lipunov24}, and were used to check for chromaticity and ensure that the transit was on target, but were not included in the fit.

Lastly, the SETI Institute \& Unistellar Citizen Science Network (UCSN) observed transits of TOI-4088, TOI-4140, TOI-6166, and TOI-7425 from various locations across Austria, France, Italy, Japan, Ukraine, the USA, and the UK. The UCSN is a global network of observers using digital consumer telescopes from Unistellar telescopes \textit{eVscopes}. These telescopes are either 85\,mm or 114\,mm-aperture Newtonian reflectors equipped with SONY CMOS IMX224, IMX347, or IMX415 sensors \citep{Marchis:2020}. Observers use a dedicated mobile application to control the telescopes and obtain transit data under the auspices of the UCSN Exoplanet Transits Program \citep{Oconner:2023}. Once an observer gathers transit data, they then upload the dataset to Unistellar and SETI Institute remote servers where it is passed through a custom Python pipeline. Raw frames are typically obtained with 3.97~s exposure times and upon upload by observers, dark subtracted when dark frames were available. Images are then plate-solved and saturated or off field-of-view frames were removed from each dataset. Raw images are then aligned and stacked in groups of 15-30 for a combined integration time of up to two minutes, increasing the signal-to-noise (S/N). We determine the optical aperture size from the target source FWHM, then perform differential aperture photometry on stacked frames using up to ten reference stars chosen for their brightness and minimal scatter. These reference stars are median combined to form a ``composite'' reference star, which we use to extract the final differential photometry as detailed in \citet{Sgro:2024} and \citet{Perrocheau:2022}. Finally, we use the \texttt{PYCHEOPS} Python package, which uses least-squares minimization, to model each transit \citep{Maxted:2022}. This model served to assess the transit quality prior to inclusion in the \exofast global fits.

\subsection{Spectroscopy}

Another important component of the exoplanet confirmation process is the collection of ground-based high-resolution spectroscopy. These observations serve multiple purposes. First, they act as probes for hidden stars: each spectrum is checked for a second set of lines that would be indicative of contaminating light from another star or set of stars, blended with the primary star. Additionally, these observations serve to place important constraints on valuable parameters such as the mass and eccentricity of the planet. For each spectroscopic observation, we derived the radial velocity (RV) motion corresponding to the orbit of the planet, providing an independent confirmation that the candidate is bona fide planet. Lastly, the RVs can be used to find evidence of additional non-transiting planetary or stellar companions. These objects, typically on much longer orbital periods, appear in the RV data as long-term linear or quadratic trends, as in many cases the baseline is not long enough to capture an entire orbit. Given a long enough RV baseline, however, an additional Keplerian body can be fit and the orbital period and minimum mass of this outer body may be obtained.

In pursuit of these goals, we obtained 623 spectroscopic observations of our 29 target systems using four high-resolution \'echelle spectrographs: the Tillinghast Reflector \'Echelle Spectrograph (TRES) at the Fred Lawrence Whipple Observatory, the CHIRON spectrograph at the Cerro Tololo Inter-American Observatory (CTIO), the NN-EXPLORE Exoplanet Investigations with Doppler Spectroscopy (NEID) spectrograph at the Kitt Peak National Observatory, and the Fiber-fed Extended Range Optical Spectrograph (FEROS), mounted on the MPG/ESO 2.2~m telescope at the European Southern Observatory (ESO)'s La Silla Observatory. These RVs, the facilities that obtained them, and the methods used to reduce them, are described in the following sections. A summary table of the RVs from each instrument is presented in Table~\ref{tab:rv_summary}. A full, machine-readable table of the RVs is available in the online journal.

\begin{table*}
\centering
\caption{Summary of RV Observations}
\label{tab:rv_summary}
\footnotesize
\setlength{\tabcolsep}{3pt}
\begin{tabular}[t]{cccccc}
\toprule
TOI & Instrument & $N_{\rm RV}$ & Median $\sigma_{\rm RV}$ & \multicolumn{2}{c}{Observation Date (UT)} \\
 &  &  & (m s$^{-1}$) & First & Last \\
\midrule
3041 & CHIRON & 21 & 36.0 & 23 Oct 2021 & 20 Apr 2026 \\
3365 & CHIRON & 19 & 29.0 & 04 Dec 2022 & 08 Mar 2025 \\
3601 & TRES & 17 & 47.6 & 03 Sep 2021 & 11 Oct 2024 \\
3788 & TRES & 22 & 46.1 & 03 Oct 2021 & 01 Nov 2023 \\
3972 & NEID & 28 & 3.0 & 21 Dec 2022 & 29 Jan 2024 \\
& TRES & 22 & 22.2 & 07 Sep 2021 & 27 Dec 2021 \\
3988 & TRES & 22 & 38.4 & 09 Sep 2021 & 21 Dec 2023 \\
3998 & TRES & 12 & 25.1 & 03 Aug 2021 & 10 Oct 2023 \\
4009 & TRES & 24 & 29.9 & 19 Sep 2021 & 08 Nov 2023 \\
4079 & TRES & 23 & 57.8 & 07 Sep 2021 & 19 Feb 2024 \\
4088 & TRES & 23 & 24.9 & 07 Nov 2021 & 21 Dec 2023 \\
4140 & TRES & 24 & 22.3 & 09 Oct 2021 & 07 Nov 2023 \\
4144 & TRES & 18 & 23.6 & 08 Oct 2021 & 21 Dec 2023 \\
5236 & TRES & 24 & 29.0 & 14 Apr 2022 & 03 May 2024 \\
5479 & FEROS & 20 & 12.6 & 31 Mar 2023 & 07 Apr 2026 \\
& TRES & 25 & 27.2 & 25 Apr 2022 & 04 May 2024 \\
\bottomrule
\end{tabular}
\panelgap
\begin{tabular}[t]{cccccc}
\toprule
TOI & Instrument & $N_{\rm RV}$ & Median $\sigma_{\rm RV}$ & \multicolumn{2}{c}{Observation Date (UT)} \\
 &  &  & (m s$^{-1}$) & First & Last \\
\midrule
5925 & TRES & 28 & 41.8 & 07 May 2023 & 05 May 2024 \\
6148 & TRES & 15 & 36.2 & 25 May 2023 & 10 Jul 2024 \\
& CHIRON & 18 & 53.5 & 04 Jun 2023 & 04 Aug 2023 \\
6166 & TRES & 21 & 30.8 & 07 May 2023 & 06 Sep 2024 \\
6171 & TRES & 19 & 30.5 & 08 Jun 2023 & 12 Jul 2024 \\
6191 & TRES & 22 & 24.1 & 20 Jul 2023 & 19 Nov 2024 \\
6184 & TRES & 44 & 23.1 & 22 Jun 2023 & 12 Jul 2024 \\
6208 & TRES & 35 & 39.6 & 09 Jul 2023 & 26 May 2024 \\
6334 & TRES & 21 & 25.1 & 01 Aug 2023 & 02 Nov 2023 \\
6417 & TRES & 16 & 23.9 & 21 Oct 2023 & 22 Dec 2023 \\
6443 & TRES & 14 & 29.7 & 20 Oct 2023 & 01 Jan 2024 \\
7219 & TRES & 9 & 29.4 & 22 May 2025 & 12 Sep 2025 \\
7266 & TRES & 10 & 58.6 & 09 Sep 2025 & 11 Nov 2025 \\
7404 & TRES & 7 & 106.6 & 25 Sep 2025 & 13 Nov 2025 \\
7425 & TRES & 8 & 28.0 & 21 Aug 2025 & 14 Nov 2025 \\
7574 & TRES & 9 & 26.9 & 30 Nov 2025 & 02 Apr 2026 \\
\bottomrule
\end{tabular}

\vspace{2mm}
\begin{minipage}{0.95\linewidth}
\footnotesize
\textbf{Note:} The full table of RVs for each system is available in machine-readable form in the online journal.
\end{minipage}
\end{table*}


\subsubsection{TRES Spectroscopy}\label{subsubsec:tres}

We used the TRES instrument \citep{gaborthesis} to obtain a total of 534 spectra of 27 of the 29 systems presented in this work. The TRES spectrograph is a high-resolution, fiber-fed \'echelle spectrograph installed on the 1.5~m Tillinghast Reflector at the Fred Lawrence Whipple Observatory in southern Arizona, USA. Each observed spectrum is run through an automated pipeline to deliver the least-squares deconvolution line profile \citep{zhou2016}, allowing us to rule out hidden stellar companions that could influence RV measurements or mimic planetary signals. Additionally, we used the Stellar Parameter Classification (SPC) tool \citep{Buchhave:2012} to estimate the metallicity, effective temperature, surface gravity, and projected rotational velocity of each star. These metallicity measurements are then used to set priors in our fits, as is discussed further in \S\ref{sec:exofast}. The spectra were reduced following the works of \cite{Buchhave:2010} and \cite{Quinn:2012}. The RVs were then derived by cross-correlating these reduced spectra with a combined, median-filtered template built from all of the observed spectra. Finally, zero point offsets were corrected using standard star observations to enable the characterization of long-term trends associated with outer planetary or stellar companions.

\subsubsection{CHIRON Spectroscopy}\label{subsubsec:chiron}

We additionally used the CHIRON spectrograph \citep{Tokovinin:2013, Paredes:2021} to collect 58 spectra of TOI-3041, TOI-3365, and TOI-6148. CHIRON is a fiber-fed \'echelle spectrograph mounted on the 1.5~m Small and Moderate Aperture Research Telescope System (SMARTS) telescope at CTIO in Chile. The spectra were observed using CHIRON's image slicer mode, which has a spectral resolution of $\sim$80,000. Each observed spectrum was bracketed with two thorium-argon calibration spectra. Then, the RVs were derived by comparing the least-squares deconvolution \citep{Donati:1997, Zhou:2021} of the observed spectra to non-rotating synthetic templates generated with the ATLAS9 models \citep{Kurucz:1992}. We then fit the line profile with a rotational broadening kernel \citep{Gray:2005} to measure the star's RV and projected rotational velocity. Finally, in the same way as for the TRES observations, we used SPC \citep{Buchhave:2012} to estimate other stellar parameters, including each star's metallicity. 

\subsubsection{NEID Spectroscopy}\label{subsubsec:neid}

We obtained 11 radial velocity measurements of TOI-3972 using 800~s exposures with NEID \citep{Schwab:2016, Halverson:2016}, a fiber-fed \citep{Kanodia:2018, Kanodia:2023}, environmentally stabilized \citep{Stefannson:2016, Robertson:2019} echelle spectrograph on the 3.5\,m WIYN telescope at Kitt Peak National Observatory, operating in its high-resolution mode ($R\sim110{,}000$; 380--930\,nm). The spectra were obtained at airmasses ranging from 1.2 to 2.2 and achieved per-exposure signal-to-noise ratios of 7--25, with a median of $\sim$19, at 5530\,\AA. We processed the data using version 1.5.2 of the NEID Data Reduction Pipeline (\texttt{NEID-DRP})\footnote{\url{https://neid.ipac.caltech.edu/docs/NEID-DRP/}}, which derives RVs through cross-correlation, and adopted the barycentric-corrected, order-reweighted velocities (\texttt{CCFRVMOD}) provided by the NExScI NEID Archive\footnote{\url{https://neid.ipac.caltech.edu/}}.

\subsubsection{FEROS Spectroscopy}\label{subsubsec:feros}

We obtained 20 high-resolution spectra of TOI-5479 with the
Fiber-fed Extended Range Optical Spectrograph \citep[FEROS;][]{Kaufer1999}, mounted on the MPG/ESO 2.2\,m telescope at La Silla Observatory, Chile, between UT 2023 March 31 and 2026 April 7, in the context of the Warm gIaNts with tEss collaboration \citep[WINE,][]{wine1,wine2,wine3,wine4}. FEROS covers the $3600$--$9200$\,\AA\ range with a resolving power of $R \simeq 48\,000$, and all observations were carried out in the simultaneous calibration mode, with the second fiber illuminated by a ThAr lamp to trace instrumental drifts during the exposures. Exposure times were 1500\,s for all but one epoch (1800\,s), yielding signal-to-noise ratios of 26--55 per resolution element at 5500\,\AA. The data were reduced with the \texttt{ceres} pipeline \citep{Brahm2017a}, which performs the bias and flat-field corrections, the optimal order-by-order extraction, and the wavelength and drift calibration, and computes precise radial velocities (RVs) and bisector spans (BIS) via cross-correlation of the extracted spectra with a G2 binary mask. The resulting RVs have a median internal precision of $12.6\,{\rm m\,s^{-1}}$ ($10.4$--$24.0\,{\rm m\,s^{-1}}$), obtained over a baseline of $\sim 3.0$\,yr spanning four observing seasons. They show a dispersion of $146\,{\rm m\,s^{-1}}$, well above the individual measurement uncertainties, indicative of a clear Doppler reflex-motion signal. The corresponding BIS values show no significant correlation with the RVs, disfavoring stellar activity or a blended eclipsing binary as the origin of the observed variations. 

\raggedbottom 

\subsection{High-Resolution Imaging}\label{subsec:hri}

Several of our target systems are located in crowded fields in which source confusion is an issue. To detect nearby stars that are blended even in the follow-up time-series photometry presented in \S\ref{subsec:followup}, we observed our target systems with two techniques of high angular resolution imaging: speckle imaging and adaptive optics (AO) imaging. These observations, which were performed through the TFOP collaboration, enable the detection of nearby sources, bound or unbound, which were previously unknown. We then used the measured magnitude differences to constrain our fits, ensuring accurate stellar and planetary parameters. A summary of our high angular resolution imaging observations is presented in Table~\ref{tab:hri}. The details of the detected stellar companions which we included in our fits are presented in Table~\ref{tab:secondarylit}. The sensitivity curves and plots for all of the observations are available for download from ExoFOP-TESS\footnote{https://exofop.ipac.caltech.edu/tess/}.

\begin{table*}
\centering
\caption{Summary of High-resolution Imaging Observations}
\label{tab:hri}
\scriptsize
\setlength{\tabcolsep}{2.5pt}
\begin{tabular}{@{}lllllllll@{\hspace{4pt}}l@{}}
\toprule
TIC ID & TOI & Telescope & Date & Instrument & Imaging Type & Filter & Pix. Scale ($\arcsec$/pix) & PSF FWHM ($\arcsec$) & Contrast \\
\midrule
371864043 & 3041 & SOAR (4.1 m) & 2022 Apr 15 & HRCam & Speckle & $I$ & 0.016 & 0.064 & $\Delta$ 6.5 mag @ 1" \\
34297761 & 3365 & SOAR (4.1 m) & 2021 Nov 20 & HRCam & Speckle & $I$ & 0.016 & 0.064 & $\Delta$ 7.2 mag @ 1" \\
191202679 & 3601 & Palomar (5 m) & 2024 Aug 20 & PHARO & AO & $Hcont$ & 0.025 & 0.081 & $\Delta$ 7.053 mag @ 0.5" \\
 &  & Palomar (5 m) & 2024 Aug 20 & PHARO & AO & $Kcont$ & 0.025 & 0.101 & $\Delta$ 6.633 mag @ 0.5" \\
 &  & Gemini (8 m) & 2024 Sep 25 & 'Alopeke & Speckle & 562 nm & 0.01 & 0.02 & $\Delta$ 5.63 mag @ 0.5" \\
 &  & Gemini (8 m) & 2024 Sep 25 & 'Alopeke & Speckle & 832 nm & 0.01 & 0.02 & $\Delta$ 6.63 mag @ 0.5" \\
312811620 & 3788 & SAI-2.5m (2.5 m) & 2021 Oct 22 & Speckle Polarimeter & Speckle & $I$ & 0.02 & 0.083 & $\Delta$ 5.9 mag @ 1" \\
 &  & Palomar (5 m) & 2025 Dec 08 & PHARO & AO & $Kcont$ & 0.025 & 0.099 & $\Delta$ 6.961 mag @ 0.5" \\
 &  & Gemini (8 m) & 2026 Apr 04 & 'Alopeke & Speckle & 562 nm & 0.01 & 0.02 & $\Delta$ 3.89 mag @ 0.5" \\
 &  & Gemini (8 m) & 2026 Apr 04 & 'Alopeke & Speckle & 832 nm & 0.01 & 0.02 & $\Delta$ 5.97 mag @ 0.5" \\
284206913 & 3972 & SAI-2.5m (2.5 m) & 2021 Jul 19 & Speckle Polarimeter & Speckle & $I$ & 0.02 & 0.083 & $\Delta$ 6.9 mag @ 1" \\
 &  & Shane (3 m) & 2022 Jan 17 & ShARCS & AO & $K_s$ & 0.033 & --- & --- \\
 &  & Shane (3 m) & 2022 Jan 17 & ShARCS & AO & $J$ & 0.033 & --- & --- \\
452938965 & 3988 & SAI-2.5m (2.5 m) & 2021 Aug 02 & Speckle Polarimeter & Speckle & $I$ & 0.02 & 0.083 & $\Delta$ 6.0 mag @ 1" \\
320281287 & 3998 & SAI-2.5m (2.5 m) & 2021 Oct 22 & Speckle Polarimeter & Speckle & $I$ & 0.02 & 0.083 & $\Delta$ 5.6 mag @ 1" \\
 &  & Palomar (5 m) & 2025 Sep 16 & PHARO & AO & $Hcont$ & 0.025 & 0.088 & $\Delta$ 7.309 mag @ 0.5" \\
 &  & Palomar (5 m) & 2025 Sep 16 & PHARO & AO & $Kcont$ & 0.025 & 0.104 & $\Delta$ 6.921 mag @ 0.5" \\
 &  & Palomar (5 m) & 2025 Sep 16 & PHARO & AO & $J$ & 0.025 & 0.097 & $\Delta$ 6.581 mag @ 0.5" \\
368734712 & 4009 & SAI-2.5m (2.5 m) & 2021 Oct 30 & Speckle Polarimeter & Speckle & $I$ & 0.02 & 0.083 & $\Delta$ 5.4 mag @ 1" \\
444558604 & 4079 & SAI-2.5m (2.5 m) & 2021 Oct 30 & Speckle Polarimeter & Speckle & $I$ & 0.02 & 0.083 & $\Delta$ 5.3 mag @ 1" \\
302381397 & 4088 & SAI-2.5m (2.5 m) & 2021 Oct 30 & Speckle Polarimeter & Speckle & $I$ & 0.02 & 0.083 & $\Delta$ 5.3 mag @ 1" \\
400103802 & 4140 & SAI-2.5m (2.5 m) & 2022 Jan 17 & Speckle Polarimeter & Speckle & $I$ & 0.02 & 0.083 & $\Delta$ 4.9 mag @ 1" \\
313194972 & 4144 & SAI-2.5m (2.5 m) & 2021 Dec 30 & Speckle Polarimeter & Speckle & $I$ & 0.02 & 0.083 & $\Delta$ 5.5 mag @ 1" \\
193765661 & 5236 & SAI-2.5m (2.5 m) & 2022 Nov 08 & Speckle Polarimeter & Speckle & $I$ & 0.021 & 0.083 & $\Delta$ 7.3 mag @ 1" \\
 &  & Palomar (5 m) & 2024 Sep 22 & PHARO & AO & $Kcont$ & 0.025 & 0.112 & $\Delta$ 6.225 mag @ 0.5" \\
67444896 & 5479 & SOAR (4.1 m) & 2022 Jun 10 & HRCam & Speckle & $I$ & 0.016 & 0.064 & $\Delta$ 6.0 mag @ 1" \\
 &  & SAI-2.5m (2.5 m) & 2023 Jan 19 & Speckle Polarimeter & Speckle & $I$ & 0.021 & 0.083 & $\Delta$ 3.6 mag @ 1" \\
 &  & SAI-2.5m (2.5 m) & 2023 Jan 19 & Speckle Polarimeter & Speckle & $I$ & 0.021 & 0.083 & $\Delta$ 5.1 mag @ 1" \\
345193111 & 5925 & SOAR (4.1 m) & 2023 Aug 31 & HRCam & Speckle & $I$ & 0.016 & 0.064 & $\Delta$ 5.2 mag @ 1" \\
 &  & SAI-2.5m (2.5 m) & 2023 Sep 23 & Speckle Polarimeter & Speckle & $I$ & 0.021 & 0.083 & $\Delta$ 6.1 mag @ 1" \\
371573539 & 6148 & SOAR (4.1 m) & 2023 Aug 31 & HRCam & Speckle & $I$ & 0.016 & 0.067 & $\Delta$ 5.1 mag @ 1" \\
 &  & SAI-2.5m (2.5 m) & 2023 Sep 01 & Speckle Polarimeter & Speckle & $I$ & 0.021 & 0.083 & $\Delta$ 5.6 mag @ 1" \\
190822775 & 6166 & SAI-2.5m (2.5 m) & 2023 Aug 28 & Speckle Polarimeter & Speckle & $I$ & 0.021 & 0.083 & $\Delta$ 5.8 mag @ 1" \\
 &  & Gemini (8 m) & 2025 Dec 10 & 'Alopeke & Speckle & 562 nm & 0.01 & 0.02 & $\Delta$ 4.86 mag @ 0.5" \\
 &  & Gemini (8 m) & 2025 Dec 10 & 'Alopeke & Speckle & 832 nm & 0.01 & 0.02 & $\Delta$ 6.17 mag @ 0.5" \\
67478724 & 6171 & Palomar (5 m) & 2023 Jun 30 & PHARO & AO & $Br\gamma$ & 0.025 & 0.097 & $\Delta$ 6.869 mag @ 0.5" \\
 &  & Palomar (5 m) & 2023 Jun 30 & PHARO & AO & $Hcont$ & 0.025 & 0.102 & $\Delta$ 6.448 mag @ 0.5" \\
 &  & SAI-2.5m (2.5 m) & 2023 Aug 28 & Speckle Polarimeter & Speckle & $I$ & 0.021 & 0.083 & $\Delta$ 6.7 mag @ 1" \\
 &  & Palomar (5 m) & 2025 Sep 16 & PHARO & AO & $Kcont$ & 0.025 & 0.105 & $\Delta$ 6.943 mag @ 0.5" \\
 &  & WIYN (3.5 m) & 2025 Dec 09 & NESSI & Speckle & 562 nm & 0.02 & 0.05 & $\Delta$ 4.49 mag @ 0.5" \\
 &  & WIYN (3.5 m) & 2025 Dec 09 & NESSI & Speckle & 832 nm & 0.02 & 0.05 & $\Delta$ 4.59 mag @ 0.5" \\
197743152 & 6191 & SAI-2.5m (2.5 m) & 2023 Aug 30 & Speckle Polarimeter & Speckle & $I$ & 0.021 & 0.083 & $\Delta$ 7.1 mag @ 1" \\
257207557 & 6184 & SOAR (4.1 m) & 2023 Aug 31 & HRCam & Speckle & $I$ & 0.016 & 0.068 & $\Delta$ 5.1 mag @ 1" \\
 &  & SAI-2.5m (2.5 m) & 2023 Sep 22 & Speckle Polarimeter & Speckle & $I$ & 0.021 & 0.083 & $\Delta$ 5.9 mag @ 1" \\
 &  & Gemini (8 m) & 2025 Oct 11 & Zorro & Speckle & 562 nm & 0.01 & 0.02 & $\Delta$ 4.02 mag @ 0.5" \\
 &  & Gemini (8 m) & 2025 Oct 11 & Zorro & Speckle & 832 nm & 0.01 & 0.02 & $\Delta$ 5.3 mag @ 0.5" \\
326475995 & 6208 & SAI-2.5m (2.5 m) & 2023 Sep 02 & Speckle Polarimeter & Speckle & $I$ & 0.021 & 0.083 & $\Delta$ 5.4 mag @ 1" \\
63718617 & 6334 & SAI-2.5m (2.5 m) & 2023 Aug 27 & Speckle Polarimeter & Speckle & $I$ & 0.021 & 0.083 & $\Delta$ 6.7 mag @ 1" \\
 &  & Palomar (5 m) & 2025 Sep 16 & PHARO & AO & $Kcont$ & 0.025 & 0.099 & $\Delta$ 6.650 mag @ 0.5" \\
 &  & Gemini (8 m) & 2025 Sep 30 & 'Alopeke & Speckle & 562 nm & 0.01 & 0.02 & $\Delta$ 5.6 mag @ 0.5" \\
 &  & Gemini (8 m) & 2025 Sep 30 & 'Alopeke & Speckle & 832 nm & 0.01 & 0.02 & $\Delta$ 6.29 mag @ 0.5" \\
95191643 & 6417 & SOAR (4.1 m) & 2024 Jan 08 & HRCam & Speckle & $I$ & 0.016 & 0.064 & $\Delta$ 6.0 mag @ 1" \\
125563127 & 6443 & SOAR (4.1 m) & 2024 Jan 08 & HRCam & Speckle & $I$ & 0.016 & 0.07 & $\Delta$ 5.3 mag @ 1" \\
242674266 & 7219 & Gemini (8 m) & 2025 Sep 14 & Zorro & Speckle & 832 nm & 0.01 & 0.02 & $\Delta$ 6.11 mag @ 0.5" \\
 &  & Gemini (8 m) & 2025 Sep 14 & Zorro & Speckle & 562 nm & 0.01 & 0.02 & $\Delta$ 4.79 mag @ 0.5" \\
125695940 & 7266 & SAI-2.5m (2.5 m) & 2025 Jul 10 & Speckle Polarimeter & Speckle & $I$ & 0.021 & 0.083 & $\Delta$ 5.7 mag @ 1" \\
 &  & Palomar (5 m) & 2025 Aug 07 & PHARO & AO & $Hcont$ & 0.025 & 0.104 & $\Delta$ 6.653 mag @ 0.5" \\
 &  & Palomar (5 m) & 2025 Aug 07 & PHARO & AO & $Kcont$ & 0.025 & 0.112 & $\Delta$ 6.252 mag @ 0.5" \\
352409708 & 7404 & Palomar (5 m) & 2025 Aug 06 & PHARO & AO & $Kcont$ & 0.025 & 0.099 & $\Delta$ 6.586 mag @ 0.5" \\
285592400 & 7425 & Maidanak NT (0.6 m) & 2025 Sep 14 & Andor iKon-L 936 & Seeing-Limited & $I$ & 0.76 & 2 & --- \\
 &  & Maidanak NT (0.6 m) & 2025 Sep 18 & Andor iKon-L 936 & Seeing-Limited & $I$ & 0.76 & 2 & --- \\
 &  & Gemini (8 m) & 2025 Oct 01 & 'Alopeke & Speckle & 832 nm & 0.01 & 0.02 & $\Delta$ 5.92 mag @ 0.5" \\
 &  & Gemini (8 m) & 2025 Oct 01 & 'Alopeke & Speckle & 562 nm & 0.01 & 0.02 & $\Delta$ 4.15 mag @ 0.5" \\
 &  & Maidanak NT (0.6 m) & 2025 Oct 11 & Andor iKon-L 936 & Seeing-Limited & $R$ & 0.77 & 2 & --- \\
159084486 & 7574 & SAI-2.5m (2.5 m) & 2026 Jan 28 & Speckle Polarimeter & Speckle & $I$ & 0.021 & 0.083 & $\Delta$ 5.9 mag @ 1" \\
\bottomrule
\end{tabular}
\end{table*}

\providecommand{\tess}{\textit{TESS}\xspace}
\begin{table*}
\centering
\caption{Observed Properties of Fitted Secondary Stars}
\label{tab:secondarylit}
\scriptsize
\setlength{\tabcolsep}{3.5pt}
\begin{tabular}{l lcccc}
\toprule
&  & TIC 605485663 & TOI-3988 B & TOI-6171 B & TOI-7266 B\\
& Classification & Background Star & Bound Companion & Bound Companion & Bound Companion\\
& Planet Host & TOI-3972 & TOI-3988 A & TOI-6171 A & TOI-7266 A\\
\midrule
$\rho$ & Angular separation ($\arcsec$) & $5.113 \pm 0.001$ & $1.000 \pm 0.011$ & $0.118 \pm 0.001$ & $0.452 \pm 0.001$ \\
$\rho_{\rm proj}$ & Projected separation (AU) & --- & $771 \pm 13$ & $57.29^{+0.76}_{-0.75}$ & $334^{+54}_{-43}$ \\
$G$ & Gaia $G$ mag. & $16.88 \pm 0.02$ & --- & --- & --- \\
$G_{\rm BP}$ & Gaia $G_{\rm BP}$ mag. & $17.241 \pm 0.034$ & --- & --- & --- \\
$G_{\rm RP}$ & Gaia $G_{\rm RP}$ mag. & $15.84 \pm 0.02$ & --- & --- & --- \\
$\Delta I$ & $I$-band contrast (mag) & --- & $5.2 \pm 0.1$ & --- & $2.9 \pm 0.1$ \\
$\Delta J$ & $J$-band contrast (mag) & $5.048 \pm 0.024$ & --- & --- & --- \\
$\Delta H$ & $H$-band contrast (mag) & --- & --- & $0.352 \pm 0.005$ & $1.707 \pm 0.007$ \\
$\Delta K$ & $K$-band contrast (mag) & $4.676 \pm 0.021$ & --- & --- & $1.703 \pm 0.007$ \\
\bottomrule
\end{tabular}
\vspace{2mm}
\begin{minipage}{\textwidth}
\footnotesize
\textbf{Notes:}
Angular separations and contrasts are reported on ExoFOP. Where a companion was detected in several filters, the separations are averaged and their scatter is included in the uncertainty.\\
\end{minipage}
\end{table*}


\subsubsection{Speckle Imaging}\label{subsubsec:speckle}


High angular resolution imaging on the 2.5-m telescope at the Caucasian Observatory of Sternberg Astronomical Institute (SAI) of Lomonosov Moscow State University was implemented with the speckle polarimeter. Observations before August 2022 were secured with the EMCCD version of the instrument. For later observations, an upgraded version was used, employing the low--noise CMOS detector Hamamatsu ORCA--quest \citep{Strakhov:2023}. For all of our observations, we used the atmospheric dispersion compensator and a wide $I_\mathrm{c}$ filter. The respective angular resolution is 0.083\arcsec. Many of the targets of this article were observed with this instrument. We detected a stellar companion in our observations of TOI-3988 and TOI-7266. In the case of TOI-3988, the separation is $1000\pm11$ mas, the position angle is $205.4\pm0.3$ deg, and the magnitude difference is 5.2. In the case of TOI-7266, the separation is $452\pm2$ mas, the position angle is $151.3\pm0.2$ deg, and the magnitude difference is 2.9. Due to their close proximity and non-detection in \gaia across multiple epochs, both of these stars are likely bound to the hosts, and we name them TOI-3988\,B and TOI-7266\,B, respectively.

The High-Resolution Camera (HRCam) at the 4.1~m Southern Astrophysical Research (SOAR) telescope is a visitor-class instrument designed for diffraction-limited imaging through speckle interferometry. Utilizing an Andor iXon EMCCD detector, the system captures high-cadence data cubes of sub-30 ms exposures to freeze atmospheric turbulence and compute the target's power spectrum and autocorrelation function \citep{Tokovinin:2018}. The instrument achieves an angular resolution of approximately 36 mas and typical contrast limits of $\Delta \textrm{mag} \approx 6$ at 1\arcsec in the Cousins $I$ band, which closely matches the \tess photometric bandpass. As part of the SOAR \tess survey, HRCam provides high-resolution astrometry and photometry to resolve stellar companions, quantify flux contamination within \tess's 21-arcsecond pixels, and measure true transit depths \citep{Ziegler:2020}. HRCam obtained observations of TOI-3041, TOI-3365, TOI-5479, TOI-5925, TOI-6148, TOI-6184, TOI-6417, and TOI-6443. No evidence of stellar companions was found in any of these observations.

TOI-3601, TOI-6166, TOI-6334, and TOI-7425 were observed by the 'Alopeke instrument on the Gemini North 8~m telescope, while TOI-6184 and TOI-7219 were observed by the Zorro instrument on the Gemini South 8~m telescope \citep{Scott:2021}. Both of these identical instruments\footnote{\url{https://www.gemini.edu/sciops/instruments/alopeke-zorro/}} provide simultaneous speckle imaging in the 562~nm and 832~nm bandpasses. To provide a reconstructed image and contrast limits, we performed a Fourier analysis using a standard reduction pipeline \citep{Howell:2011}. Around TOI-6166, 'Alopeke found tentative evidence of a source with $\Delta$mag = 5.25 in the 832~nm filter with a separation of 1.233 arcsec and a position angle of 155.7 deg east of north. However, as this source was not detected in the 562~nm filter or in the SAI speckle observation, we chose not to include it in our analysis. No other evidence of stellar companions was detected.

Finally, TOI-6171 was observed by the NN-EXPLORE Exoplanet Stellar Speckle Imager \citep[NESSI;][]{Scott:2018}. NESSI, like 'Alopeke and Zorro, is a dual-channel speckle imager that observes simultaneously in filters centered at 562~nm and 832~nm. It is mounted on the 3.5~m WIYN telescope along with the NEID instrument. The data reduction was also performed following the procedure described in \cite{Howell:2011}. No companion was found in these observations.

\subsubsection{Adaptive Optics Imaging}\label{subsubsec:ao}


Observations of TOI-3601, TOI-3788, TOI-3998, TOI-5236, TOI-6171, TOI-6334, TOI-7266, and TOI-7404 were made with the PHARO instrument \citep{Hayward:2001} on the Palomar Hale telescope (5~m) behind the P3K natural guide-star AO system \citep{Dekany:2013}. The pixel scale for PHARO is $0.025\arcsec$. The Palomar data were collected in a 3--point quincunx dither pattern in the $K_{cont}$ and $H_{cont}$ filters. The reduced science frames were combined into a single mosaic image with final resolutions of $\sim 0.10$\arcsec and $\sim 0.11$\arcsec, respectively. The sensitivities of the final combined AO images were determined by injecting simulated sources azimuthally around the primary target every $20^\circ $ at separations of integer multiples of the central source's FWHM \citep{Furlan:2017}. The brightness of each injected source was scaled until standard aperture photometry detected it with $5\sigma $ significance.  The final $5\sigma $ limit at each separation was determined from the average of all of the determined limits at that separation and the uncertainty on the limit was set by the rms dispersion of the azimuthal slices at a given radial distance.  

We discovered a stellar companion to TOI-6171\,A located $0.118 \pm 0.001 \arcsec$ to the north ($4.86 \pm 1.72^\circ$ E of N) in the $Hcont$ filter. This companion, hereafter TOI-6171\,B, is fainter than TOI-6171\,A by $\Delta Hcont = 0.352 \pm 0.005$ mag. This companion is not in the \gaia catalogue, so this observation establishes its discovery. Additionally, a nearby stellar companion to TOI-7266\,A was discovered located $0.456\pm0.001$\arcsec\ to the southeast ($152.2^\circ\pm0.1^\circ$ E of N) in both near-infrared images.  The companion is fainter than the primary target by $\Delta Hcont = 1.707\pm0.007$mag and $\Delta Kcont = 1.703\pm0.007$mag. These observations corroborate SAI's detection of the same source in the $I$ filter.


Finally, TOI-3972 was observed with the ShARCS instrument on the Shane 3~m telescope at Lick Observatory \citep{Kupke:2012, Gavel:2014, McGurk:2014}. The observations were collected in the $J$ filter ($\lambda_0 = 1.238 \mu \mathrm{m}$, $\Delta \lambda = 0.271 \mu \mathrm{m}$) and the $K_s$ filter ($\lambda_0 = 2.150 \mu \mathrm{m}$, $\Delta \lambda = 0.320 \mu \mathrm{m}$). The data were reduced using the publicly available \texttt{SImMER} pipeline\footnote{\url{https://github.com/arjunsavel/SImMER}} \citep{Savel:2020, Savel:2022}. In these data, we detected three unbound background stars, one with a separation of $\sim 5.113 \arcsec$ to the southwest ($\sim227.9^{\circ}$ E of N), another with a separation of $\sim 9.434 \arcsec$ to the north ($\sim0.4^{\circ}$ E of N), and one with a separation of $\sim 9.572 \arcsec$ to the southwest ($\sim194.8^{\circ}$ E of N). Each of these has been observed by \gaia and has astrometric observations consistent with being chance alignments, rather than bound companions. All three are also registered in the TIC, with identifiers TIC 605485663, TIC 284206920, and TIC 284206903 (nearest to furthest from TOI-3972).

\subsection{Archival Photometry and Astrometry}\label{subsec:archival}

To measure the SED of each of our host stars, we downloaded archival photometry from \gaia Data Release 3 (\textit{Gaia }DR3; \citealt{GaiaDR3}), the Two Micron All-Sky Survey (2MASS; \citealt{Cutri:2003, Skrutskie:2006}), and the Wide-field Infrared Survey Explorer (WISE; \citealt{Wright:2010, Cutri:2012}) using the VizieR service\footnote{\url{https://vizier.cds.unistra.fr/viz-bin/VizieR}} \citep{Ochsenbein:2000}, along with the \texttt{astroquery} package \citep{Ginsburg:2019}. From \gaia, we retrieved the magnitudes from the wider $G$ filter, along with the narrower $G_{\rm BP}$ and $G_{\rm RP}$ filters, spanning the entire visual wavelength range. From 2MASS, we retrieved the near-infrared $J$, $H$, and $K_{\rm s}$ magnitudes. Finally, from WISE, we chose to only use the $W1$, $W2$, and $W3$ bandpasses, as the $W4$ bandpass has a known discrepancy causing sources to appear too bright in this wavelength range \citep{Wright:2010}. These observations span a total wavelength range of $0.33 - 17 \mu$m. To ensure that the uncertainties on these observations are not underestimated, we adopted a systematic floor on the uncertainties of each measurement as described in \cite{Eastman:2019}.

We also obtained astrometric information on each star using \gaia DR3 to determine the distance to each star and whether they have nearby bound companions. A correction was applied to the parallax measurements following the work of \cite{Lindegren:2021} to account for systematic biases. The \gaia renormalised unit weight error (RUWE) is a measure of the quality of the single-source astrometric source, and an elevated RUWE is often a good indicator that a star has a binary companion \citep{Castro-Ginard:2024}. \cite{Castro-Ginard:2024} developed a position-varying lower RUWE threshold for an astrometric solution to be indicative of a binary system. We used the Python package \texttt{GaiaUnlimited}\footnote{\url{https://github.com/gaia-unlimited/gaiaunlimited}} to estimate this threshold for each host star in this work and found that three of them have an elevated RUWE: TOI-3788 with a RUWE of 1.429 ($>$ RUWE threshold of 1.219), TOI-4009 with a RUWE of 1.338 ($> 1.196$), and TOI-7266 with a RUWE of 3.257 ($> 1.239$). While both TOI-3788 and TOI-4009 show no evidence of a nearby companion in the high-resolution imaging, TOI-7266 has a bright, nearby companion detected in the high angular resolution imaging that is very likely influencing \gaia's astrometric solution, elevating the uncertainties. However, the parallax measurement remains suitable for the purposes of this work. Additionally, TOI-4079 has a distant \gaia-detected, bright companion, with a separation of $11.5\arcsec$ and $\Delta T = -0.584$ mag. However, due to the companion's large separation, it was not included in our global fits. The archival photometry and astrometry are included in Table~\ref{tab:lit}.

\renewcommand{\arraystretch}{\litstretch}
\providecommand{\bjdtdb}{\ensuremath{\rm {BJD_{TDB}}}}
\providecommand{\feh}{\ensuremath{\left[{\rm Fe}/{\rm H}\right]}}
\providecommand{\teff}{\ensuremath{T_{\rm eff}}}
\providecommand{\teq}{\ensuremath{T_{\rm eq}}}
\providecommand{\ecosw}{\ensuremath{e\cos{\omega_*}}}
\providecommand{\esinw}{\ensuremath{e\sin{\omega_*}}}
\providecommand\msun{M$_\odot$\xspace}
\providecommand{\rsun}{R$_\odot$\xspace}
\providecommand{\lsun}{L$_\odot$\xspace}
\providecommand{\mj}{\ensuremath{\,M_{\rm J}}}
\providecommand{\rj}{\ensuremath{\,R_{\rm J}}}
\providecommand{\me}{\ensuremath{\,M_{\rm E}}}
\providecommand{\re}{\ensuremath{\,R_{\rm E}}}
\providecommand{\fave}{\langle F \rangle}
\providecommand{\fluxcgs}{10$^9$ erg s$^{-1}$ cm$^{-2}$}
\providecommand{\tess}{\textit{TESS}\xspace}
\begin{table*}
\centering
\caption{Measured Properties from Literature}
\label{tab:lit}
\scriptsize
\setlength{\tabcolsep}{3pt}
\begin{tabular}{llcccccc}
\toprule
&  & TOI-3041 & TOI-3365 & TOI-3601 & TOI-3788 & TOI-3972 & Source \\
\multicolumn{8}{l}{\textbf{Other identifiers}:} \\
& \tess Input Catalog & TIC 371864043 & TIC 34297761 & TIC 191202679 & TIC 312811620 & TIC 284206913 & \\
& TYCHO-2 & --- & TYC 8181-2918-1 & --- & TYC 3768-1773-1 & TYC 4016-89-1 & \\
& 2MASS & J09571802-6846449 & J09455937-4616036 & J00344830+4523148 & J06174379+5422528 & J00371289+6112186 & \\
& Gaia DR3 & 5243264296502760704 & 5411470059225900800 & 389013055062728064 & 996225841925467264 & 427232186629403264 & \\
\midrule
\multicolumn{8}{l}{\textbf{Astrometric Parameters}:} \\
$\alpha_{J2000}\ddagger$ & Right Ascension (h:m:s) & 09:57:18.012 & 09:45:59.370 & 00:34:48.297 & 06:17:43.798 & 00:37:12.901  & 1 \\
$\delta_{J2000}\ddagger$ & Declination (d:m:s) & -68:46:44.995 & -46:16:03.645 & 45:23:14.799 & 54:22:52.878 & 61:12:18.676  & 1 \\
$\mu_{\alpha}$ & Gaia DR3 proper motion in RA (mas yr$^{-1}$)& $-25.220 \pm 0.012$ & $-42.398 \pm 0.014$ & $-4.844 \pm 0.013$ & $11.324 \pm 0.028$ & $147.884 \pm 0.013$  & 1 \\
$\mu_{\delta}$ & Gaia DR3 proper motion in Dec (mas yr$^{-1}$)& $14.087 \pm 0.013$ & $8.805 \pm 0.016$ & $-6.875 \pm 0.013$ & $-11.425 \pm 0.025$ & $-21.092 \pm 0.015$  & 1 \\
$\pi$ & Gaia DR3 Parallax (mas) & $2.0792 \pm 0.0108$ & $3.7029 \pm 0.0156$ & $1.3719 \pm 0.0160$ & $1.8955 \pm 0.0258$ & $6.0930 \pm 0.0138$  & 1 \\
$v\sin{i_\star}$ & Projected rotational velocity (km s$^{-1}$) & $6.20 \pm 0.03$ & $5.55 \pm 0.02$ & $15.3 \pm 0.2$ & $12.6 \pm 0.1$ & $2.13 \pm 0.12$  & 2 \\
\multicolumn{8}{l}{\textbf{Photometric Parameters}:} \\
$G$ & Gaia $G$ mag. & $11.654 \pm 0.020$ & $11.293 \pm 0.020$ & $12.791 \pm 0.020$ & $11.706 \pm 0.020$ & $11.271 \pm 0.020$  & 1 \\
$G_{\rm BP}$ & Gaia $G_{\rm BP}$ mag. & $11.951 \pm 0.020$ & $11.632 \pm 0.020$ & $13.069 \pm 0.020$ & $11.981 \pm 0.020$ & $11.684 \pm 0.020$  & 1 \\
$G_{\rm RP}$ & Gaia $G_{\rm RP}$ mag. & $11.190 \pm 0.020$ & $10.794 \pm 0.020$ & $12.357 \pm 0.020$ & $11.268 \pm 0.020$ & $10.697 \pm 0.020$  & 1 \\
$T$ & TESS mag. & $11.2553 \pm 0.0060$ & $10.8467 \pm 0.0060$ & $12.4226 \pm 0.0070$ & $11.3321 \pm 0.0070$ & $10.7486 \pm 0.0060$  & 3 \\
$J$ & 2MASS $J$ mag. & $10.668 \pm 0.023$ & $10.242 \pm 0.022$ & $11.894 \pm 0.022$ & $10.813 \pm 0.022$ & $10.037 \pm 0.024$  & 4 \\
$H$ & 2MASS $H$ mag. & $10.407 \pm 0.022$ & $9.950 \pm 0.024$ & $11.673 \pm 0.024$ & $10.560 \pm 0.028$ & $9.685 \pm 0.031$  & 4 \\
$K$ & 2MASS $K$ mag. & $10.345 \pm 0.021$ & $9.888 \pm 0.021$ & $11.595 \pm 0.020$ & $10.492 \pm 0.020$ & $9.583 \pm 0.021$  & 4 \\
$W1$ & WISE $W1$ mag. & $10.315 \pm 0.030$ & $9.828 \pm 0.030$ & $11.576 \pm 0.030$ & $10.464 \pm 0.030$ & $9.529 \pm 0.030$  & 5 \\
$W2$ & WISE $W2$ mag. & $10.353 \pm 0.030$ & $9.884 \pm 0.030$ & $11.602 \pm 0.030$ & $10.482 \pm 0.030$ & $9.594 \pm 0.030$  & 5 \\
$W3$ & WISE $W3$ mag. & $10.313 \pm 0.041$ & $9.861 \pm 0.048$ & $11.543 \pm 0.130$ & $10.407 \pm 0.072$ & $9.613 \pm 0.037$  & 5 \\
\bottomrule
\end{tabular}
\end{table*}

\providecommand{\bjdtdb}{\ensuremath{\rm {BJD_{TDB}}}}
\providecommand{\feh}{\ensuremath{\left[{\rm Fe}/{\rm H}\right]}}
\providecommand{\teff}{\ensuremath{T_{\rm eff}}}
\providecommand{\teq}{\ensuremath{T_{\rm eq}}}
\providecommand{\ecosw}{\ensuremath{e\cos{\omega_*}}}
\providecommand{\esinw}{\ensuremath{e\sin{\omega_*}}}
\providecommand\msun{M$_\odot$\xspace}
\providecommand{\rsun}{R$_\odot$\xspace}
\providecommand{\lsun}{L$_\odot$\xspace}
\providecommand{\mj}{\ensuremath{\,M_{\rm J}}}
\providecommand{\rj}{\ensuremath{\,R_{\rm J}}}
\providecommand{\me}{\ensuremath{\,M_{\rm E}}}
\providecommand{\re}{\ensuremath{\,R_{\rm E}}}
\providecommand{\fave}{\langle F \rangle}
\providecommand{\fluxcgs}{10$^9$ erg s$^{-1}$ cm$^{-2}$}
\providecommand{\tess}{\textit{TESS}\xspace}
\begin{table*}
\centering
\contcaption{}
\scriptsize
\setlength{\tabcolsep}{3pt}
\begin{tabular}{llcccccc}
\toprule
&  & TOI-3988 & TOI-3998 & TOI-4009 & TOI-4079 & TOI-4088 & Source \\
\multicolumn{8}{l}{\textbf{Other identifiers}:} \\
& \tess Input Catalog & TIC 452938965 & TIC 320281287 & TIC 368734712 & TIC 444558604 & TIC 302381397 & \\
& TYCHO-2 & --- & --- & --- & --- & --- & \\
& 2MASS & J00340202+5850394 & J00303281+5619534 & J00510713+7056451 & J01195535+6432169 & J06401948+7633113 & \\
& Gaia DR3 & 428028954600689280 & 421665531052612096 & 536125894094897536 & 524829576048693120 & 1115757466129413760 & \\
\midrule
\multicolumn{8}{l}{\textbf{Astrometric Parameters}:} \\
$\alpha_{J2000}\ddagger$ & Right Ascension (h:m:s) & 00:34:02.017 & 00:30:32.802 & 00:51:07.155 & 01:19:55.359 & 06:40:19.503  & 1 \\
$\delta_{J2000}\ddagger$ & Declination (d:m:s) & 58:50:39.558 & 56:19:53.345 & 70:56:44.979 & 64:32:16.983 & 76:33:11.280  & 1 \\
$\mu_{\alpha}$ & Gaia DR3 proper motion in RA (mas yr$^{-1}$)& $3.040 \pm 0.010$ & $40.382 \pm 0.008$ & $8.952 \pm 0.013$ & $17.947 \pm 0.008$ & $-11.732 \pm 0.010$  & 1 \\
$\mu_{\delta}$ & Gaia DR3 proper motion in Dec (mas yr$^{-1}$)& $-6.725 \pm 0.011$ & $-4.035 \pm 0.011$ & $-4.002 \pm 0.015$ & $0.738 \pm 0.011$ & $-14.979 \pm 0.013$  & 1 \\
$\pi$ & Gaia DR3 Parallax (mas) & $1.2794 \pm 0.0122$ & $2.4433 \pm 0.0109$ & $1.5330 \pm 0.0129$ & $2.7511 \pm 0.0110$ & $2.2683 \pm 0.0119$  & 1 \\
$v\sin{i_\star}$ & Projected rotational velocity (km s$^{-1}$) & $7.63 \pm 0.15$ & $4.69 \pm 0.06$ & $9.53 \pm 0.13$ & $18.1 \pm 0.1$ & $4.02 \pm 0.13$  & 2 \\
\multicolumn{8}{l}{\textbf{Photometric Parameters}:} \\
$G$ & Gaia $G$ mag. & $12.765 \pm 0.020$ & $12.591 \pm 0.020$ & $12.664 \pm 0.020$ & $12.678 \pm 0.020$ & $12.734 \pm 0.020$  & 1 \\
$G_{\rm BP}$ & Gaia $G_{\rm BP}$ mag. & $13.178 \pm 0.020$ & $12.966 \pm 0.020$ & $13.055 \pm 0.020$ & $13.111 \pm 0.020$ & $13.112 \pm 0.020$  & 1 \\
$G_{\rm RP}$ & Gaia $G_{\rm RP}$ mag. & $12.164 \pm 0.020$ & $12.039 \pm 0.020$ & $12.103 \pm 0.020$ & $12.078 \pm 0.020$ & $12.185 \pm 0.020$  & 1 \\
$T$ & TESS mag. & $12.2454 \pm 0.0630$ & $12.1134 \pm 0.0060$ & $12.1693 \pm 0.0140$ & $12.1484 \pm 0.0300$ & $12.2523 \pm 0.0090$  & 3 \\
$J$ & 2MASS $J$ mag. & $11.440 \pm 0.023$ & $11.347 \pm 0.023$ & $11.446 \pm 0.020$ & $11.325 \pm 0.025$ & $11.602 \pm 0.022$  & 4 \\
$H$ & 2MASS $H$ mag. & $11.165 \pm 0.028$ & $11.024 \pm 0.023$ & $11.169 \pm 0.027$ & $11.084 \pm 0.030$ & $11.214 \pm 0.028$  & 4 \\
$K$ & 2MASS $K$ mag. & $11.069 \pm 0.026$ & $10.906 \pm 0.020$ & $11.088 \pm 0.025$ & $10.931 \pm 0.020$ & $11.177 \pm 0.023$  & 4 \\
$W1$ & WISE $W1$ mag. & $10.954 \pm 0.030$ & $10.804 \pm 0.030$ & $11.023 \pm 0.030$ & $10.785 \pm 0.030$ & $11.095 \pm 0.030$  & 5 \\
$W2$ & WISE $W2$ mag. & $10.976 \pm 0.030$ & $10.830 \pm 0.030$ & $11.057 \pm 0.030$ & $10.798 \pm 0.030$ & $11.136 \pm 0.030$  & 5 \\
$W3$ & WISE $W3$ mag. & $10.894 \pm 0.067$ & $10.720 \pm 0.068$ & $11.345 \pm 0.137$ & $10.618 \pm 0.080$ & $11.135 \pm 0.126$  & 5 \\
\bottomrule
\end{tabular}
\end{table*}

\providecommand{\bjdtdb}{\ensuremath{\rm {BJD_{TDB}}}}
\providecommand{\feh}{\ensuremath{\left[{\rm Fe}/{\rm H}\right]}}
\providecommand{\teff}{\ensuremath{T_{\rm eff}}}
\providecommand{\teq}{\ensuremath{T_{\rm eq}}}
\providecommand{\ecosw}{\ensuremath{e\cos{\omega_*}}}
\providecommand{\esinw}{\ensuremath{e\sin{\omega_*}}}
\providecommand\msun{M$_\odot$\xspace}
\providecommand{\rsun}{R$_\odot$\xspace}
\providecommand{\lsun}{L$_\odot$\xspace}
\providecommand{\mj}{\ensuremath{\,M_{\rm J}}}
\providecommand{\rj}{\ensuremath{\,R_{\rm J}}}
\providecommand{\me}{\ensuremath{\,M_{\rm E}}}
\providecommand{\re}{\ensuremath{\,R_{\rm E}}}
\providecommand{\fave}{\langle F \rangle}
\providecommand{\fluxcgs}{10$^9$ erg s$^{-1}$ cm$^{-2}$}
\providecommand{\tess}{\textit{TESS}\xspace}
\begin{table*}
\centering
\contcaption{}
\scriptsize
\setlength{\tabcolsep}{3pt}
\begin{tabular}{llcccccc}
\toprule
&  & TOI-4140 & TOI-4144 & TOI-5236 & TOI-5479 & TOI-5925 & Source \\
\multicolumn{8}{l}{\textbf{Other identifiers}:} \\
& \tess Input Catalog & TIC 400103802 & TIC 313194972 & TIC 193765661 & TIC 67444896 & TIC 345193111 & \\
& TYCHO-2 & --- & --- & TYC 3164-167-1 & --- & --- & \\
& 2MASS & J06275465+7400292 & J05462827+7502442 & J20250880+4313597 & J11253524+1246006 & J20550360+1204572 & \\
& Gaia DR3 & 1114282814878271360 & 502876967407491328 & 2068978419000426240 & 3965507334046321792 & 1757765531106021632 & \\
\midrule
\multicolumn{8}{l}{\textbf{Astrometric Parameters}:} \\
$\alpha_{J2000}\ddagger$ & Right Ascension (h:m:s) & 06:27:54.664 & 05:46:28.279 & 20:25:08.791 & 11:25:35.240 & 20:55:03.594  & 1 \\
$\delta_{J2000}\ddagger$ & Declination (d:m:s) & 74:00:29.303 & 75:02:44.166 & 43:13:59.764 & 12:46:00.610 & 12:04:57.318  & 1 \\
$\mu_{\alpha}$ & Gaia DR3 proper motion in RA (mas yr$^{-1}$)& $6.819 \pm 0.010$ & $-1.920 \pm 0.009$ & $-11.582 \pm 0.011$ & $-5.490 \pm 0.019$ & $3.842 \pm 0.015$  & 1 \\
$\mu_{\delta}$ & Gaia DR3 proper motion in Dec (mas yr$^{-1}$)& $-2.190 \pm 0.011$ & $-14.547 \pm 0.011$ & $-10.581 \pm 0.013$ & $-3.035 \pm 0.017$ & $-7.518 \pm 0.011$  & 1 \\
$\pi$ & Gaia DR3 Parallax (mas) & $3.2623 \pm 0.0116$ & $3.8371 \pm 0.0097$ & $1.7219 \pm 0.0120$ & $1.6361 \pm 0.0192$ & $1.9092 \pm 0.0175$  & 1 \\
$v\sin{i_\star}$ & Projected rotational velocity (km s$^{-1}$) & $3.86 \pm 0.09$ & $3.35 \pm 0.17$ & $8.30 \pm 0.18$ & $4.17 \pm 0.08$ & $8.49 \pm 0.07$  & 2 \\
\multicolumn{8}{l}{\textbf{Photometric Parameters}:} \\
$G$ & Gaia $G$ mag. & $12.625 \pm 0.020$ & $12.428 \pm 0.020$ & $11.762 \pm 0.020$ & $12.702 \pm 0.020$ & $12.627 \pm 0.020$  & 1 \\
$G_{\rm BP}$ & Gaia $G_{\rm BP}$ mag. & $13.021 \pm 0.020$ & $12.854 \pm 0.020$ & $12.087 \pm 0.020$ & $13.027 \pm 0.020$ & $12.916 \pm 0.020$  & 1 \\
$G_{\rm RP}$ & Gaia $G_{\rm RP}$ mag. & $12.068 \pm 0.020$ & $11.843 \pm 0.020$ & $11.271 \pm 0.020$ & $12.215 \pm 0.020$ & $12.180 \pm 0.020$  & 1 \\
$T$ & TESS mag. & $12.1311 \pm 0.0060$ & $11.9039 \pm 0.0060$ & $11.3335 \pm 0.0200$ & $12.2773 \pm 0.0070$ & $12.2493 \pm 0.0110$  & 3 \\
$J$ & 2MASS $J$ mag. & $11.445 \pm 0.021$ & $11.183 \pm 0.023$ & $10.602 \pm 0.023$ & $11.732 \pm 0.022$ & $11.653 \pm 0.023$  & 4 \\
$H$ & 2MASS $H$ mag. & $11.070 \pm 0.029$ & $10.815 \pm 0.030$ & $10.300 \pm 0.024$ & $11.451 \pm 0.022$ & $11.479 \pm 0.030$  & 4 \\
$K$ & 2MASS $K$ mag. & $11.018 \pm 0.025$ & $10.721 \pm 0.025$ & $10.267 \pm 0.022$ & $11.383 \pm 0.021$ & $11.403 \pm 0.023$  & 4 \\
$W1$ & WISE $W1$ mag. & $11.001 \pm 0.030$ & $10.678 \pm 0.030$ & $9.874 \pm 0.030$ & $11.291 \pm 0.030$ & $11.332 \pm 0.030$  & 5 \\
$W2$ & WISE $W2$ mag. & $11.055 \pm 0.030$ & $10.740 \pm 0.030$ & $9.906 \pm 0.030$ & $11.347 \pm 0.030$ & $11.369 \pm 0.030$  & 5 \\
$W3$ & WISE $W3$ mag. & $11.177 \pm 0.151$ & $10.611 \pm 0.085$ & $9.717 \pm 0.059$ & $11.183 \pm 0.164$ & $11.638 \pm 0.182$  & 5 \\
\bottomrule
\end{tabular}
\end{table*}

\providecommand{\bjdtdb}{\ensuremath{\rm {BJD_{TDB}}}}
\providecommand{\feh}{\ensuremath{\left[{\rm Fe}/{\rm H}\right]}}
\providecommand{\teff}{\ensuremath{T_{\rm eff}}}
\providecommand{\teq}{\ensuremath{T_{\rm eq}}}
\providecommand{\ecosw}{\ensuremath{e\cos{\omega_*}}}
\providecommand{\esinw}{\ensuremath{e\sin{\omega_*}}}
\providecommand\msun{M$_\odot$\xspace}
\providecommand{\rsun}{R$_\odot$\xspace}
\providecommand{\lsun}{L$_\odot$\xspace}
\providecommand{\mj}{\ensuremath{\,M_{\rm J}}}
\providecommand{\rj}{\ensuremath{\,R_{\rm J}}}
\providecommand{\me}{\ensuremath{\,M_{\rm E}}}
\providecommand{\re}{\ensuremath{\,R_{\rm E}}}
\providecommand{\fave}{\langle F \rangle}
\providecommand{\fluxcgs}{10$^9$ erg s$^{-1}$ cm$^{-2}$}
\providecommand{\tess}{\textit{TESS}\xspace}
\begin{table*}
\centering
\contcaption{}
\scriptsize
\setlength{\tabcolsep}{3pt}
\begin{tabular}{llcccccc}
\toprule
&  & TOI-6148 & TOI-6166 & TOI-6171 & TOI-6184 & TOI-6191 & Source \\
\multicolumn{8}{l}{\textbf{Other identifiers}:} \\
& \tess Input Catalog & TIC 371573539 & TIC 190822775 & TIC 67478724 & TIC 257207557 & TIC 197743152 & \\
& TYCHO-2 & TYC 1689-1490-1 & --- & TYC 2756-665-1 & TYC 1153-1229-1 & TYC 3615-419-1 & \\
& 2MASS & J22080862+1930236 & J21435567+4634164 & J23183728+3218059 & J22503322+0926366 & J22281033+4934256 & \\
& Gaia DR3 & 1778172295119486592 & 1974640218262906880 & 1910379604762707712 & 2716415258209633792 & 1988134971147656960 & \\
\midrule
\multicolumn{8}{l}{\textbf{Astrometric Parameters}:} \\
$\alpha_{J2000}\ddagger$ & Right Ascension (h:m:s) & 22:08:08.623 & 21:43:55.675 & 23:18:37.290 & 22:50:33.234 & 22:28:10.342  & 1 \\
$\delta_{J2000}\ddagger$ & Declination (d:m:s) & 19:30:23.487 & 46:34:16.421 & 32:18:05.968 & 09:26:36.687 & 49:34:25.601  & 1 \\
$\mu_{\alpha}$ & Gaia DR3 proper motion in RA (mas yr$^{-1}$)& $-14.831 \pm 0.017$ & $-2.874 \pm 0.010$ & $5.052 \pm 0.015$ & $-9.422 \pm 0.018$ & $6.289 \pm 0.011$  & 1 \\
$\mu_{\delta}$ & Gaia DR3 proper motion in Dec (mas yr$^{-1}$)& $-11.831 \pm 0.018$ & $-15.702 \pm 0.010$ & $-1.576 \pm 0.015$ & $-27.637 \pm 0.016$ & $-29.793 \pm 0.010$  & 1 \\
$\pi$ & Gaia DR3 Parallax (mas) & $2.0125 \pm 0.0156$ & $2.3728 \pm 0.0102$ & $2.0178 \pm 0.0180$ & $2.9674 \pm 0.0164$ & $3.2346 \pm 0.0108$  & 1 \\
$v\sin{i_\star}$ & Projected rotational velocity (km s$^{-1}$) & $12.1 \pm 0.1$ & $5.57 \pm 0.61$ & $7.75 \pm 0.04$ & $3.47 \pm 0.08$ & $4.22 \pm 0.07$  & 2 \\
\multicolumn{8}{l}{\textbf{Photometric Parameters}:} \\
$G$ & Gaia $G$ mag. & $11.742 \pm 0.020$ & $12.560 \pm 0.020$ & $11.496 \pm 0.020$ & $12.377 \pm 0.020$ & $11.883 \pm 0.020$  & 1 \\
$G_{\rm BP}$ & Gaia $G_{\rm BP}$ mag. & $12.035 \pm 0.020$ & $12.892 \pm 0.020$ & $11.835 \pm 0.020$ & $12.770 \pm 0.020$ & $12.254 \pm 0.020$  & 1 \\
$G_{\rm RP}$ & Gaia $G_{\rm RP}$ mag. & $11.291 \pm 0.020$ & $12.059 \pm 0.020$ & $10.993 \pm 0.020$ & $11.821 \pm 0.020$ & $11.348 \pm 0.020$  & 1 \\
$T$ & TESS mag. & $11.3510 \pm 0.0070$ & $12.1280 \pm 0.0060$ & $11.0537 \pm 0.0080$ & $11.8836 \pm 0.0060$ & $11.4105 \pm 0.0060$  & 3 \\
$J$ & 2MASS $J$ mag. & $10.789 \pm 0.027$ & $11.475 \pm 0.021$ & $10.425 \pm 0.021$ & $11.187 \pm 0.023$ & $10.754 \pm 0.020$  & 4 \\
$H$ & 2MASS $H$ mag. & $10.592 \pm 0.031$ & $11.204 \pm 0.023$ & $10.130 \pm 0.020$ & $10.831 \pm 0.022$ & $10.414 \pm 0.020$  & 4 \\
$K$ & 2MASS $K$ mag. & $10.517 \pm 0.020$ & $11.152 \pm 0.020$ & $10.075 \pm 0.020$ & $10.752 \pm 0.023$ & $10.379 \pm 0.020$  & 4 \\
$W1$ & WISE $W1$ mag. & $10.470 \pm 0.030$ & $11.066 \pm 0.030$ & $10.041 \pm 0.030$ & $10.725 \pm 0.030$ & $10.317 \pm 0.030$  & 5 \\
$W2$ & WISE $W2$ mag. & $10.535 \pm 0.030$ & $11.115 \pm 0.030$ & $10.091 \pm 0.030$ & $10.770 \pm 0.030$ & $10.376 \pm 0.030$  & 5 \\
$W3$ & WISE $W3$ mag. & $10.445 \pm 0.078$ & $10.833 \pm 0.091$ & $10.013 \pm 0.047$ & $10.752 \pm 0.114$ & $10.412 \pm 0.065$  & 5 \\
\bottomrule
\end{tabular}
\end{table*}

\providecommand{\bjdtdb}{\ensuremath{\rm {BJD_{TDB}}}}
\providecommand{\feh}{\ensuremath{\left[{\rm Fe}/{\rm H}\right]}}
\providecommand{\teff}{\ensuremath{T_{\rm eff}}}
\providecommand{\teq}{\ensuremath{T_{\rm eq}}}
\providecommand{\ecosw}{\ensuremath{e\cos{\omega_*}}}
\providecommand{\esinw}{\ensuremath{e\sin{\omega_*}}}
\providecommand\msun{M$_\odot$\xspace}
\providecommand{\rsun}{R$_\odot$\xspace}
\providecommand{\lsun}{L$_\odot$\xspace}
\providecommand{\mj}{\ensuremath{\,M_{\rm J}}}
\providecommand{\rj}{\ensuremath{\,R_{\rm J}}}
\providecommand{\me}{\ensuremath{\,M_{\rm E}}}
\providecommand{\re}{\ensuremath{\,R_{\rm E}}}
\providecommand{\fave}{\langle F \rangle}
\providecommand{\fluxcgs}{10$^9$ erg s$^{-1}$ cm$^{-2}$}
\providecommand{\tess}{\textit{TESS}\xspace}
\begin{table*}
\centering
\contcaption{}
\scriptsize
\setlength{\tabcolsep}{3pt}
\begin{tabular}{llcccccc}
\toprule
&  & TOI-6208 & TOI-6334 & TOI-6417 & TOI-6443 & TOI-7219 & Source \\
\multicolumn{8}{l}{\textbf{Other identifiers}:} \\
& \tess Input Catalog & TIC 326475995 & TIC 63718617 & TIC 95191643 & TIC 125563127 & TIC 242674266 & \\
& TYCHO-2 & --- & TYC 2318-162-1 & --- & TYC 6543-2384-1 & --- & \\
& 2MASS & J22072113+5331105 & J02152881+3521259 & J07432357-0831234 & J07393705-2527556 & J19354858-0936501 & \\
& Gaia DR3 & 2005283023221079552 & 326490808114021504 & 3041838535020669184 & 5614234537560392320 & 4194211383330904576 & \\
\midrule
\multicolumn{8}{l}{\textbf{Astrometric Parameters}:} \\
$\alpha_{J2000}\ddagger$ & Right Ascension (h:m:s) & 22:07:21.136 & 02:15:28.815 & 07:43:23.579 & 07:39:37.059 & 19:35:48.587  & 1 \\
$\delta_{J2000}\ddagger$ & Declination (d:m:s) & 53:31:10.439 & 35:21:26.124 & -08:31:23.441 & -25:27:55.646 & -09:36:50.112  & 1 \\
$\mu_{\alpha}$ & Gaia DR3 proper motion in RA (mas yr$^{-1}$)& $17.134 \pm 0.011$ & $5.107 \pm 0.023$ & $4.732 \pm 0.011$ & $-5.887 \pm 0.008$ & $3.137 \pm 0.015$  & 1 \\
$\mu_{\delta}$ & Gaia DR3 proper motion in Dec (mas yr$^{-1}$)& $4.437 \pm 0.010$ & $-4.629 \pm 0.023$ & $-8.344 \pm 0.011$ & $2.387 \pm 0.010$ & $-3.847 \pm 0.011$  & 1 \\
$\pi$ & Gaia DR3 Parallax (mas) & $1.4470 \pm 0.0097$ & $1.3079 \pm 0.0200$ & $1.7131 \pm 0.0136$ & $1.5898 \pm 0.0105$ & $3.0363 \pm 0.0132$  & 1 \\
$v\sin{i_\star}$ & Projected rotational velocity (km s$^{-1}$) & $11.6 \pm 0.1$ & $8.01 \pm 0.03$ & $5.89 \pm 0.07$ & $7.83 \pm 0.13$ & $6.59 \pm 0.10$  & 2 \\
\multicolumn{8}{l}{\textbf{Photometric Parameters}:} \\
$G$ & Gaia $G$ mag. & $12.427 \pm 0.020$ & $11.912 \pm 0.020$ & $12.094 \pm 0.020$ & $12.161 \pm 0.020$ & $12.500 \pm 0.020$  & 1 \\
$G_{\rm BP}$ & Gaia $G_{\rm BP}$ mag. & $12.774 \pm 0.020$ & $12.244 \pm 0.020$ & $12.388 \pm 0.020$ & $12.429 \pm 0.020$ & $12.978 \pm 0.020$  & 1 \\
$G_{\rm RP}$ & Gaia $G_{\rm RP}$ mag. & $11.910 \pm 0.020$ & $11.415 \pm 0.020$ & $11.639 \pm 0.020$ & $11.742 \pm 0.020$ & $11.849 \pm 0.020$  & 1 \\
$T$ & TESS mag. & $11.9806 \pm 0.0060$ & $11.4808 \pm 0.0080$ & $11.7024 \pm 0.0070$ & $11.8052 \pm 0.0120$ & $11.9146 \pm 0.0060$  & 3 \\
$J$ & 2MASS $J$ mag. & $11.307 \pm 0.024$ & $10.880 \pm 0.020$ & $11.098 \pm 0.022$ & $11.281 \pm 0.025$ & $11.075 \pm 0.027$  & 4 \\
$H$ & 2MASS $H$ mag. & $11.092 \pm 0.031$ & $10.611 \pm 0.022$ & $10.857 \pm 0.022$ & $11.089 \pm 0.027$ & $10.716 \pm 0.023$  & 4 \\
$K$ & 2MASS $K$ mag. & $10.975 \pm 0.020$ & $10.562 \pm 0.020$ & $10.809 \pm 0.021$ & $10.973 \pm 0.023$ & $10.592 \pm 0.021$  & 4 \\
$W1$ & WISE $W1$ mag. & $10.894 \pm 0.030$ & $10.474 \pm 0.030$ & $10.791 \pm 0.030$ & $10.934 \pm 0.030$ & $10.548 \pm 0.030$  & 5 \\
$W2$ & WISE $W2$ mag. & $10.928 \pm 0.030$ & $10.511 \pm 0.030$ & $10.816 \pm 0.030$ & $10.962 \pm 0.030$ & $10.569 \pm 0.030$  & 5 \\
$W3$ & WISE $W3$ mag. & $11.243 \pm 0.143$ & $10.631 \pm 0.105$ & $10.683 \pm 0.090$ & $10.822 \pm 0.109$ & $10.649 \pm 0.096$  & 5 \\
\bottomrule
\end{tabular}
\end{table*}

\providecommand{\bjdtdb}{\ensuremath{\rm {BJD_{TDB}}}}
\providecommand{\feh}{\ensuremath{\left[{\rm Fe}/{\rm H}\right]}}
\providecommand{\teff}{\ensuremath{T_{\rm eff}}}
\providecommand{\teq}{\ensuremath{T_{\rm eq}}}
\providecommand{\ecosw}{\ensuremath{e\cos{\omega_*}}}
\providecommand{\esinw}{\ensuremath{e\sin{\omega_*}}}
\providecommand\msun{M$_\odot$\xspace}
\providecommand{\rsun}{R$_\odot$\xspace}
\providecommand{\lsun}{L$_\odot$\xspace}
\providecommand{\mj}{\ensuremath{\,M_{\rm J}}}
\providecommand{\rj}{\ensuremath{\,R_{\rm J}}}
\providecommand{\me}{\ensuremath{\,M_{\rm E}}}
\providecommand{\re}{\ensuremath{\,R_{\rm E}}}
\providecommand{\fave}{\langle F \rangle}
\providecommand{\fluxcgs}{10$^9$ erg s$^{-1}$ cm$^{-2}$}
\providecommand{\tess}{\textit{TESS}\xspace}
\begin{table*}
\centering
\contcaption{}
\scriptsize
\setlength{\tabcolsep}{3pt}
\begin{tabular}{llccccc}
\toprule
&  & TOI-7266 & TOI-7404 & TOI-7425 & TOI-7574 & Source \\
\multicolumn{7}{l}{\textbf{Other identifiers}:} \\
& \tess Input Catalog & TIC 125695940 & TIC 352409708 & TIC 285592400 & TIC 159084486 & \\
& TYCHO-2 & TYC 1112-2111-1 & --- & --- & TYC 129-1-1 & \\
& 2MASS & J21043215+1238329 & J01023683+4055376 & J00523174+6420034 & J05534033+0538354 & \\
& Gaia DR3 & 1757473855585097088 & 374416144850764032 & 524207836587213184 & 3320263294677241472 & \\
\midrule
\multicolumn{7}{l}{\textbf{Astrometric Parameters}:} \\
$\alpha_{J2000}\ddagger$ & Right Ascension (h:m:s) & 21:04:32.151 & 01:02:36.838 & 00:52:31.738 & 05:53:40.336  & 1 \\
$\delta_{J2000}\ddagger$ & Declination (d:m:s) & 12:38:32.916 & 40:55:37.705 & 64:20:03.332 & 05:38:35.387  & 1 \\
$\mu_{\alpha}$ & Gaia DR3 proper motion in RA (mas yr$^{-1}$)& $5.011 \pm 0.041$ & $-1.899 \pm 0.015$ & $1.689 \pm 0.008$ & $-12.999 \pm 0.021$  & 1 \\
$\mu_{\delta}$ & Gaia DR3 proper motion in Dec (mas yr$^{-1}$)& $3.077 \pm 0.034$ & $-6.047 \pm 0.010$ & $-4.940 \pm 0.012$ & $-12.163 \pm 0.018$  & 1 \\
$\pi$ & Gaia DR3 Parallax (mas) & $1.6263 \pm 0.0449$ & $0.9826 \pm 0.0192$ & $1.9254 \pm 0.0100$ & $3.7531 \pm 0.0189$  & 1 \\
$v\sin{i_\star}$ & Projected rotational velocity (km s$^{-1}$) & $14.5 \pm 0.3$ & $37.4 \pm 0.4$ & $8.46 \pm 0.06$ & $3.36 \pm 0.09$  & 2 \\
\multicolumn{7}{l}{\textbf{Photometric Parameters}:} \\
$G$ & Gaia $G$ mag. & $12.119 \pm 0.020$ & $12.649 \pm 0.020$ & $12.594 \pm 0.020$ & $12.255 \pm 0.020$  & 1 \\
$G_{\rm BP}$ & Gaia $G_{\rm BP}$ mag. & $12.358 \pm 0.020$ & $12.909 \pm 0.020$ & $13.024 \pm 0.020$ & $12.633 \pm 0.020$  & 1 \\
$G_{\rm RP}$ & Gaia $G_{\rm RP}$ mag. & $11.600 \pm 0.020$ & $12.239 \pm 0.020$ & $11.997 \pm 0.020$ & $11.715 \pm 0.020$  & 1 \\
$T$ & TESS mag. & $11.6790 \pm 0.0060$ & $12.3047 \pm 0.0070$ & $12.0637 \pm 0.0060$ & $11.7771 \pm 0.0060$  & 3 \\
$J$ & 2MASS $J$ mag. & $11.066 \pm 0.021$ & $11.835 \pm 0.031$ & $11.277 \pm 0.024$ & $11.091 \pm 0.021$  & 4 \\
$H$ & 2MASS $H$ mag. & $10.809 \pm 0.023$ & $11.563 \pm 0.032$ & $10.951 \pm 0.028$ & $10.760 \pm 0.024$  & 4 \\
$K$ & 2MASS $K$ mag. & $10.719 \pm 0.022$ & $11.505 \pm 0.020$ & $10.911 \pm 0.023$ & $10.696 \pm 0.025$  & 4 \\
$W1$ & WISE $W1$ mag. & $10.688 \pm 0.030$ & $11.475 \pm 0.030$ & $10.823 \pm 0.030$ & $10.641 \pm 0.030$  & 5 \\
$W2$ & WISE $W2$ mag. & $10.710 \pm 0.030$ & $11.499 \pm 0.030$ & $10.851 \pm 0.030$ & $10.680 \pm 0.030$  & 5 \\
$W3$ & WISE $W3$ mag. & $10.715 \pm 0.114$ & $11.449 \pm 0.142$ & $10.851 \pm 0.121$ & $11.015 \pm 0.130$  & 5 \\
\bottomrule
\end{tabular}
\vspace{2mm}
\begin{minipage}{\textwidth}
\textbf{Notes:}
\footnotesize
The uncertainties of the photometric measurements have a systematic floor applied that is usually larger than the reported catalog errors.\\
$\ddagger$ Right Ascension and Declination are in epoch J2000. Coordinates are from Vizier where Gaia RA and Dec have been precessed and corrected from epoch J2016.\\
Sources: (1) \cite{GaiaDR3}; (2) \S\ref{subsubsec:tres} \& \S\ref{subsubsec:chiron}; (3) \cite{Stassun:2019}; (4) \cite{Cutri:2003, Skrutskie:2006}; (5) \cite{Wright:2010, Cutri:2012}
\end{minipage}
\end{table*}
\renewcommand{\arraystretch}{\bodystretch}

\section{\exofast Global Fits}\label{sec:exofast}

To determine the stellar and planetary parameters and their uncertainties for each system, we used the \texttt{IDL}-based open-source fitting software \exofast\footnote{\url{https://github.com/jdeast/EXOFASTv2}} \citep{Eastman:2013, Eastman:2019}. \exofast is a differential evolution Markov Chain Monte Carlo (MCMC) code that has the functionality to simultaneously fit the SED of the host star and any nearby stars using MESA Isochrones and Stellar Tracks (MIST; \citealt{Paxton:2011}), the radial velocity modulation induced by the planet's orbit, and the transits. We performed a global fit of each of the systems using the same methodology as the first two papers in the MEEP series \citep{Schulte:2024, Schulte:2025} to ensure that the final sample generated by the MEEP survey is self-consistent. 

To properly account for the contamination of nearby stars, we performed multi-component SED fits for all systems that had a nearby star with a separation of less than 6.5 arcsec and a similar brightness to the target star in the \tess bandpass, such that $\Delta T < 6$\,mag. We chose these constraints to select for situations in which the nearby star meaningfully contributes flux to the SED (0.4\% at $\Delta T = 6$\,mag) and is blended in the WISE photometry, which has the largest point-spread functions of all of the ground-based photometry we used \citep{Wright:2010}. These conditions were met for four of our target systems: TOI-3972, TOI-3988, TOI-6171, and TOI-7266. For each of these systems, we included the magnitude difference measured by the high-resolution imaging observations to place an additional constraint on the SED of the nearby stars. These magnitude differences are listed in Table~\ref{tab:secondarylit}. The stellar parameters for each of these stars, TIC 605485663, TOI-3988\,B, TOI-6171\,B, and TOI-7266\,B, are presented in Table~\ref{tab:secondarymedian}.

\providecommand{\bjdtdb}{\ensuremath{\rm {BJD_{TDB}}}}
\providecommand{\feh}{\ensuremath{\left[{\rm Fe}/{\rm H}\right]}}
\providecommand{\teff}{\ensuremath{T_{\rm eff}}}
\providecommand{\teq}{\ensuremath{T_{\rm eq}}}
\providecommand{\ecosw}{\ensuremath{e\cos{\omega_\star}}}
\providecommand{\esinw}{\ensuremath{e\sin{\omega_\star}}}
\providecommand{\msun}{\ensuremath{\,M_\Sun}}
\providecommand{\rsun}{\ensuremath{\,R_\Sun}}
\providecommand{\lsun}{\ensuremath{\,L_\Sun}}
\providecommand{\mj}{\ensuremath{\,M_{\rm J}}}
\providecommand{\rj}{\ensuremath{\,R_{\rm J}}}
\providecommand{\me}{\ensuremath{\,M_{\rm E}}}
\providecommand{\re}{\ensuremath{\,R_{\rm E}}}
\providecommand{\fave}{\langle F \rangle}
\providecommand{\fluxcgs}{10$^9$ erg s$^{-1}$ cm$^{-2}$}
\providecommand{\tess}{\textit{TESS}\xspace}
\begin{table*}
\centering
\caption{Median Values and 68\% Confidence Intervals for the Fitted Stellar Parameters of Secondary Stars}
\label{tab:secondarymedian}
\scriptsize
\setlength{\tabcolsep}{3.5pt}
\begin{tabular}{llcccc}
\toprule
&  & TIC 605485663 & TOI-3988 B & TOI-6171 B & TOI-7266 B\\
& Classification & Background Star & Bound Companion & Bound Companion & Bound Companion\\
& Planet Host & TOI-3972 & TOI-3988 A & TOI-6171 A & TOI-7266 A\\
\midrule
\multicolumn{6}{l}{\textbf{Stellar Parameters}:} \\
$M_\star$ & Mass (\msun) & $0.94^{+0.25}_{-0.13}$ & $0.564^{+0.040}_{-0.045}$ & $1.108^{+0.086}_{-0.056}$ & $0.957^{+0.110}_{-0.097}$ \\
$R_\star$ & Radius (\rsun) & $1.96^{+0.58}_{-0.38}$ & $0.527^{+0.038}_{-0.043}$ & $1.387^{+0.054}_{-0.051}$ & $0.918^{+0.130}_{-0.093}$ \\
$L_\star$ & Luminosity (\lsun) & $4.8^{+3.4}_{-1.9}$ & $0.0569^{+0.0093}_{-0.0094}$ & $2.00^{+0.18}_{-0.17}$ & $0.67^{+0.43}_{-0.25}$ \\
$\rho_\star$ & Density (cgs) & $0.18^{+0.18}_{-0.10}$ & $5.44^{+1.10}_{-0.80}$ & $0.589^{+0.083}_{-0.073}$ & $1.74^{+0.44}_{-0.43}$ \\
$\log{g}$ & Surface gravity (cgs) & $3.84^{+0.21}_{-0.24}$ & $4.746^{+0.044}_{-0.039}$ & $4.201^{+0.044}_{-0.043}$ & $4.492^{+0.053}_{-0.068}$ \\
$T_{\rm eff}$ & Effective temperature (K) & $5990^{+1000}_{-830}$ & $3880^{+120}_{-110}$ & $5830 \pm 160$ & $5450^{+370}_{-350}$ \\
$[{\rm Fe/H}]$ & Metallicity (dex) & $-1.0^{+1.1}_{-2.3}$ & $0.13^{+0.13}_{-0.15}$ & $0.21^{+0.11}_{-0.12}$ & $0.13^{+0.14}_{-0.13}$ \\
$[{\rm Fe/H}]_{0}^{*}$ & Initial metallicity (dex) & $-0.86^{+0.97}_{-2.00}$ & $0.10 \pm 0.11$ & $0.239^{+0.091}_{-0.093}$ & $0.12^{+0.12}_{-0.11}$ \\
Age$^{*}$ & Age (Gyr) & $9.0^{+3.3}_{-4.6}$ & $2.03^{+1.60}_{-0.70}$ & $6.8^{+1.8}_{-2.5}$ & $3.5^{+2.6}_{-1.4}$ \\
EEP & Equivalent evolutionary phase & $461^{+14}_{-21}$ & $272^{+12}_{-11}$ & $428^{+12}_{-30}$ & $339^{+10}_{-12}$ \\
$A_V^{*}$ & V-band extinction (mag) & $1.54^{+0.74}_{-0.64}$ & $1.00 \pm 0.25$ & $0.183^{+0.048}_{-0.076}$ & $0.21^{+0.13}_{-0.14}$ \\
$d^{*}$ & Distance (pc) & $2960^{+680}_{-450}$ & $770.7^{+9.5}_{-9.3}$ & $485.5^{+4.9}_{-4.8}$ & $660^{+120}_{-89}$ \\
\bottomrule
\end{tabular}
\begin{flushleft}
\textbf{Note:} *When the secondary star is a bound companion, the initial metallicity, age, \textit{V}-band extinction, and distance are fixed to those of the primary star.
\end{flushleft}
\end{table*}

One drawback of MCMC is that the computational cost and runtime of the fits varies widely depending on the number of parameters being fit and the degeneracies between those parameters. As in the previous MEEP papers, we adopted a strict set of convergence criteria: A fit is considered to have converged if the Gelman-Rubin statistic \citep{Ford:2006} is smaller than 1.01 (indicating that most chains are in agreement), and the number of independent draws is larger than 1000 (indicating that the chains are well-mixed) \citep{Ford:2006, Eastman:2019}. However, some of the parameters in one of our fits were not able to meet these strict criteria. TOI-3972 has a fainter ($\Delta T = 5.48$\,mag) unbound background star which we chose to include to properly account for its influence on the SED of the target system. While all of TOI-3972's parameters met our convergence criteria, the background star's \teff reached a Gelman-Rubin statistic of 1.12 and had 808 independent draws before the fit reached its 6.5-day runtime limit. This reflects the relatively poor constraints on the background star's parameters due to our limited photometric data. However, as the chains and posteriors for all parameters were well-behaved and met the less conservative convergence criteria offered by \cite{brooks_general_1998}, we chose to adopt this fit. The results of our \exofast fits are presented in Table~\ref{tab:median}.

\renewcommand{\arraystretch}{\medianstretch}
\providecommand{\bjdtdb}{\ensuremath{\rm {BJD_{TDB}}}}
\providecommand{\feh}{\ensuremath{\left[{\rm Fe}/{\rm H}\right]}}
\providecommand{\teff}{\ensuremath{T_{\rm eff}}}
\providecommand{\teq}{\ensuremath{T_{\rm eq}}}
\providecommand{\ecosw}{\ensuremath{e\cos{\omega_\star}}}
\providecommand{\esinw}{\ensuremath{e\sin{\omega_\star}}}
\providecommand{\msun}{\ensuremath{\,M_\Sun}}
\providecommand{\rsun}{\ensuremath{\,R_\Sun}}
\providecommand{\lsun}{\ensuremath{\,L_\Sun}}
\providecommand{\mj}{\ensuremath{\,M_{\rm J}}}
\providecommand{\rj}{\ensuremath{\,R_{\rm J}}}
\providecommand{\me}{\ensuremath{\,M_{\rm E}}}
\providecommand{\re}{\ensuremath{\,R_{\rm E}}}
\providecommand{\fave}{\langle F \rangle}
\providecommand{\fluxcgs}{10$^9$ erg s$^{-1}$ cm$^{-2}$}
\providecommand{\tess}{\textit{TESS}\xspace}
\begin{table*}
\centering
\caption{Median Values and 68\% Confidence Intervals for Fitted Stellar and Planetary Parameters}
\label{tab:median}
\scriptsize
\setlength{\tabcolsep}{3.5pt}
\begin{tabular}{llccccc}
\toprule
&  & TOI-3041 & TOI-3365 & TOI-3601 & TOI-3788 & TOI-3972\\
\midrule
\multicolumn{7}{l}{\textbf{Priors}:} \\
$\pi$ & Gaia Parallax (mas)& $\mathcal{G}$[2.1072, 0.01472] & $\mathcal{G}$[3.729, 0.01853] & $\mathcal{G}$[1.3968, 0.01887] & $\mathcal{G}$[1.9462, 0.02767] & $\mathcal{G}$[6.1319, 0.01704] \\
$[{\rm Fe/H}]$ & Metallicity (dex)& $\mathcal{G}$[-0.0087268, 0.17593] & $\mathcal{G}$[0.031413, 0.045772] & $\mathcal{G}$[0.2385, 0.10733] & $\mathcal{G}$[0.091131, 0.13002] & $\mathcal{G}$[0.28009, 0.028306] \\
$A_V$ & V-band extinction (mag)& $\mathcal{U}$[0, 0.69097] & $\mathcal{U}$[0, 1.6348] & $\mathcal{U}$[0, 0.27671] & $\mathcal{U}$[0, 0.51904] & $\mathcal{U}$[0, 3.4193] \\
$D_T$ & Dilution in \tess& $\mathcal{G}$[0, 0.016388] & $\mathcal{G}$[0, 0.019179] & $\mathcal{G}$[0, 0.0013418] & $\mathcal{G}$[0, 0.003419] & $\mathcal{G}$[0, 0.01603] \\
\hline
\multicolumn{7}{l}{\textbf{Stellar Parameters}:} \\
$M_\star$ & Mass (\msun) & $1.414^{+0.077}_{-0.078}$ & $1.042^{+0.070}_{-0.061}$ & $1.315^{+0.069}_{-0.075}$ & $1.431^{+0.085}_{-0.087}$ & $0.939^{+0.046}_{-0.039}$ \\
$R_\star$ & Radius (\rsun) & $1.729^{+0.056}_{-0.050}$ & $1.313^{+0.035}_{-0.034}$ & $1.516^{+0.061}_{-0.057}$ & $1.753^{+0.070}_{-0.063}$ & $0.940^{+0.021}_{-0.019}$ \\
$L_\star$ & Luminosity (\lsun) & $4.48^{+0.50}_{-0.39}$ & $1.88^{+0.19}_{-0.16}$ & $3.15^{+0.23}_{-0.24}$ & $4.77^{+0.61}_{-0.50}$ & $0.643^{+0.049}_{-0.037}$ \\
$\rho_\star$ & Density (cgs) & $0.386^{+0.040}_{-0.041}$ & $0.650^{+0.060}_{-0.054}$ & $0.531^{+0.075}_{-0.070}$ & $0.375^{+0.051}_{-0.050}$ & $1.601^{+0.090}_{-0.096}$ \\
$\log{g}$ & Surface gravity (cgs) & $4.113^{+0.033}_{-0.037}$ & $4.220^{+0.031}_{-0.030}$ & $4.195^{+0.042}_{-0.047}$ & $4.106^{+0.042}_{-0.047}$ & $4.466^{+0.019}_{-0.020}$ \\
$T_{\rm eff}$ & Effective temperature (K) & $6390^{+190}_{-160}$ & $5900^{+140}_{-130}$ & $6240^{+150}_{-160}$ & $6440^{+230}_{-210}$ & $5334^{+85}_{-77}$ \\
$[{\rm Fe/H}]$ & Metallicity (dex) & $0.094^{+0.120}_{-0.097}$ & $0.043^{+0.044}_{-0.040}$ & $0.246^{+0.094}_{-0.099}$ & $0.081^{+0.100}_{-0.090}$ & $0.284 \pm 0.028$ \\
$[{\rm Fe/H}]_{0}$ & Initial metallicity (dex) & $0.211^{+0.100}_{-0.096}$ & $0.101 \pm 0.045$ & $0.298^{+0.082}_{-0.085}$ & $0.202^{+0.097}_{-0.090}$ & $0.281^{+0.039}_{-0.040}$ \\
Age & Age (Gyr) & $2.26^{+0.91}_{-0.73}$ & $7.8^{+2.8}_{-2.5}$ & $2.7^{+1.5}_{-1.2}$ & $2.15^{+0.99}_{-0.78}$ & $7.6^{+3.4}_{-3.2}$ \\
EEP & Equivalent evolutionary phase & $364^{+28}_{-18}$ & $427^{+11}_{-18}$ & $363^{+41}_{-23}$ & $363^{+32}_{-18}$ & $367^{+24}_{-23}$ \\
$A_V$ & V-band extinction (mag) & $0.31^{+0.12}_{-0.10}$ & $0.23 \pm 0.11$ & $0.170^{+0.073}_{-0.094}$ & $0.25^{+0.13}_{-0.12}$ & $0.084^{+0.110}_{-0.061}$ \\
$d$ & Distance (pc) & $474.3 \pm 3.3$ & $268.2 \pm 1.3$ & $716.3^{+9.9}_{-9.5}$ & $513.2^{+7.3}_{-7.1}$ & $163.06 \pm 0.45$ \\
\multicolumn{7}{l}{\textbf{Planetary Parameters}:} \\
$P$ & Period (days) & $2.95995978^{+0.00000073}_{-0.00000074}$ & $5.3928630 \pm 0.0000016$ & $4.0517742^{+0.0000052}_{-0.0000053}$ & $3.0923706 \pm 0.0000016$ & $10.5113703 \pm 0.0000025$ \\
$R_{\rm P}$ & Radius (\rj) & $1.373^{+0.047}_{-0.043}$ & $1.083^{+0.034}_{-0.033}$ & $1.118^{+0.055}_{-0.053}$ & $1.380^{+0.059}_{-0.055}$ & $1.046^{+0.026}_{-0.024}$ \\
$M_{\rm P}$ & Mass (\mj) & $1.57 \pm 0.11$ & $5.20^{+0.25}_{-0.22}$ & $2.91 \pm 0.26$ & $1.26 \pm 0.17$ & $2.945^{+0.110}_{-0.099}$ \\
$T_C$ & Time of conjunction (\bjdtdb) & $2459359.13587 \pm 0.00022$ & $2459992.69333 \pm 0.00022$ & $2459879.37084^{+0.00071}_{-0.00072}$ & $2458863.84393 \pm 0.00072$ & $2459892.72756 \pm 0.00011$ \\
$T_0$ & Optimal conjunction time (\bjdtdb) & $2459702.49148 \pm 0.00017$ & $2460052.01553^{+0.00021}_{-0.00022}$ & $2459944.19941^{+0.00067}_{-0.00068}$ & $2459837.94087^{+0.00033}_{-0.00032}$ & $2460123.978108 \pm 0.000090$ \\
$a$ & Semi-major axis (AU) & $0.04530^{+0.00081}_{-0.00085}$ & $0.0611^{+0.0013}_{-0.0012}$ & $0.05454^{+0.00094}_{-0.00110}$ & $0.04682^{+0.00091}_{-0.00097}$ & $0.0921^{+0.0015}_{-0.0013}$ \\
$i$ & Inclination (Degrees) & $84.47^{+0.51}_{-0.54}$ & $86.56 \pm 0.22$ & $84.74^{+0.50}_{-0.54}$ & $82.38^{+0.61}_{-0.79}$ & $89.32^{+0.38}_{-0.28}$ \\
$e$ & Eccentricity & $0.033^{+0.039}_{-0.023}$ & $0.120^{+0.016}_{-0.015}$ & $0.058^{+0.060}_{-0.040}$ & $0.052^{+0.058}_{-0.036}$ & $0.278^{+0.015}_{-0.012}$ \\
$\omega_\star$ & Argument of periastron (Degrees) & $97^{+78}_{-85}$ & $-132.9^{+6.4}_{-8.2}$ & $56^{+55}_{-78}$ & $78^{+85}_{-95}$ & $111.5^{+2.2}_{-2.8}$ \\
$\teq$ & Equilibrium temperature (K) & $1903^{+46}_{-38}$ & $1318^{+28}_{-26}$ & $1587^{+26}_{-27}$ & $1902^{+48}_{-44}$ & $821^{+14}_{-12}$ \\
$\tau_{\rm circ}$ & Tidal circularization timescale (Gyr) & $0.133^{+0.026}_{-0.025}$ & $13.7^{+2.2}_{-1.9}$ & $2.42^{+0.79}_{-0.63}$ & $0.122^{+0.039}_{-0.035}$ & $75 \pm 13$ \\
$K$ & RV semi-amplitude (m/s) & $175^{+12}_{-11}$ & $587.5^{+9.9}_{-9.6}$ & $309 \pm 25$ & $137 \pm 18$ & $295.5^{+5.8}_{-4.8}$ \\
$\dot{\gamma}$ & RV slope (m/s/day) & ---& $-0.384 \pm 0.019$ & $-0.173^{+0.065}_{-0.066}$ & $0.156^{+0.072}_{-0.071}$ & ---\\
$R_{\rm P}/R_\star$ & Radius of planet in stellar radii  & $0.08159^{+0.00073}_{-0.00071}$ & $0.08475^{+0.00086}_{-0.00087}$ & $0.0758 \pm 0.0014$ & $0.08086 \pm 0.00082$ & $0.11433^{+0.00061}_{-0.00051}$ \\
$a/R_\star$ & Semi-major axis in stellar radii  & $5.63^{+0.19}_{-0.21}$ & $10.01^{+0.30}_{-0.28}$ & $7.73^{+0.35}_{-0.36}$ & $5.74^{+0.25}_{-0.27}$ & $21.09^{+0.39}_{-0.43}$ \\
Depth & \tess flux decrement at mid-transit & $0.00718 \pm 0.00012$ & $0.00766 \pm 0.00014$ & $0.00598 \pm 0.00016$ & $0.006622^{+0.000084}_{-0.000085}$ & $0.01613^{+0.00019}_{-0.00018}$ \\
$\tau$ & Ingress/egress transit duration (days) & $0.01608^{+0.00093}_{-0.00092}$ & $0.0210 \pm 0.0012$ & $0.0171^{+0.0025}_{-0.0024}$ & $0.0208^{+0.0020}_{-0.0019}$ & $0.01407^{+0.00055}_{-0.00042}$ \\
$T_{14}$ & Total transit duration (days) & $0.15591^{+0.00091}_{-0.00090}$ & $0.1626 \pm 0.0011$ & $0.1343^{+0.0027}_{-0.0026}$ & $0.1327^{+0.0018}_{-0.0017}$ & $0.13294^{+0.00049}_{-0.00042}$ \\
$b$ & Transit impact parameter & $0.531^{+0.034}_{-0.039}$ & $0.650^{+0.021}_{-0.025}$ & $0.688^{+0.043}_{-0.058}$ & $0.745^{+0.023}_{-0.027}$ & $0.182^{+0.073}_{-0.100}$ \\
$\rho_{\rm P}$ & Density (cgs) & $0.750^{+0.087}_{-0.083}$ & $5.08^{+0.51}_{-0.46}$ & $2.58^{+0.48}_{-0.43}$ & $0.59^{+0.12}_{-0.11}$ & $3.20^{+0.21}_{-0.22}$ \\
$\log{g_{\rm P}}$ & Surface gravity (cgs) & $3.314^{+0.038}_{-0.040}$ & $4.042 \pm 0.030$ & $3.761^{+0.057}_{-0.062}$ & $3.214^{+0.068}_{-0.074}$ & $3.825^{+0.020}_{-0.022}$ \\
$M_{\rm P}/M_\star$ & Mass ratio  & $0.001061^{+0.000078}_{-0.000073}$ & $0.00476 \pm 0.00013$ & $0.00212^{+0.00018}_{-0.00017}$ & $0.00084 \pm 0.00011$ & $0.002987^{+0.000072}_{-0.000069}$ \\
$d/R_\star$ & Separation at mid-transit  & $5.55^{+0.29}_{-0.43}$ & $10.82^{+0.47}_{-0.45}$ & $7.48^{+0.54}_{-0.68}$ & $5.64^{+0.42}_{-0.58}$ & $15.47^{+0.51}_{-0.59}$ \\
\bottomrule
\end{tabular}
\end{table*}
\providecommand{\bjdtdb}{\ensuremath{\rm {BJD_{TDB}}}}
\providecommand{\feh}{\ensuremath{\left[{\rm Fe}/{\rm H}\right]}}
\providecommand{\teff}{\ensuremath{T_{\rm eff}}}
\providecommand{\teq}{\ensuremath{T_{\rm eq}}}
\providecommand{\ecosw}{\ensuremath{e\cos{\omega_\star}}}
\providecommand{\esinw}{\ensuremath{e\sin{\omega_\star}}}
\providecommand{\msun}{\ensuremath{\,M_\Sun}}
\providecommand{\rsun}{\ensuremath{\,R_\Sun}}
\providecommand{\lsun}{\ensuremath{\,L_\Sun}}
\providecommand{\mj}{\ensuremath{\,M_{\rm J}}}
\providecommand{\rj}{\ensuremath{\,R_{\rm J}}}
\providecommand{\me}{\ensuremath{\,M_{\rm E}}}
\providecommand{\re}{\ensuremath{\,R_{\rm E}}}
\providecommand{\fave}{\langle F \rangle}
\providecommand{\fluxcgs}{10$^9$ erg s$^{-1}$ cm$^{-2}$}
\providecommand{\tess}{\textit{TESS}\xspace}
\begin{table*}
\centering
\contcaption{}
\scriptsize
\setlength{\tabcolsep}{3.5pt}
\begin{tabular}{llccccc}
\toprule
&  & TOI-3988 A$^{*}$ & TOI-3998 A & TOI-4009$^{*}$ & TOI-4079 & TOI-4088\\
\midrule
\multicolumn{7}{l}{\textbf{Priors}:} \\
$\pi$ & Gaia Parallax (mas)& $\mathcal{G}$[1.2974, 0.01577] & $\mathcal{G}$[2.4617, 0.01479] & $\mathcal{G}$[1.5491, 0.01632] & $\mathcal{G}$[2.7679, 0.01487] & $\mathcal{G}$[2.2859, 0.01554] \\
$[{\rm Fe/H}]$ & Metallicity (dex)& $\mathcal{G}$[-0.12645, 0.14352] & $\mathcal{G}$[0.060407, 0.045522] & $\mathcal{G}$[0.42639, 0.030641] & $\mathcal{G}$[0.15198, 0.069865] & $\mathcal{G}$[0.20851, 0.081489] \\
$A_V$ & V-band extinction (mag)& $\mathcal{U}$[0, 3.2255] & $\mathcal{U}$[0, 1.3378] & $\mathcal{U}$[0, 1.9224] & $\mathcal{U}$[0, 5.2863] & $\mathcal{U}$[0, 0.3997] \\
$D_T$ & Dilution in \tess& $\mathcal{G}$[0, 0.018844] & $\mathcal{G}$[0, 0.058236] & $\mathcal{G}$[0, 0.022828] & $\mathcal{G}$[0, 0.063041] & $\mathcal{G}$[0, 0.0030068] \\
\hline
\multicolumn{7}{l}{\textbf{Stellar Parameters}:} \\
$M_\star$ & Mass (\msun) & $1.53^{+0.12}_{-0.20}$ & $1.146^{+0.069}_{-0.085}$ & $1.408^{+0.077}_{-0.080}$ & $1.116^{+0.054}_{-0.063}$ & $1.032^{+0.066}_{-0.059}$ \\
$R_\star$ & Radius (\rsun) & $2.124^{+0.089}_{-0.090}$ & $1.198^{+0.045}_{-0.040}$ & $1.777^{+0.064}_{-0.063}$ & $1.093^{+0.039}_{-0.036}$ & $1.169^{+0.037}_{-0.035}$ \\
$L_\star$ & Luminosity (\lsun) & $7.5^{+2.0}_{-1.5}$ & $1.95^{+0.30}_{-0.27}$ & $3.83^{+0.60}_{-0.52}$ & $1.61^{+0.24}_{-0.21}$ & $1.33 \pm 0.11$ \\
$\rho_\star$ & Density (cgs) & $0.222^{+0.045}_{-0.037}$ & $0.94^{+0.11}_{-0.12}$ & $0.351^{+0.047}_{-0.040}$ & $1.20 \pm 0.12$ & $0.912^{+0.100}_{-0.095}$ \\
$\log{g}$ & Surface gravity (cgs) & $3.964^{+0.062}_{-0.068}$ & $4.341^{+0.036}_{-0.048}$ & $4.085 \pm 0.040$ & $4.409^{+0.028}_{-0.035}$ & $4.316^{+0.036}_{-0.037}$ \\
$T_{\rm eff}$ & Effective temperature (K) & $6550^{+460}_{-370}$ & $6240^{+220}_{-240}$ & $6050^{+250}_{-230}$ & $6210 \pm 190$ & $5730^{+130}_{-140}$ \\
$[{\rm Fe/H}]$ & Metallicity (dex) & $0.01^{+0.11}_{-0.12}$ & $0.049^{+0.045}_{-0.046}$ & $0.423 \pm 0.029$ & $0.113^{+0.068}_{-0.067}$ & $0.188^{+0.078}_{-0.077}$ \\
$[{\rm Fe/H}]_{0}$ & Initial metallicity (dex) & $0.10 \pm 0.11$ & $0.081^{+0.049}_{-0.051}$ & $0.441^{+0.038}_{-0.044}$ & $0.108^{+0.066}_{-0.067}$ & $0.213^{+0.072}_{-0.071}$ \\
Age & Age (Gyr) & $2.03^{+1.60}_{-0.70}$ & $2.7^{+3.0}_{-1.6}$ & $2.9^{+1.3}_{-1.0}$ & $2.1^{+2.4}_{-1.3}$ & $7.5^{+3.3}_{-3.0}$ \\
EEP & Equivalent evolutionary phase & $386^{+61}_{-28}$ & $351^{+52}_{-26}$ & $385^{+26}_{-33}$ & $339^{+31}_{-27}$ & $408^{+18}_{-31}$ \\
$A_V$ & V-band extinction (mag) & $1.00 \pm 0.25$ & $0.74^{+0.17}_{-0.19}$ & $0.54^{+0.18}_{-0.19}$ & $0.90 \pm 0.17$ & $0.24^{+0.10}_{-0.12}$ \\
$d$ & Distance (pc) & $770.7^{+9.5}_{-9.3}$ & $406.1 \pm 2.4$ & $645.3^{+6.8}_{-6.7}$ & $361.3^{+2.0}_{-1.9}$ & $437.3^{+3.0}_{-2.9}$ \\
\multicolumn{7}{l}{\textbf{Planetary Parameters}:} \\
$P$ & Period (days) & $2.8744224 \pm 0.0000027$ & $1.81450944 \pm 0.00000049$ & $4.3271526 \pm 0.0000038$ & $1.98722964^{+0.00000053}_{-0.00000054}$ & $2.72844245 \pm 0.00000097$ \\
$R_{\rm P}$ & Radius (\rj) & $1.522^{+0.071}_{-0.070}$ & $1.287^{+0.049}_{-0.044}$ & $1.422^{+0.065}_{-0.062}$ & $1.338^{+0.049}_{-0.046}$ & $1.144^{+0.039}_{-0.038}$ \\
$M_{\rm P}$ & Mass (\mj) & $1.15 \pm 0.15$ & $1.05 \pm 0.12$ & $0.72 \pm 0.11$ & $5.18^{+0.40}_{-0.39}$ & $0.798^{+0.073}_{-0.071}$ \\
$T_C$ & Time of conjunction (\bjdtdb) & $2459904.65119^{+0.00064}_{-0.00066}$ & $2459880.38562^{+0.00027}_{-0.00026}$ & $2459882.41240^{+0.00072}_{-0.00078}$ & $2459882.62774 \pm 0.00015$ & $2459912.55438^{+0.00032}_{-0.00028}$ \\
$T_0$ & Optimal conjunction time (\bjdtdb) & $2459893.15388^{+0.00056}_{-0.00057}$ & $2459669.90252 \pm 0.00012$ & $2460046.84469 \pm 0.00051$ & $2459904.48745 \pm 0.00015$ & $2459735.20568 \pm 0.00019$ \\
$a$ & Semi-major axis (AU) & $0.0456^{+0.0012}_{-0.0020}$ & $0.03047^{+0.00060}_{-0.00077}$ & $0.0583^{+0.0010}_{-0.0011}$ & $0.03214^{+0.00051}_{-0.00062}$ & $0.03863^{+0.00081}_{-0.00075}$ \\
$i$ & Inclination (Degrees) & $83.6^{+1.2}_{-1.1}$ & $81.61^{+0.63}_{-0.87}$ & $82.71^{+0.54}_{-0.56}$ & $88.63^{+0.93}_{-0.97}$ & $84.27^{+0.40}_{-0.44}$ \\
$e$ & Eccentricity & $0.073^{+0.079}_{-0.051}$ & $0.082^{+0.051}_{-0.042}$ & $0.047^{+0.052}_{-0.033}$ & $0.063^{+0.042}_{-0.038}$ & $0.057^{+0.055}_{-0.040}$ \\
$\omega_\star$ & Argument of periastron (Degrees) & $-105^{+48}_{-86}$ & $58^{+38}_{-36}$ & $-158^{+95}_{-100}$ & $115^{+41}_{-51}$ & $10^{+65}_{-62}$ \\
$\teq$ & Equilibrium temperature (K) & $2164^{+100}_{-90}$ & $1886^{+58}_{-57}$ & $1614^{+49}_{-47}$ & $1749^{+55}_{-52}$ & $1520^{+28}_{-30}$ \\
$\tau_{\rm circ}$ & Tidal circularization timescale (Gyr) & $0.048^{+0.017}_{-0.013}$ & $0.0120^{+0.0035}_{-0.0036}$ & $0.257^{+0.084}_{-0.069}$ & $0.074 \pm 0.017$ & $0.092^{+0.021}_{-0.019}$ \\
$K$ & RV semi-amplitude (m/s) & $125^{+15}_{-14}$ & $159 \pm 17$ & $71 \pm 10$ & $779 \pm 52$ & $113.1^{+9.5}_{-9.2}$ \\
$\dot{\gamma}$ & RV slope (m/s/day) & $-0.100^{+0.055}_{-0.054}$ & $-0.060^{+0.047}_{-0.049}$ & $-0.064 \pm 0.051$ & ---& ---\\
$R_{\rm P}/R_\star$ & Radius of planet in stellar radii  & $0.0737 \pm 0.0010$ & $0.11035 \pm 0.00069$ & $0.0822^{+0.0016}_{-0.0015}$ & $0.12574^{+0.00094}_{-0.00092}$ & $0.10055^{+0.00080}_{-0.00082}$ \\
$a/R_\star$ & Semi-major axis in stellar radii  & $4.59^{+0.29}_{-0.27}$ & $5.47^{+0.21}_{-0.24}$ & $7.03^{+0.30}_{-0.28}$ & $6.32^{+0.20}_{-0.22}$ & $7.11 \pm 0.26$ \\
Depth & \tess flux decrement at mid-transit & $0.00581 \pm 0.00012$ & $0.01228^{+0.00012}_{-0.00011}$ & $0.00588^{+0.00019}_{-0.00018}$ & $0.01801^{+0.00032}_{-0.00031}$ & $0.01056 \pm 0.00011$ \\
$\tau$ & Ingress/egress transit duration (days) & $0.0189^{+0.0029}_{-0.0026}$ & $0.01719^{+0.00077}_{-0.00076}$ & $0.0408^{+0.0050}_{-0.0039}$ & $0.01225^{+0.00058}_{-0.00025}$ & $0.0175^{+0.0012}_{-0.0011}$ \\
$T_{14}$ & Total transit duration (days) & $0.1964 \pm 0.0026$ & $0.08264^{+0.00065}_{-0.00064}$ & $0.1205^{+0.0025}_{-0.0024}$ & $0.10698^{+0.00058}_{-0.00051}$ & $0.1033 \pm 0.0010$ \\
$b$ & Transit impact parameter & $0.546^{+0.075}_{-0.110}$ & $0.749^{+0.011}_{-0.012}$ & $0.8966^{+0.0078}_{-0.0081}$ & $0.143^{+0.110}_{-0.097}$ & $0.703^{+0.020}_{-0.023}$ \\
$\rho_{\rm P}$ & Density (cgs) & $0.404^{+0.090}_{-0.075}$ & $0.613 \pm 0.097$ & $0.311^{+0.071}_{-0.061}$ & $2.68^{+0.34}_{-0.32}$ & $0.659^{+0.092}_{-0.083}$ \\
$\log{g_{\rm P}}$ & Surface gravity (cgs) & $3.090^{+0.072}_{-0.076}$ & $3.198^{+0.054}_{-0.065}$ & $2.947^{+0.076}_{-0.083}$ & $3.855^{+0.040}_{-0.044}$ & $3.178^{+0.047}_{-0.049}$ \\
$M_{\rm P}/M_\star$ & Mass ratio  & $0.000730^{+0.000091}_{-0.000086}$ & $0.000879^{+0.000097}_{-0.000094}$ & $0.000494^{+0.000072}_{-0.000071}$ & $0.00445 \pm 0.00031$ & $0.000736^{+0.000064}_{-0.000061}$ \\
$d/R_\star$ & Separation at mid-transit  & $4.76^{+0.68}_{-0.47}$ & $5.14^{+0.44}_{-0.50}$ & $7.06^{+0.57}_{-0.49}$ & $6.03^{+0.41}_{-0.45}$ & $7.04^{+0.45}_{-0.52}$ \\
\bottomrule
\end{tabular}
\end{table*}
\providecommand{\bjdtdb}{\ensuremath{\rm {BJD_{TDB}}}}
\providecommand{\feh}{\ensuremath{\left[{\rm Fe}/{\rm H}\right]}}
\providecommand{\teff}{\ensuremath{T_{\rm eff}}}
\providecommand{\teq}{\ensuremath{T_{\rm eq}}}
\providecommand{\ecosw}{\ensuremath{e\cos{\omega_\star}}}
\providecommand{\esinw}{\ensuremath{e\sin{\omega_\star}}}
\providecommand{\msun}{\ensuremath{\,M_\Sun}}
\providecommand{\rsun}{\ensuremath{\,R_\Sun}}
\providecommand{\lsun}{\ensuremath{\,L_\Sun}}
\providecommand{\mj}{\ensuremath{\,M_{\rm J}}}
\providecommand{\rj}{\ensuremath{\,R_{\rm J}}}
\providecommand{\me}{\ensuremath{\,M_{\rm E}}}
\providecommand{\re}{\ensuremath{\,R_{\rm E}}}
\providecommand{\fave}{\langle F \rangle}
\providecommand{\fluxcgs}{10$^9$ erg s$^{-1}$ cm$^{-2}$}
\providecommand{\tess}{\textit{TESS}\xspace}
\begin{table*}
\centering
\contcaption{}
\scriptsize
\setlength{\tabcolsep}{3.5pt}
\begin{tabular}{llccccc}
\toprule
&  & TOI-4140 & TOI-4144 & TOI-5236$^{*}$ & TOI-5479$^{*}$ & TOI-5925\\
\midrule
\multicolumn{7}{l}{\textbf{Priors}:} \\
$\pi$ & Gaia Parallax (mas)& $\mathcal{G}$[3.2795, 0.01532] & $\mathcal{G}$[3.852, 0.01393] & $\mathcal{G}$[1.7673, 0.01562] & $\mathcal{G}$[1.6643, 0.02165] & $\mathcal{G}$[1.9353, 0.02016] \\
$[{\rm Fe/H}]$ & Metallicity (dex)& $\mathcal{G}$[0.18458, 0.044439] & $\mathcal{G}$[0.32924, 0.050307] & $\mathcal{G}$[0.38605, 0.069701] & $\mathcal{G}$[0.21657, 0.079744] & $\mathcal{G}$[-0.014113, 0.10144] \\
$A_V$ & V-band extinction (mag)& $\mathcal{U}$[0, 0.3997] & $\mathcal{U}$[0, 0.65375] & $\mathcal{U}$[0, 7.2625] & $\mathcal{U}$[0, 0.11004] & $\mathcal{U}$[0, 0.50731] \\
$D_T$ & Dilution in \tess& $\mathcal{G}$[0, 0.00067541] & $\mathcal{G}$[0, 0.0036516] & $\mathcal{G}$[0, 0.038616] & --- & $\mathcal{G}$[0, 0.0031289] \\
\hline
\multicolumn{7}{l}{\textbf{Stellar Parameters}:} \\
$M_\star$ & Mass (\msun) & $0.934 \pm 0.043$ & $0.940^{+0.045}_{-0.042}$ & $1.59^{+0.14}_{-0.18}$ & $1.126^{+0.090}_{-0.072}$ & $1.153^{+0.064}_{-0.073}$ \\
$R_\star$ & Radius (\rsun) & $0.896^{+0.028}_{-0.024}$ & $0.906^{+0.027}_{-0.026}$ & $2.47^{+0.14}_{-0.13}$ & $1.457^{+0.055}_{-0.053}$ & $1.202^{+0.042}_{-0.039}$ \\
$L_\star$ & Luminosity (\lsun) & $0.680^{+0.057}_{-0.044}$ & $0.644^{+0.049}_{-0.047}$ & $7.1^{+2.0}_{-1.5}$ & $2.213^{+0.100}_{-0.097}$ & $2.03^{+0.29}_{-0.22}$ \\
$\rho_\star$ & Density (cgs) & $1.84^{+0.17}_{-0.18}$ & $1.78 \pm 0.17$ & $0.146^{+0.030}_{-0.025}$ & $0.516^{+0.074}_{-0.066}$ & $0.94^{+0.10}_{-0.11}$ \\
$\log{g}$ & Surface gravity (cgs) & $4.505^{+0.028}_{-0.034}$ & $4.497^{+0.030}_{-0.032}$ & $3.848^{+0.060}_{-0.065}$ & $4.165^{+0.044}_{-0.048}$ & $4.341^{+0.034}_{-0.041}$ \\
$T_{\rm eff}$ & Effective temperature (K) & $5540^{+110}_{-100}$ & $5430 \pm 110$ & $6000^{+350}_{-330}$ & $5830 \pm 110$ & $6290^{+210}_{-190}$ \\
$[{\rm Fe/H}]$ & Metallicity (dex) & $0.181 \pm 0.044$ & $0.331^{+0.048}_{-0.049}$ & $0.376^{+0.062}_{-0.067}$ & $0.218^{+0.076}_{-0.078}$ & $0.017^{+0.083}_{-0.075}$ \\
$[{\rm Fe/H}]_{0}$ & Initial metallicity (dex) & $0.171 \pm 0.052$ & $0.311^{+0.053}_{-0.054}$ & $0.385^{+0.064}_{-0.071}$ & $0.250^{+0.066}_{-0.068}$ & $0.055^{+0.075}_{-0.070}$ \\
Age & Age (Gyr) & $5.4^{+4.1}_{-2.9}$ & $6.3^{+4.1}_{-3.3}$ & $2.40^{+1.30}_{-0.77}$ & $6.6^{+2.7}_{-2.2}$ & $2.5^{+2.4}_{-1.4}$ \\
EEP & Equivalent evolutionary phase & $347^{+31}_{-21}$ & $351^{+34}_{-20}$ & $408^{+51}_{-24}$ & $431^{+15}_{-29}$ & $349^{+45}_{-25}$ \\
$A_V$ & V-band extinction (mag) & $0.156^{+0.110}_{-0.092}$ & $0.27^{+0.10}_{-0.11}$ & $0.60^{+0.33}_{-0.32}$ & $0.056^{+0.037}_{-0.038}$ & $0.21^{+0.15}_{-0.14}$ \\
$d$ & Distance (pc) & $305.0 \pm 1.4$ & $259.67^{+0.95}_{-0.94}$ & $566.1^{+5.1}_{-5.0}$ & $601.1 \pm 7.8$ & $516.3^{+5.4}_{-5.3}$ \\
\multicolumn{7}{l}{\textbf{Planetary Parameters}:} \\
$P$ & Period (days) & $4.2511241 \pm 0.0000013$ & $16.9020425^{+0.0000066}_{-0.0000065}$ & $10.997518 \pm 0.000020$ & $15.130540 \pm 0.000016$ & $6.1613677 \pm 0.0000078$ \\
$R_{\rm P}$ & Radius (\rj) & $0.936^{+0.029}_{-0.025}$ & $1.010^{+0.031}_{-0.029}$ & $1.379^{+0.092}_{-0.084}$ & $1.055^{+0.044}_{-0.041}$ & $1.184^{+0.048}_{-0.045}$ \\
$M_{\rm P}$ & Mass (\mj) & $0.354^{+0.072}_{-0.067}$ & $1.78^{+0.19}_{-0.20}$ & $1.61^{+0.20}_{-0.18}$ & $2.50^{+0.29}_{-0.27}$ & $3.12 \pm 0.19$ \\
$T_C$ & Time of conjunction (\bjdtdb) & $2459910.46580 \pm 0.00016$ & $2459744.98925 \pm 0.00032$ & $2459807.1602 \pm 0.0016$ & $2459570.00224^{+0.00096}_{-0.00100}$ & $2459814.46094^{+0.00074}_{-0.00073}$ \\
$T_0$ & Optimal conjunction time (\bjdtdb) & $2459744.67224 \pm 0.00015$ & $2460167.54118 \pm 0.00028$ & $2460302.0491 \pm 0.0013$ & $2460205.48565^{+0.00063}_{-0.00064}$ & $2460313.53142 \pm 0.00036$ \\
$a$ & Semi-major axis (AU) & $0.05020^{+0.00075}_{-0.00078}$ & $0.1263^{+0.0020}_{-0.0019}$ & $0.1130^{+0.0031}_{-0.0044}$ & $0.1246^{+0.0032}_{-0.0027}$ & $0.0690^{+0.0012}_{-0.0015}$ \\
$i$ & Inclination (Degrees) & $89.57^{+0.30}_{-0.42}$ & $89.78^{+0.15}_{-0.17}$ & $86.71 \pm 0.48$ & $88.66^{+0.88}_{-0.87}$ & $85.67^{+0.32}_{-0.40}$ \\
$e$ & Eccentricity & $0.067^{+0.066}_{-0.044}$ & $0.205 \pm 0.034$ & $0.118^{+0.083}_{-0.075}$ & $0.594^{+0.032}_{-0.036}$ & $0.307^{+0.041}_{-0.035}$ \\
$\omega_\star$ & Argument of periastron (Degrees) & $37^{+91}_{-41}$ & $-98 \pm 11$ & $-85^{+36}_{-18}$ & $164.4^{+5.8}_{-6.1}$ & $44.3^{+7.4}_{-7.7}$ \\
$\teq$ & Equilibrium temperature (K) & $1128^{+21}_{-18}$ & $701^{+13}_{-12}$ & $1355^{+69}_{-60}$ & $961 \pm 14$ & $1266^{+37}_{-32}$ \\
$\tau_{\rm circ}$ & Tidal circularization timescale (Gyr) & $0.69^{+0.22}_{-0.19}$ & $626^{+94}_{-100}$ & $33^{+15}_{-11}$ & $15.5^{+12.0}_{-6.7}$ & $4.0^{+1.6}_{-1.3}$ \\
$K$ & RV semi-amplitude (m/s) & $46.8^{+9.4}_{-8.8}$ & $149 \pm 16$ & $109 \pm 10$ & $235^{+27}_{-26}$ & $330^{+17}_{-16}$ \\
$\dot{\gamma}$ & RV slope (m/s/day) & $-0.127 \pm 0.037$ & ---& $0.038^{+0.032}_{-0.031}$ & $-0.075^{+0.064}_{-0.062}$ & $-0.230^{+0.099}_{-0.100}$ \\
$R_{\rm P}/R_\star$ & Radius of planet in stellar radii  & $0.10735^{+0.00046}_{-0.00044}$ & $0.11441^{+0.00067}_{-0.00056}$ & $0.0573 \pm 0.0013$ & $0.07436^{+0.00110}_{-0.00096}$ & $0.1012 \pm 0.0015$ \\
$a/R_\star$ & Semi-major axis in stellar radii  & $12.07^{+0.35}_{-0.41}$ & $29.98^{+0.95}_{-0.96}$ & $9.78^{+0.62}_{-0.60}$ & $18.42^{+0.84}_{-0.83}$ & $12.35^{+0.44}_{-0.48}$ \\
Depth & \tess flux decrement at mid-transit & $0.01387 \pm 0.00015$ & $0.01592 \pm 0.00019$ & $0.00352 \pm 0.00012$ & $0.00642 \pm 0.00016$ & $0.01059 \pm 0.00023$ \\
$\tau$ & Ingress/egress transit duration (days) & $0.01175^{+0.00030}_{-0.00010}$ & $0.02519^{+0.00130}_{-0.00048}$ & $0.0299^{+0.0053}_{-0.0049}$ & $0.01374^{+0.00190}_{-0.00075}$ & $0.0179^{+0.0020}_{-0.0018}$ \\
$T_{14}$ & Total transit duration (days) & $0.12018^{+0.00050}_{-0.00047}$ & $0.2414^{+0.0013}_{-0.0011}$ & $0.3412^{+0.0056}_{-0.0053}$ & $0.1882^{+0.0020}_{-0.0018}$ & $0.1065^{+0.0017}_{-0.0016}$ \\
$b$ & Transit impact parameter & $0.087^{+0.087}_{-0.061}$ & $0.138^{+0.110}_{-0.095}$ & $0.632^{+0.062}_{-0.089}$ & $0.24 \pm 0.16$ & $0.698^{+0.033}_{-0.038}$ \\
$\rho_{\rm P}$ & Density (cgs) & $0.53^{+0.12}_{-0.11}$ & $2.14^{+0.32}_{-0.30}$ & $0.76^{+0.19}_{-0.16}$ & $2.64^{+0.47}_{-0.43}$ & $2.33^{+0.32}_{-0.29}$ \\
$\log{g_{\rm P}}$ & Surface gravity (cgs) & $3.000^{+0.084}_{-0.095}$ & $3.636^{+0.052}_{-0.059}$ & $3.321^{+0.074}_{-0.076}$ & $3.746^{+0.059}_{-0.064}$ & $3.742^{+0.041}_{-0.044}$ \\
$M_{\rm P}/M_\star$ & Mass ratio  & $0.000363^{+0.000073}_{-0.000068}$ & $0.00181 \pm 0.00019$ & $0.000980^{+0.000094}_{-0.000093}$ & $0.00211 \pm 0.00022$ & $0.00259 \pm 0.00013$ \\
$d/R_\star$ & Separation at mid-transit  & $11.69^{+0.70}_{-0.83}$ & $35.9 \pm 2.1$ & $10.8^{+1.4}_{-1.2}$ & $10.3^{+1.1}_{-1.0}$ & $9.21^{+0.78}_{-0.83}$ \\
\bottomrule
\end{tabular}
\end{table*}
\providecommand{\bjdtdb}{\ensuremath{\rm {BJD_{TDB}}}}
\providecommand{\feh}{\ensuremath{\left[{\rm Fe}/{\rm H}\right]}}
\providecommand{\teff}{\ensuremath{T_{\rm eff}}}
\providecommand{\teq}{\ensuremath{T_{\rm eq}}}
\providecommand{\ecosw}{\ensuremath{e\cos{\omega_\star}}}
\providecommand{\esinw}{\ensuremath{e\sin{\omega_\star}}}
\providecommand{\msun}{\ensuremath{\,M_\Sun}}
\providecommand{\rsun}{\ensuremath{\,R_\Sun}}
\providecommand{\lsun}{\ensuremath{\,L_\Sun}}
\providecommand{\mj}{\ensuremath{\,M_{\rm J}}}
\providecommand{\rj}{\ensuremath{\,R_{\rm J}}}
\providecommand{\me}{\ensuremath{\,M_{\rm E}}}
\providecommand{\re}{\ensuremath{\,R_{\rm E}}}
\providecommand{\fave}{\langle F \rangle}
\providecommand{\fluxcgs}{10$^9$ erg s$^{-1}$ cm$^{-2}$}
\providecommand{\tess}{\textit{TESS}\xspace}
\begin{table*}
\centering
\contcaption{}
\scriptsize
\setlength{\tabcolsep}{3.5pt}
\begin{tabular}{llccccc}
\toprule
&  & TOI-6148 & TOI-6166 & TOI-6171 A$^{*}$ & TOI-6184 & TOI-6191\\
\midrule
\multicolumn{7}{l}{\textbf{Priors}:} \\
$\pi$ & Gaia Parallax (mas)& $\mathcal{G}$[2.0626, 0.033] & $\mathcal{G}$[2.3905, 0.01428] & $\mathcal{G}$[2.0628, 0.02059] & $\mathcal{G}$[2.9896, 0.01921] & $\mathcal{G}$[3.2727, 0.01472] \\
$[{\rm Fe/H}]$ & Metallicity (dex)& $\mathcal{G}$[0.322, 0.08] & $\mathcal{G}$[0.13786, 0.10097] & $\mathcal{G}$[0.19524, 0.12273] & $\mathcal{G}$[0.21948, 0.058682] & $\mathcal{G}$[0.26101, 0.060506] \\
$A_V$ & V-band extinction (mag)& $\mathcal{U}$[0, 0.15841] & $\mathcal{U}$[0, 1.4608] & $\mathcal{U}$[0, 0.25163] & $\mathcal{U}$[0, 0.27833] & $\mathcal{U}$[0, 0.90376] \\
$D_T$ & Dilution in \tess& $\mathcal{G}$[0, 0.00044813] & $\mathcal{G}$[0, 0.031507] & $\mathcal{G}$[0, 0.014052] & $\mathcal{G}$[0, 0.00032891] & $\mathcal{G}$[0, 0.027341] \\
\hline
\multicolumn{7}{l}{\textbf{Stellar Parameters}:} \\
$M_\star$ & Mass (\msun) & $1.374^{+0.063}_{-0.064}$ & $1.070^{+0.071}_{-0.075}$ & $1.149^{+0.120}_{-0.067}$ & $0.989^{+0.052}_{-0.040}$ & $1.066^{+0.068}_{-0.066}$ \\
$R_\star$ & Radius (\rsun) & $1.662^{+0.068}_{-0.065}$ & $1.135^{+0.041}_{-0.036}$ & $1.648^{+0.066}_{-0.064}$ & $1.114^{+0.033}_{-0.032}$ & $1.182^{+0.044}_{-0.038}$ \\
$L_\star$ & Luminosity (\lsun) & $3.53 \pm 0.19$ & $1.48^{+0.20}_{-0.17}$ & $2.86 \pm 0.22$ & $1.067^{+0.044}_{-0.052}$ & $1.37^{+0.13}_{-0.11}$ \\
$\rho_\star$ & Density (cgs) & $0.421^{+0.058}_{-0.053}$ & $1.03^{+0.12}_{-0.13}$ & $0.365^{+0.063}_{-0.052}$ & $1.010^{+0.100}_{-0.092}$ & $0.91 \pm 0.12$ \\
$\log{g}$ & Surface gravity (cgs) & $4.133^{+0.040}_{-0.041}$ & $4.358^{+0.037}_{-0.045}$ & $4.067^{+0.057}_{-0.050}$ & $4.340^{+0.033}_{-0.030}$ & $4.322^{+0.041}_{-0.046}$ \\
$T_{\rm eff}$ & Effective temperature (K) & $6130 \pm 130$ & $5980^{+190}_{-180}$ & $5850 \pm 150$ & $5552^{+89}_{-92}$ & $5740^{+160}_{-150}$ \\
$[{\rm Fe/H}]$ & Metallicity (dex) & $0.332^{+0.071}_{-0.077}$ & $0.117^{+0.095}_{-0.092}$ & $0.21^{+0.10}_{-0.11}$ & $0.257^{+0.054}_{-0.055}$ & $0.272 \pm 0.059$ \\
$[{\rm Fe/H}]_{0}$ & Initial metallicity (dex) & $0.376^{+0.065}_{-0.076}$ & $0.135^{+0.086}_{-0.084}$ & $0.239^{+0.091}_{-0.093}$ & $0.276^{+0.054}_{-0.055}$ & $0.282 \pm 0.058$ \\
Age & Age (Gyr) & $2.77^{+1.10}_{-0.90}$ & $4.6^{+3.7}_{-2.6}$ & $6.8^{+1.8}_{-2.5}$ & $9.3^{+2.8}_{-3.1}$ & $6.4^{+3.7}_{-2.9}$ \\
EEP & Equivalent evolutionary phase & $373^{+27}_{-24}$ & $377^{+35}_{-37}$ & $447.5^{+8.1}_{-39.0}$ & $411^{+13}_{-24}$ & $403^{+23}_{-40}$ \\
$A_V$ & V-band extinction (mag) & $0.094^{+0.046}_{-0.058}$ & $0.29^{+0.16}_{-0.15}$ & $0.183^{+0.048}_{-0.076}$ & $0.215^{+0.042}_{-0.063}$ & $0.20^{+0.12}_{-0.11}$ \\
$d$ & Distance (pc) & $483.9^{+7.7}_{-7.6}$ & $418.3 \pm 2.5$ & $485.5^{+4.9}_{-4.8}$ & $334.6 \pm 2.1$ & $305.5 \pm 1.4$ \\
\multicolumn{7}{l}{\textbf{Planetary Parameters}:} \\
$P$ & Period (days) & $8.367809 \pm 0.000014$ & $1.80211763 \pm 0.00000074$ & $3.7477271 \pm 0.0000044$ & $3.6152434 \pm 0.0000023$ & $3.7752310 \pm 0.0000026$ \\
$R_{\rm P}$ & Radius (\rj) & $1.070^{+0.055}_{-0.050}$ & $1.331^{+0.049}_{-0.044}$ & $0.954^{+0.039}_{-0.038}$ & $1.212^{+0.040}_{-0.038}$ & $1.067^{+0.043}_{-0.038}$ \\
$M_{\rm P}$ & Mass (\mj) & $7.51 \pm 0.38$ & $0.94 \pm 0.11$ & $1.34^{+0.19}_{-0.17}$ & $0.67 \pm 0.10$ & $0.411^{+0.100}_{-0.099}$ \\
$T_C$ & Time of conjunction (\bjdtdb) & $2459849.6217 \pm 0.0011$ & $2459850.64269 \pm 0.00019$ & $2459848.43151 \pm 0.00064$ & $2459847.82127^{+0.00034}_{-0.00033}$ & $2459880.41938^{+0.00063}_{-0.00055}$ \\
$T_0$ & Optimal conjunction time (\bjdtdb) & $2460393.53007^{+0.00056}_{-0.00057}$ & $2460093.92872 \pm 0.00016$ & $2460249.43861 \pm 0.00042$ & $2460205.73060 \pm 0.00021$ & $2460265.49299 \pm 0.00021$ \\
$a$ & Semi-major axis (AU) & $0.0898^{+0.0013}_{-0.0014}$ & $0.02965^{+0.00064}_{-0.00071}$ & $0.04947^{+0.00170}_{-0.00098}$ & $0.04594^{+0.00078}_{-0.00063}$ & $0.0485 \pm 0.0010$ \\
$i$ & Inclination (Degrees) & $83.91^{+0.58}_{-0.61}$ & $88.4^{+1.1}_{-1.2}$ & $88.61^{+0.96}_{-1.20}$ & $85.99 \pm 0.29$ & $84.01^{+0.49}_{-0.64}$ \\
$e$ & Eccentricity & $0.460^{+0.028}_{-0.029}$ & $0.085^{+0.053}_{-0.050}$ & $0.101^{+0.058}_{-0.057}$ & $0.041^{+0.046}_{-0.029}$ & $0.088^{+0.072}_{-0.058}$ \\
$\omega_\star$ & Argument of periastron (Degrees) & $108.7^{+3.9}_{-3.8}$ & $37^{+40}_{-32}$ & $-81^{+31}_{-35}$ & $-17^{+97}_{-110}$ & $46^{+78}_{-46}$ \\
$\teq$ & Equilibrium temperature (K) & $1273 \pm 17$ & $1785^{+51}_{-45}$ & $1622^{+33}_{-32}$ & $1318^{+15}_{-17}$ & $1367^{+28}_{-25}$ \\
$\tau_{\rm circ}$ & Tidal circularization timescale (Gyr) & $18.9^{+9.1}_{-6.2}$ & $0.0083^{+0.0025}_{-0.0024}$ & $1.50^{+0.35}_{-0.31}$ & $0.196^{+0.049}_{-0.044}$ & $0.259^{+0.110}_{-0.096}$ \\
$K$ & RV semi-amplitude (m/s) & $680^{+30}_{-29}$ & $151 \pm 16$ & $159 \pm 19$ & $89 \pm 14$ & $51^{+13}_{-12}$ \\
$\dot{\gamma}$ & RV slope (m/s/day) & ---& $-0.273^{+0.068}_{-0.072}$ & ---& $0.17 \pm 0.11$ & ---\\
$R_{\rm P}/R_\star$ & Radius of planet in stellar radii  & $0.0662 \pm 0.0011$ & $0.1204 \pm 0.0011$ & $0.05948^{+0.00059}_{-0.00055}$ & $0.11178^{+0.00091}_{-0.00096}$ & $0.09272^{+0.00097}_{-0.00093}$ \\
$a/R_\star$ & Semi-major axis in stellar radii  & $11.61 \pm 0.51$ & $5.62^{+0.21}_{-0.24}$ & $6.48^{+0.35}_{-0.32}$ & $8.87^{+0.29}_{-0.28}$ & $8.83^{+0.36}_{-0.40}$ \\
Depth & \tess flux decrement at mid-transit & $0.00458 \pm 0.00012$ & $0.01654^{+0.00034}_{-0.00033}$ & $0.004051^{+0.000086}_{-0.000085}$ & $0.01366 \pm 0.00017$ & $0.00755 \pm 0.00014$ \\
$\tau$ & Ingress/egress transit duration (days) & $0.0130^{+0.0016}_{-0.0015}$ & $0.01199^{+0.00063}_{-0.00027}$ & $0.01212^{+0.00110}_{-0.00035}$ & $0.0188 \pm 0.0012$ & $0.0274^{+0.0017}_{-0.0016}$ \\
$T_{14}$ & Total transit duration (days) & $0.1174 \pm 0.0019$ & $0.10857^{+0.00071}_{-0.00060}$ & $0.2089^{+0.0014}_{-0.0012}$ & $0.1203^{+0.0011}_{-0.0010}$ & $0.08535^{+0.00098}_{-0.00097}$ \\
$b$ & Transit impact parameter & $0.681^{+0.037}_{-0.047}$ & $0.145^{+0.110}_{-0.098}$ & $0.17^{+0.15}_{-0.12}$ & $0.622^{+0.026}_{-0.029}$ & $0.8765^{+0.0058}_{-0.0061}$ \\
$\rho_{\rm P}$ & Density (cgs) & $7.6^{+1.3}_{-1.1}$ & $0.494^{+0.081}_{-0.075}$ & $1.91^{+0.41}_{-0.33}$ & $0.468^{+0.090}_{-0.084}$ & $0.42^{+0.12}_{-0.11}$ \\
$\log{g_{\rm P}}$ & Surface gravity (cgs) & $4.211^{+0.047}_{-0.050}$ & $3.119^{+0.057}_{-0.062}$ & $3.561^{+0.072}_{-0.071}$ & $3.055^{+0.070}_{-0.080}$ & $2.95^{+0.10}_{-0.13}$ \\
$M_{\rm P}/M_\star$ & Mass ratio  & $0.00523^{+0.00023}_{-0.00022}$ & $0.000844^{+0.000092}_{-0.000093}$ & $0.00110 \pm 0.00013$ & $0.000647 \pm 0.000099$ & $0.000368^{+0.000091}_{-0.000088}$ \\
$d/R_\star$ & Separation at mid-transit  & $6.38^{+0.48}_{-0.45}$ & $5.35^{+0.42}_{-0.48}$ & $7.03^{+0.74}_{-0.70}$ & $8.87^{+0.55}_{-0.50}$ & $8.40^{+0.75}_{-0.82}$ \\
\bottomrule
\end{tabular}
\end{table*}
\providecommand{\bjdtdb}{\ensuremath{\rm {BJD_{TDB}}}}
\providecommand{\feh}{\ensuremath{\left[{\rm Fe}/{\rm H}\right]}}
\providecommand{\teff}{\ensuremath{T_{\rm eff}}}
\providecommand{\teq}{\ensuremath{T_{\rm eq}}}
\providecommand{\ecosw}{\ensuremath{e\cos{\omega_\star}}}
\providecommand{\esinw}{\ensuremath{e\sin{\omega_\star}}}
\providecommand{\msun}{\ensuremath{\,M_\Sun}}
\providecommand{\rsun}{\ensuremath{\,R_\Sun}}
\providecommand{\lsun}{\ensuremath{\,L_\Sun}}
\providecommand{\mj}{\ensuremath{\,M_{\rm J}}}
\providecommand{\rj}{\ensuremath{\,R_{\rm J}}}
\providecommand{\me}{\ensuremath{\,M_{\rm E}}}
\providecommand{\re}{\ensuremath{\,R_{\rm E}}}
\providecommand{\fave}{\langle F \rangle}
\providecommand{\fluxcgs}{10$^9$ erg s$^{-1}$ cm$^{-2}$}
\providecommand{\tess}{\textit{TESS}\xspace}
\begin{table*}
\centering
\contcaption{}
\scriptsize
\setlength{\tabcolsep}{3.5pt}
\begin{tabular}{llccccc}
\toprule
&  & TOI-6208$^{*}$ & TOI-6334$^{*}$ & TOI-6417 & TOI-6443 & TOI-7219\\
\midrule
\multicolumn{7}{l}{\textbf{Priors}:} \\
$\pi$ & Gaia Parallax (mas)& $\mathcal{G}$[1.4631, 0.01393] & $\mathcal{G}$[1.3496, 0.02236] & $\mathcal{G}$[1.7436, 0.01688] & $\mathcal{G}$[1.6177, 0.0145] & $\mathcal{G}$[3.0571, 0.01656] \\
$[{\rm Fe/H}]$ & Metallicity (dex)& $\mathcal{G}$[0.15394, 0.10752] & $\mathcal{G}$[0.10601, 0.1123] & $\mathcal{G}$[0.10655, 0.084444] & $\mathcal{G}$[0.28276, 0.088468] & $\mathcal{G}$[0.20103, 0.083148] \\
$A_V$ & V-band extinction (mag)& $\mathcal{U}$[0, 2.5733] & $\mathcal{U}$[0, 0.27833] & $\mathcal{U}$[0, 0.38068] & $\mathcal{U}$[0, 3.9128] & $\mathcal{U}$[0, 0.99924] \\
$D_T$ & Dilution in \tess& $\mathcal{G}$[0, 0.052059] & $\mathcal{G}$[0, 0.0026859] & $\mathcal{G}$[0, 0.01357] & $\mathcal{G}$[0, 0.032125] & $\mathcal{G}$[0, 0.0097473] \\
\hline
\multicolumn{7}{l}{\textbf{Stellar Parameters}:} \\
$M_\star$ & Mass (\msun) & $1.47^{+0.13}_{-0.18}$ & $1.480^{+0.160}_{-0.082}$ & $1.395^{+0.073}_{-0.079}$ & $1.66 \pm 0.12$ & $1.055^{+0.069}_{-0.067}$ \\
$R_\star$ & Radius (\rsun) & $1.957^{+0.080}_{-0.078}$ & $2.572^{+0.096}_{-0.090}$ & $1.695^{+0.066}_{-0.062}$ & $1.632^{+0.053}_{-0.051}$ & $1.166^{+0.038}_{-0.036}$ \\
$L_\star$ & Luminosity (\lsun) & $5.47^{+1.30}_{-0.92}$ & $7.92^{+0.48}_{-0.56}$ & $4.23^{+0.29}_{-0.37}$ & $6.5^{+1.9}_{-1.3}$ & $1.37^{+0.15}_{-0.14}$ \\
$\rho_\star$ & Density (cgs) & $0.274^{+0.051}_{-0.044}$ & $0.125^{+0.017}_{-0.015}$ & $0.403^{+0.057}_{-0.051}$ & $0.536^{+0.064}_{-0.059}$ & $0.94^{+0.11}_{-0.10}$ \\
$\log{g}$ & Surface gravity (cgs) & $4.020^{+0.059}_{-0.066}$ & $3.793^{+0.048}_{-0.043}$ & $4.123^{+0.042}_{-0.044}$ & $4.231^{+0.039}_{-0.040}$ & $4.328^{+0.038}_{-0.039}$ \\
$T_{\rm eff}$ & Effective temperature (K) & $6300^{+410}_{-310}$ & $6030^{+120}_{-140}$ & $6350^{+160}_{-170}$ & $7210^{+490}_{-410}$ & $5780^{+160}_{-170}$ \\
$[{\rm Fe/H}]$ & Metallicity (dex) & $0.16^{+0.11}_{-0.10}$ & $0.14^{+0.11}_{-0.10}$ & $0.116^{+0.078}_{-0.077}$ & $0.226^{+0.082}_{-0.085}$ & $0.226 \pm 0.083$ \\
$[{\rm Fe/H}]_{0}$ & Initial metallicity (dex) & $0.24 \pm 0.10$ & $0.181^{+0.096}_{-0.100}$ & $0.223^{+0.077}_{-0.081}$ & $0.329^{+0.073}_{-0.081}$ & $0.241 \pm 0.076$ \\
Age & Age (Gyr) & $2.4^{+2.0}_{-1.1}$ & $2.87^{+0.49}_{-0.86}$ & $2.39^{+1.00}_{-0.83}$ & $0.34^{+0.50}_{-0.21}$ & $6.5^{+3.6}_{-3.0}$ \\
EEP & Equivalent evolutionary phase & $385^{+62}_{-36}$ & $452.5^{+7.0}_{-50.0}$ & $366^{+31}_{-20}$ & $308^{+23}_{-28}$ & $401^{+22}_{-40}$ \\
$A_V$ & V-band extinction (mag) & $0.54^{+0.25}_{-0.23}$ & $0.217^{+0.045}_{-0.089}$ & $0.281^{+0.070}_{-0.110}$ & $0.67 \pm 0.25$ & $0.75^{+0.13}_{-0.15}$ \\
$d$ & Distance (pc) & $683.7^{+6.6}_{-6.4}$ & $741 \pm 12$ & $573.0^{+5.6}_{-5.5}$ & $618.3^{+5.6}_{-5.5}$ & $327.1 \pm 1.8$ \\
\multicolumn{7}{l}{\textbf{Planetary Parameters}:} \\
$P$ & Period (days) & $5.2266134 \pm 0.0000062$ & $3.8497745 \pm 0.0000039$ & $3.3458228 \pm 0.0000021$ & $9.9872665^{+0.0000098}_{-0.0000100}$ & $2.9069477^{+0.0000029}_{-0.0000028}$ \\
$R_{\rm P}$ & Radius (\rj) & $1.482^{+0.069}_{-0.065}$ & $1.528^{+0.060}_{-0.055}$ & $1.259^{+0.054}_{-0.052}$ & $0.909^{+0.032}_{-0.031}$ & $1.257^{+0.070}_{-0.053}$ \\
$M_{\rm P}$ & Mass (\mj) & $0.87^{+0.13}_{-0.12}$ & $0.94^{+0.14}_{-0.13}$ & $1.10 \pm 0.11$ & $5.05^{+0.31}_{-0.30}$ & $2.08^{+0.26}_{-0.25}$ \\
$T_C$ & Time of conjunction (\bjdtdb) & $2459872.10373^{+0.00076}_{-0.00075}$ & $2459905.99685^{+0.00070}_{-0.00072}$ & $2459987.10169 \pm 0.00039$ & $2459971.0695^{+0.0011}_{-0.0010}$ & $2460532.39246^{+0.00047}_{-0.00046}$ \\
$T_0$ & Optimal conjunction time (\bjdtdb) & $2460227.51364^{+0.00060}_{-0.00061}$ & $2459990.69226 \pm 0.00068$ & $2460137.66401^{+0.00036}_{-0.00037}$ & $2460989.77205 \pm 0.00030$ & $2460433.55643 \pm 0.00040$ \\
$a$ & Semi-major axis (AU) & $0.0670^{+0.0019}_{-0.0028}$ & $0.0548^{+0.0019}_{-0.0010}$ & $0.04892^{+0.00084}_{-0.00094}$ & $0.1075^{+0.0025}_{-0.0026}$ & $0.04060^{+0.00087}_{-0.00088}$ \\
$i$ & Inclination (Degrees) & $87.9^{+1.4}_{-1.2}$ & $88.0^{+1.4}_{-1.8}$ & $86.6^{+1.7}_{-1.1}$ & $85.75^{+0.34}_{-0.38}$ & $83.27^{+0.43}_{-0.49}$ \\
$e$ & Eccentricity & $0.146^{+0.068}_{-0.073}$ & $0.056^{+0.057}_{-0.039}$ & $0.052^{+0.056}_{-0.037}$ & $0.416^{+0.029}_{-0.027}$ & $0.035^{+0.043}_{-0.025}$ \\
$\omega_\star$ & Argument of periastron (Degrees) & $-7^{+34}_{-31}$ & $-155^{+74}_{-80}$ & $113^{+78}_{-61}$ & $138.3^{+4.8}_{-5.5}$ & $-70^{+110}_{-120}$ \\
$\teq$ & Equilibrium temperature (K) & $1649^{+70}_{-58}$ & $1986 \pm 34$ & $1803^{+29}_{-33}$ & $1355^{+74}_{-63}$ & $1494^{+34}_{-36}$ \\
$\tau_{\rm circ}$ & Tidal circularization timescale (Gyr) & $0.46^{+0.19}_{-0.15}$ & $0.145^{+0.037}_{-0.035}$ & $0.232^{+0.074}_{-0.062}$ & $106^{+44}_{-34}$ & $0.205^{+0.064}_{-0.061}$ \\
$K$ & RV semi-amplitude (m/s) & $81.0^{+9.9}_{-10.0}$ & $92 \pm 13$ & $119 \pm 11$ & $372^{+17}_{-15}$ & $283 \pm 33$ \\
$\dot{\gamma}$ & RV slope (m/s/day) & ---& ---& $-1.05^{+0.44}_{-0.42}$ & $0.86^{+0.48}_{-0.50}$ & ---\\
$R_{\rm P}/R_\star$ & Radius of planet in stellar radii  & $0.0779^{+0.0017}_{-0.0016}$ & $0.06104^{+0.00059}_{-0.00054}$ & $0.07631^{+0.00099}_{-0.00092}$ & $0.05721 \pm 0.00052$ & $0.1105^{+0.0037}_{-0.0023}$ \\
$a/R_\star$ & Semi-major axis in stellar radii  & $7.34^{+0.43}_{-0.42}$ & $4.60^{+0.20}_{-0.19}$ & $6.20^{+0.28}_{-0.27}$ & $14.16 \pm 0.54$ & $7.49 \pm 0.28$ \\
Depth & \tess flux decrement at mid-transit & $0.00679^{+0.00028}_{-0.00027}$ & $0.004267^{+0.000083}_{-0.000081}$ & $0.00645 \pm 0.00014$ & $0.003369^{+0.000059}_{-0.000058}$ & $0.01060^{+0.00030}_{-0.00027}$ \\
$\tau$ & Ingress/egress transit duration (days) & $0.0184^{+0.0028}_{-0.0013}$ & $0.01690^{+0.00140}_{-0.00046}$ & $0.0137^{+0.0017}_{-0.0015}$ & $0.0126^{+0.0011}_{-0.0010}$ & $0.0338^{+0.0091}_{-0.0043}$ \\
$T_{14}$ & Total transit duration (days) & $0.2374^{+0.0027}_{-0.0021}$ & $0.2824^{+0.0019}_{-0.0017}$ & $0.1688^{+0.0018}_{-0.0016}$ & $0.1297^{+0.0014}_{-0.0013}$ & $0.0848^{+0.0018}_{-0.0017}$ \\
$b$ & Transit impact parameter & $0.27^{+0.16}_{-0.18}$ & $0.17^{+0.15}_{-0.11}$ & $0.35^{+0.11}_{-0.19}$ & $0.681^{+0.028}_{-0.033}$ & $0.879^{+0.014}_{-0.011}$ \\
$\rho_{\rm P}$ & Density (cgs) & $0.331^{+0.075}_{-0.062}$ & $0.325^{+0.062}_{-0.056}$ & $0.68^{+0.12}_{-0.11}$ & $8.35^{+1.00}_{-0.95}$ & $1.29 \pm 0.26$ \\
$\log{g_{\rm P}}$ & Surface gravity (cgs) & $2.993^{+0.074}_{-0.078}$ & $2.997^{+0.067}_{-0.074}$ & $3.233^{+0.059}_{-0.063}$ & $4.181^{+0.038}_{-0.040}$ & $3.510^{+0.066}_{-0.081}$ \\
$M_{\rm P}/M_\star$ & Mass ratio  & $0.000575^{+0.000073}_{-0.000074}$ & $0.000596^{+0.000083}_{-0.000084}$ & $0.000753 \pm 0.000072$ & $0.00291^{+0.00014}_{-0.00013}$ & $0.00188^{+0.00023}_{-0.00022}$ \\
$d/R_\star$ & Separation at mid-transit  & $7.33^{+0.90}_{-0.79}$ & $4.63^{+0.42}_{-0.33}$ & $6.02^{+0.45}_{-0.56}$ & $9.17^{+0.72}_{-0.73}$ & $7.51^{+0.50}_{-0.48}$ \\
\bottomrule
\end{tabular}
\end{table*}
\providecommand{\bjdtdb}{\ensuremath{\rm {BJD_{TDB}}}}
\providecommand{\feh}{\ensuremath{\left[{\rm Fe}/{\rm H}\right]}}
\providecommand{\teff}{\ensuremath{T_{\rm eff}}}
\providecommand{\teq}{\ensuremath{T_{\rm eq}}}
\providecommand{\ecosw}{\ensuremath{e\cos{\omega_\star}}}
\providecommand{\esinw}{\ensuremath{e\sin{\omega_\star}}}
\providecommand{\msun}{\ensuremath{\,M_\Sun}}
\providecommand{\rsun}{\ensuremath{\,R_\Sun}}
\providecommand{\lsun}{\ensuremath{\,L_\Sun}}
\providecommand{\mj}{\ensuremath{\,M_{\rm J}}}
\providecommand{\rj}{\ensuremath{\,R_{\rm J}}}
\providecommand{\me}{\ensuremath{\,M_{\rm E}}}
\providecommand{\re}{\ensuremath{\,R_{\rm E}}}
\providecommand{\fave}{\langle F \rangle}
\providecommand{\fluxcgs}{10$^9$ erg s$^{-1}$ cm$^{-2}$}
\providecommand{\tess}{\textit{TESS}\xspace}
\begin{table*}
\centering
\contcaption{}
\scriptsize
\setlength{\tabcolsep}{3.5pt}
\begin{tabular}{llcccc}
\toprule
&  & TOI-7266 A & TOI-7404$^{*}$ & TOI-7425$^{*}$ & TOI-7574\\
\midrule
\multicolumn{6}{l}{\textbf{Priors}:} \\
$\pi$ & Gaia Parallax (mas)& $\mathcal{G}$[0.6613, 0.51311] & $\mathcal{G}$[1.009, 0.02165] & $\mathcal{G}$[1.9413, 0.01414] & $\mathcal{G}$[3.7757, 0.02138] \\
$[{\rm Fe/H}]$ & Metallicity (dex)& $\mathcal{G}$[0.030781, 0.148] & $\mathcal{G}$[0.26381, 0.11981] & $\mathcal{G}$[0.39238, 0.042848] & $\mathcal{G}$[0.072416, 0.051784] \\
$A_V$ & V-band extinction (mag)& $\mathcal{U}$[0, 0.409] & $\mathcal{U}$[0, 0.18569] & $\mathcal{U}$[0, 5.12] & $\mathcal{U}$[0, 3.4706] \\
$D_T$ & Dilution in \tess& $\mathcal{G}$[0, 0.014552] & $\mathcal{G}$[0, 0.0086174] & $\mathcal{G}$[0, 0.019646] & $\mathcal{G}$[0, 0.0010978] \\
\hline
\multicolumn{6}{l}{\textbf{Stellar Parameters}:} \\
$M_\star$ & Mass (\msun) & $1.33^{+0.19}_{-0.18}$ & $1.536^{+0.073}_{-0.093}$ & $1.295^{+0.075}_{-0.120}$ & $0.913^{+0.048}_{-0.041}$ \\
$R_\star$ & Radius (\rsun) & $1.91^{+0.30}_{-0.23}$ & $2.139^{+0.095}_{-0.087}$ & $1.606^{+0.055}_{-0.050}$ & $0.922^{+0.030}_{-0.028}$ \\
$L_\star$ & Luminosity (\lsun) & $4.8^{+2.2}_{-1.4}$ & $6.43^{+0.45}_{-0.41}$ & $2.89^{+0.38}_{-0.35}$ & $0.720^{+0.077}_{-0.051}$ \\
$\rho_\star$ & Density (cgs) & $0.266^{+0.100}_{-0.079}$ & $0.220^{+0.033}_{-0.030}$ & $0.439^{+0.049}_{-0.056}$ & $1.64^{+0.18}_{-0.16}$ \\
$\log{g}$ & Surface gravity (cgs) & $3.994^{+0.089}_{-0.095}$ & $3.962^{+0.043}_{-0.048}$ & $4.138^{+0.035}_{-0.049}$ & $4.469 \pm 0.034$ \\
$T_{\rm eff}$ & Effective temperature (K) & $6200^{+250}_{-230}$ & $6280 \pm 140$ & $5930 \pm 200$ & $5540^{+140}_{-120}$ \\
$[{\rm Fe/H}]$ & Metallicity (dex) & $0.05^{+0.13}_{-0.12}$ & $0.20^{+0.10}_{-0.11}$ & $0.393 \pm 0.041$ & $0.075 \pm 0.051$ \\
$[{\rm Fe/H}]_{0}$ & Initial metallicity (dex) & $0.12^{+0.12}_{-0.11}$ & $0.26 \pm 0.10$ & $0.401^{+0.045}_{-0.047}$ & $0.089 \pm 0.056$ \\
Age & Age (Gyr) & $3.5^{+2.6}_{-1.4}$ & $2.22^{+0.71}_{-0.50}$ & $4.0^{+2.9}_{-1.4}$ & $7.5 \pm 3.8$ \\
EEP & Equivalent evolutionary phase & $411^{+41}_{-29}$ & $390^{+21}_{-20}$ & $397^{+48}_{-36}$ & $365^{+28}_{-26}$ \\
$A_V$ & V-band extinction (mag) & $0.21^{+0.13}_{-0.14}$ & $0.090^{+0.062}_{-0.060}$ & $0.67^{+0.15}_{-0.17}$ & $0.149^{+0.140}_{-0.100}$ \\
$d$ & Distance (pc) & $660^{+120}_{-89}$ & $990 \pm 21$ & $515.3^{+3.8}_{-3.7}$ & $264.9 \pm 1.5$ \\
\multicolumn{6}{l}{\textbf{Planetary Parameters}:} \\
$P$ & Period (days) & $3.8789280^{+0.0000060}_{-0.0000061}$ & $1.4634427^{+0.0000016}_{-0.0000017}$ & $1.28157053 \pm 0.00000031$ & $4.4797048^{+0.0000088}_{-0.0000087}$ \\
$R_{\rm P}$ & Radius (\rj) & $1.42^{+0.24}_{-0.19}$ & $1.233^{+0.066}_{-0.059}$ & $1.792^{+0.063}_{-0.060}$ & $1.143^{+0.042}_{-0.040}$ \\
$M_{\rm P}$ & Mass (\mj) & $2.17^{+0.42}_{-0.40}$ & $8.7^{+1.4}_{-1.5}$ & $1.17 \pm 0.15$ & $2.64 \pm 0.31$ \\
$T_C$ & Time of conjunction (\bjdtdb) & $2460552.58533^{+0.00077}_{-0.00075}$ & $2460608.4588^{+0.0010}_{-0.0011}$ & $2460633.58710^{+0.00047}_{-0.00041}$ & $2460683.83080^{+0.00059}_{-0.00058}$ \\
$T_0$ & Optimal conjunction time (\bjdtdb) & $2460447.85435^{+0.00064}_{-0.00066}$ & $2460033.32597^{+0.00073}_{-0.00074}$ & $2459909.49980 \pm 0.00018$ & $2460939.17431^{+0.00030}_{-0.00031}$ \\
$a$ & Semi-major axis (AU) & $0.0531^{+0.0024}_{-0.0025}$ & $0.02916^{+0.00045}_{-0.00060}$ & $0.02517^{+0.00048}_{-0.00083}$ & $0.05164^{+0.00089}_{-0.00078}$ \\
$i$ & Inclination (Degrees) & $85.6^{+2.4}_{-2.1}$ & $82.7^{+4.1}_{-2.9}$ & $78.45^{+0.86}_{-1.20}$ & $87.70^{+0.46}_{-0.31}$ \\
$e$ & Eccentricity & $0.103^{+0.100}_{-0.070}$ & $0.044^{+0.066}_{-0.031}$ & $0.050^{+0.056}_{-0.035}$ & $0.077^{+0.061}_{-0.048}$ \\
$\omega_\star$ & Argument of periastron (Degrees) & $20^{+77}_{-70}$ & $-90 \pm 120$ & $63^{+90}_{-85}$ & $144^{+52}_{-45}$ \\
$\teq$ & Equilibrium temperature (K) & $1790^{+140}_{-120}$ & $2599^{+45}_{-42}$ & $2290^{+56}_{-58}$ & $1128^{+26}_{-20}$ \\
$\tau_{\rm circ}$ & Tidal circularization timescale (Gyr) & $0.39^{+0.38}_{-0.22}$ & $0.060^{+0.025}_{-0.021}$ & $0.00064^{+0.00017}_{-0.00018}$ & $2.32^{+0.65}_{-0.59}$ \\
$K$ & RV semi-amplitude (m/s) & $234 \pm 38$ & $1160^{+170}_{-190}$ & $182^{+21}_{-20}$ & $346^{+38}_{-39}$ \\
$\dot{\gamma}$ & RV slope (m/s/day) & $-2.1 \pm 1.4$ & $8.1^{+7.4}_{-6.6}$ & ---& ---\\
$R_{\rm P}/R_\star$ & Radius of planet in stellar radii  & $0.0767^{+0.0016}_{-0.0014}$ & $0.05922^{+0.00110}_{-0.00097}$ & $0.11462^{+0.00085}_{-0.00086}$ & $0.1275^{+0.0018}_{-0.0022}$ \\
$a/R_\star$ & Semi-major axis in stellar radii  & $5.96^{+0.68}_{-0.66}$ & $2.92 \pm 0.14$ & $3.37^{+0.12}_{-0.15}$ & $12.05 \pm 0.42$ \\
Depth & \tess flux decrement at mid-transit & $0.00648 \pm 0.00015$ & $0.00390 \pm 0.00011$ & $0.01390^{+0.00019}_{-0.00018}$ & $0.01841^{+0.00043}_{-0.00044}$ \\
$\tau$ & Ingress/egress transit duration (days) & $0.0177^{+0.0044}_{-0.0032}$ & $0.0107^{+0.0020}_{-0.0013}$ & $0.0192 \pm 0.0011$ & $0.0166^{+0.0018}_{-0.0020}$ \\
$T_{14}$ & Total transit duration (days) & $0.1990^{+0.0038}_{-0.0030}$ & $0.1616^{+0.0022}_{-0.0020}$ & $0.11036^{+0.00099}_{-0.00098}$ & $0.1172 \pm 0.0018$ \\
$b$ & Transit impact parameter & $0.45^{+0.14}_{-0.24}$ & $0.37^{+0.15}_{-0.21}$ & $0.661^{+0.020}_{-0.023}$ & $0.470^{+0.067}_{-0.110}$ \\
$\rho_{\rm P}$ & Density (cgs) & $0.91^{+0.48}_{-0.34}$ & $5.7 \pm 1.4$ & $0.252^{+0.045}_{-0.041}$ & $2.18^{+0.38}_{-0.34}$ \\
$\log{g_{\rm P}}$ & Surface gravity (cgs) & $3.42^{+0.13}_{-0.15}$ & $4.150^{+0.080}_{-0.096}$ & $2.955^{+0.063}_{-0.067}$ & $3.698^{+0.059}_{-0.063}$ \\
$M_{\rm P}/M_\star$ & Mass ratio  & $0.00157 \pm 0.00027$ & $0.00541^{+0.00084}_{-0.00091}$ & $0.00088 \pm 0.00010$ & $0.00275 \pm 0.00031$ \\
$d/R_\star$ & Separation at mid-transit  & $5.8^{+1.0}_{-1.1}$ & $2.93^{+0.21}_{-0.24}$ & $3.31^{+0.21}_{-0.33}$ & $11.60^{+0.85}_{-1.00}$ \\
\bottomrule
\end{tabular}
\begin{flushleft}
\textbf{Notes:} The priors for each system are labeled as $\mathcal{G}$[mean, standard deviation] if they are Gaussian priors and $\mathcal{U}$[lower limit, upper limit] if they are uniform priors.
$^*$Star has bimodal mass and age posteriors. See Table~\ref{tab:bimodal} for each individual solution.
\end{flushleft}
\end{table*}
\renewcommand{\arraystretch}{\bodystretch}

\subsection{Priors and Constraints}\label{subsec:priors}

We applied several Gaussian priors, limits, and starting values to our global fits to account for past observations and ensure that each MCMC converged. We enforced Gaussian priors on the parallax of each star that we fit using the corrected \gaia DR3 \citep{Lindegren:2021, GaiaDR3} observations (see \S\ref{subsec:archival}). Then, we placed Gaussian priors on the metallicity of each primary star using a weighted average of the spectroscopic metallicities determined by SPC (see \S\ref{subsubsec:tres} \& \S\ref{subsubsec:chiron}). The weight used was the signal-to-noise per resolution element of the corresponding spectrum. In the case of TOI-3972, we did not place a prior on the metallicity of the unbound background star that was included in our multi-component SED fit.

While the SPOC PDCSAP lightcurves have been corrected for crowding, we chose to allow the dilution factor to float as a free parameter for our \tess lightcurves to account for possible over-correction or under-correction and ensure that the ground-based and \tess transits agree on the radius of each planet. However, we placed a Gaussian prior on the dilution of each star centered at zero, with a width equal to 10\% of the dilution factor $D$, defined as $D = C/(1 + C)$, where $C$ is the contamination ratio from the \tess Input Catalog (TIC) v8.2 \citep{Stassun:2018, Stassun:2019}, which we downloaded using MAST. We fit for dilution for every system in this paper except for TOI-5479. TOI-5479 is in a sparse field and has a very low contamination ratio of $4.9 \times 10^{-8}$, indicating that contamination from other sources in the \tess data is negligible. In addition to the aforementioned Gaussian priors, we placed uniform priors on the visual extinction $A_V$ of each target star utilizing conservative upper limits from the \cite{Schlafly:2011} dust maps. 

In all of our initial fits, we allowed an RV linear slope to float as a free parameter. After this initial fit converged, we checked to see if the median slope was consistent with zero within $1\sigma$. If it was, we removed the slope from the fit to ensure that we did not inflate our errors. Additionally, we allowed the eccentricity to float as a free parameter in our fits, in line with the previous installments of the MEEP survey \citep{Schulte:2024, Schulte:2025}. Eccentricity is parameterized as $\sqrt{e} \cos \omega_\star$ and $\sqrt{e} \sin \omega_\star$ in \exofast fits, but even with this parameterization, the fitted eccentricity is biased towards positive values, often larger than their true values when the true values are very close to zero (the so-called Lucy-Sweeney bias; \citealt{Lucy:1971}). \cite{Eastman:2019} argues that an orbital eccentricity is significant if the median eccentricity is $2.45\sigma$ greater than zero. In this article, seven of our planets meet these requirements and are discussed further in \S\ref{subsec:eccentricity}.

TOI-3988\,A, TOI-6171\,A, and TOI-7266\,A all have likely bound stellar companions that were detected in our high angular resolution imaging (\S\ref{subsec:hri}). While modeling these systems, we fixed the initial metallicity, age, visual extinction, and distance of the companion star to those of the primary star to account for the likely scenario that these binary pairs formed together from the same nebula. The current surface metallicity of each star was allowed to be different to account for the differences in evolution and gravitational settling of each star in these pairs. Finally, we adopted starting values for the mass, radius, and effective temperature of each star from the TIC. Starting values for the transit epoch $T_C$, orbital period $P$, and the ratio of planetary and stellar radii $R_{\rm P}/R_*$ were retrieved from ExoFOP-TESS\footnote{\url{https://exofop.ipac.caltech.edu/tess/}}. All of the priors used to constrain our fits are listed at the top of Table~\ref{tab:median}.

\subsection{Bimodal Posteriors}\label{subsec:bimodal}

\begin{figure*}
    \centering
    \includegraphics[width=0.33\linewidth]{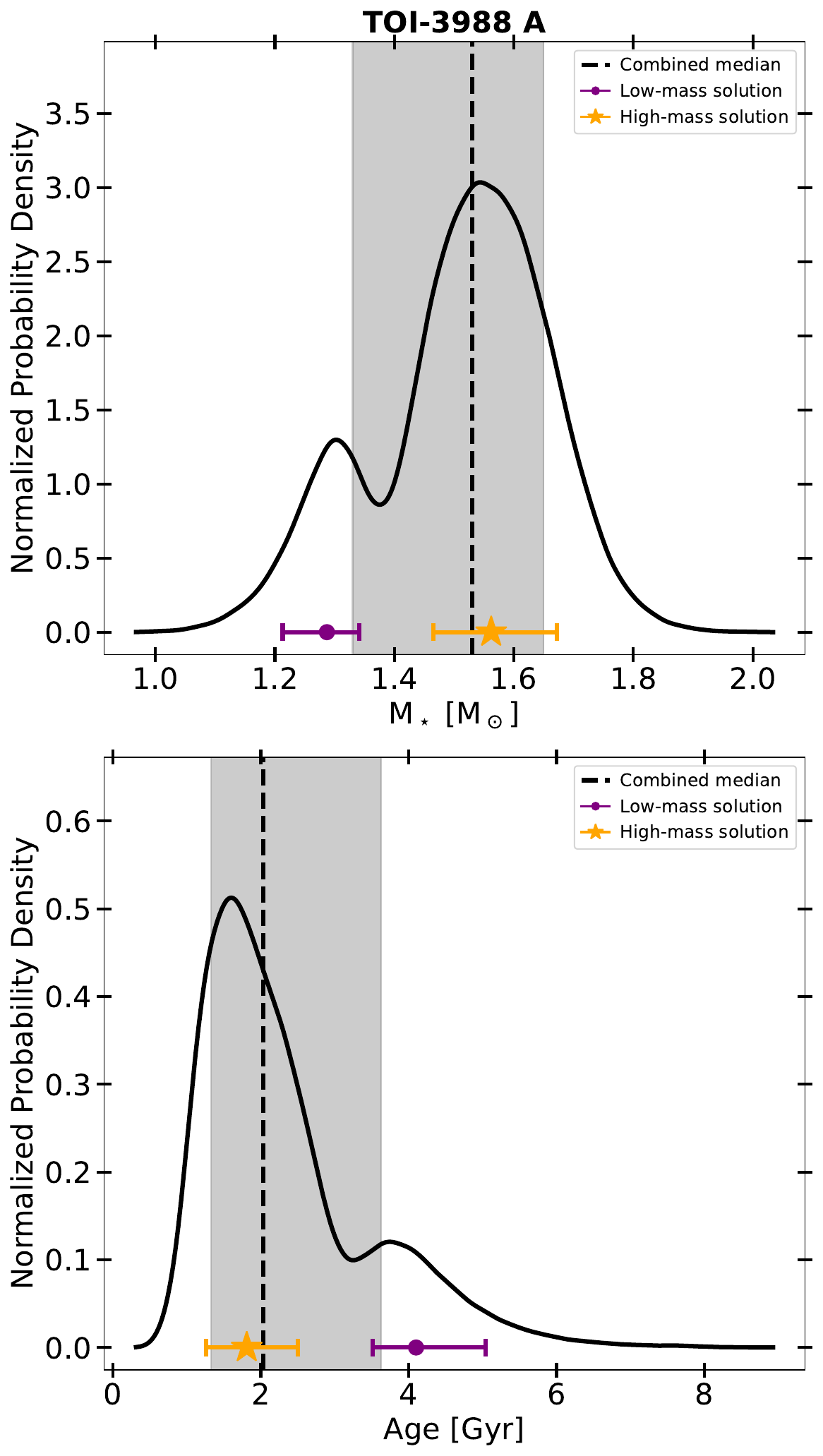}
    \includegraphics[width=0.33\linewidth]{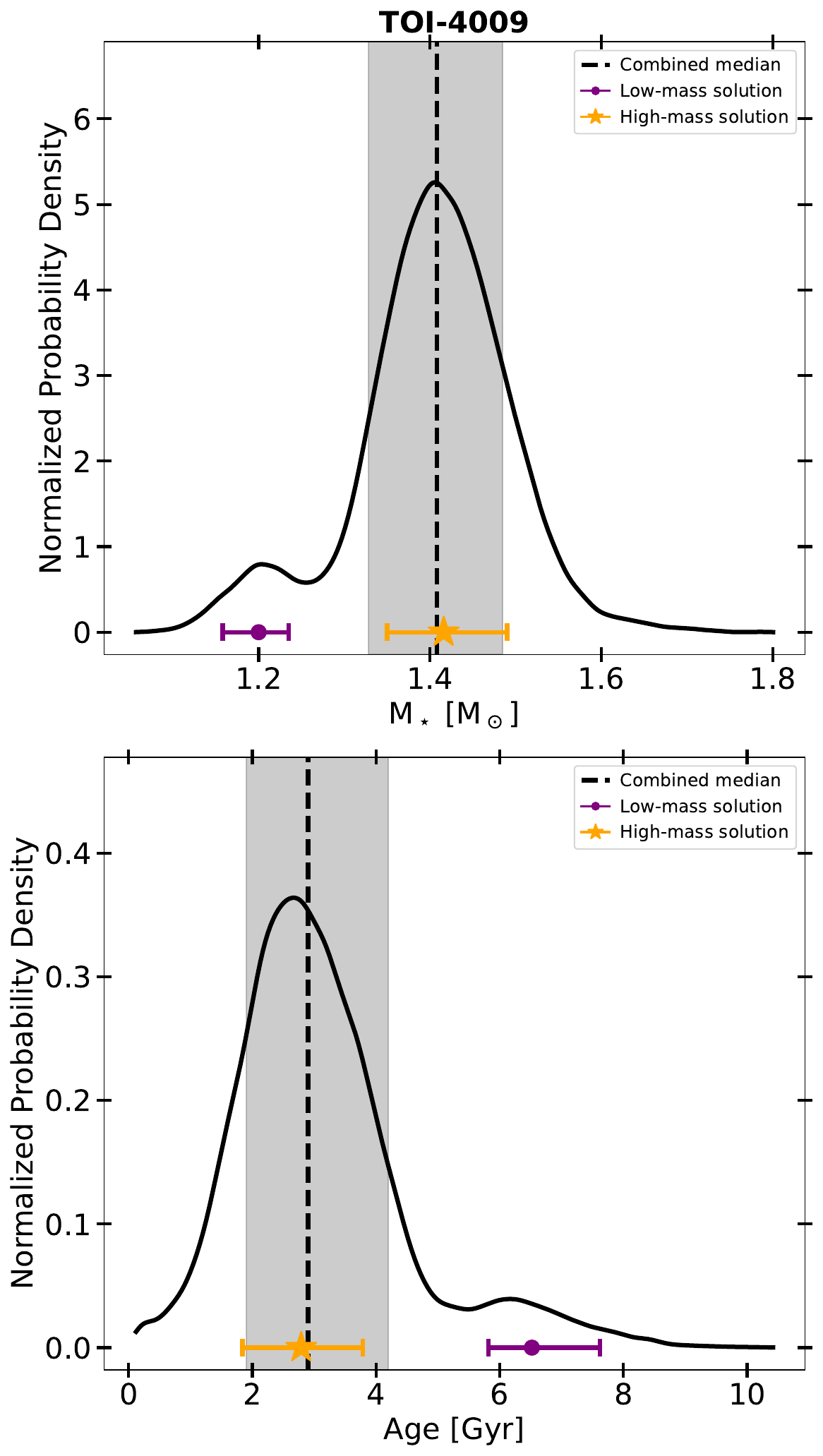}
    \includegraphics[width=0.33\linewidth]{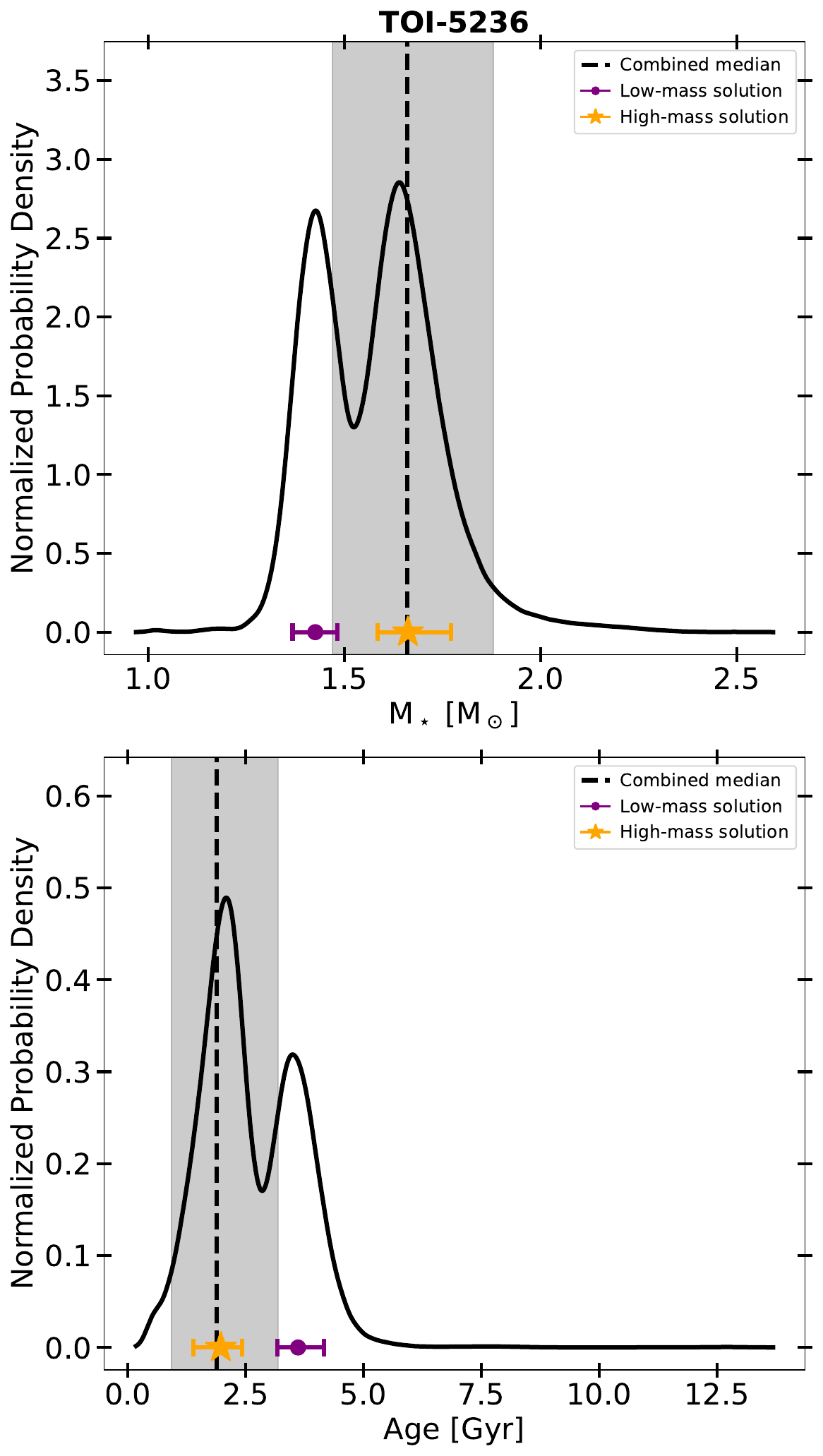}
    \caption{Bimodal posterior distributions in stellar mass and age for TOI-3988\,A, TOI-4009, and TOI-5236. In each frame, the solid black line represents a kernel density estimation of the posterior distribution for that parameter. The dashed black line and grey shaded region represents the median and 68\% confidence interval reported in Table~\ref{tab:median}. The purple marker and error bars represent the median and 68\% confidence interval of the low-mass solution, while the orange marker and error bars represent those of the high-mass solution, as reported in Table~\ref{tab:bimodal}. The solution with the highest probability is marked with a star, while the one with the lowest probability is marked with a circle.}
    \label{fig:bimodal_1}
\end{figure*}

\begin{figure*}
    \centering
    \includegraphics[width=0.33\linewidth]{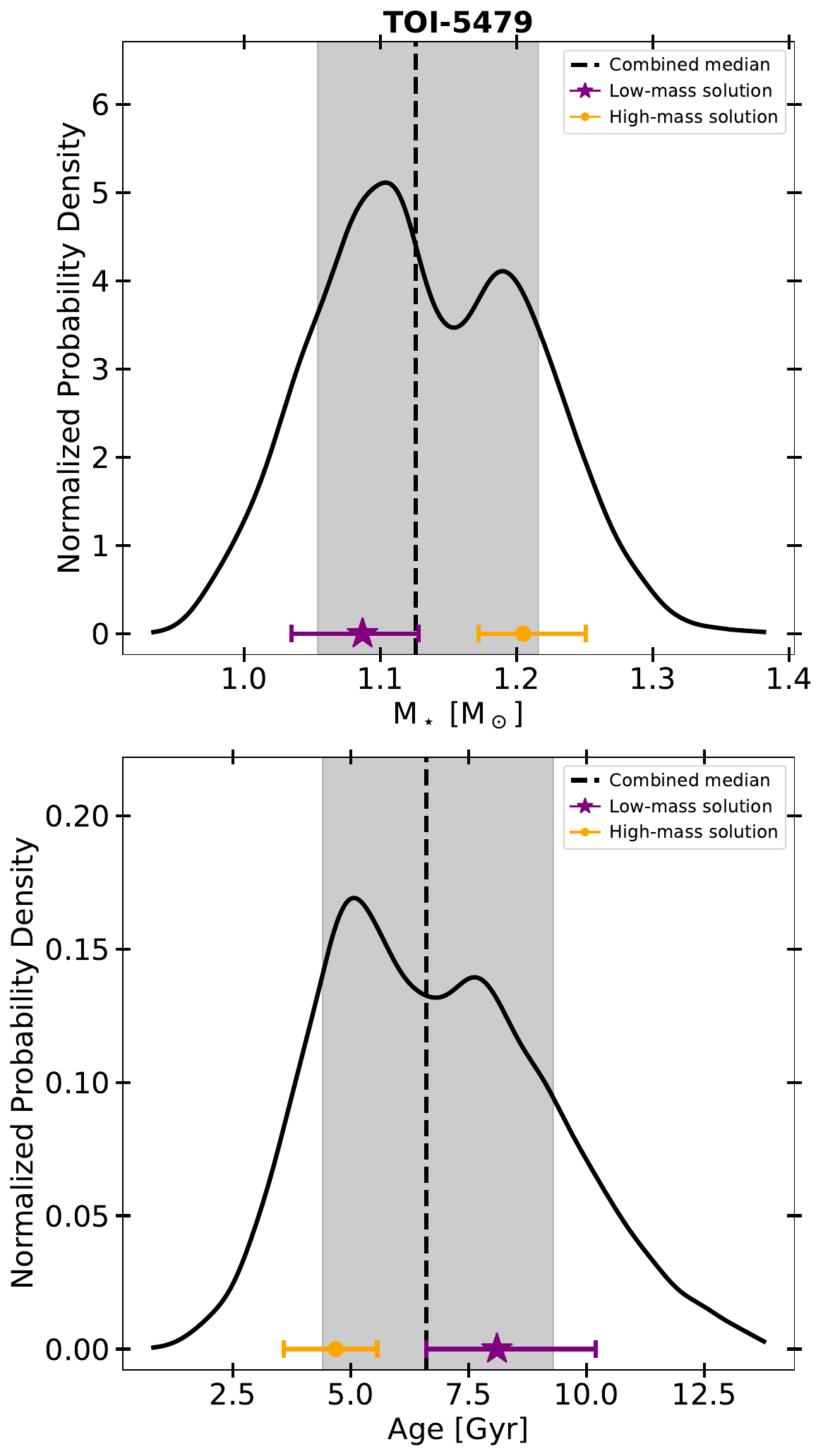}
    \includegraphics[width=0.33\linewidth]{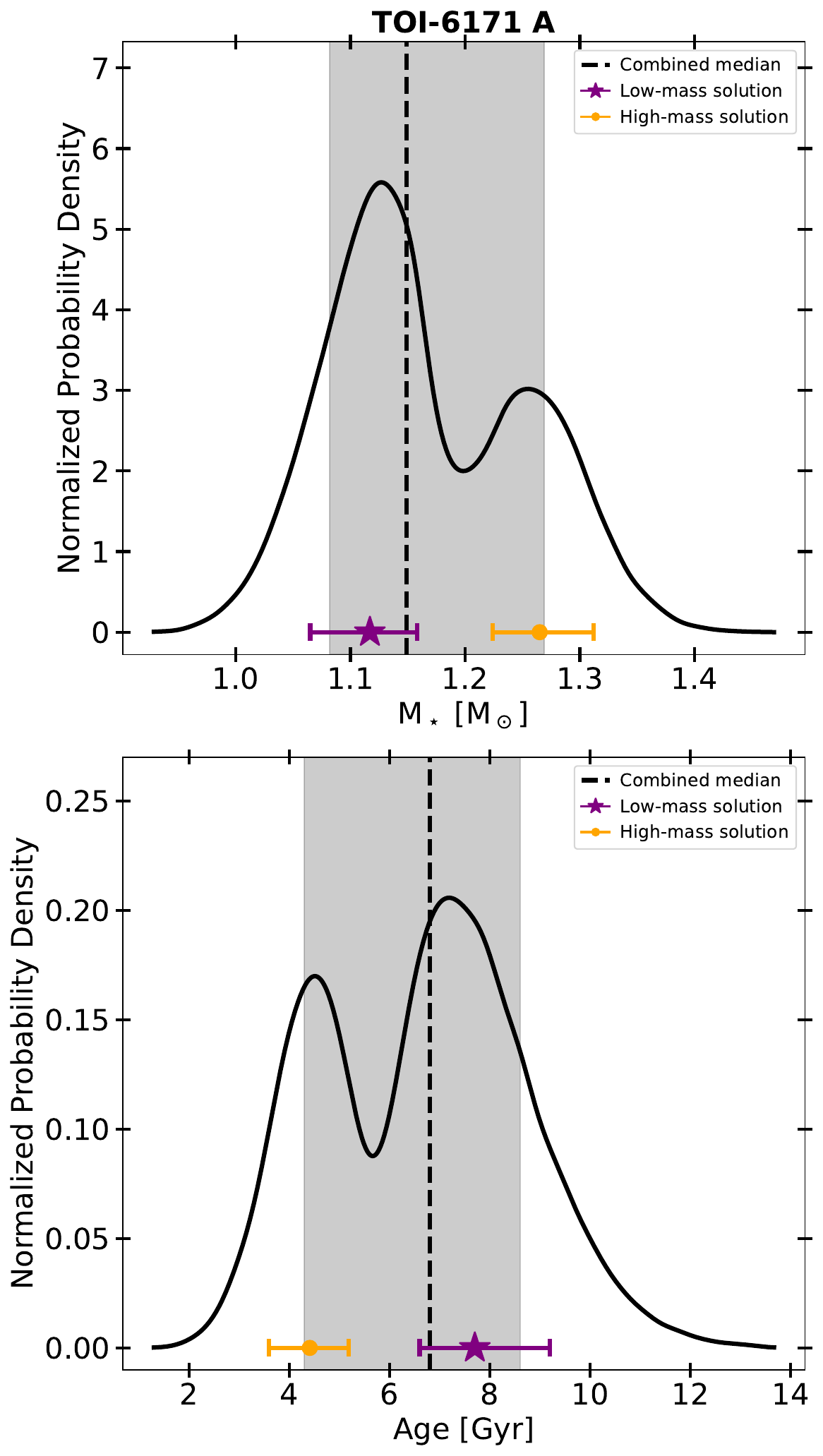}
    \includegraphics[width=0.33\linewidth]{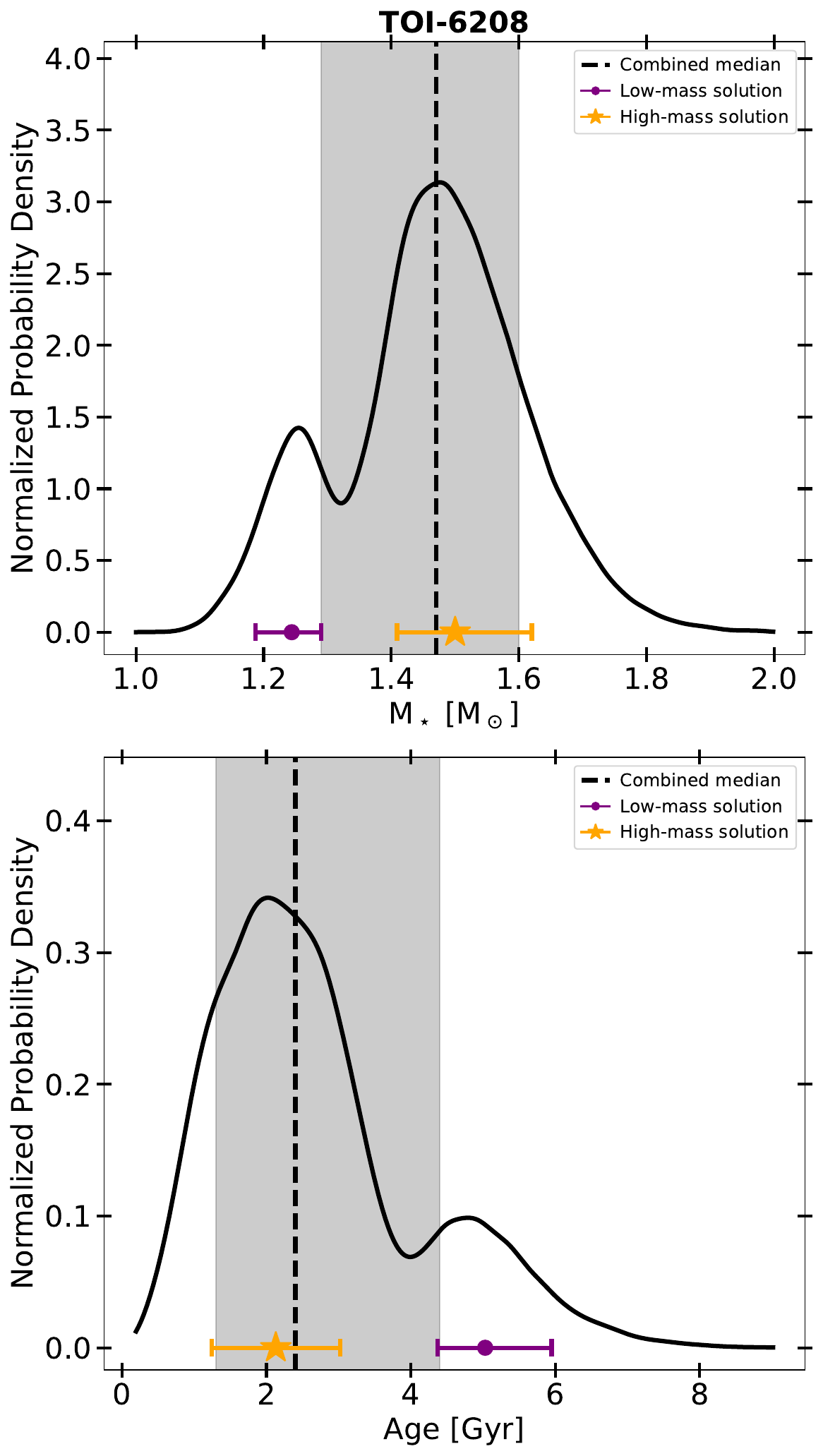}
    \caption{Same as Figure~\ref{fig:bimodal_1}, but for TOI-5479, TOI-6171, and TOI-6208.}
    \label{fig:bimodal_2}
\end{figure*}

\begin{figure*}
    \centering
    \includegraphics[width=0.33\linewidth]{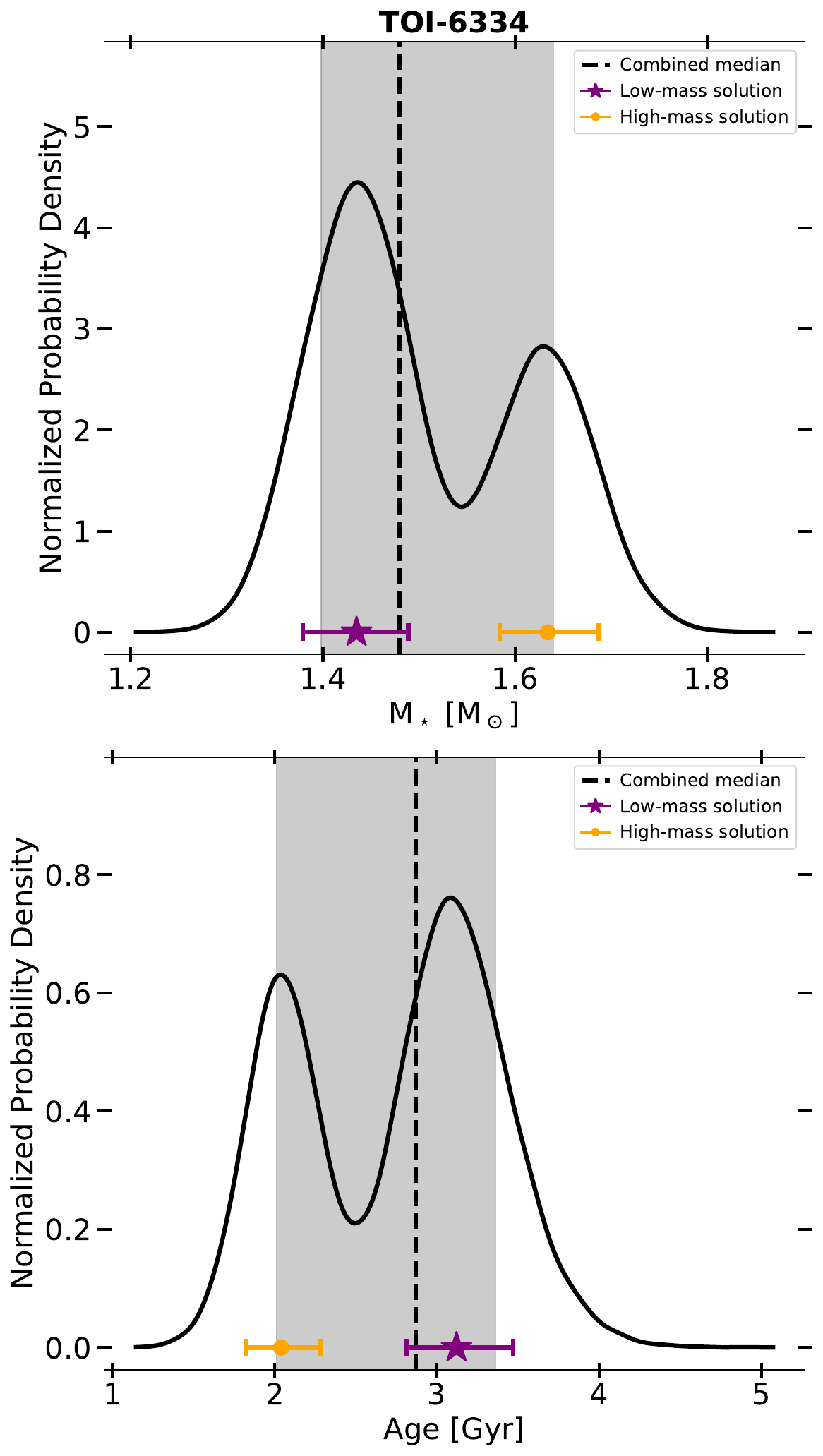}
    \includegraphics[width=0.33\linewidth]{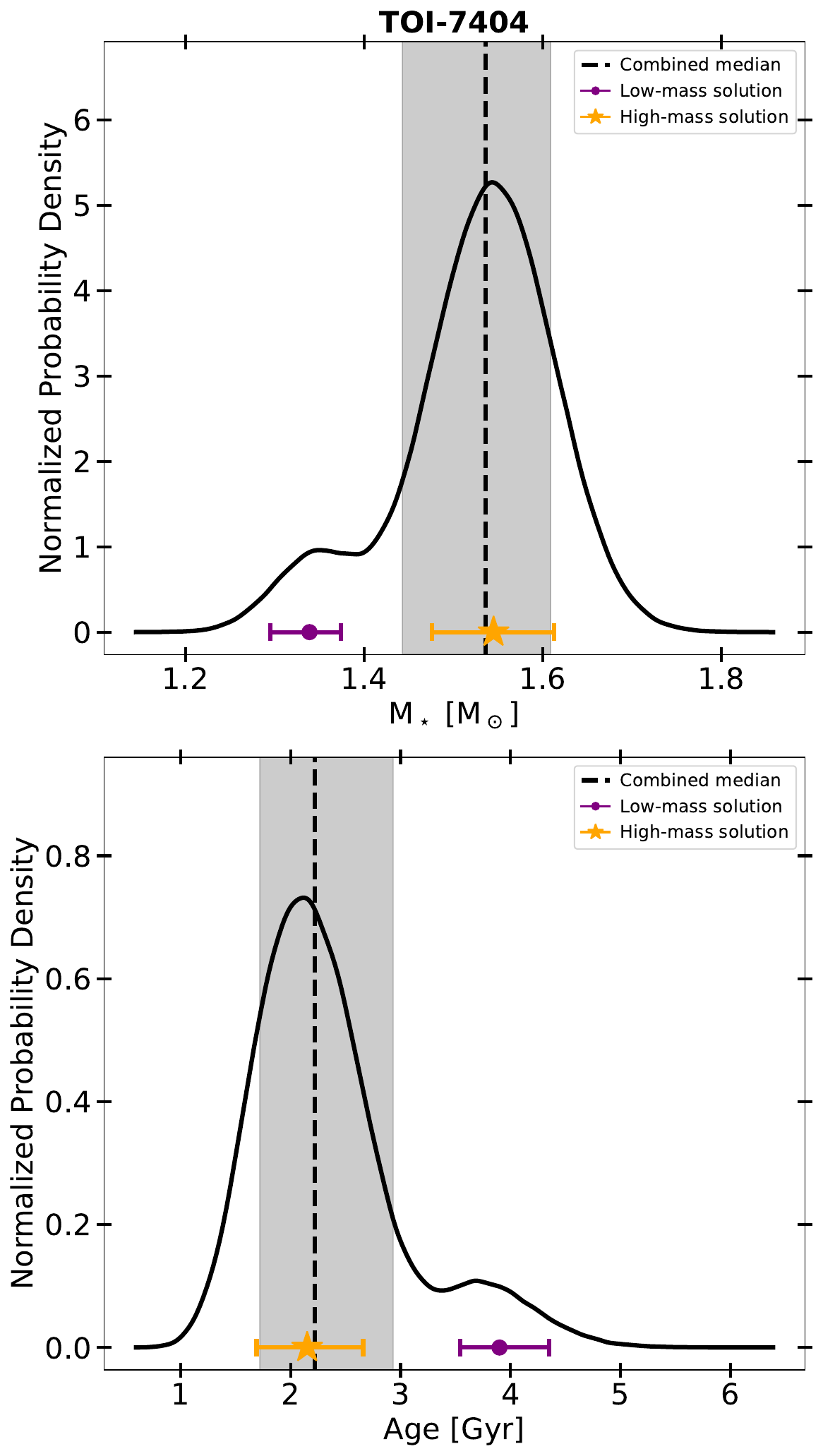}
    \includegraphics[width=0.33\linewidth]{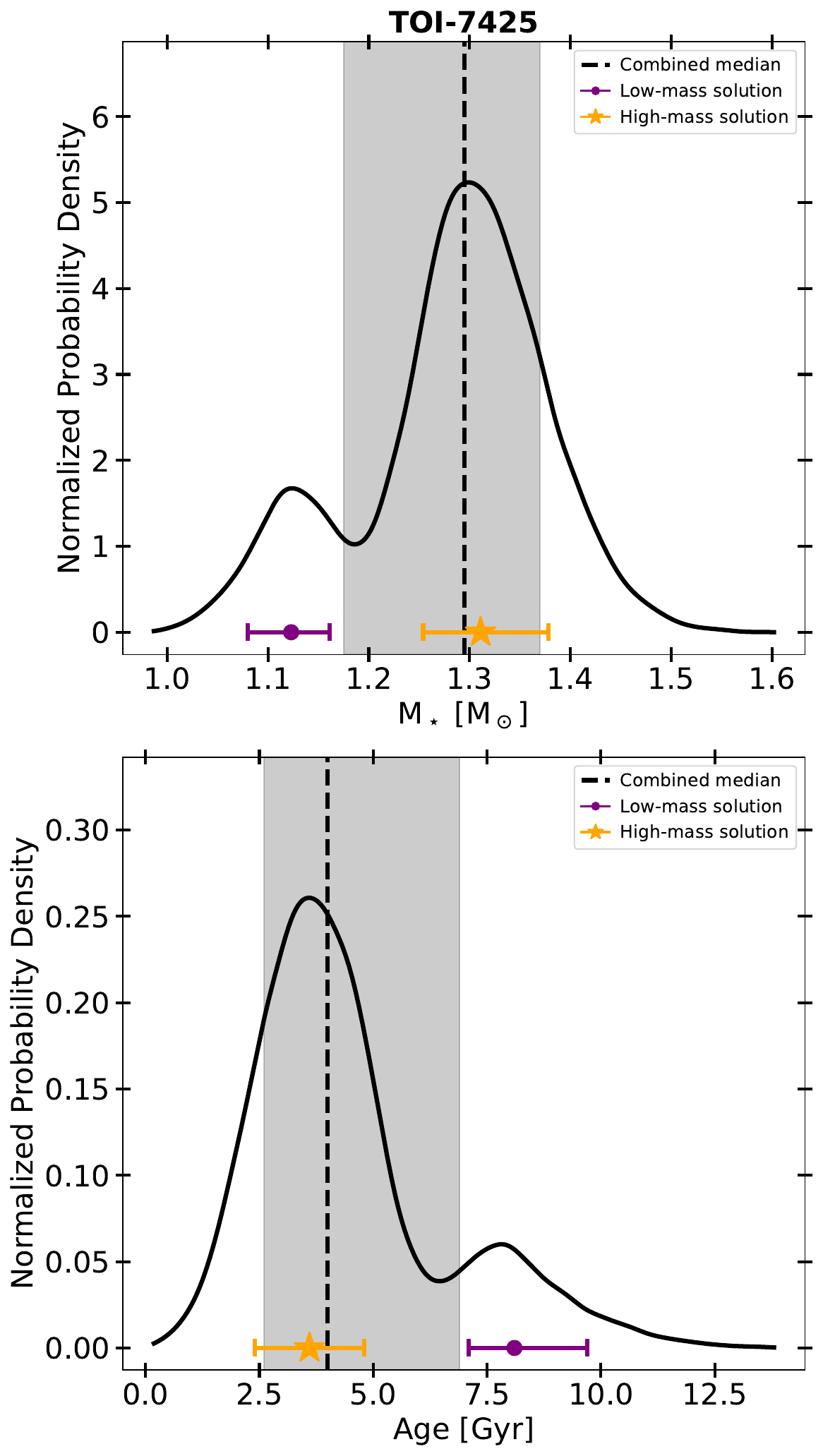}
    \caption{Same as Figure~\ref{fig:bimodal_1}, but for TOI-6334, TOI-7404, and TOI-7425.}
    \label{fig:bimodal_3}
\end{figure*}

Nine of the host stars in this work (TOI-3988\,A, TOI-4009, TOI-5236, TOI-5479, TOI-6171\,A, TOI-6208, TOI-6334, TOI-7404, and TOI-7425) have bimodal stellar mass and age posteriors. These bimodal posteriors occur because the star's observables place it near the subgiant branch, making it difficult for the MCMC and MIST models to discern whether the star is a subgiant star with a smaller mass, or a main sequence star with a larger mass \citep{Stassun:2018}. This degeneracy primarily arises from the star's position on the color-magnitude diagram: uncertain extinction and reddening allows for two stars with quite different ages and masses to explain the available data. In some cases, such as TOI-3988\,A, both solutions are nearly equally favored. TOI-5236's higher-mass dwarf solution is slightly favored with a probability of 60.4\% over its lower-mass subgiant solution. In others, such as TOI-4009, the difference is much starker. TOI-4009's higher-mass dwarf solution is favored with 92.4\% probability over it's lower-mass subgiant solution. In most cases, the medians and 68\% confidence intervals presented in Table~\ref{tab:median} are representative of the most likely parameter space. However, to represent the bimodal nature of these posteriors along with the fact that only one can be the ``true'' solution, we split each bimodal posterior and present each solution, along with their probabilities, in Table~\ref{tab:bimodal}. To split each solution, we found the stellar mass value associated with the local minimum between each peak in the posterior. Then, we used \exofast's built-in script \texttt{splitpdf.pro} to separate MCMC samples with stellar masses above and below the selected cutoff. The probability of each solution is determined by summing the number of samples on each side of the cutoff. Each bimodal stellar mass and age posterior is shown in Figures~\ref{fig:bimodal_1}--\ref{fig:bimodal_3}, demonstrating the overlap between the median solution and the split solutions. Of the nine host stars with bimodal posteriors, three of them favor the subgiant solution.

\renewcommand{\arraystretch}{\bimodalstretch}
\providecommand{\bjdtdb}{\ensuremath{\rm {BJD_{TDB}}}}
\providecommand{\feh}{\ensuremath{\left[{\rm Fe}/{\rm H}\right]}}
\providecommand{\teff}{\ensuremath{T_{\rm eff}}}
\providecommand{\teq}{\ensuremath{T_{\rm eq}}}
\providecommand{\ecosw}{\ensuremath{e\cos{\omega_\star}}}
\providecommand{\esinw}{\ensuremath{e\sin{\omega_\star}}}
\providecommand{\msun}{\ensuremath{\,M_\Sun}}
\providecommand{\rsun}{\ensuremath{\,R_\Sun}}
\providecommand{\lsun}{\ensuremath{\,L_\Sun}}
\providecommand{\mj}{\ensuremath{\,M_{\rm J}}}
\providecommand{\rj}{\ensuremath{\,R_{\rm J}}}
\providecommand{\me}{\ensuremath{\,M_{\rm E}}}
\providecommand{\re}{\ensuremath{\,R_{\rm E}}}
\providecommand{\fave}{\langle F \rangle}
\providecommand{\fluxcgs}{10$^9$ erg s$^{-1}$ cm$^{-2}$}
\providecommand{\tess}{\textit{TESS}\xspace}
\begin{table*}
\centering
\caption{Median Values and 68\% Confidence Intervals for Solutions which are Bimodal in Mass}
\label{tab:bimodal}
\scriptsize
\setlength{\tabcolsep}{3.5pt}
\begin{tabular}{llcccc}
\toprule
&  & \multicolumn{2}{c}{TOI-3988 A} & \multicolumn{2}{c}{TOI-4009}\\
&  & Low-mass solution & High-mass solution & Low-mass solution & High-mass solution\\
&  & 19.7\% probability & 80.3\% probability & 7.6\% probability & 92.4\% probability\\
\midrule
\multicolumn{6}{l}{\textbf{Stellar Parameters}:} \\
$M_\star$ & Mass (\msun) & $1.287^{+0.054}_{-0.074}$ & $1.562^{+0.110}_{-0.097}$ & $1.200^{+0.035}_{-0.042}$ & $1.416^{+0.074}_{-0.066}$ \\
$R_\star$ & Radius (\rsun) & $2.160^{+0.085}_{-0.081}$ & $2.114^{+0.089}_{-0.090}$ & $1.795^{+0.066}_{-0.061}$ & $1.776^{+0.063}_{-0.064}$ \\
$L_\star$ & Luminosity (\lsun) & $6.0^{+1.1}_{-1.0}$ & $7.9^{+1.9}_{-1.3}$ & $3.20^{+0.38}_{-0.34}$ & $3.88^{+0.59}_{-0.49}$ \\
$\rho_\star$ & Density (cgs) & $0.179^{+0.022}_{-0.021}$ & $0.232^{+0.042}_{-0.031}$ & $0.292 \pm 0.029$ & $0.355^{+0.046}_{-0.036}$ \\
$\log{g}$ & Surface gravity (cgs) & $3.877^{+0.034}_{-0.041}$ & $3.979^{+0.055}_{-0.047}$ & $4.008^{+0.028}_{-0.031}$ & $4.089^{+0.039}_{-0.033}$ \\
$T_{\rm eff}$ & Effective temperature (K) & $6150^{+260}_{-300}$ & $6650^{+440}_{-330}$ & $5760 \pm 170$ & $6070^{+250}_{-220}$ \\
$[{\rm Fe/H}]$ & Metallicity (dex) & $-0.011^{+0.095}_{-0.110}$ & $0.01^{+0.11}_{-0.12}$ & $0.425 \pm 0.030$ & $0.422 \pm 0.029$ \\
$[{\rm Fe/H}]_{0}$ & Initial metallicity (dex) & $0.039^{+0.083}_{-0.094}$ & $0.12^{+0.10}_{-0.11}$ & $0.406^{+0.035}_{-0.037}$ & $0.444^{+0.036}_{-0.043}$ \\
Age & Age (Gyr) & $4.10^{+0.94}_{-0.59}$ & $1.81^{+0.69}_{-0.55}$ & $6.52^{+1.10}_{-0.70}$ & $2.79^{+1.00}_{-0.95}$ \\
EEP & Equivalent evolutionary phase & $453.1^{+5.6}_{-6.6}$ & $378^{+23}_{-24}$ & $451.9^{+4.4}_{-3.7}$ & $382^{+23}_{-31}$ \\
$A_V$ & V-band extinction (mag) & $0.72^{+0.18}_{-0.23}$ & $1.06^{+0.23}_{-0.20}$ & $0.31^{+0.15}_{-0.16}$ & $0.56^{+0.17}_{-0.18}$ \\
$d$ & Distance (pc) & $769.3^{+9.4}_{-9.2}$ & $771.0^{+9.5}_{-9.2}$ & $644.3^{+6.7}_{-6.6}$ & $645.4^{+6.8}_{-6.7}$ \\
\multicolumn{6}{l}{\textbf{Planetary Parameters}:} \\
$P$ & Period (days) & $2.8744223^{+0.0000027}_{-0.0000028}$ & $2.8744225 \pm 0.0000027$ & $4.3271527^{+0.0000037}_{-0.0000038}$ & $4.3271526 \pm 0.0000038$ \\
$R_{\rm P}$ & Radius (\rj) & $1.551^{+0.067}_{-0.065}$ & $1.514^{+0.070}_{-0.069}$ & $1.452^{+0.065}_{-0.061}$ & $1.420^{+0.064}_{-0.062}$ \\
$M_{\rm P}$ & Mass (\mj) & $1.04 \pm 0.13$ & $1.18 \pm 0.14$ & $0.646^{+0.100}_{-0.096}$ & $0.73 \pm 0.11$ \\
$T_C$ & Time of conjunction (\bjdtdb) & $2459904.65113^{+0.00065}_{-0.00068}$ & $2459904.65121^{+0.00063}_{-0.00066}$ & $2459882.41235^{+0.00082}_{-0.00087}$ & $2459882.41241^{+0.00072}_{-0.00077}$ \\
$T_0$ & Optimal conjunction time (\bjdtdb) & $2459893.15381^{+0.00057}_{-0.00056}$ & $2459893.15390 \pm 0.00056$ & $2460046.84467^{+0.00051}_{-0.00052}$ & $2460046.84470 \pm 0.00051$ \\
$a$ & Semi-major axis (AU) & $0.04305^{+0.00059}_{-0.00084}$ & $0.04591^{+0.00100}_{-0.00097}$ & $0.05523^{+0.00053}_{-0.00065}$ & $0.05836^{+0.00100}_{-0.00092}$ \\
$i$ & Inclination (Degrees) & $82.43^{+1.10}_{-0.89}$ & $83.81^{+1.10}_{-0.91}$ & $81.83^{+0.53}_{-0.67}$ & $82.75^{+0.53}_{-0.49}$ \\
$e$ & Eccentricity & $0.053^{+0.066}_{-0.037}$ & $0.079^{+0.078}_{-0.055}$ & $0.066^{+0.054}_{-0.045}$ & $0.046^{+0.051}_{-0.032}$ \\
$\omega_\star$ & Argument of periastron (Degrees) & $150 \pm 100$ & $-102^{+39}_{-68}$ & $114^{+57}_{-66}$ & $-147^{+94}_{-99}$ \\
$\teq$ & Equilibrium temperature (K) & $2099^{+81}_{-84}$ & $2180^{+100}_{-84}$ & $1584^{+42}_{-40}$ & $1617^{+49}_{-47}$ \\
$\tau_{\rm circ}$ & Tidal circularization timescale (Gyr) & $0.0375^{+0.0110}_{-0.0098}$ & $0.051^{+0.017}_{-0.013}$ & $0.180^{+0.060}_{-0.051}$ & $0.263^{+0.083}_{-0.066}$ \\
$K$ & RV semi-amplitude (m/s) & $125 \pm 15$ & $125 \pm 14$ & $70^{+11}_{-10}$ & $71 \pm 10$ \\
$\dot{\gamma}$ & RV slope (m/s/day) & $-0.103^{+0.057}_{-0.055}$ & $-0.099^{+0.054}_{-0.053}$ & $-0.062^{+0.054}_{-0.053}$ & $-0.065 \pm 0.051$ \\
$R_{\rm P}/R_\star$ & Radius of planet in stellar radii  & $0.07381^{+0.00099}_{-0.00110}$ & $0.0736 \pm 0.0010$ & $0.0830^{+0.0017}_{-0.0015}$ & $0.0821^{+0.0016}_{-0.0015}$ \\
$a/R_\star$ & Semi-major axis in stellar radii  & $4.28^{+0.17}_{-0.18}$ & $4.66^{+0.27}_{-0.22}$ & $6.61^{+0.21}_{-0.23}$ & $7.06^{+0.29}_{-0.25}$ \\
Depth & \tess flux decrement at mid-transit & $0.00588^{+0.00013}_{-0.00012}$ & $0.00580 \pm 0.00012$ & $0.00580^{+0.00019}_{-0.00018}$ & $0.00589^{+0.00019}_{-0.00018}$ \\
$\tau$ & Ingress/egress transit duration (days) & $0.0197^{+0.0025}_{-0.0026}$ & $0.0187^{+0.0030}_{-0.0025}$ & $0.0429^{+0.0060}_{-0.0043}$ & $0.0406^{+0.0049}_{-0.0038}$ \\
$T_{14}$ & Total transit duration (days) & $0.1974^{+0.0023}_{-0.0025}$ & $0.1961^{+0.0026}_{-0.0025}$ & $0.1220^{+0.0023}_{-0.0025}$ & $0.1204 \pm 0.0024$ \\
$b$ & Transit impact parameter & $0.567^{+0.059}_{-0.093}$ & $0.539^{+0.080}_{-0.110}$ & $0.8993^{+0.0079}_{-0.0082}$ & $0.8964^{+0.0077}_{-0.0080}$ \\
$\rho_{\rm P}$ & Density (cgs) & $0.345^{+0.064}_{-0.057}$ & $0.420^{+0.088}_{-0.072}$ & $0.261^{+0.056}_{-0.050}$ & $0.316^{+0.070}_{-0.060}$ \\
$\log{g_{\rm P}}$ & Surface gravity (cgs) & $3.029^{+0.061}_{-0.068}$ & $3.105^{+0.067}_{-0.068}$ & $2.879^{+0.073}_{-0.081}$ & $2.953^{+0.074}_{-0.080}$ \\
$M_{\rm P}/M_\star$ & Mass ratio  & $0.000777^{+0.000096}_{-0.000093}$ & $0.000720^{+0.000085}_{-0.000083}$ & $0.000515^{+0.000077}_{-0.000076}$ & $0.000493 \pm 0.000071$ \\
$d/R_\star$ & Separation at mid-transit  & $4.26^{+0.32}_{-0.39}$ & $4.89^{+0.65}_{-0.43}$ & $6.34^{+0.44}_{-0.51}$ & $7.10^{+0.56}_{-0.43}$ \\
\bottomrule
\end{tabular}
\end{table*}
\providecommand{\bjdtdb}{\ensuremath{\rm {BJD_{TDB}}}}
\providecommand{\feh}{\ensuremath{\left[{\rm Fe}/{\rm H}\right]}}
\providecommand{\teff}{\ensuremath{T_{\rm eff}}}
\providecommand{\teq}{\ensuremath{T_{\rm eq}}}
\providecommand{\ecosw}{\ensuremath{e\cos{\omega_\star}}}
\providecommand{\esinw}{\ensuremath{e\sin{\omega_\star}}}
\providecommand{\msun}{\ensuremath{\,M_\Sun}}
\providecommand{\rsun}{\ensuremath{\,R_\Sun}}
\providecommand{\lsun}{\ensuremath{\,L_\Sun}}
\providecommand{\mj}{\ensuremath{\,M_{\rm J}}}
\providecommand{\rj}{\ensuremath{\,R_{\rm J}}}
\providecommand{\me}{\ensuremath{\,M_{\rm E}}}
\providecommand{\re}{\ensuremath{\,R_{\rm E}}}
\providecommand{\fave}{\langle F \rangle}
\providecommand{\fluxcgs}{10$^9$ erg s$^{-1}$ cm$^{-2}$}
\providecommand{\tess}{\textit{TESS}\xspace}
\begin{table*}
\centering
\contcaption{}
\scriptsize
\setlength{\tabcolsep}{3.5pt}
\begin{tabular}{llcccc}
\toprule
&  & \multicolumn{2}{c}{TOI-5236} & \multicolumn{2}{c}{TOI-5479}\\
&  & Low-mass solution & High-mass solution & Low-mass solution & High-mass solution\\
&  & 39.6\% probability & 60.4\% probability & 60.9\% probability & 39.1\% probability\\
\midrule
\multicolumn{6}{l}{\textbf{Stellar Parameters}:} \\
$M_\star$ & Mass (\msun) & $1.426^{+0.056}_{-0.058}$ & $1.662^{+0.110}_{-0.077}$ & $1.087^{+0.041}_{-0.052}$ & $1.205^{+0.046}_{-0.033}$ \\
$R_\star$ & Radius (\rsun) & $2.49 \pm 0.14$ & $2.46^{+0.14}_{-0.12}$ & $1.460^{+0.056}_{-0.052}$ & $1.452^{+0.054}_{-0.051}$ \\
$L_\star$ & Luminosity (\lsun) & $5.98^{+1.10}_{-0.90}$ & $7.9^{+2.1}_{-1.3}$ & $2.200^{+0.100}_{-0.094}$ & $2.236^{+0.098}_{-0.100}$ \\
$\rho_\star$ & Density (cgs) & $0.130^{+0.021}_{-0.018}$ & $0.157^{+0.027}_{-0.022}$ & $0.489^{+0.064}_{-0.057}$ & $0.556^{+0.066}_{-0.057}$ \\
$\log{g}$ & Surface gravity (cgs) & $3.798^{+0.043}_{-0.042}$ & $3.876^{+0.050}_{-0.043}$ & $4.143 \pm 0.039$ & $4.196^{+0.035}_{-0.032}$ \\
$T_{\rm eff}$ & Effective temperature (K) & $5740^{+240}_{-270}$ & $6150^{+360}_{-230}$ & $5820 \pm 110$ & $5860 \pm 110$ \\
$[{\rm Fe/H}]$ & Metallicity (dex) & $0.372^{+0.068}_{-0.069}$ & $0.378^{+0.059}_{-0.065}$ & $0.203^{+0.074}_{-0.076}$ & $0.239^{+0.072}_{-0.073}$ \\
$[{\rm Fe/H}]_{0}$ & Initial metallicity (dex) & $0.351^{+0.064}_{-0.068}$ & $0.406^{+0.056}_{-0.063}$ & $0.237^{+0.065}_{-0.069}$ & $0.269^{+0.064}_{-0.061}$ \\
Age & Age (Gyr) & $3.61^{+0.55}_{-0.44}$ & $1.97^{+0.45}_{-0.58}$ & $8.1^{+2.1}_{-1.5}$ & $4.68^{+0.88}_{-1.10}$ \\
EEP & Equivalent evolutionary phase & $457.3^{+8.6}_{-7.1}$ & $397^{+12}_{-22}$ & $441.2^{+8.2}_{-10.0}$ & $407^{+13}_{-23}$ \\
$A_V$ & V-band extinction (mag) & $0.37^{+0.25}_{-0.24}$ & $0.74^{+0.30}_{-0.25}$ & $0.052^{+0.038}_{-0.036}$ & $0.063^{+0.034}_{-0.039}$ \\
$d$ & Distance (pc) & $565.8 \pm 5.0$ & $566.3^{+5.1}_{-4.9}$ & $600.5 \pm 7.7$ & $601.8^{+7.8}_{-7.7}$ \\
\multicolumn{6}{l}{\textbf{Planetary Parameters}:} \\
$P$ & Period (days) & $10.997519 \pm 0.000020$ & $10.997517 \pm 0.000020$ & $15.130540 \pm 0.000016$ & $15.130539^{+0.000016}_{-0.000015}$ \\
$R_{\rm P}$ & Radius (\rj) & $1.391^{+0.092}_{-0.088}$ & $1.371^{+0.090}_{-0.081}$ & $1.058^{+0.045}_{-0.041}$ & $1.049^{+0.042}_{-0.039}$ \\
$M_{\rm P}$ & Mass (\mj) & $1.50 \pm 0.14$ & $1.69^{+0.18}_{-0.16}$ & $2.44 \pm 0.26$ & $2.61^{+0.28}_{-0.27}$ \\
$T_C$ & Time of conjunction (\bjdtdb) & $2459807.1603 \pm 0.0016$ & $2459807.1601 \pm 0.0016$ & $2459570.00220^{+0.00099}_{-0.00100}$ & $2459570.00229^{+0.00094}_{-0.00098}$ \\
$T_0$ & Optimal conjunction time (\bjdtdb) & $2460302.0492 \pm 0.0013$ & $2460302.0491 \pm 0.0013$ & $2460205.48565^{+0.00063}_{-0.00064}$ & $2460205.48565^{+0.00062}_{-0.00064}$ \\
$a$ & Semi-major axis (AU) & $0.1090^{+0.0014}_{-0.0015}$ & $0.1147^{+0.0025}_{-0.0018}$ & $0.1232^{+0.0015}_{-0.0020}$ & $0.1275^{+0.0016}_{-0.0012}$ \\
$i$ & Inclination (Degrees) & $86.43^{+0.44}_{-0.42}$ & $86.86^{+0.44}_{-0.39}$ & $88.49^{+0.97}_{-0.84}$ & $88.88^{+0.76}_{-0.79}$ \\
$e$ & Eccentricity & $0.090^{+0.073}_{-0.060}$ & $0.138 \pm 0.080$ & $0.598^{+0.031}_{-0.036}$ & $0.589^{+0.033}_{-0.036}$ \\
$\omega_\star$ & Argument of periastron (Degrees) & $-81^{+50}_{-25}$ & $-86^{+27}_{-15}$ & $163.4^{+5.8}_{-6.1}$ & $165.9^{+5.6}_{-5.8}$ \\
$\teq$ & Equilibrium temperature (K) & $1319^{+52}_{-47}$ & $1379^{+70}_{-54}$ & $966^{+12}_{-11}$ & $952 \pm 11$ \\
$\tau_{\rm circ}$ & Tidal circularization timescale (Gyr) & $30.8^{+12.0}_{-9.0}$ & $35^{+16}_{-12}$ & $13.7^{+9.8}_{-5.7}$ & $18.8^{+13.0}_{-7.8}$ \\
$K$ & RV semi-amplitude (m/s) & $108.8^{+10.0}_{-9.9}$ & $110^{+11}_{-10}$ & $236^{+27}_{-26}$ & $233 \pm 27$ \\
$\dot{\gamma}$ & RV slope (m/s/day) & $0.038 \pm 0.031$ & $0.039^{+0.032}_{-0.031}$ & $-0.075^{+0.063}_{-0.062}$ & $-0.075^{+0.065}_{-0.063}$ \\
$R_{\rm P}/R_\star$ & Radius of planet in stellar radii  & $0.0574 \pm 0.0013$ & $0.0572 \pm 0.0013$ & $0.07444^{+0.00110}_{-0.00099}$ & $0.07425^{+0.00100}_{-0.00090}$ \\
$a/R_\star$ & Semi-major axis in stellar radii  & $9.40^{+0.49}_{-0.46}$ & $10.02^{+0.55}_{-0.50}$ & $18.10^{+0.76}_{-0.74}$ & $18.90^{+0.72}_{-0.67}$ \\
Depth & \tess flux decrement at mid-transit & $0.00356 \pm 0.00011$ & $0.00349 \pm 0.00011$ & $0.00643^{+0.00015}_{-0.00016}$ & $0.00641^{+0.00016}_{-0.00015}$ \\
$\tau$ & Ingress/egress transit duration (days) & $0.0302^{+0.0053}_{-0.0045}$ & $0.0296^{+0.0054}_{-0.0051}$ & $0.01390^{+0.00200}_{-0.00089}$ & $0.01353^{+0.00160}_{-0.00058}$ \\
$T_{14}$ & Total transit duration (days) & $0.3422^{+0.0056}_{-0.0050}$ & $0.3405^{+0.0056}_{-0.0053}$ & $0.1884^{+0.0020}_{-0.0018}$ & $0.1880^{+0.0018}_{-0.0017}$ \\
$b$ & Transit impact parameter & $0.636^{+0.060}_{-0.078}$ & $0.629^{+0.063}_{-0.097}$ & $0.26^{+0.16}_{-0.17}$ & $0.21^{+0.16}_{-0.14}$ \\
$\rho_{\rm P}$ & Density (cgs) & $0.69^{+0.16}_{-0.13}$ & $0.81^{+0.19}_{-0.15}$ & $2.54^{+0.45}_{-0.41}$ & $2.79^{+0.47}_{-0.41}$ \\
$\log{g_{\rm P}}$ & Surface gravity (cgs) & $3.283^{+0.067}_{-0.068}$ & $3.346^{+0.068}_{-0.069}$ & $3.731^{+0.058}_{-0.063}$ & $3.768^{+0.055}_{-0.058}$ \\
$M_{\rm P}/M_\star$ & Mass ratio  & $0.001007 \pm 0.000092$ & $0.000963 \pm 0.000090$ & $0.00215 \pm 0.00022$ & $0.00206 \pm 0.00022$ \\
$d/R_\star$ & Separation at mid-transit  & $10.07^{+1.10}_{-0.92}$ & $11.3^{+1.3}_{-1.2}$ & $9.95^{+1.00}_{-0.94}$ & $10.79^{+1.10}_{-0.91}$ \\
\bottomrule
\end{tabular}
\end{table*}
\providecommand{\bjdtdb}{\ensuremath{\rm {BJD_{TDB}}}}
\providecommand{\feh}{\ensuremath{\left[{\rm Fe}/{\rm H}\right]}}
\providecommand{\teff}{\ensuremath{T_{\rm eff}}}
\providecommand{\teq}{\ensuremath{T_{\rm eq}}}
\providecommand{\ecosw}{\ensuremath{e\cos{\omega_\star}}}
\providecommand{\esinw}{\ensuremath{e\sin{\omega_\star}}}
\providecommand{\msun}{\ensuremath{\,M_\Sun}}
\providecommand{\rsun}{\ensuremath{\,R_\Sun}}
\providecommand{\lsun}{\ensuremath{\,L_\Sun}}
\providecommand{\mj}{\ensuremath{\,M_{\rm J}}}
\providecommand{\rj}{\ensuremath{\,R_{\rm J}}}
\providecommand{\me}{\ensuremath{\,M_{\rm E}}}
\providecommand{\re}{\ensuremath{\,R_{\rm E}}}
\providecommand{\fave}{\langle F \rangle}
\providecommand{\fluxcgs}{10$^9$ erg s$^{-1}$ cm$^{-2}$}
\providecommand{\tess}{\textit{TESS}\xspace}
\begin{table*}
\centering
\contcaption{}
\scriptsize
\setlength{\tabcolsep}{3.5pt}
\begin{tabular}{llcccc}
\toprule
&  & \multicolumn{2}{c}{TOI-6171 A} & \multicolumn{2}{c}{TOI-6208}\\
&  & Low-mass solution & High-mass solution & Low-mass solution & High-mass solution\\
&  & 64.1\% probability & 35.9\% probability & 18.6\% probability & 81.4\% probability\\
\midrule
\multicolumn{6}{l}{\textbf{Stellar Parameters}:} \\
$M_\star$ & Mass (\msun) & $1.117^{+0.041}_{-0.052}$ & $1.265^{+0.047}_{-0.041}$ & $1.244^{+0.046}_{-0.057}$ & $1.500^{+0.120}_{-0.091}$ \\
$R_\star$ & Radius (\rsun) & $1.660^{+0.063}_{-0.066}$ & $1.627^{+0.062}_{-0.057}$ & $1.990^{+0.075}_{-0.076}$ & $1.949^{+0.079}_{-0.077}$ \\
$L_\star$ & Luminosity (\lsun) & $2.84 \pm 0.22$ & $2.90 \pm 0.20$ & $4.54^{+0.63}_{-0.54}$ & $5.72^{+1.30}_{-0.88}$ \\
$\rho_\star$ & Density (cgs) & $0.342^{+0.047}_{-0.037}$ & $0.415^{+0.046}_{-0.042}$ & $0.222^{+0.026}_{-0.024}$ & $0.285^{+0.049}_{-0.036}$ \\
$\log{g}$ & Surface gravity (cgs) & $4.044^{+0.038}_{-0.036}$ & $4.118^{+0.032}_{-0.031}$ & $3.934^{+0.033}_{-0.036}$ & $4.033^{+0.054}_{-0.044}$ \\
$T_{\rm eff}$ & Effective temperature (K) & $5820 \pm 150$ & $5900 \pm 130$ & $5980^{+210}_{-200}$ & $6380^{+390}_{-280}$ \\
$[{\rm Fe/H}]$ & Metallicity (dex) & $0.185^{+0.100}_{-0.099}$ & $0.266^{+0.089}_{-0.091}$ & $0.126^{+0.100}_{-0.095}$ & $0.16 \pm 0.10$ \\
$[{\rm Fe/H}]_{0}$ & Initial metallicity (dex) & $0.212^{+0.087}_{-0.088}$ & $0.285^{+0.079}_{-0.076}$ & $0.152^{+0.087}_{-0.082}$ & $0.254^{+0.098}_{-0.096}$ \\
Age & Age (Gyr) & $7.7^{+1.5}_{-1.1}$ & $4.41^{+0.77}_{-0.82}$ & $5.03^{+0.92}_{-0.66}$ & $2.13 \pm 0.89$ \\
EEP & Equivalent evolutionary phase & $451.8^{+5.0}_{-6.4}$ & $410^{+11}_{-19}$ & $452.0 \pm 5.5$ & $375^{+27}_{-28}$ \\
$A_V$ & V-band extinction (mag) & $0.169^{+0.056}_{-0.079}$ & $0.205^{+0.034}_{-0.057}$ & $0.31 \pm 0.16$ & $0.60^{+0.23}_{-0.20}$ \\
$d$ & Distance (pc) & $485.4^{+4.9}_{-4.8}$ & $485.7^{+4.9}_{-4.8}$ & $683.2^{+6.5}_{-6.3}$ & $683.8^{+6.6}_{-6.4}$ \\
\multicolumn{6}{l}{\textbf{Planetary Parameters}:} \\
$P$ & Period (days) & $3.7477271^{+0.0000044}_{-0.0000045}$ & $3.7477271 \pm 0.0000044$ & $5.2266136^{+0.0000062}_{-0.0000063}$ & $5.2266134 \pm 0.0000062$ \\
$R_{\rm P}$ & Radius (\rj) & $0.961^{+0.037}_{-0.038}$ & $0.942^{+0.036}_{-0.034}$ & $1.505^{+0.068}_{-0.064}$ & $1.477^{+0.068}_{-0.064}$ \\
$M_{\rm P}$ & Mass (\mj) & $1.30^{+0.16}_{-0.15}$ & $1.44 \pm 0.18$ & $0.787^{+0.094}_{-0.100}$ & $0.90 \pm 0.12$ \\
$T_C$ & Time of conjunction (\bjdtdb) & $2459848.43152^{+0.00063}_{-0.00064}$ & $2459848.43149^{+0.00064}_{-0.00063}$ & $2459872.10377^{+0.00078}_{-0.00076}$ & $2459872.10373^{+0.00076}_{-0.00074}$ \\
$T_0$ & Optimal conjunction time (\bjdtdb) & $2460249.43861^{+0.00042}_{-0.00043}$ & $2460249.43860^{+0.00043}_{-0.00041}$ & $2460227.51361^{+0.00061}_{-0.00062}$ & $2460227.51364 \pm 0.00060$ \\
$a$ & Semi-major axis (AU) & $0.04902^{+0.00059}_{-0.00078}$ & $0.05108^{+0.00062}_{-0.00055}$ & $0.06341^{+0.00077}_{-0.00098}$ & $0.0675^{+0.0017}_{-0.0014}$ \\
$i$ & Inclination (Degrees) & $88.5^{+1.0}_{-1.2}$ & $88.74^{+0.87}_{-1.10}$ & $87.3^{+1.7}_{-1.2}$ & $88.0^{+1.3}_{-1.1}$ \\
$e$ & Eccentricity & $0.079^{+0.054}_{-0.047}$ & $0.137^{+0.048}_{-0.042}$ & $0.147^{+0.060}_{-0.064}$ & $0.146^{+0.069}_{-0.075}$ \\
$\omega_\star$ & Argument of periastron (Degrees) & $-80^{+38}_{-43}$ & $-82^{+21}_{-25}$ & $19^{+30}_{-29}$ & $-12^{+29}_{-28}$ \\
$\teq$ & Equilibrium temperature (K) & $1632^{+30}_{-31}$ & $1606^{+28}_{-29}$ & $1614^{+49}_{-46}$ & $1658^{+70}_{-59}$ \\
$\tau_{\rm circ}$ & Tidal circularization timescale (Gyr) & $1.43^{+0.32}_{-0.28}$ & $1.66^{+0.33}_{-0.34}$ & $0.35^{+0.13}_{-0.10}$ & $0.49^{+0.19}_{-0.15}$ \\
$K$ & RV semi-amplitude (m/s) & $158^{+19}_{-18}$ & $162 \pm 20$ & $80.8^{+9.6}_{-10.0}$ & $81.0^{+9.9}_{-10.0}$ \\
$\dot{\gamma}$ & RV slope (m/s/day) & ---& ---& ---& ---\\
$R_{\rm P}/R_\star$ & Radius of planet in stellar radii  & $0.05947^{+0.00059}_{-0.00055}$ & $0.05948^{+0.00058}_{-0.00055}$ & $0.0778^{+0.0017}_{-0.0016}$ & $0.0779 \pm 0.0016$ \\
$a/R_\star$ & Semi-major axis in stellar radii  & $6.33^{+0.28}_{-0.24}$ & $6.76 \pm 0.24$ & $6.84 \pm 0.26$ & $7.44^{+0.40}_{-0.33}$ \\
Depth & \tess flux decrement at mid-transit & $0.004055 \pm 0.000086$ & $0.004043^{+0.000086}_{-0.000085}$ & $0.00688^{+0.00028}_{-0.00026}$ & $0.00677 \pm 0.00027$ \\
$\tau$ & Ingress/egress transit duration (days) & $0.01213^{+0.00110}_{-0.00036}$ & $0.01210^{+0.00110}_{-0.00033}$ & $0.0189^{+0.0030}_{-0.0016}$ & $0.0184^{+0.0027}_{-0.0012}$ \\
$T_{14}$ & Total transit duration (days) & $0.2089^{+0.0014}_{-0.0012}$ & $0.2089^{+0.0014}_{-0.0012}$ & $0.2383^{+0.0029}_{-0.0022}$ & $0.2373^{+0.0026}_{-0.0020}$ \\
$b$ & Transit impact parameter & $0.17^{+0.15}_{-0.12}$ & $0.17^{+0.15}_{-0.12}$ & $0.30^{+0.15}_{-0.20}$ & $0.26^{+0.16}_{-0.17}$ \\
$\rho_{\rm P}$ & Density (cgs) & $1.81^{+0.35}_{-0.29}$ & $2.13^{+0.39}_{-0.34}$ & $0.284^{+0.054}_{-0.048}$ & $0.343^{+0.073}_{-0.060}$ \\
$\log{g_{\rm P}}$ & Surface gravity (cgs) & $3.541^{+0.064}_{-0.065}$ & $3.602^{+0.062}_{-0.067}$ & $2.933^{+0.062}_{-0.069}$ & $3.006^{+0.070}_{-0.072}$ \\
$M_{\rm P}/M_\star$ & Mass ratio  & $0.00111 \pm 0.00013$ & $0.00108 \pm 0.00013$ & $0.000607^{+0.000073}_{-0.000077}$ & $0.000569^{+0.000070}_{-0.000072}$ \\
$d/R_\star$ & Separation at mid-transit  & $6.73^{+0.60}_{-0.53}$ & $7.60^{+0.55}_{-0.52}$ & $6.45 \pm 0.57$ & $7.50^{+0.86}_{-0.65}$ \\
\bottomrule
\end{tabular}
\end{table*}
\providecommand{\bjdtdb}{\ensuremath{\rm {BJD_{TDB}}}}
\providecommand{\feh}{\ensuremath{\left[{\rm Fe}/{\rm H}\right]}}
\providecommand{\teff}{\ensuremath{T_{\rm eff}}}
\providecommand{\teq}{\ensuremath{T_{\rm eq}}}
\providecommand{\ecosw}{\ensuremath{e\cos{\omega_\star}}}
\providecommand{\esinw}{\ensuremath{e\sin{\omega_\star}}}
\providecommand{\msun}{\ensuremath{\,M_\Sun}}
\providecommand{\rsun}{\ensuremath{\,R_\Sun}}
\providecommand{\lsun}{\ensuremath{\,L_\Sun}}
\providecommand{\mj}{\ensuremath{\,M_{\rm J}}}
\providecommand{\rj}{\ensuremath{\,R_{\rm J}}}
\providecommand{\me}{\ensuremath{\,M_{\rm E}}}
\providecommand{\re}{\ensuremath{\,R_{\rm E}}}
\providecommand{\fave}{\langle F \rangle}
\providecommand{\fluxcgs}{10$^9$ erg s$^{-1}$ cm$^{-2}$}
\providecommand{\tess}{\textit{TESS}\xspace}
\begin{table*}
\centering
\contcaption{}
\scriptsize
\setlength{\tabcolsep}{3.5pt}
\begin{tabular}{llcccc}
\toprule
&  & \multicolumn{2}{c}{TOI-6334} & \multicolumn{2}{c}{TOI-7404}\\
&  & Low-mass solution & High-mass solution & Low-mass solution & High-mass solution\\
&  & 63.8\% probability & 36.2\% probability & 9.6\% probability & 90.4\% probability\\
\midrule
\multicolumn{6}{l}{\textbf{Stellar Parameters}:} \\
$M_\star$ & Mass (\msun) & $1.435^{+0.054}_{-0.056}$ & $1.634^{+0.053}_{-0.050}$ & $1.339^{+0.035}_{-0.044}$ & $1.545^{+0.068}_{-0.069}$ \\
$R_\star$ & Radius (\rsun) & $2.577^{+0.100}_{-0.087}$ & $2.565^{+0.089}_{-0.095}$ & $2.174^{+0.089}_{-0.090}$ & $2.135^{+0.095}_{-0.086}$ \\
$L_\star$ & Luminosity (\lsun) & $7.81^{+0.50}_{-0.58}$ & $8.07^{+0.44}_{-0.48}$ & $6.28^{+0.43}_{-0.38}$ & $6.45^{+0.45}_{-0.41}$ \\
$\rho_\star$ & Density (cgs) & $0.118 \pm 0.012$ & $0.137^{+0.015}_{-0.012}$ & $0.183^{+0.023}_{-0.020}$ & $0.223^{+0.031}_{-0.028}$ \\
$\log{g}$ & Surface gravity (cgs) & $3.772^{+0.028}_{-0.033}$ & $3.833^{+0.029}_{-0.024}$ & $3.889^{+0.034}_{-0.033}$ & $3.967 \pm 0.041$ \\
$T_{\rm eff}$ & Effective temperature (K) & $6010^{+130}_{-140}$ & $6070 \pm 110$ & $6200^{+130}_{-120}$ & $6290 \pm 140$ \\
$[{\rm Fe/H}]$ & Metallicity (dex) & $0.116^{+0.100}_{-0.094}$ & $0.188 \pm 0.095$ & $0.129^{+0.100}_{-0.095}$ & $0.20 \pm 0.10$ \\
$[{\rm Fe/H}]_{0}$ & Initial metallicity (dex) & $0.146^{+0.092}_{-0.091}$ & $0.236^{+0.081}_{-0.083}$ & $0.160^{+0.083}_{-0.077}$ & $0.271^{+0.100}_{-0.097}$ \\
Age & Age (Gyr) & $3.12^{+0.35}_{-0.31}$ & $2.04^{+0.24}_{-0.22}$ & $3.90^{+0.45}_{-0.36}$ & $2.15^{+0.51}_{-0.46}$ \\
EEP & Equivalent evolutionary phase & $456.6^{+4.5}_{-5.4}$ & $403.6^{+5.9}_{-7.5}$ & $449.6^{+5.4}_{-5.2}$ & $387^{+17}_{-19}$ \\
$A_V$ & V-band extinction (mag) & $0.202^{+0.055}_{-0.096}$ & $0.235^{+0.032}_{-0.064}$ & $0.061^{+0.063}_{-0.044}$ & $0.093^{+0.060}_{-0.061}$ \\
$d$ & Distance (pc) & $740 \pm 12$ & $743^{+13}_{-12}$ & $986^{+21}_{-20}$ & $990 \pm 21$ \\
\multicolumn{6}{l}{\textbf{Planetary Parameters}:} \\
$P$ & Period (days) & $3.8497745 \pm 0.0000039$ & $3.8497745^{+0.0000040}_{-0.0000039}$ & $1.4634426 \pm 0.0000017$ & $1.4634427^{+0.0000016}_{-0.0000017}$ \\
$R_{\rm P}$ & Radius (\rj) & $1.531^{+0.063}_{-0.055}$ & $1.523^{+0.054}_{-0.057}$ & $1.262^{+0.062}_{-0.061}$ & $1.229^{+0.065}_{-0.058}$ \\
$M_{\rm P}$ & Mass (\mj) & $0.91 \pm 0.13$ & $0.99 \pm 0.14$ & $7.9^{+1.3}_{-1.6}$ & $8.7^{+1.3}_{-1.5}$ \\
$T_C$ & Time of conjunction (\bjdtdb) & $2459905.99686^{+0.00070}_{-0.00072}$ & $2459905.99684^{+0.00070}_{-0.00071}$ & $2460608.4587 \pm 0.0011$ & $2460608.4588 \pm 0.0010$ \\
$T_0$ & Optimal conjunction time (\bjdtdb) & $2459990.69226 \pm 0.00068$ & $2459990.69227 \pm 0.00068$ & $2460033.32591^{+0.00075}_{-0.00076}$ & $2460033.32597^{+0.00073}_{-0.00074}$ \\
$a$ & Semi-major axis (AU) & $0.05423^{+0.00067}_{-0.00072}$ & $0.05663^{+0.00060}_{-0.00058}$ & $0.02786^{+0.00024}_{-0.00031}$ & $0.02921^{+0.00042}_{-0.00044}$ \\
$i$ & Inclination (Degrees) & $87.7^{+1.6}_{-1.9}$ & $88.4^{+1.1}_{-1.6}$ & $79.9^{+4.1}_{-2.1}$ & $83.0^{+4.0}_{-2.8}$ \\
$e$ & Eccentricity & $0.050^{+0.052}_{-0.035}$ & $0.069^{+0.061}_{-0.046}$ & $0.054^{+0.088}_{-0.040}$ & $0.043^{+0.063}_{-0.031}$ \\
$\omega_\star$ & Argument of periastron (Degrees) & $171^{+72}_{-92}$ & $-133^{+58}_{-36}$ & $70^{+110}_{-130}$ & $-90 \pm 120$ \\
$\teq$ & Equilibrium temperature (K) & $1997^{+31}_{-35}$ & $1970^{+26}_{-29}$ & $2642^{+43}_{-40}$ & $2595^{+42}_{-40}$ \\
$\tau_{\rm circ}$ & Tidal circularization timescale (Gyr) & $0.136^{+0.035}_{-0.033}$ & $0.160^{+0.035}_{-0.033}$ & $0.044^{+0.017}_{-0.016}$ & $0.062^{+0.024}_{-0.021}$ \\
$K$ & RV semi-amplitude (m/s) & $93 \pm 13$ & $92 \pm 13$ & $1150^{+190}_{-230}$ & $1160^{+170}_{-190}$ \\
$\dot{\gamma}$ & RV slope (m/s/day) & ---& ---& $8.4^{+8.7}_{-7.3}$ & $8.1^{+7.3}_{-6.6}$ \\
$R_{\rm P}/R_\star$ & Radius of planet in stellar radii  & $0.06106^{+0.00060}_{-0.00055}$ & $0.06101^{+0.00057}_{-0.00053}$ & $0.0597 \pm 0.0011$ & $0.05917^{+0.00100}_{-0.00096}$ \\
$a/R_\star$ & Semi-major axis in stellar radii  & $4.52^{+0.14}_{-0.17}$ & $4.75^{+0.17}_{-0.14}$ & $2.75^{+0.11}_{-0.10}$ & $2.94 \pm 0.13$ \\
Depth & \tess flux decrement at mid-transit & $0.004272^{+0.000083}_{-0.000081}$ & $0.004260 \pm 0.000081$ & $0.00392 \pm 0.00011$ & $0.00390 \pm 0.00011$ \\
$\tau$ & Ingress/egress transit duration (days) & $0.01701^{+0.00150}_{-0.00054}$ & $0.01675^{+0.00120}_{-0.00036}$ & $0.0121^{+0.0021}_{-0.0022}$ & $0.0106^{+0.0019}_{-0.0012}$ \\
$T_{14}$ & Total transit duration (days) & $0.2826^{+0.0020}_{-0.0018}$ & $0.2822^{+0.0019}_{-0.0017}$ & $0.1627 \pm 0.0025$ & $0.1615^{+0.0022}_{-0.0020}$ \\
$b$ & Transit impact parameter & $0.18^{+0.15}_{-0.12}$ & $0.141^{+0.140}_{-0.098}$ & $0.48^{+0.10}_{-0.21}$ & $0.36^{+0.14}_{-0.21}$ \\
$\rho_{\rm P}$ & Density (cgs) & $0.313^{+0.057}_{-0.053}$ & $0.346^{+0.063}_{-0.057}$ & $4.8 \pm 1.2$ & $5.8^{+1.4}_{-1.3}$ \\
$\log{g_{\rm P}}$ & Surface gravity (cgs) & $2.982^{+0.064}_{-0.072}$ & $3.023^{+0.064}_{-0.071}$ & $4.089^{+0.083}_{-0.098}$ & $4.155^{+0.078}_{-0.091}$ \\
$M_{\rm P}/M_\star$ & Mass ratio  & $0.000608^{+0.000083}_{-0.000084}$ & $0.000577^{+0.000079}_{-0.000081}$ & $0.00566^{+0.00092}_{-0.00120}$ & $0.00539^{+0.00081}_{-0.00089}$ \\
$d/R_\star$ & Separation at mid-transit  & $4.50^{+0.27}_{-0.30}$ & $4.91^{+0.37}_{-0.29}$ & $2.72^{+0.20}_{-0.27}$ & $2.95^{+0.21}_{-0.22}$ \\
\bottomrule
\end{tabular}
\end{table*}
\providecommand{\bjdtdb}{\ensuremath{\rm {BJD_{TDB}}}}
\providecommand{\feh}{\ensuremath{\left[{\rm Fe}/{\rm H}\right]}}
\providecommand{\teff}{\ensuremath{T_{\rm eff}}}
\providecommand{\teq}{\ensuremath{T_{\rm eq}}}
\providecommand{\ecosw}{\ensuremath{e\cos{\omega_\star}}}
\providecommand{\esinw}{\ensuremath{e\sin{\omega_\star}}}
\providecommand{\msun}{\ensuremath{\,M_\Sun}}
\providecommand{\rsun}{\ensuremath{\,R_\Sun}}
\providecommand{\lsun}{\ensuremath{\,L_\Sun}}
\providecommand{\mj}{\ensuremath{\,M_{\rm J}}}
\providecommand{\rj}{\ensuremath{\,R_{\rm J}}}
\providecommand{\me}{\ensuremath{\,M_{\rm E}}}
\providecommand{\re}{\ensuremath{\,R_{\rm E}}}
\providecommand{\fave}{\langle F \rangle}
\providecommand{\fluxcgs}{10$^9$ erg s$^{-1}$ cm$^{-2}$}
\providecommand{\tess}{\textit{TESS}\xspace}
\begin{table*}
\centering
\contcaption{}
\scriptsize
\setlength{\tabcolsep}{3.5pt}
\begin{tabular}{llcc}
\toprule
&  & \multicolumn{2}{c}{TOI-7425}\\
&  & Low-mass solution & High-mass solution\\
&  & 17.3\% probability & 82.7\% probability\\
\midrule
\multicolumn{4}{l}{\textbf{Stellar Parameters}:} \\
$M_\star$ & Mass (\msun) & $1.123^{+0.038}_{-0.043}$ & $1.311^{+0.067}_{-0.057}$ \\
$R_\star$ & Radius (\rsun) & $1.621^{+0.059}_{-0.062}$ & $1.604^{+0.053}_{-0.048}$ \\
$L_\star$ & Luminosity (\lsun) & $2.54^{+0.29}_{-0.28}$ & $2.95^{+0.36}_{-0.32}$ \\
$\rho_\star$ & Density (cgs) & $0.370^{+0.045}_{-0.037}$ & $0.449^{+0.045}_{-0.042}$ \\
$\log{g}$ & Surface gravity (cgs) & $4.068^{+0.033}_{-0.032}$ & $4.146 \pm 0.031$ \\
$T_{\rm eff}$ & Effective temperature (K) & $5730^{+160}_{-170}$ & $5970^{+190}_{-180}$ \\
$[{\rm Fe/H}]$ & Metallicity (dex) & $0.386 \pm 0.042$ & $0.394 \pm 0.041$ \\
$[{\rm Fe/H}]_{0}$ & Initial metallicity (dex) & $0.382^{+0.042}_{-0.043}$ & $0.405^{+0.045}_{-0.046}$ \\
Age & Age (Gyr) & $8.1^{+1.6}_{-1.0}$ & $3.6 \pm 1.2$ \\
EEP & Equivalent evolutionary phase & $451.2^{+4.9}_{-4.5}$ & $390^{+24}_{-34}$ \\
$A_V$ & V-band extinction (mag) & $0.50^{+0.14}_{-0.16}$ & $0.70^{+0.14}_{-0.15}$ \\
$d$ & Distance (pc) & $515.0^{+3.8}_{-3.7}$ & $515.4^{+3.8}_{-3.7}$ \\
\multicolumn{4}{l}{\textbf{Planetary Parameters}:} \\
$P$ & Period (days) & $1.28157053 \pm 0.00000031$ & $1.28157053 \pm 0.00000031$ \\
$R_{\rm P}$ & Radius (\rj) & $1.808^{+0.068}_{-0.069}$ & $1.789^{+0.061}_{-0.058}$ \\
$M_{\rm P}$ & Mass (\mj) & $1.06^{+0.12}_{-0.11}$ & $1.19^{+0.15}_{-0.14}$ \\
$T_C$ & Time of conjunction (\bjdtdb) & $2460633.58719^{+0.00053}_{-0.00048}$ & $2460633.58708^{+0.00045}_{-0.00040}$ \\
$T_0$ & Optimal conjunction time (\bjdtdb) & $2459909.49980 \pm 0.00018$ & $2459909.49980 \pm 0.00018$ \\
$a$ & Semi-major axis (AU) & $0.02400^{+0.00027}_{-0.00031}$ & $0.02528^{+0.00042}_{-0.00037}$ \\
$i$ & Inclination (Degrees) & $77.0^{+1.0}_{-1.1}$ & $78.63^{+0.77}_{-0.85}$ \\
$e$ & Eccentricity & $0.095 \pm 0.050$ & $0.043^{+0.049}_{-0.030}$ \\
$\omega_\star$ & Argument of periastron (Degrees) & $75^{+38}_{-34}$ & $53^{+110}_{-90}$ \\
$\teq$ & Equilibrium temperature (K) & $2268^{+57}_{-59}$ & $2295^{+55}_{-57}$ \\
$\tau_{\rm circ}$ & Tidal circularization timescale (Gyr) & $0.00047^{+0.00016}_{-0.00013}$ & $0.00067^{+0.00016}_{-0.00015}$ \\
$K$ & RV semi-amplitude (m/s) & $180^{+19}_{-18}$ & $183 \pm 21$ \\
$\dot{\gamma}$ & RV slope (m/s/day) & ---& ---\\
$R_{\rm P}/R_\star$ & Radius of planet in stellar radii  & $0.11467^{+0.00088}_{-0.00090}$ & $0.11461^{+0.00084}_{-0.00085}$ \\
$a/R_\star$ & Semi-major axis in stellar radii  & $3.18^{+0.12}_{-0.11}$ & $3.39 \pm 0.11$ \\
Depth & \tess flux decrement at mid-transit & $0.01400 \pm 0.00018$ & $0.01388 \pm 0.00018$ \\
$\tau$ & Ingress/egress transit duration (days) & $0.0194 \pm 0.0011$ & $0.0192 \pm 0.0010$ \\
$T_{14}$ & Total transit duration (days) & $0.11073^{+0.00100}_{-0.00098}$ & $0.11029 \pm 0.00096$ \\
$b$ & Transit impact parameter & $0.661^{+0.021}_{-0.024}$ & $0.661^{+0.020}_{-0.023}$ \\
$\rho_{\rm P}$ & Density (cgs) & $0.222^{+0.039}_{-0.033}$ & $0.257^{+0.044}_{-0.039}$ \\
$\log{g_{\rm P}}$ & Surface gravity (cgs) & $2.906^{+0.058}_{-0.059}$ & $2.964^{+0.060}_{-0.062}$ \\
$M_{\rm P}/M_\star$ & Mass ratio  & $0.000906^{+0.000095}_{-0.000090}$ & $0.00087 \pm 0.00010$ \\
$d/R_\star$ & Separation at mid-transit  & $2.92^{+0.27}_{-0.24}$ & $3.36^{+0.20}_{-0.24}$ \\
\bottomrule
\end{tabular}
\end{table*}
\renewcommand{\arraystretch}{\bodystretch}

\subsection{Automation}\label{subsec:automation}

To efficiently set up and analyze the results of the \exofast global fits presented in this work, we developed several tools to automate parts of our workflow. These tools are hosted on GitHub\footnote{\url{https://github.com/jackschulte/exofast_tools}} and include the \tess lightcurve pipeline mentioned in \S\ref{subsec:TESS} along with several other pipelines used to generate priors, download and reformat RVs, and generate the scripts required to run an \exofast fit. Additionally, the code used to construct the figures in Appendix\,\ref{sec:systemfigures} is included in a separate GitHub repository\footnote{\url{https://github.com/jackschulte/system_figure_pipeline}} and the code used to construct most of the tables in this article is included in a third repository\footnote{\url{https://github.com/jackschulte/LaTeX_table_pipeline}}.

\section{Results and Discussion}\label{sec:discussion}

\begin{figure}
    \centering
    \includegraphics[width=0.95\linewidth]{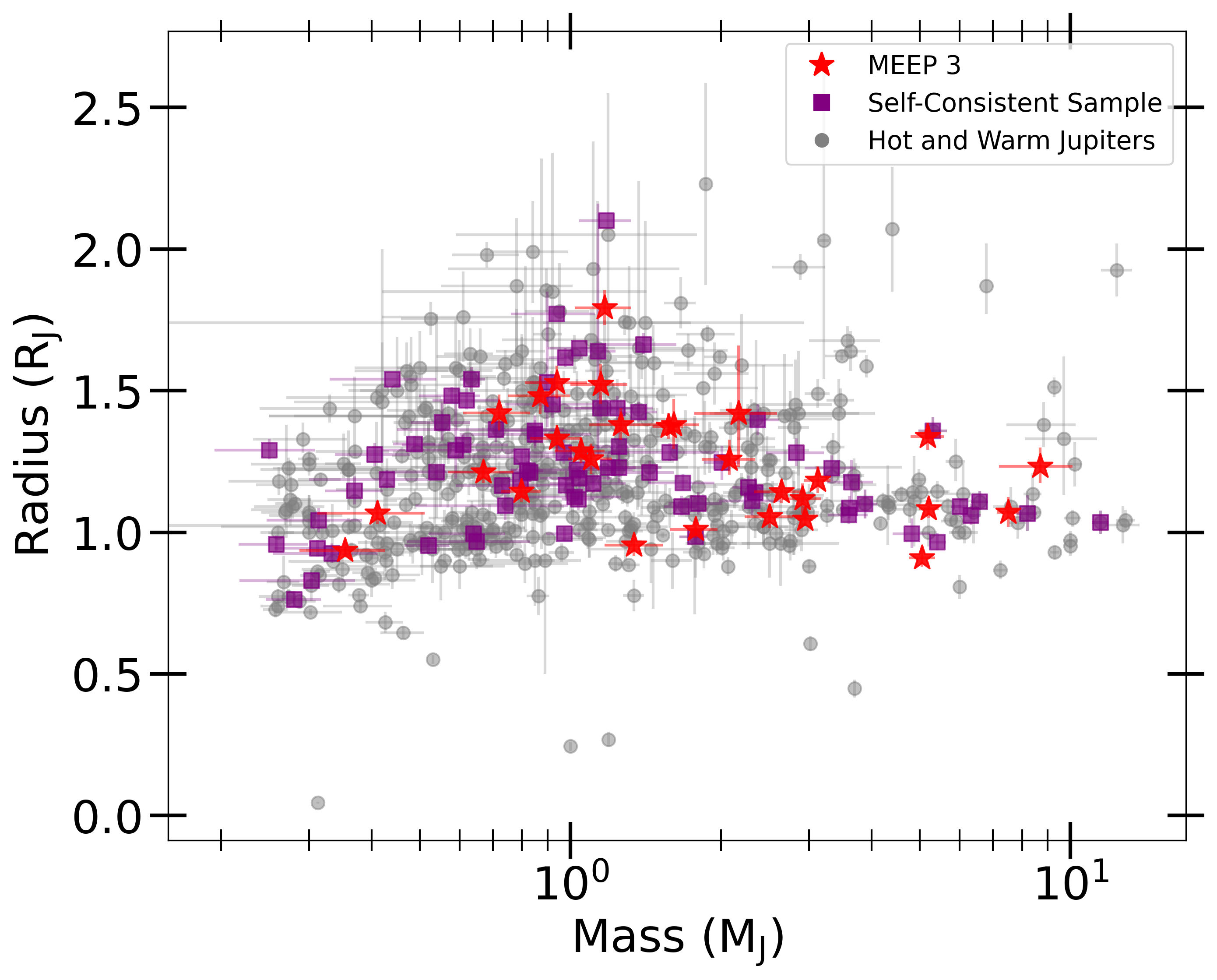}
    \caption{Mass-radius diagram of the HJ and WJ systems discovered by the MEEP survey, compared to those systems in the NASA Exoplanet Archive (NEA; retrieved \nearetrievaldate). Systems discovered in the MEEP survey are represented as red stars. Those discovered following a similar methodology to this work (\citealt{Rodriguez:2019, Rodriguez:2021, Ikwut-Ukwa:2022, Yee:2022, Yee:2023a, Rodriguez:2023, RodriguezMartinez:2025, Yee:2025}) are represented with purple squares, while those from the NASA Exoplanet Archive are represented as grey circles.}
    \label{fig:mass-radius}
\end{figure}

In this article, we present the discovery and characterization of 29 giant planets detected by \tess. These planets span a large mass range (0.35\,\mj < $M_{\rm P}$ < 8.7\,\mj) and have orbital periods between 1.28 days and 16.9 days. In Figure~\ref{fig:mass-radius}, we compare the mass-radius distribution of planets discovered by the MEEP survey (\citealt{Schulte:2024, Schulte:2025}, this work) to those discovered by our larger collaboration (labeled as ``Self-Consistent Sample''; \citealt{Rodriguez:2019, Rodriguez:2021, Ikwut-Ukwa:2022, Yee:2022, Yee:2023a, Rodriguez:2023, RodriguezMartinez:2025, Yee:2025}) and to those discovered by other teams. We find that the mass-radius distribution of the planets discovered by our survey and related surveys is representative of the greater underlying population of hot and warm Jupiters. In this work and the others included in the self-consistent sample, we do well to fill out the higher-mass part of this diagram, discovering 12 super-Jupiters with masses exceeding 5 \mj.

\begin{figure*}
    \includegraphics[width=0.47\textwidth]{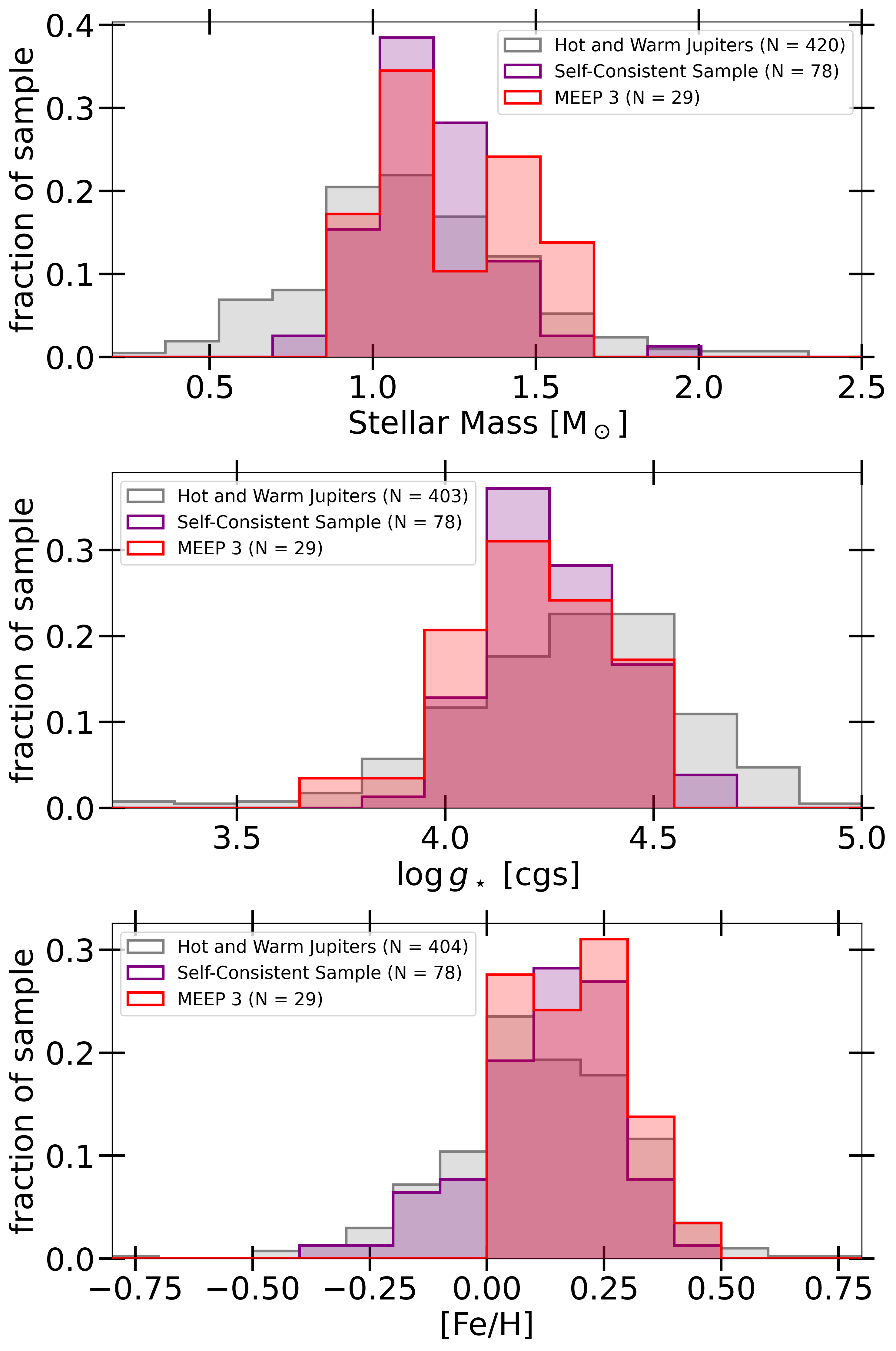}
    \includegraphics[width=0.47\textwidth]{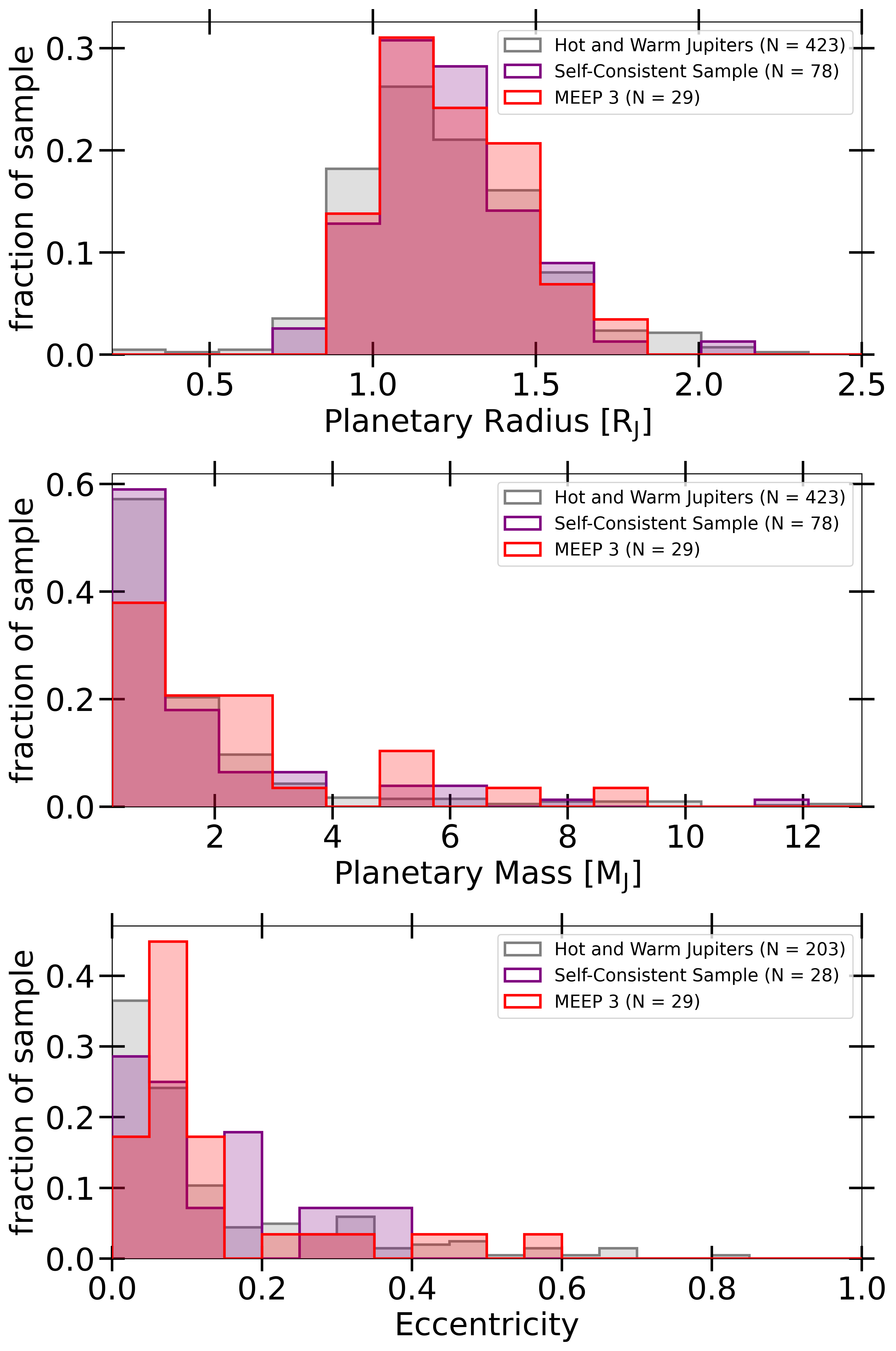}
    \caption{Comparison of host star parameters and planetary parameters among hot and warm Jupiters ($P < 25$ days, $M_{\rm P} = [0.25, 13]$\,\mj). Red histograms correspond to the planets and host stars in this work, while purple corresponds to those discovered by \citealt{Rodriguez:2019, Rodriguez:2021, Rodriguez:2023, Ikwut-Ukwa:2022, Yee:2022, Yee:2023a, Yee:2025, RodriguezMartinez:2025}, and grey corresponds to hot and warm Jupiters retrieved from the NEA (retrieved \nearetrievaldate). Systems in which the orbital eccentricity was fixed to 0 are excluded from the eccentricity histogram. Including the systems discovered in this work in the self-consistent sample, we find that the self-consistent sample is statistically distinct from the literature sample in terms of all three stellar parameters.}
    \label{fig:histograms}
\end{figure*}

In addition, we investigated the similarities and differences between the HJ and WJ systems in the self-consistent sample and those from the literature in terms of several key parameters. Specifically, we performed Kolmogorov-Smirnov (K-S) tests \citep{Massey:1951} of host star mass, $\log{g_\star}$, and metallicity, as well as the planet's radius, mass, and eccentricity to test whether the self-consistent sample is drawn from the same underlying population as the literature hot and warm Jupiter systems. Since the orbital period range of the self-consistent sample is tied to the length of a \tess sector and is therefore much smaller than the literature sample ($<\sim25$\,days, compared to $< 200$\,days), we chose to limit the WJ sample to only those with periods smaller than 25 days for the purpose of this analysis. Histograms of each of these parameters are shown in Figure~\ref{fig:histograms}. Each K-S test was performed including the systems from this work in the ``self-consistent sample,'' although these are separated in the histograms to show this article's contribution to the distribution of each parameter. We find that the K-S tests performed on the three host star parameters all had $p$-values smaller than $5 \times 10^{-2}$ ($p = 6.8 \times 10^{-4}$,  $2.8 \times 10^{-4}$, and $3.0 \times 10^{-3}$, for stellar mass, $\log{g_\star}$, and metallicity, respectively), indicating that the distribution of the self-consistent sample's \textit{host star parameters} is likely to be distinct from that of the literature sample. None of the \textit{planetary parameters} had $p$-values above $5 \times 10^{-2}$, indicating that the planets in the self-consistent sample are drawn from the underlying population as those in the literature sample. To ensure that the differences in the host star parameters are not the result of the samples having different selection functions, we performed all of the K-S tests two more times, constraining both samples to only those with FGK hosts (4000\,K $<$ \teff $<$ 7500\,K) and only those with HJs ($P<10$\,days). In both cases, the $p$-values did not change above or below $5 \times 10^{-2}$ for any of the host star parameters, so neither affected our interpretation. The $p$-value of the orbital eccentricity did fall to $3.4 \times 10^{-2}$ when comparing to the literature HJs; however, with only 50 systems with measured nonzero eccentricities in the self-consistent sample and 162 in the literature sample, we argue that this is unlikely to be significant.

One likely explanation for the discrepancy among host star parameters is that the global fits performed on all stars in the self-consistent sample provide a better constraint on the stellar density due to the transit shape, affecting the distributions of both $\log{ g_\star}$ and stellar mass. The difference in host star metallicity distributions, while not as stark, may point to differences in the literature as to how metallicity is measured. All host star metallicities in the self-consistent sample are influenced by the inclusion of spectroscopic metallicity priors, most of which are from dedicated pipelines run on observations from the TRES and CHIRON spectrographs. While both the self-consistent sample and the literature sample report a preference for metal-rich hosts, consistent with prior findings \citep{Fischer:2005}, the self-consistent sample has systematically larger host star metallicities. This is especially true for the host stars in this paper, all of which are more metal-rich than the Sun. An ongoing effort to reanalyze all HJs in the literature (Brennom et al. in prep) will further elucidate these differences.

\subsection{Planets with Significant Eccentricity}\label{subsec:eccentricity}

\begin{figure}
    \centering
    \includegraphics[width=\linewidth]{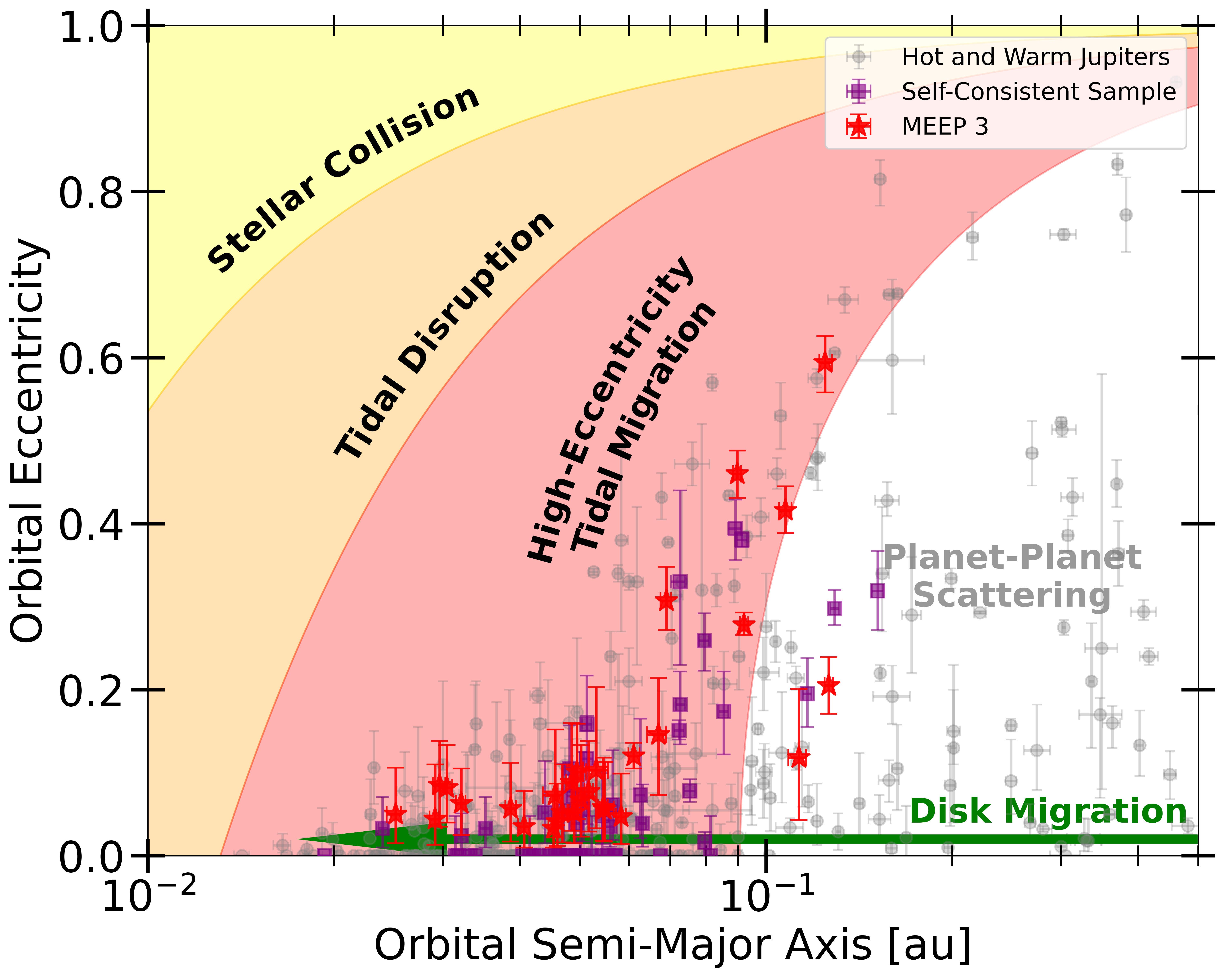}
    \caption{Semi-major axis vs. orbital eccentricity distribution for the planets presented in this work, those included in the self-consistent sample, and the literature HJs and WJs. This figure is an updated version of the one presented in \citealt{Schulte:2025}, which was adapted from \citealt{Dawson:2018}. The region labeled ``Stellar Collision'' represents the parameter space corresponding to a giant planet colliding with a star with a radius of 1 R$_\odot$. The region labeled ``Tidal Disruption'' corresponds to a Jupiter-like planet encountering the Roche limit of a Sun-like star. The region labeled ``High-Eccentricity Tidal Migration'' corresponds to the tidal migration of a giant planet governed by Equation 11 in \citealt{Dawson:2018}. The red stars represent the planets presented in this work, while the purple squares represent those in the self-consistent sample defined in \S\ref{sec:discussion}. The grey circles represent literature giant planets with $P < 200$ days, masses between 0.25\,\mj and 13\,\mj, and reported eccentricities (NEA accessed on \nearetrievaldate).}
    \label{fig:a_ecc}
\end{figure}

In this article, seven of the hot and warm Jupiters have orbital eccentricities that are significant. These planets, TOI-3365\,b, TOI-3972\,b, TOI-4144\,b, TOI-5479\,b, TOI-5925\,b, TOI-6148\,b, and TOI-6443\,b, each have $> 6 \sigma$ orbital eccentricities. Four of them, TOI-3365\,b, TOI-5925\,b, TOI-6148\,b, and TOI-6443\,b, are HJs with orbital periods $<10$ days, while TOI-3972\,b, TOI-4144\,b, and TOI-5479\,b, are WJs with $P > 10$ days. TOI-5479\,b has the largest eccentricity in our sample, with $e = 0.594^{+0.032}_{-0.036}$. Due to tides from their host stars, HJs on shorter orbital periods are typically circular, while HJs and WJs with larger periastron distances are less affected by host star tides, allowing larger orbital eccentricities to remain after many Gyr. This is one of the most discerning differences between the HJ and WJ populations, as $\sim13\%$ of HJs have measured non-zero eccentricities with $>2.45\sigma$ (95\% confidence in light of the Lucy-Sweeney bias; \citealt{Lucy:1971, Eastman:2019}), whereas $\sim 74\%$ of WJs have significant eccentricities. This difference is apparent in Figure~\ref{fig:a_ecc}, where the HJ eccentricity distribution matches expectations from the theory of high-eccentricity tidal migration, as was argued by \cite{Bonomo:2017}. The WJ population instead appears to be consistent with the expectations from planet-planet scattering, in agreement with the work of \cite{dong_planet-planet_2026}. However, further work (both observational and theoretical) is required to determine the frequency of each migration mechanism \citep{kawai_identifying_2025}.

TOI-3972\,b, a WJ which has an orbital eccentricity of $0.278^{+0.015}_{-0.012}$, had its spin-orbit alignment angle measured by \cite{wang_single-star_2024} via the Rossiter-McLaughlin effect, and was determined to have an aligned orbit. The authors argue that this is typical of WJs, which have a tendency to be aligned, agnostic to their orbital eccentricities. This is in contrast with HJs, which are frequently misaligned. The absence of a misaligned orbit implies that TOI-3972\,b's eccentricity may be primordial, or that it was excited without misaligning the orbit. \cite{Zink:2023} discusses an avenue through which HJs may undergo coplanar high-eccentricity tidal migration, but with TOI-3972\,b's 10.5-day period, its tidal circularization timescale would be $\sim 75$ Gyr, indicating that it will not become a HJ within the lifetime of its host star.

\subsection{Tentative Outer Companions}\label{subsec:companions}

16 of the 29 systems presented in this article show at least tentative evidence for an outer companion in the RV observations, by meeting the $>1\sigma$ significance constraints on the fitted RV linear slope, as defined in \S\ref{sec:exofast}. Of these, seven (TOI-3365, TOI-3601, TOI-3788, TOI-4140, TOI-5925, TOI-6166, and TOI-6417) have RV slopes with at least $2\sigma$ significance, and only three of these have RV slopes with $>3\sigma$ significance: TOI-4140 ($3.4\sigma$), TOI-6166 ($4\sigma$), and TOI-3365 ($20\sigma$). TOI-3365 stands out as having clear evidence of a distant outer companion, with an RV slope of $-0.384 \pm 0.019$\, m\,s$^{-1}$day$^{-1}$. Notably, its planet also has significant orbital eccentricity ($0.12^{+0.016}_{-0.015}$). Over an 825~day baseline, there is a 317\,m\,s$^{-1}$ drift and no evidence of turnover, implying that the outer companion is likely orbiting TOI-3365 on a much longer orbital period. Given that its orbital period is likely to be at least 825 days, its minimum mass would be at least 2.4\,\mj, but is likely even larger. Without more RV data, it is impossible to say whether the object is a planet, brown dwarf, or star. Regardless, it is possible that TOI-3365\,b underwent HEM, triggered via vZKL oscillations caused by a mutual inclination between TOI-3365\,b and this distant outer companion.

Continued RV monitoring of long-term trends in planet-hosting systems often requires a sustained, multi-year commitment, but it has the potential to yield useful information about the migration histories of these systems. Systems containing HJs and massive outer companions are compelling candidates for studying HEM through vZKL oscillations or planet-planet scattering, which are believed to be the primary mechanisms shaping the population of HJs \citep{Bonomo:2017}. As RV baselines can be short and sparsely sampled, dedicated follow-up campaigns are necessary to identify which HJ systems contain outer companions \citep[e.g.,][]{Knutson:2014}. Furthermore, the upcoming release of \gaia Data Release 4 will yield epoch astrometry and provide constraints on the mutual inclinations between the inner giant planets and the outer companions, further clarifying the picture and elucidating the likelihood of vZKL-driven HEM.

\section{Summary}\label{sec:summary}

In this article, the third in the MEEP series, we presented the discovery of 29 new transiting HJ and WJ systems detected by the \tess spacecraft, spanning a wide range in mass ($0.35 - 8.7$\,\mj), radius ($0.91 - 1.8$ \rj), and orbital eccentricity ($0 - 0.6$). We discovered three companion stars, TOI-3988\,B, TOI-6171\,B, and TOI-7266\,B, and found compelling evidence of an outer planetary or stellar companion in the TOI-3365 system. Finally, we performed two-sample Kolmogorov-Smirnov tests on host star and planetary parameters to compare the 78 HJ and WJ systems in our self-consistent sample to the 470 systems in the literature and found significant differences between the samples in terms of the host star mass, surface gravity, and metallicity. We interpret these differences as being signatures of the different analysis techniques, such as our joint-fitting strategy. Over the coming years, as \tess completes its all-sky survey of the solar neighborhood and new ground-based and space-based observatories detect thousands of new transiting exoplanets, the efficient discovery of HJ and WJ systems will enable a deeper understanding of how these enigmatic systems form and evolve.

\section*{Acknowledgements}

This paper made use of data collected by the TESS mission and are publicly available from the Mikulski Archive for Space Telescopes (MAST) operated by the Space Telescope Science Institute (STScI). We acknowledge the use of public TESS data from pipelines at the TESS Science Office and at the TESS Science Processing Operations Center. Resources supporting this work were provided by the NASA High-End Computing (HEC) Program through the NASA Advanced Supercomputing (NAS) Division at Ames Research Center for the production of the SPOC data products.

Research reported in this publication was supported in part by funding provided by the National Aeronautics and Space Administration (NASA), under award number 80NSSC25M7087, Michigan Space Grant Consortium (MSGC).

This work was supported in part through computational resources and services provided by the Institute for Cyber-Enabled Research at Michigan State University.

This research is also based in part on observations obtained at the Southern Astrophysical Research (SOAR) telescope, which is a joint project of the Minist\'{e}rio da Ci\^{e}ncia, Tecnologia e Inova\c{c}\~{o}es (MCTI/LNA) do Brasil, the US National Science Foundation’s NOIRLab, the University of North Carolina at Chapel Hill (UNC), and Michigan State University (MSU).


Funding for the TESS mission is provided by NASA's Science Mission Directorate. KAC acknowledges support from the TESS mission via subaward s3449 from MIT and NASA grants 80NSSC24K1889 and 80NSSC26K0081.


This work makes use of observations from the LCOGT network. Part of the LCOGT telescope time was granted by NOIRLab through the Mid-Scale Innovations Program (MSIP). MSIP is funded by NSF.


This paper is based on observations made with the Las Cumbres Observatory’s education network telescopes that were upgraded through generous support from the Gordon and Betty Moore Foundation.


This paper is based on observations made with the MuSCAT instruments, developed by the Astrobiology Center (ABC) in Japan, the University of Tokyo, and Las Cumbres Observatory (LCOGT). MuSCAT3 was developed with financial support by JSPS KAKENHI (JP18H05439) and JST PRESTO (JPMJPR1775), and is located at the Faulkes Telescope North on Maui, HI (USA), operated by LCOGT. MuSCAT4 was developed with financial support provided by the Heising-Simons Foundation (grant 2022-3611), JST grant number JPMJCR1761, and the ABC in Japan, and is located at the Faulkes Telescope South at Siding Spring Observatory (Australia), operated by LCOGT.

In this study, observational data obtained within the scope of projects numbered 22BT100-1958 and 25ATUG100-3012, carried out using the TUG100 telescope at the TUG (T\"UB\.{I}TAK National Observatory, Antalya) site of the T\"urkiye National Observatories, have been utilised. We express our gratitude for the invaluable support provided by the T\"urkiye National Observatories, the observation team, and all staff members.

The work of I.A.S., V.K., V.L., V.S., and A.K. was conducted under the state assignment of Lomonosov Moscow State University. MASTER was partially supported by Lomonosov MSU Development Programme before 2018.

DRC acknowledges partial support from NASA Grant 18-2XRP18\_2-0007. This research has made use of the Exoplanet Follow-up Observation Program (ExoFOP; DOI: 10.26134/ExoFOP5) website, which is operated by the California Institute of Technology, under contract with the National Aeronautics and Space Administration under the Exoplanet Exploration Program. Based on observations obtained at the Hale Telescope, Palomar Observatory, as part of a collaborative agreement between the Caltech Optical Observatories and the Jet Propulsion Laboratory operated by Caltech for NASA.

Funding for KB was provided by the European Union (ERC AdG SUBSTELLAR, GA 101054354).

We acknowledge financial support from the Agencia Estatal de Investigaci\'on of the Ministerio de Ciencia e Innovaci\'on MCIN/AEI/10.13039/501100011033 and the ERDF “A way of making Europe” through projects PID2021-125627OB-C32 and PID2024-158486OB-C32. This work is supported by the ERC Grant (ERC Advanced Grant SPEAR, GA 101200674). Funded by the European Union. Views and opinions expressed are, however, those of the authors only and do not necessarily reflect those of the European Union or the European Research Council Executive Agency (ERCEA). Neither the European Union nor the granting authority can be held responsible for them.

This work is partly supported by JSPS KAKENHI Grant Numbers
JP24H00017, JP25K24620, JP26H01402, JP26K00755. JP24K00689,
JP24K17082, JP24K17083, JP24H00248, and JSPS Grant-in-Aid for JSPS
Fellows Grant Number JP24KJ0241 and JP25KJ0091.
This article is based on observations made with the MuSCAT2
instrument, developed by ABC, at Telescopio Carlos Sánchez operated on
the island of Tenerife by the IAC in the Spanish Observatorio del
Teide.

R.B. acknowledges support from FONDECYT Project 1241963.

This research has made use of the NASA Exoplanet Archive, which is operated by the California Institute of Technology, under contract with the National Aeronautics and Space Administration under the Exoplanet Exploration Program.

L.A.S. and T.M.E. were partially supported during this work by the NASA Exoplanets Research Program via grant 80NSSC24K0165, which supports the UNITE (Unistellar Network Investigating TESS Exoplanets) program, under the auspices of which the Unistellar data were collected. This program is also made possible in part by the kind contributions of the Gordon and Betty Moore Foundation.

This work is partly supported by JSPS Grant-in-Aid for JSPS Fellows Grant Number JP25KJ0091.

A.J.\ acknowledges support from Fondecyt project 1251439.

\section*{Data Availability}

The \tess observations of each system, which are presented in \S\ref{subsec:TESS} are publicly available and can be accessed through MAST\footnote{\url{https://archive.stsci.edu/}}. The ground-based time-series photometry presented in \S\ref{subsec:followup} and the high-resolution imaging observations presented in \S\ref{subsec:hri} are available for download on ExoFOP-TESS\footnote{\url{https://exofop.ipac.caltech.edu/tess/}}. The archival photometry and astrometry discussed in \S\ref{subsec:archival} are accessible from the VizieR service\footnote{\url{https://vizier.cds.unistra.fr/viz-bin/VizieR}} \citep{Ochsenbein:2000}. Finally, the pipelines we constructed to perform the discoveries presented in this article are hosted on JS's GitHub account\footnote{\url{https://github.com/jackschulte}}, with each repository listed in \S\ref{subsec:automation}.



\bibliographystyle{mnras}
\bibliography{main} 




\appendix

\section{\exofast Fit Results for Each System}\label{sec:systemfigures}

In this article, we collected many ground-based spectroscopic and photometric observations to complement the \tess data of each target system. Here, we present these data in the form of 29 figures, one for each system, summarizing the data used to constrain our \exofast fits, along with the fit results themselves. Each figure shows the transits, the RVs, the stellar SEDs, and the MIST evolutionary track that describe each system. Our two-component SED fits are shown with in terms of both of their components, while the systems with bimodal stellar mass and age have both solutions plotted on the best-fit MIST track. All \textit{eVscope} transit observations are differentiated by the initials of their citizen scientist observers to ensure that these plots may be compared against the files hosted on ExoFOP-TESS.

\begin{figure*}
    \centering
    \includegraphics[width=\textwidth,height=0.8\textheight,keepaspectratio]{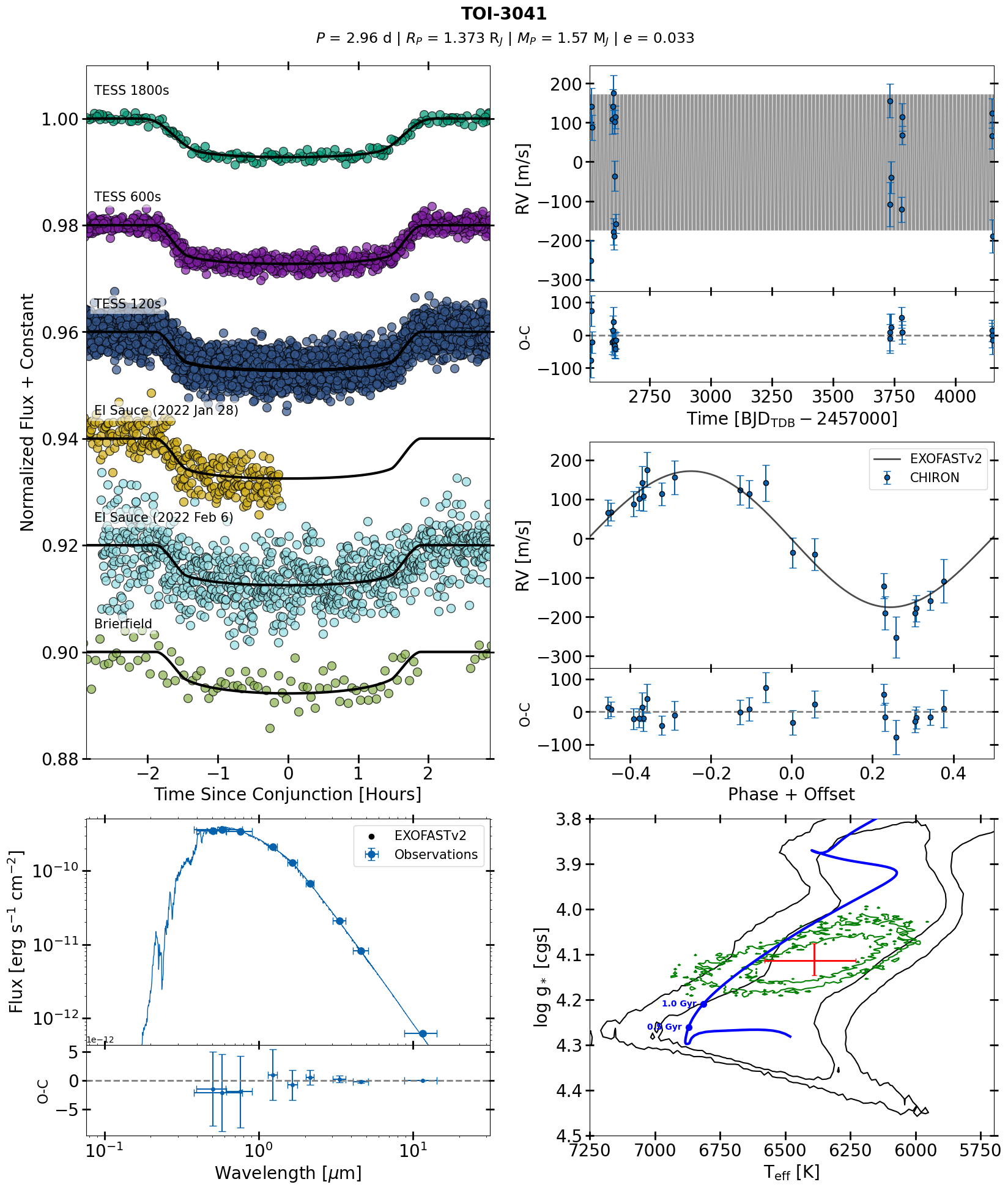}
    \caption{Photometric and radial velocity observations of the TOI-3041 system.
    \textbf{Upper left:} Phase-folded, unbinned, transits of TOI-3041\,b shown in comparison to the best-fit time of conjunction with an arbitrary normalized flux offset. Multiple \tess sectors in the same cadence are stacked on top of each other.
    \textbf{Bottom left:} The spectral energy distribution of the target star compared to the best-fit \exofast model. Residuals are shown on a linear scale, using the same units as the primary y-axis.
    \textbf{Upper right:} RV observations versus time, including any significant long-term trend. The residuals are shown in the subpanel below in the same units.
    \textbf{Middle right:} RV observations phase-folded using the best-fit ephemeris from the \exofast global fit. The phase is shifted so that the transit occurs at Phase + Offset = 0. The residuals are shown in the subpanel below in the same units.
    \textbf{Bottom right:} The evolutionary track and current evolutionary stage of the planet according to the best-fit MESA Isochrones and Stellar Tracks (MIST) model. The blue line indicates the best-fit MIST track, while the black contours show the 1$\sigma$ and 2$\sigma$ constraints on the star's current \teff and $\log{ g}$ from the MIST isochrone alone. The green contours represent the 1$\sigma$ and 2$\sigma$ constraints on the star's \teff and $\log{g}$ from the \exofast global fit, combining constraints from observations of the star and planet. The red cross indicates the median and 68\% confidence interval reported in Table \ref{tab:median}.}
    \label{fig:toi3041}
\end{figure*}

\begin{figure*}
    \centering    
    \includegraphics[width=\textwidth,height=\textheight,keepaspectratio]{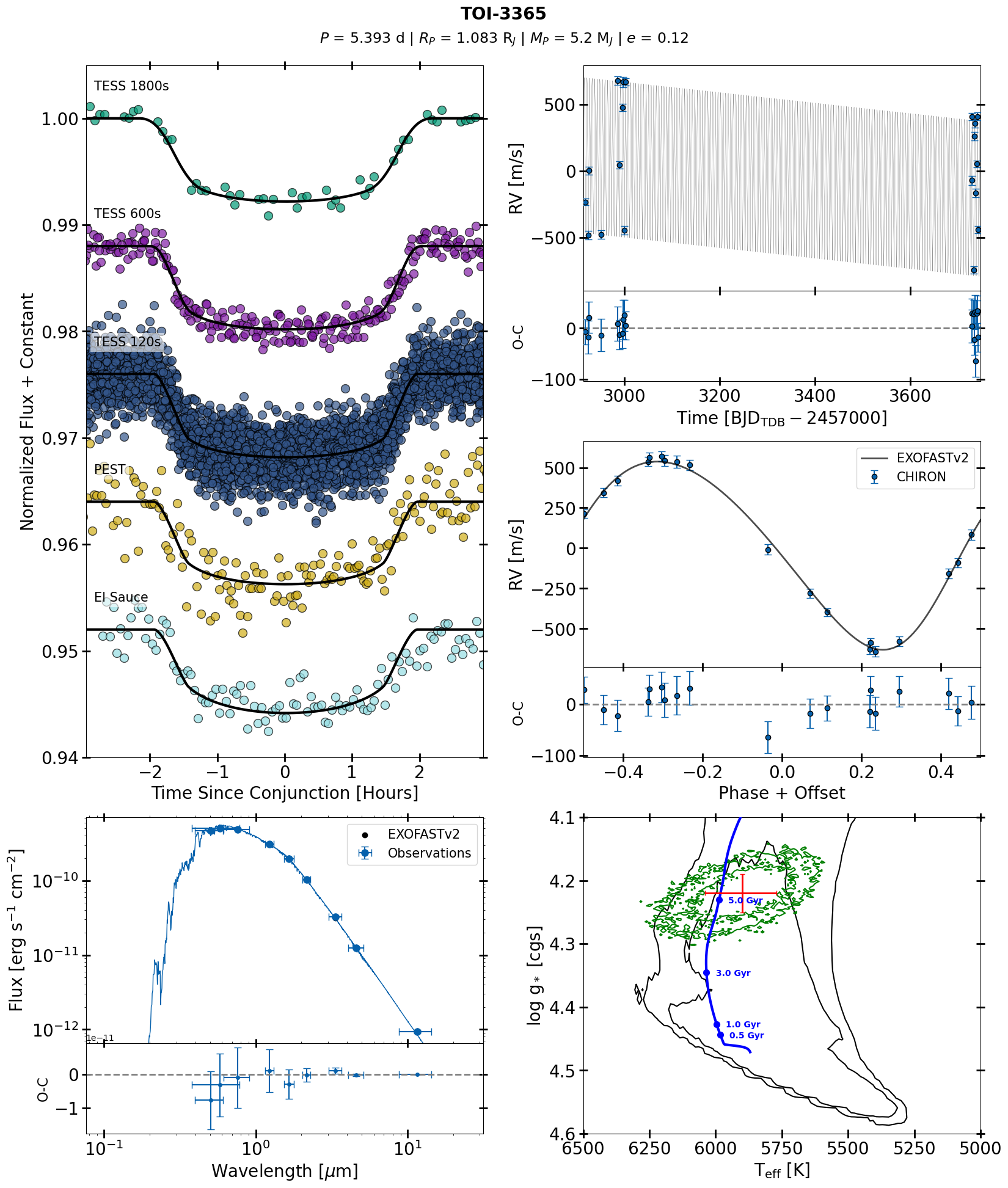}
    \caption{Same as Figure~\ref{fig:toi3041}, but for TOI-3365.}
    \label{fig:toi3365}
\end{figure*}

\begin{figure*}
    \centering    
    \includegraphics[width=\textwidth,height=\textheight,keepaspectratio]{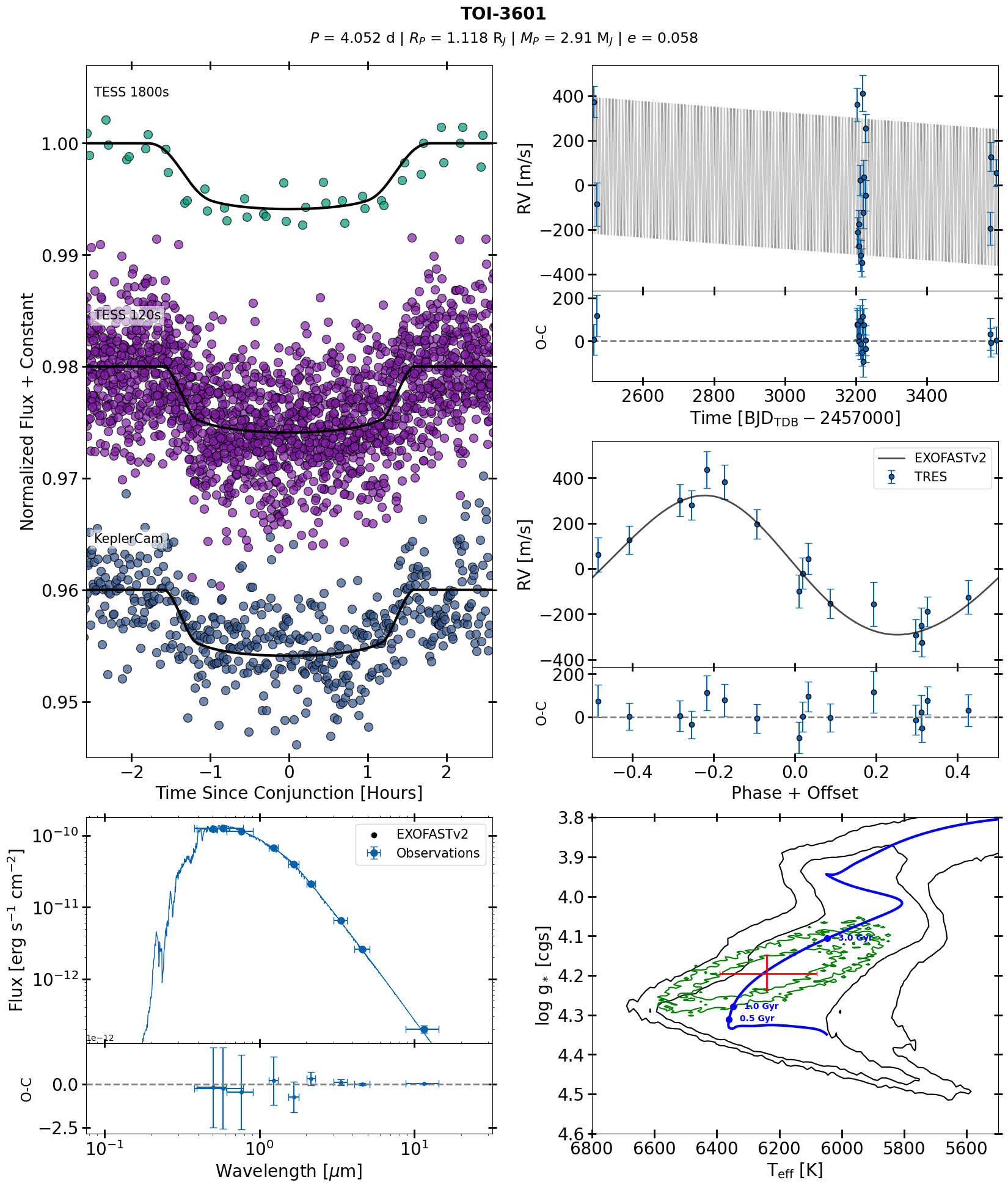}
    \caption{Same as Figure~\ref{fig:toi3041}, but for TOI-3601.}
    \label{fig:toi3601}
\end{figure*}

\begin{figure*}
    \centering    
    \includegraphics[width=\textwidth,height=\textheight,keepaspectratio]{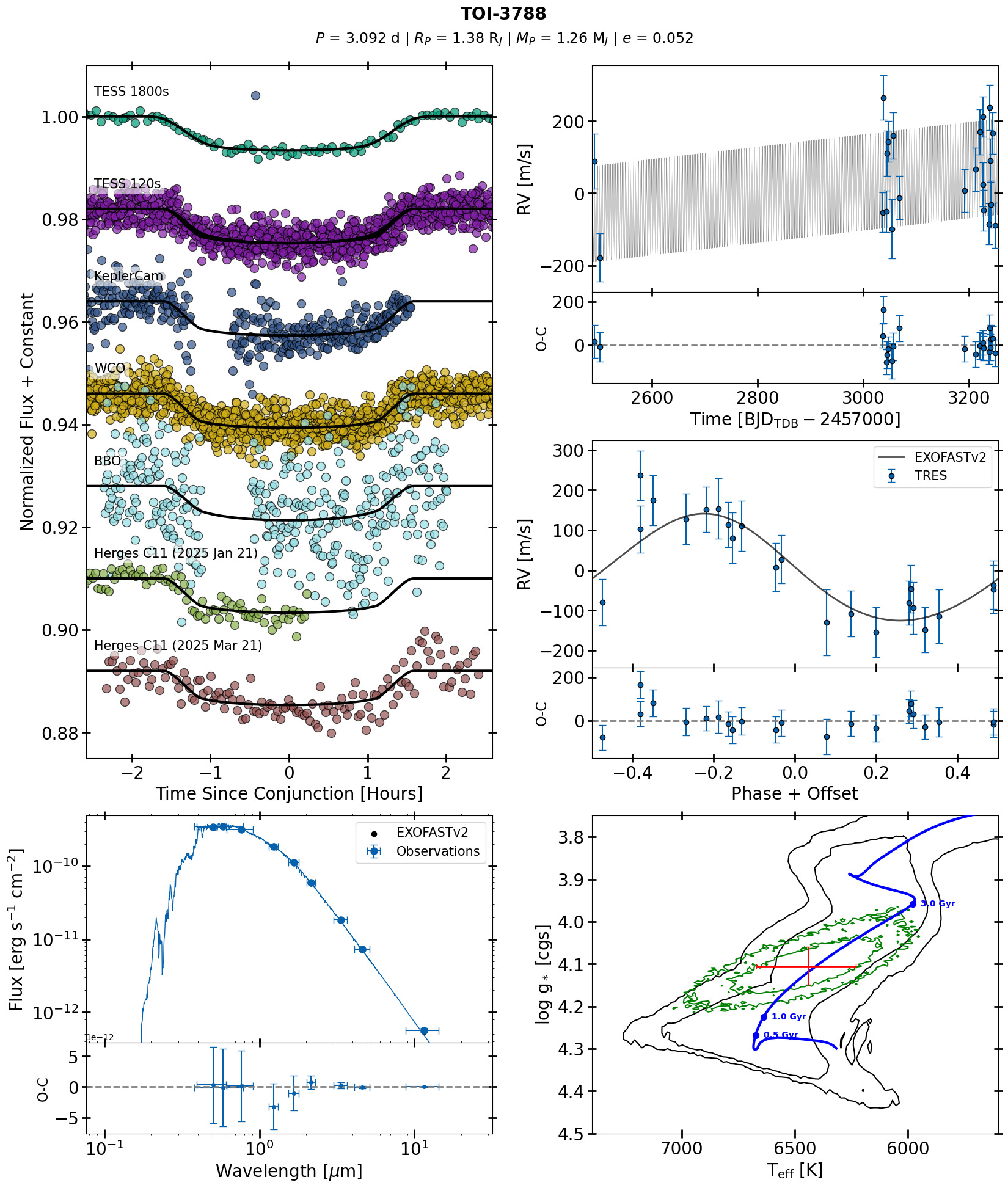}
    \caption{Same as Figure~\ref{fig:toi3041}, but for TOI-3788.}
    \label{fig:toi3788}
\end{figure*}

\begin{figure*}
    \centering    
    \includegraphics[width=\textwidth,height=\textheight,keepaspectratio]{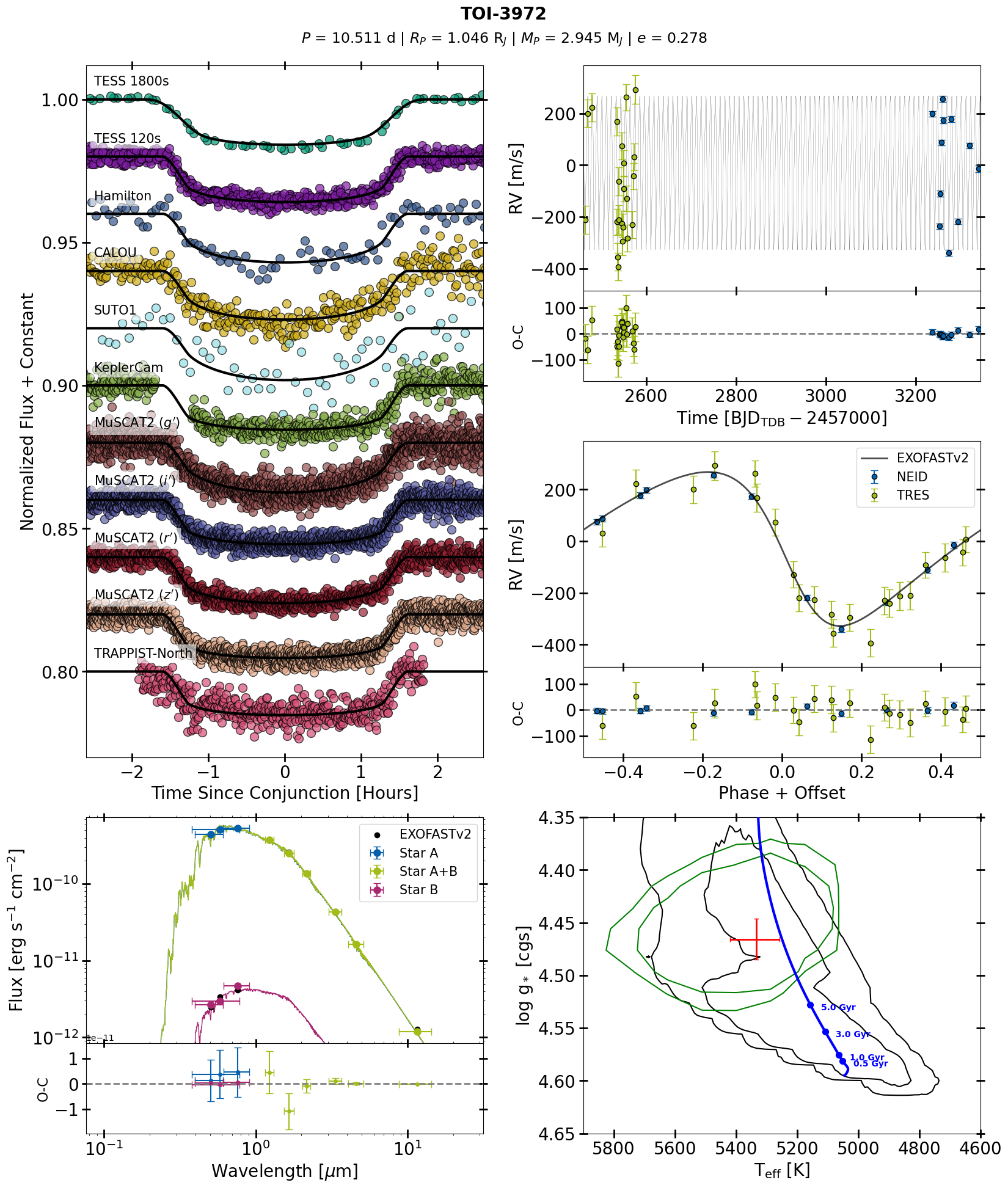}
    \caption{Same as Figure~\ref{fig:toi3041}, but for TOI-3972. In the SED plot in the \textbf{bottom left}, blue circles represent the unblended \gaia observations of the target star TOI-3972, the magenta circles represent the unblended \gaia observations of the unbound background star TIC 605485663, and the green circles represent the blended WISE and 2MASS observations of both stars. Corresponding model atmospheres are shown for each star, as well as the summed atmosphere.}
    \label{fig:toi3972}
\end{figure*}

\begin{figure*}
    \centering    
    \includegraphics[width=\textwidth,height=\textheight,keepaspectratio]{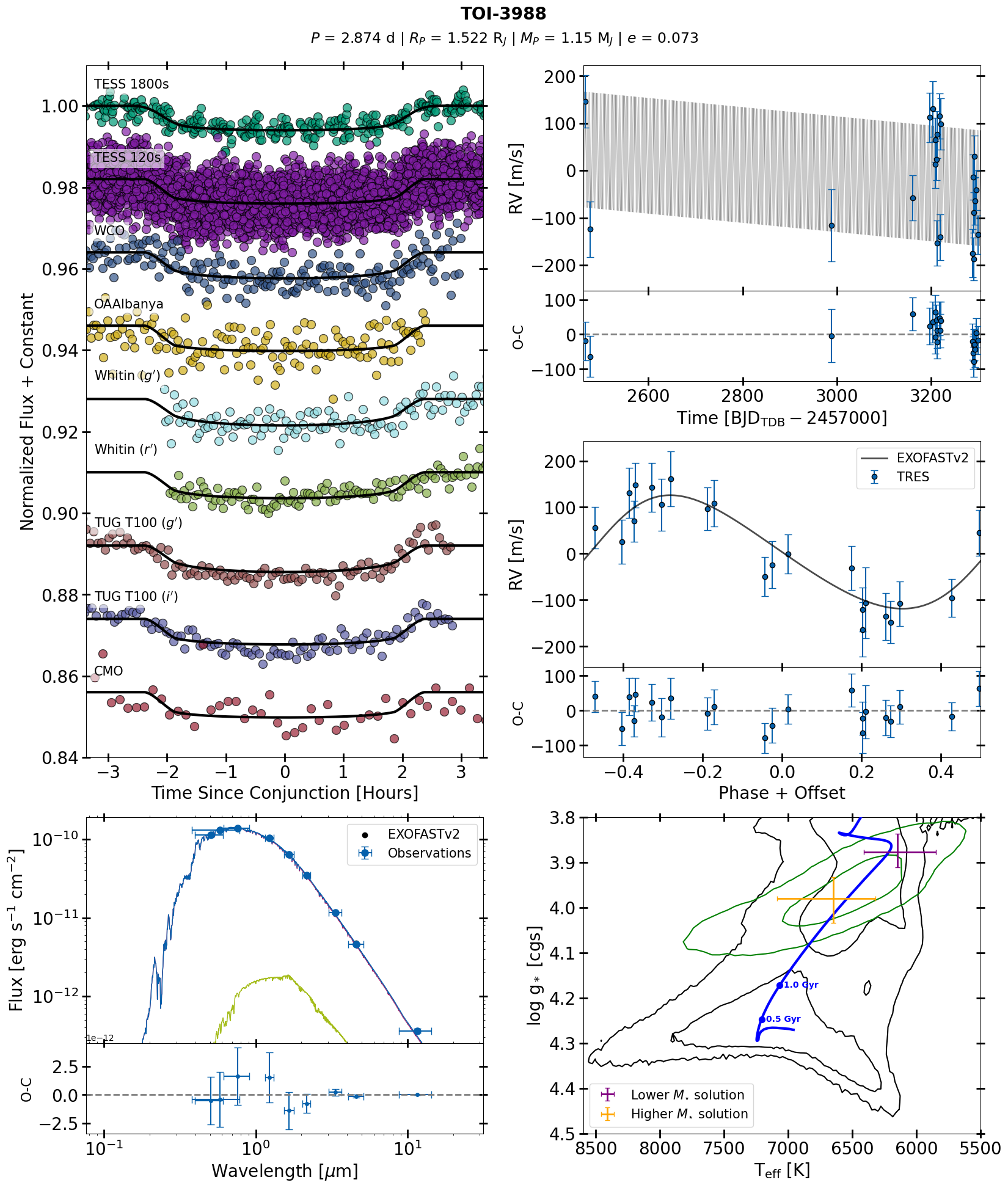}
    \caption{Same as Figure~\ref{fig:toi3041}, but for TOI-3988. In the SED plot in the \textbf{bottom left}, model atmospheres are shown for TOI-3988\,A (magenta), TOI-3988\,B (green), as well as the summed atmosphere (blue). All \gaia, 2MASS, and WISE observations were blended. In the evolutionary plot in the \textbf{bottom right}, the purple cross represents the median and 68\% confidence interval of the lower-mass subgiant solution, while the orange cross represents that of the higher-mass dwarf solution. Both  of these are listed in Table \ref{tab:bimodal}.}
    \label{fig:toi3988}
\end{figure*}

\begin{figure*}
    \centering    
    \includegraphics[width=\textwidth,height=\textheight,keepaspectratio]{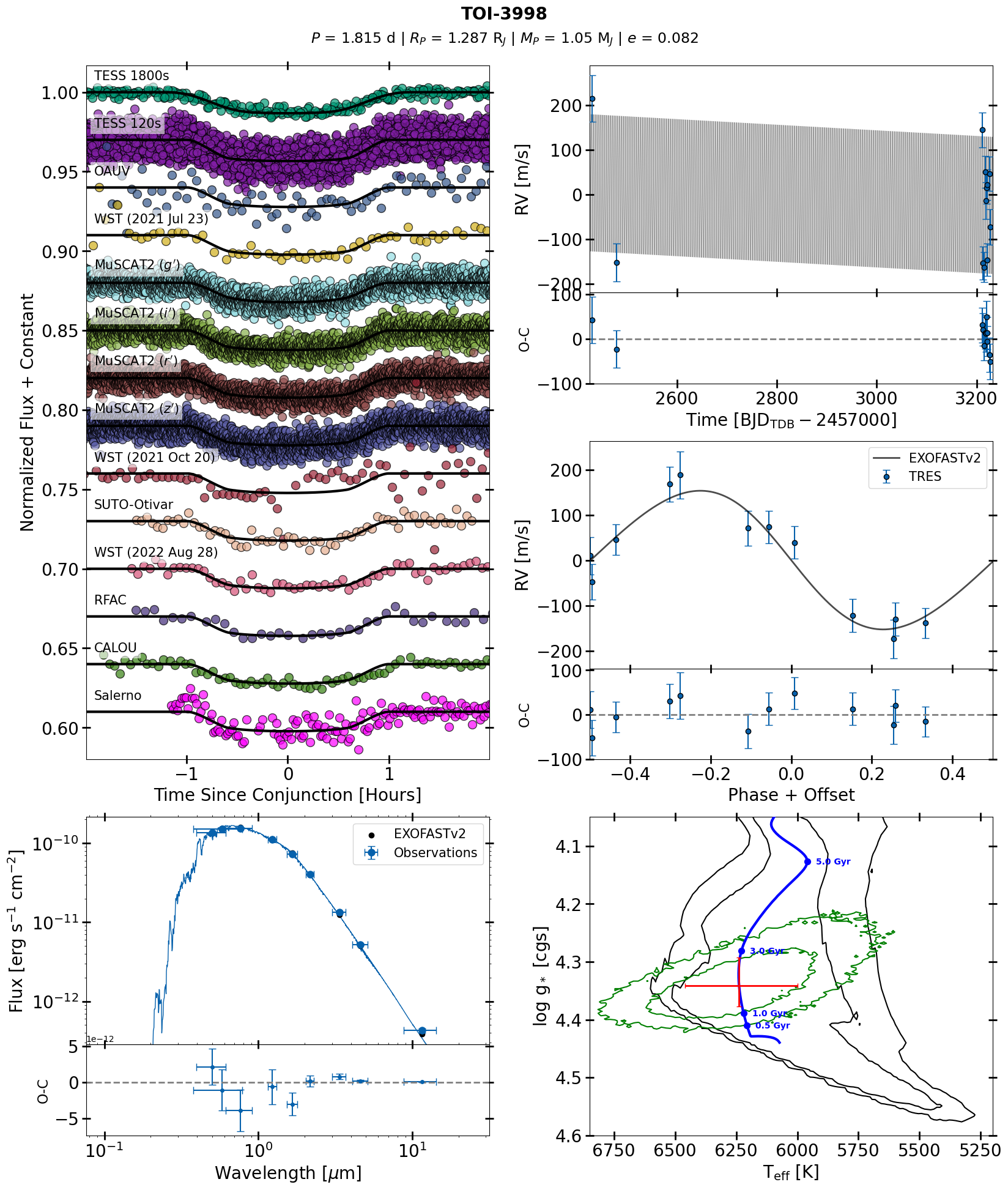}
    \caption{Same as Figure~\ref{fig:toi3041}, but for TOI-3998.}
    \label{fig:toi3998}
\end{figure*}

\begin{figure*}
    \centering    
    \includegraphics[width=\textwidth,height=\textheight,keepaspectratio]{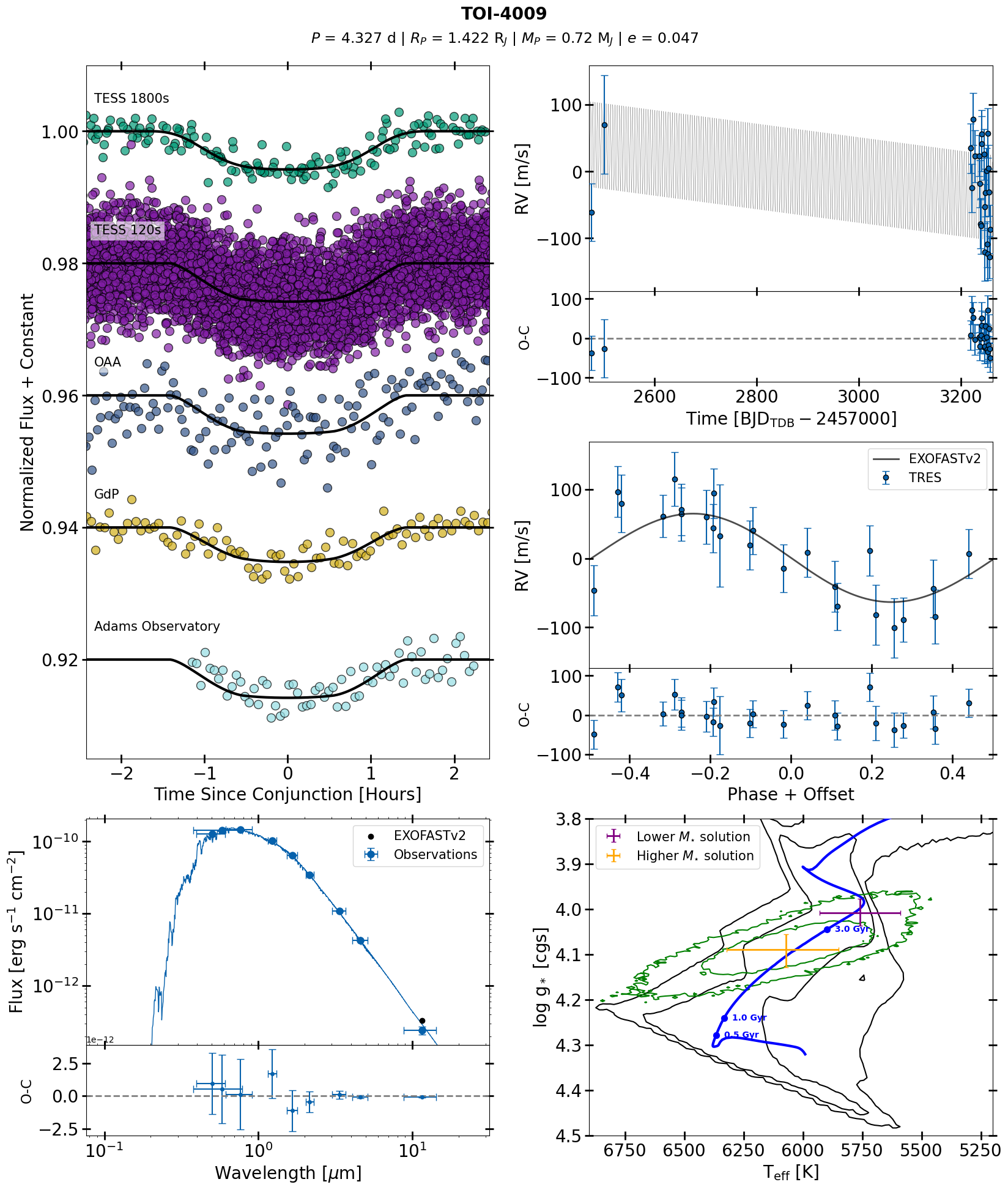}
    \caption{Same as Figure~\ref{fig:toi3041}, but for TOI-4009. In the evolutionary plot in the \textbf{bottom right}, the purple cross represents the median and 68\% confidence interval of the lower-mass subgiant solution, while the orange cross represents that of the higher-mass dwarf solution. Both  of these are listed in Table \ref{tab:bimodal}.}
    \label{fig:toi4009}
\end{figure*}

\begin{figure*}
    \centering    
    \includegraphics[width=\textwidth,height=\textheight,keepaspectratio]{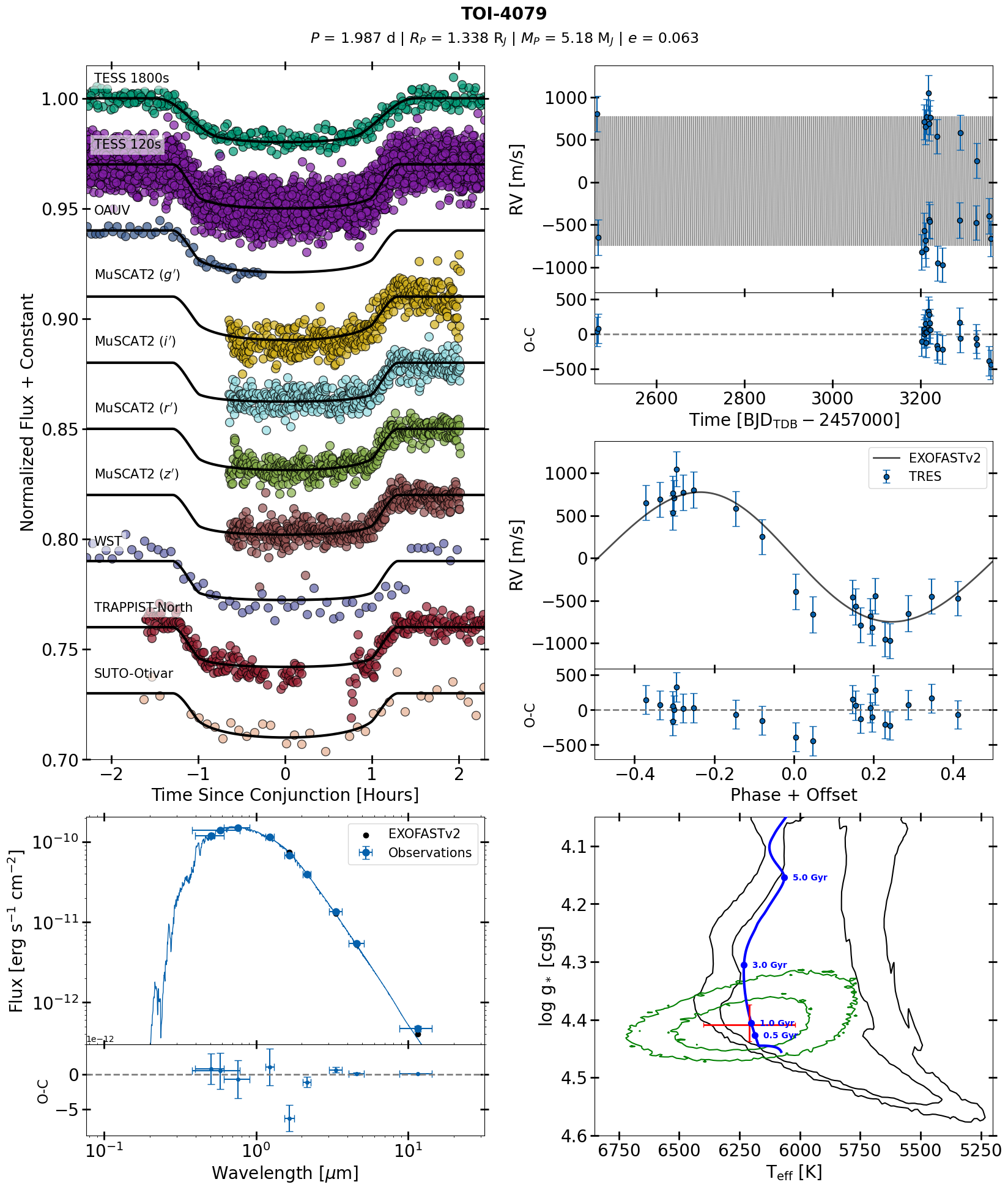}
    \caption{Same as Figure~\ref{fig:toi3041}, but for TOI-4079.}
    \label{fig:toi4079}
\end{figure*}

\begin{figure*}
    \centering    
    \includegraphics[width=\textwidth,height=\textheight,keepaspectratio]{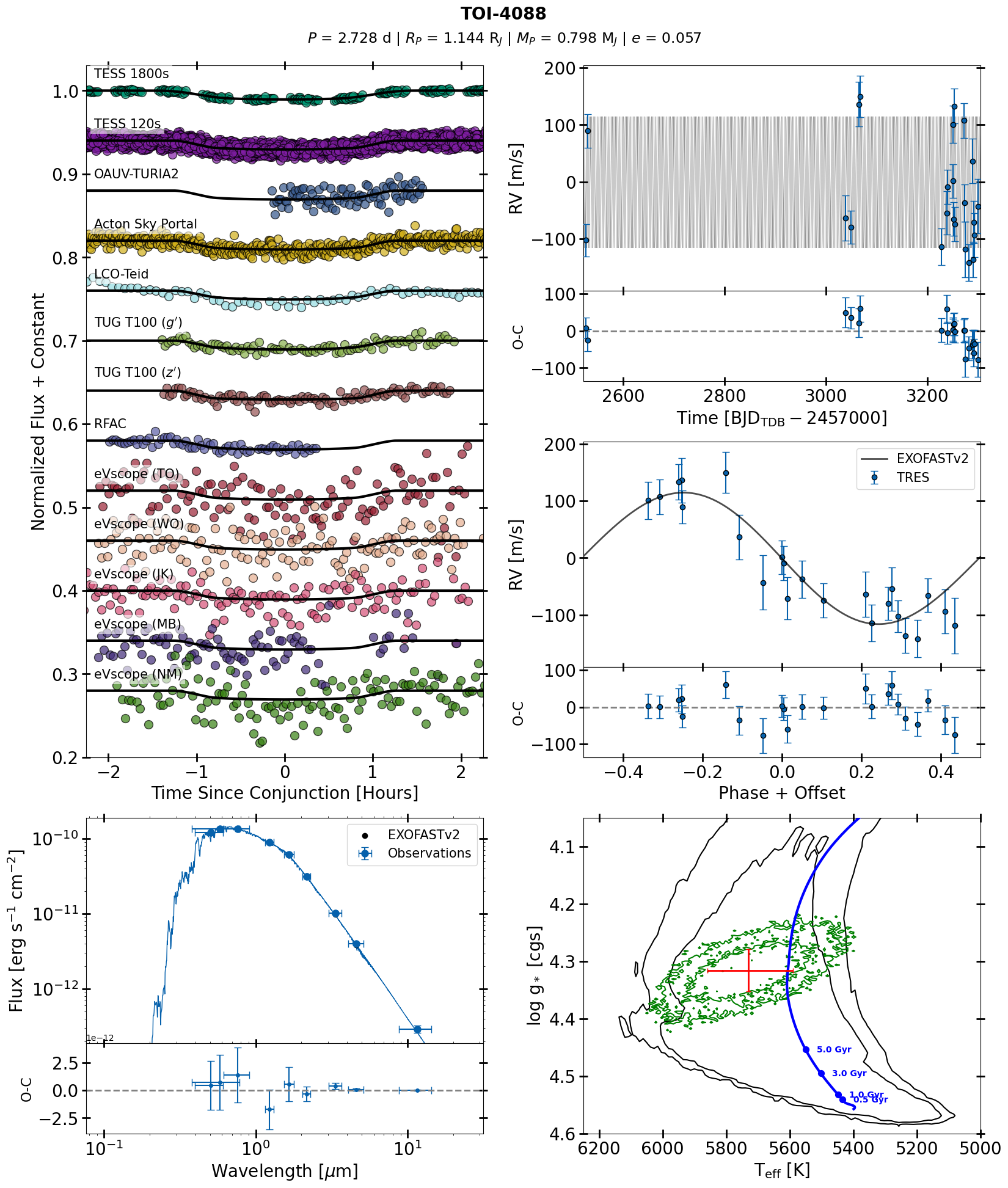}
    \caption{Same as Figure~\ref{fig:toi3041}, but for TOI-4088.}
    \label{fig:toi4088}
\end{figure*}

\begin{figure*}
    \centering    
    \includegraphics[width=\textwidth,height=\textheight,keepaspectratio]{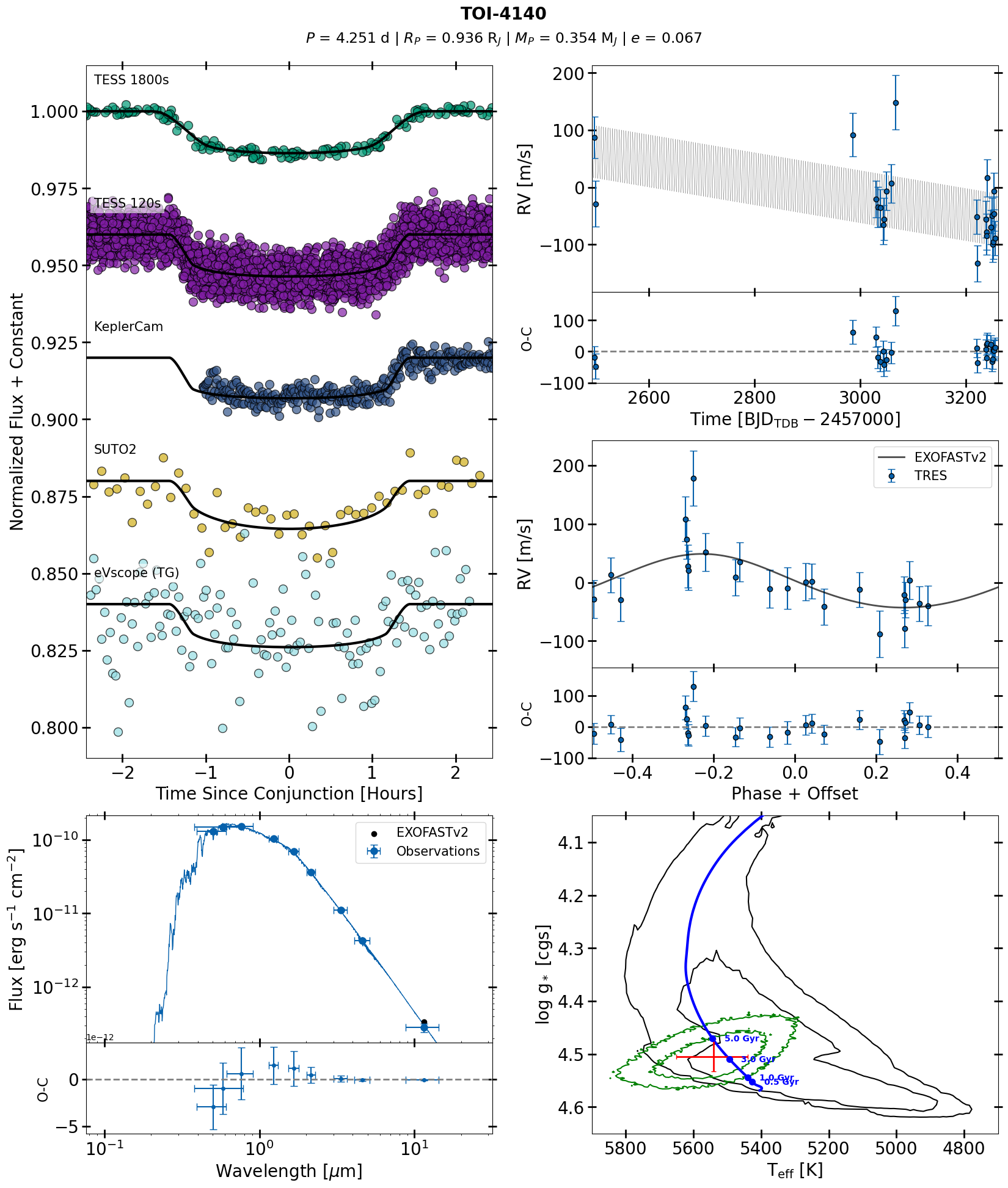}
    \caption{Same as Figure~\ref{fig:toi3041}, but for TOI-4140.}
    \label{fig:toi4140}
\end{figure*}

\begin{figure*}
    \centering    
    \includegraphics[width=\textwidth,height=\textheight,keepaspectratio]{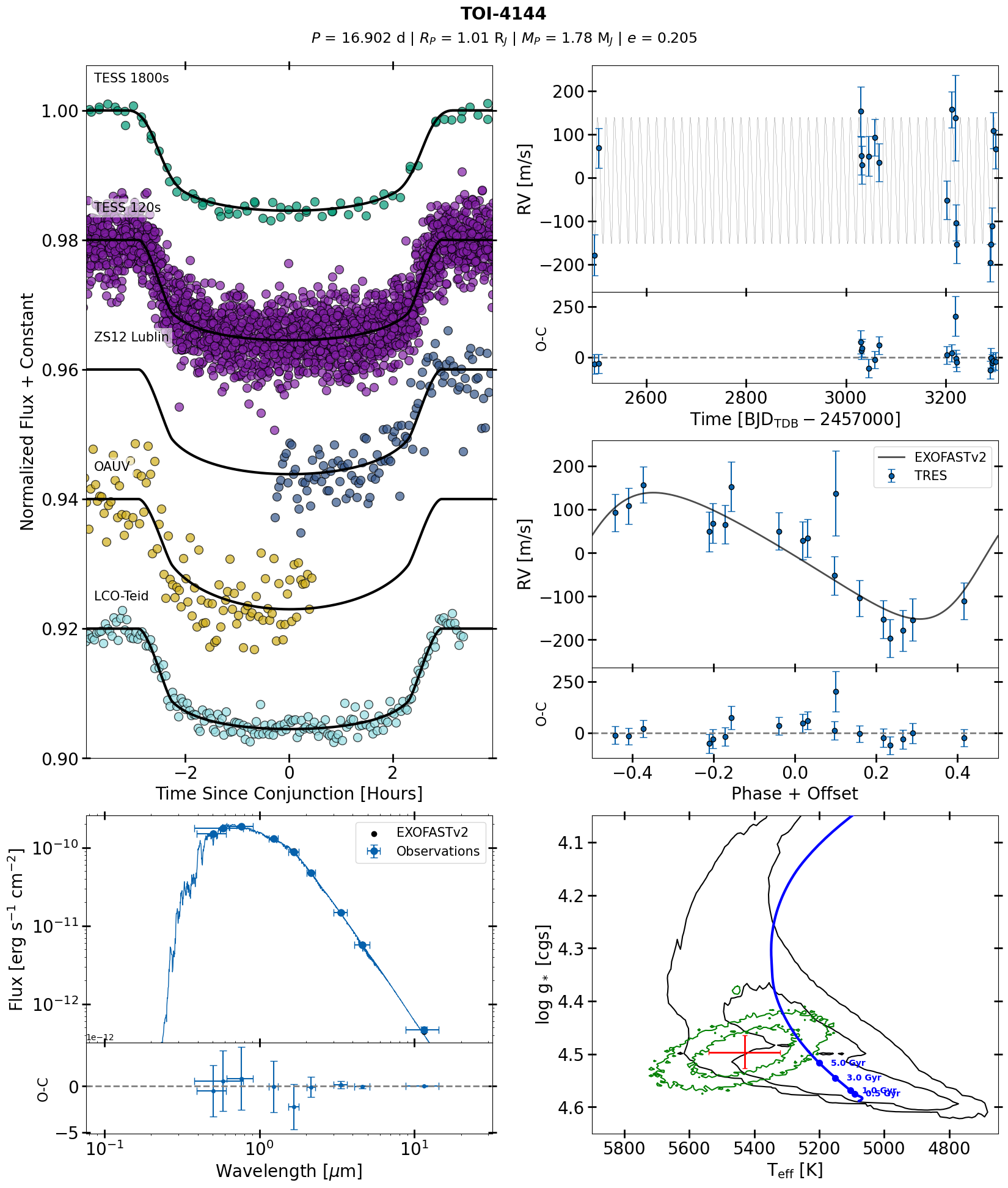}
    \caption{Same as Figure~\ref{fig:toi3041}, but for TOI-4144.}
    \label{fig:toi4144}
\end{figure*}

\begin{figure*}
    \centering    
    \includegraphics[width=\textwidth,height=\textheight,keepaspectratio]{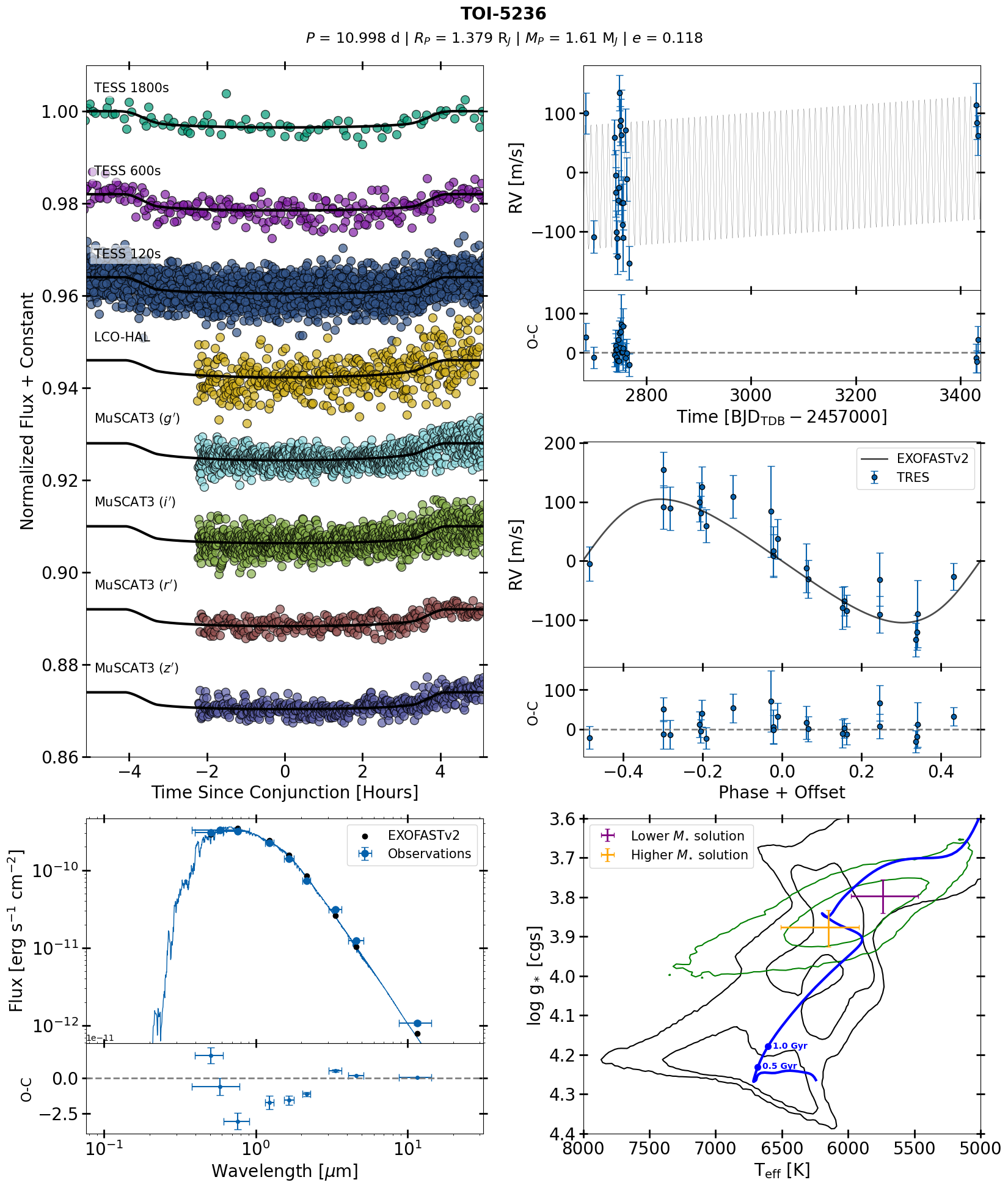}
    \caption{Same as Figure~\ref{fig:toi3041}, but for TOI-5236. In the evolutionary plot in the \textbf{bottom right}, the purple cross represents the median and 68\% confidence interval of the lower-mass subgiant solution, while the orange cross represents that of the higher-mass dwarf solution. Both  of these are listed in Table \ref{tab:bimodal}.}
    \label{fig:toi5236}
\end{figure*}

\begin{figure*}
    \centering    
    \includegraphics[width=\textwidth,height=\textheight,keepaspectratio]{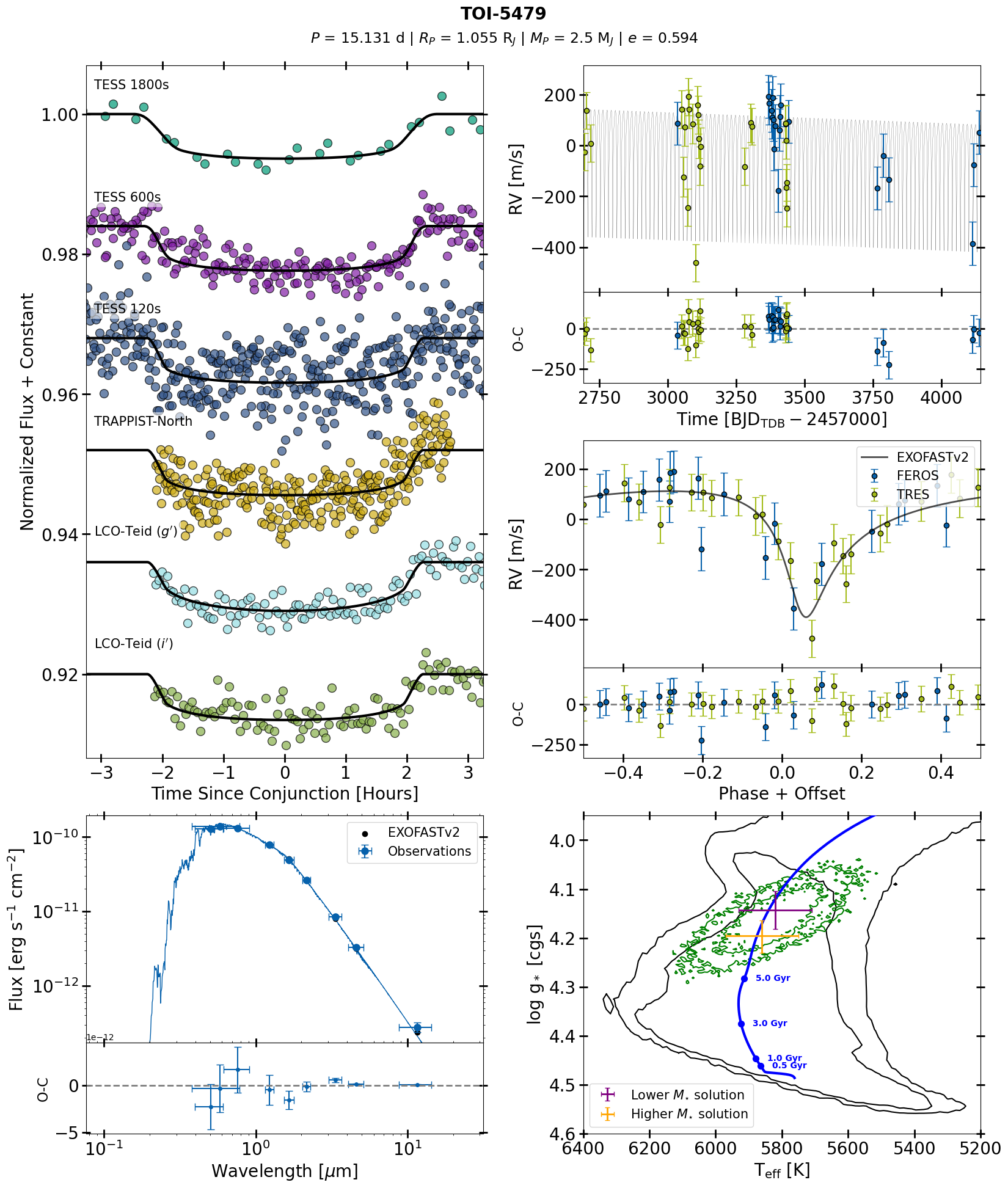}
    \caption{Same as Figure~\ref{fig:toi3041}, but for TOI-5479. In the evolutionary plot in the \textbf{bottom right}, the purple cross represents the median and 68\% confidence interval of the lower-mass subgiant solution, while the orange cross represents that of the higher-mass dwarf solution. Both  of these are listed in Table \ref{tab:bimodal}.}
    \label{fig:toi5479}
\end{figure*}

\begin{figure*}
    \centering    
    \includegraphics[width=\textwidth,height=\textheight,keepaspectratio]{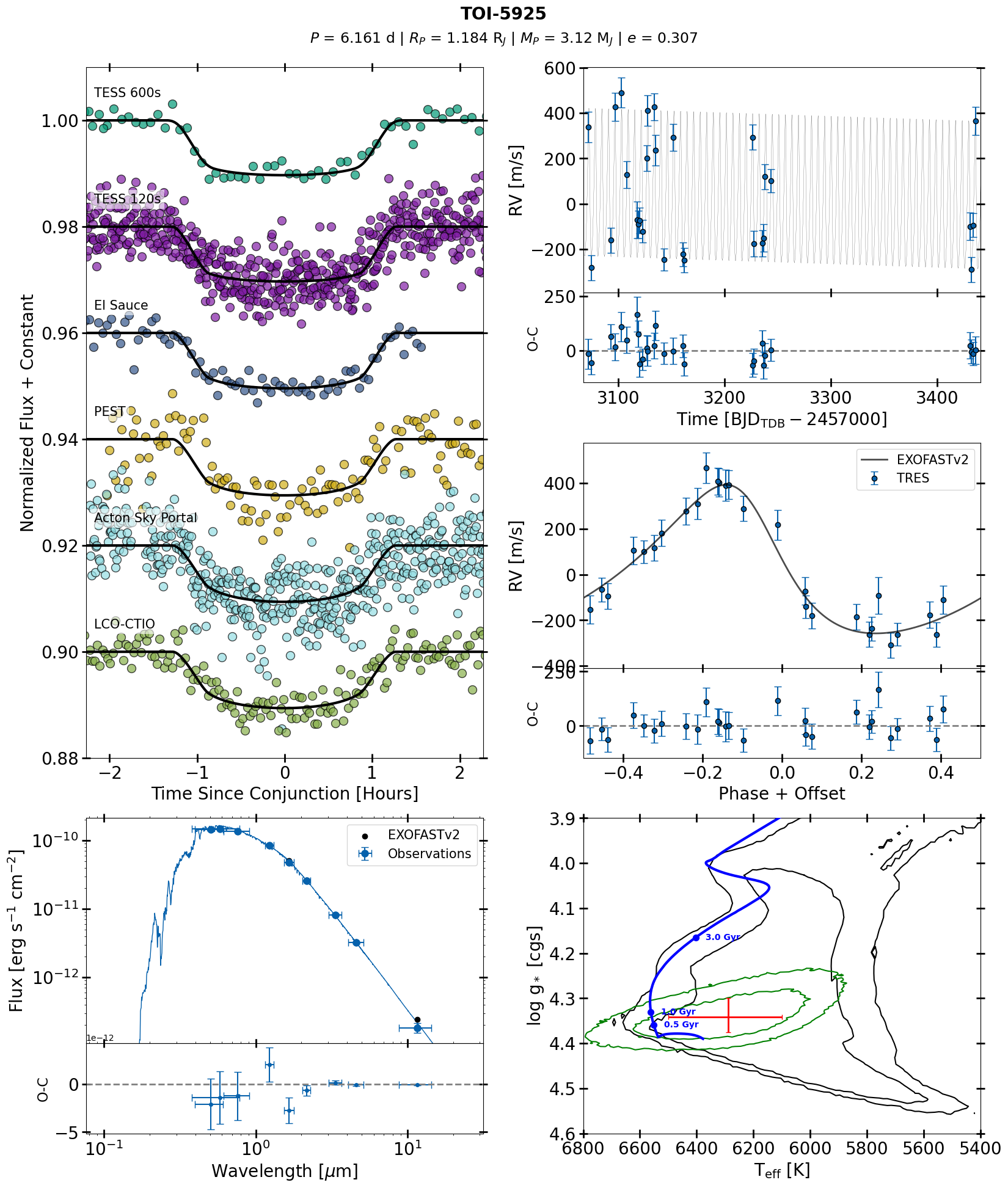}
    \caption{Same as Figure~\ref{fig:toi3041}, but for TOI-5925.}
    \label{fig:toi5925}
\end{figure*}

\begin{figure*}
    \centering    
    \includegraphics[width=\textwidth,height=\textheight,keepaspectratio]{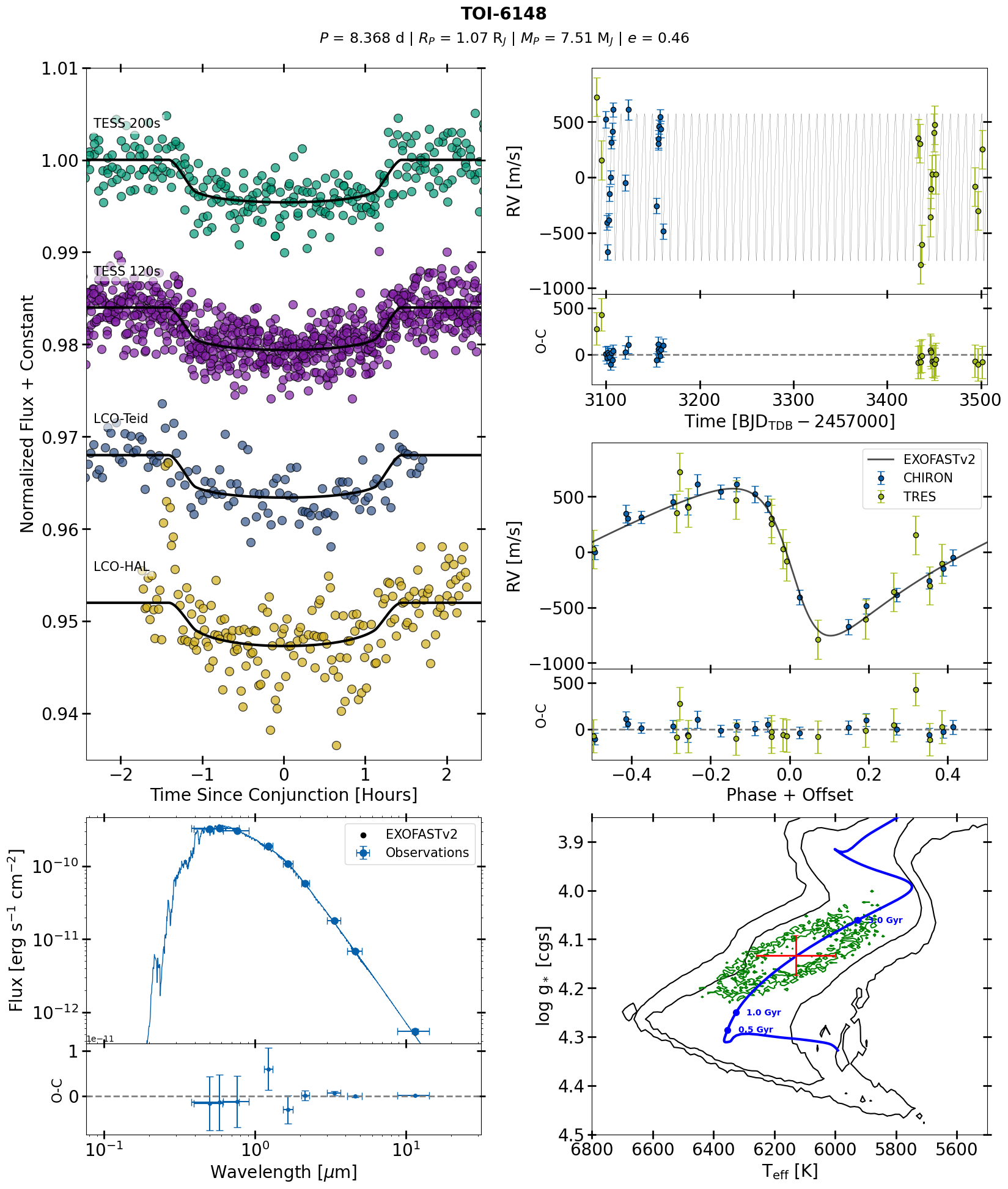}
    \caption{Same as Figure~\ref{fig:toi3041}, but for TOI-6148.}
    \label{fig:toi6148}
\end{figure*}

\begin{figure*}
    \centering    
    \includegraphics[width=\textwidth,height=\textheight,keepaspectratio]{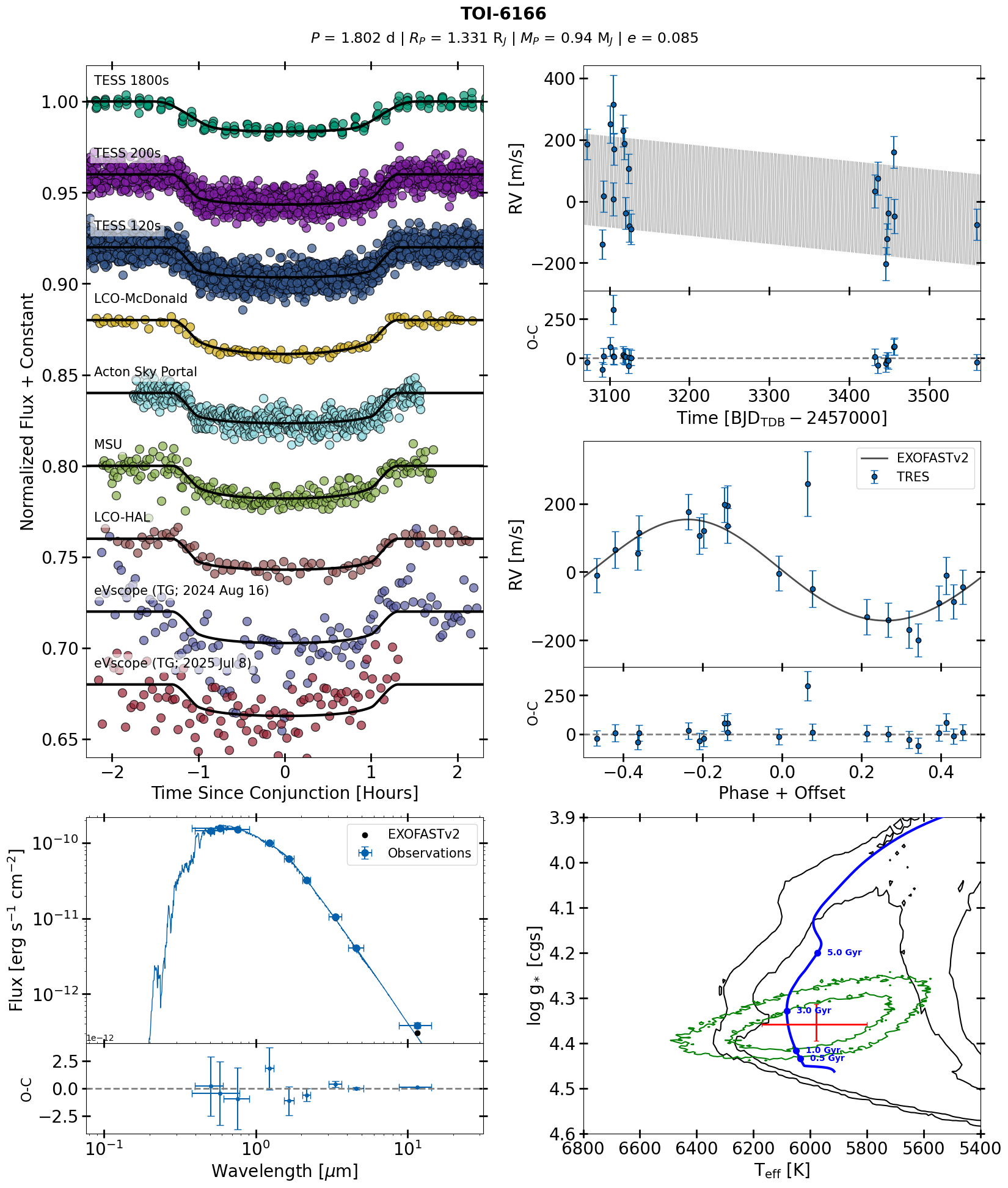}
    \caption{Same as Figure~\ref{fig:toi3041}, but for TOI-6166.}
    \label{fig:toi6166}
\end{figure*}

\begin{figure*}
    \centering    
    \includegraphics[width=\textwidth,height=\textheight,keepaspectratio]{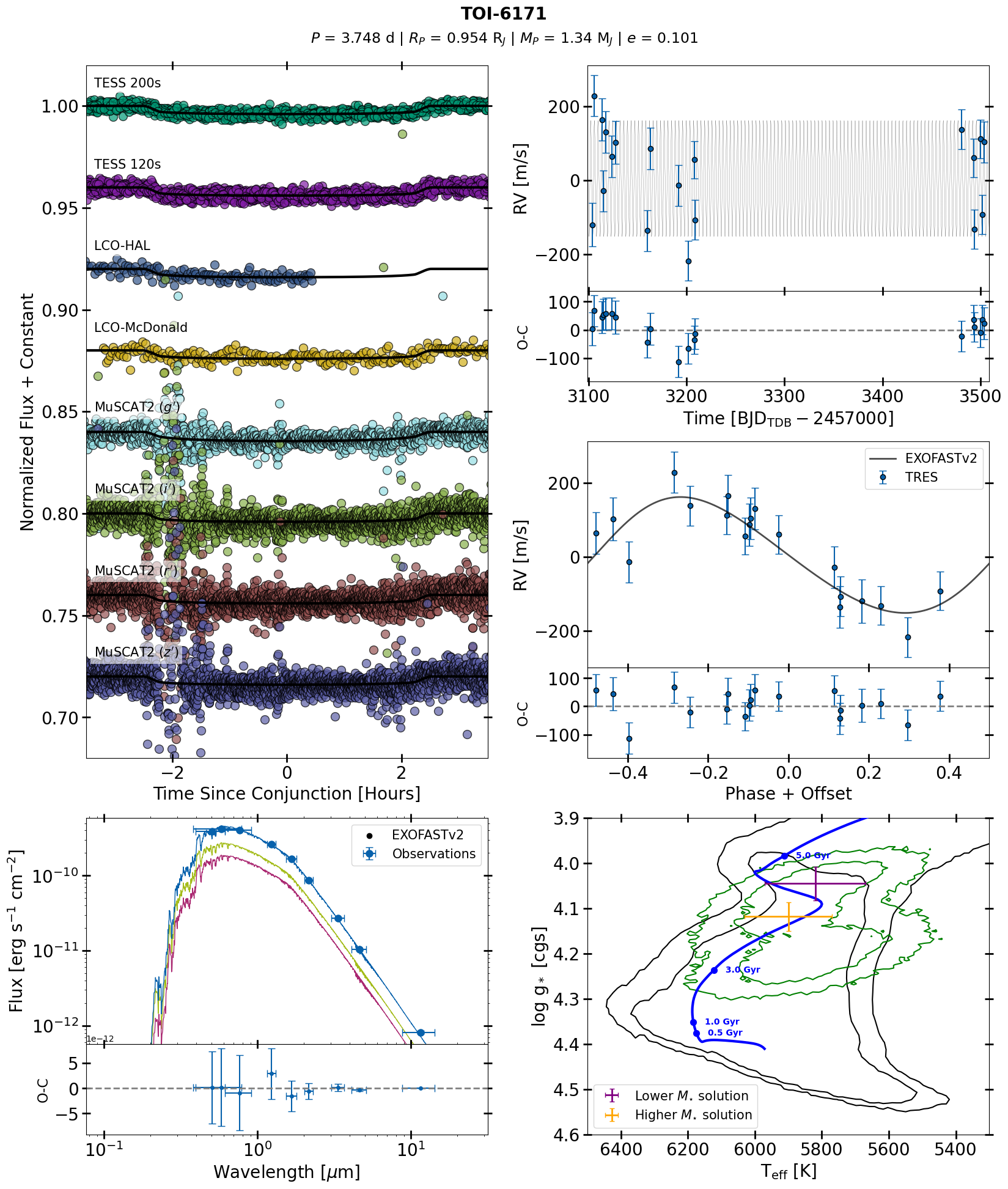}
    \caption{Same as Figure~\ref{fig:toi3041}, but for TOI-6171. In the SED plot in the \textbf{bottom left}, model atmospheres are shown for TOI-6171\,A (green), TOI-6171\,B (magenta), as well as the summed atmosphere (blue). All \gaia, 2MASS, and WISE observations were blended. In the evolutionary plot in the \textbf{bottom right}, the purple cross represents the median and 68\% confidence interval of the lower-mass subgiant solution, while the orange cross represents that of the higher-mass dwarf solution. Both  of these are listed in Table \ref{tab:bimodal}.}
    \label{fig:toi6171}
\end{figure*}

\begin{figure*}
    \centering    
    \includegraphics[width=\textwidth,height=\textheight,keepaspectratio]{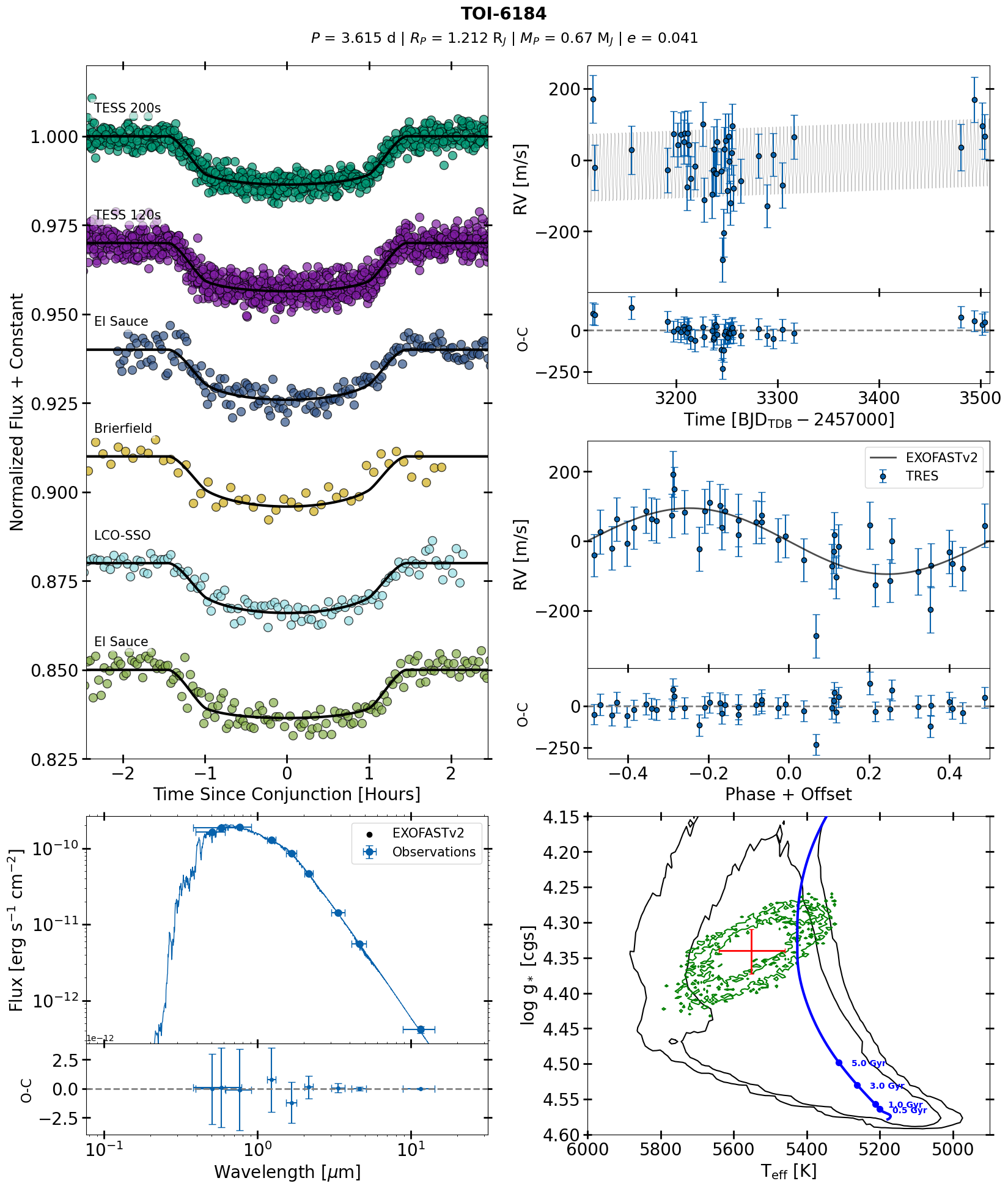}
    \caption{Same as Figure~\ref{fig:toi3041}, but for TOI-6184.}
    \label{fig:toi6184}
\end{figure*}

\begin{figure*}
    \centering    
    \includegraphics[width=\textwidth,height=\textheight,keepaspectratio]{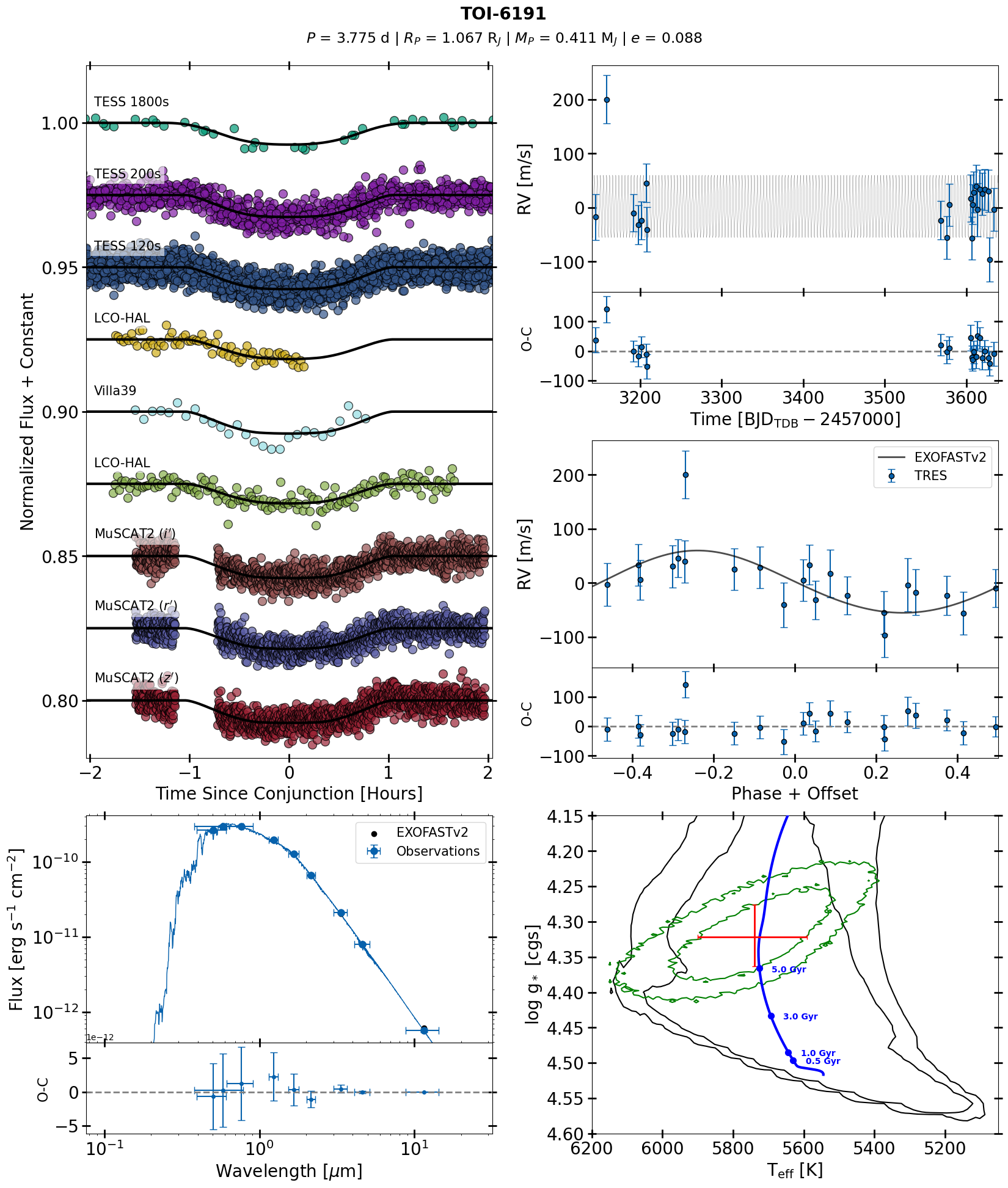}
    \caption{Same as Figure~\ref{fig:toi3041}, but for TOI-6191.}
    \label{fig:toi6191}
\end{figure*}

\begin{figure*}
    \centering    
    \includegraphics[width=\textwidth,height=\textheight,keepaspectratio]{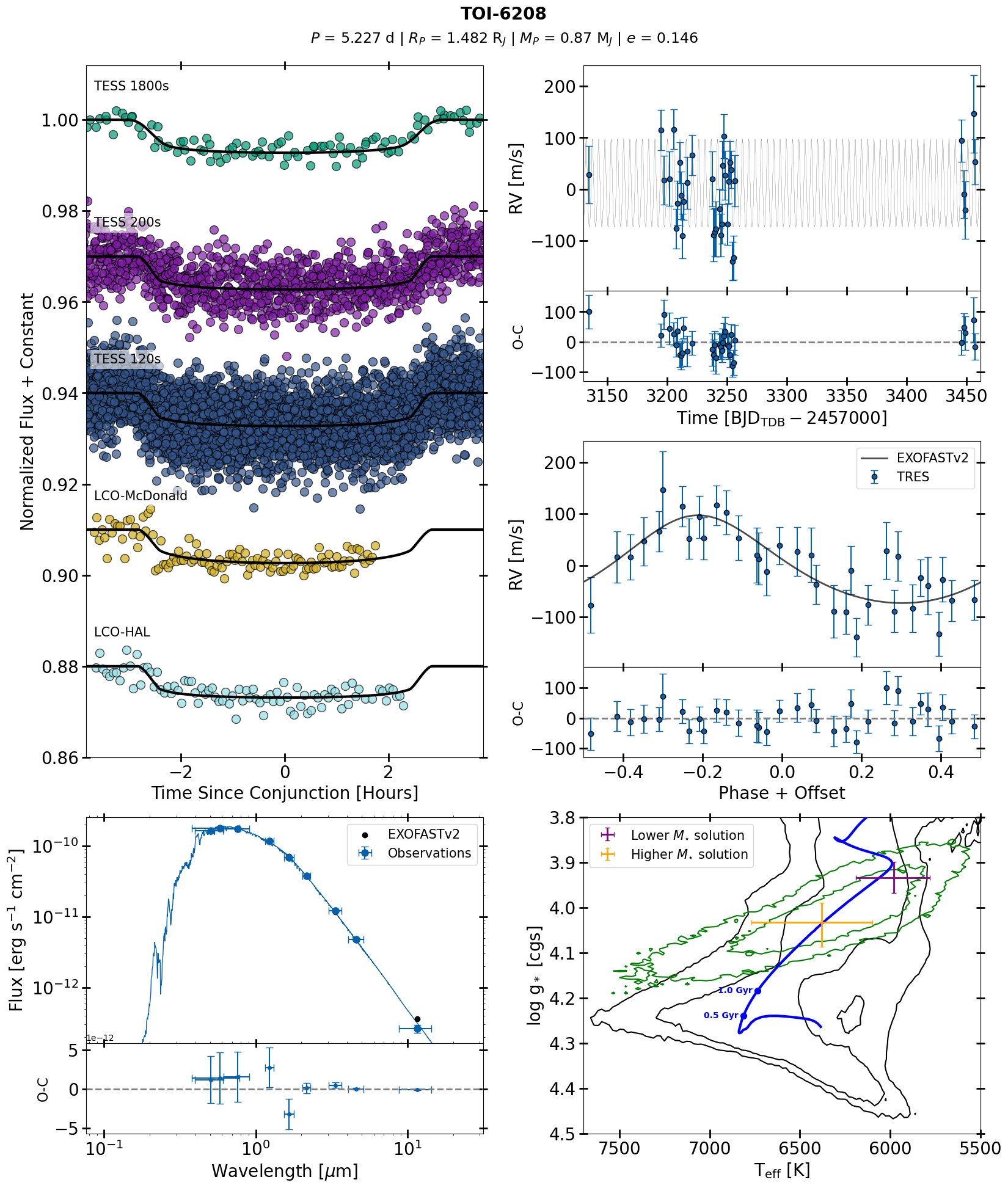}
    \caption{Same as Figure~\ref{fig:toi3041}, but for TOI-6208. In the evolutionary plot in the \textbf{bottom right}, the purple cross represents the median and 68\% confidence interval of the lower-mass subgiant solution, while the orange cross represents that of the higher-mass dwarf solution. Both  of these are listed in Table \ref{tab:bimodal}.}
    \label{fig:toi6208}
\end{figure*}

\begin{figure*}
    \centering    
    \includegraphics[width=\textwidth,height=\textheight,keepaspectratio]{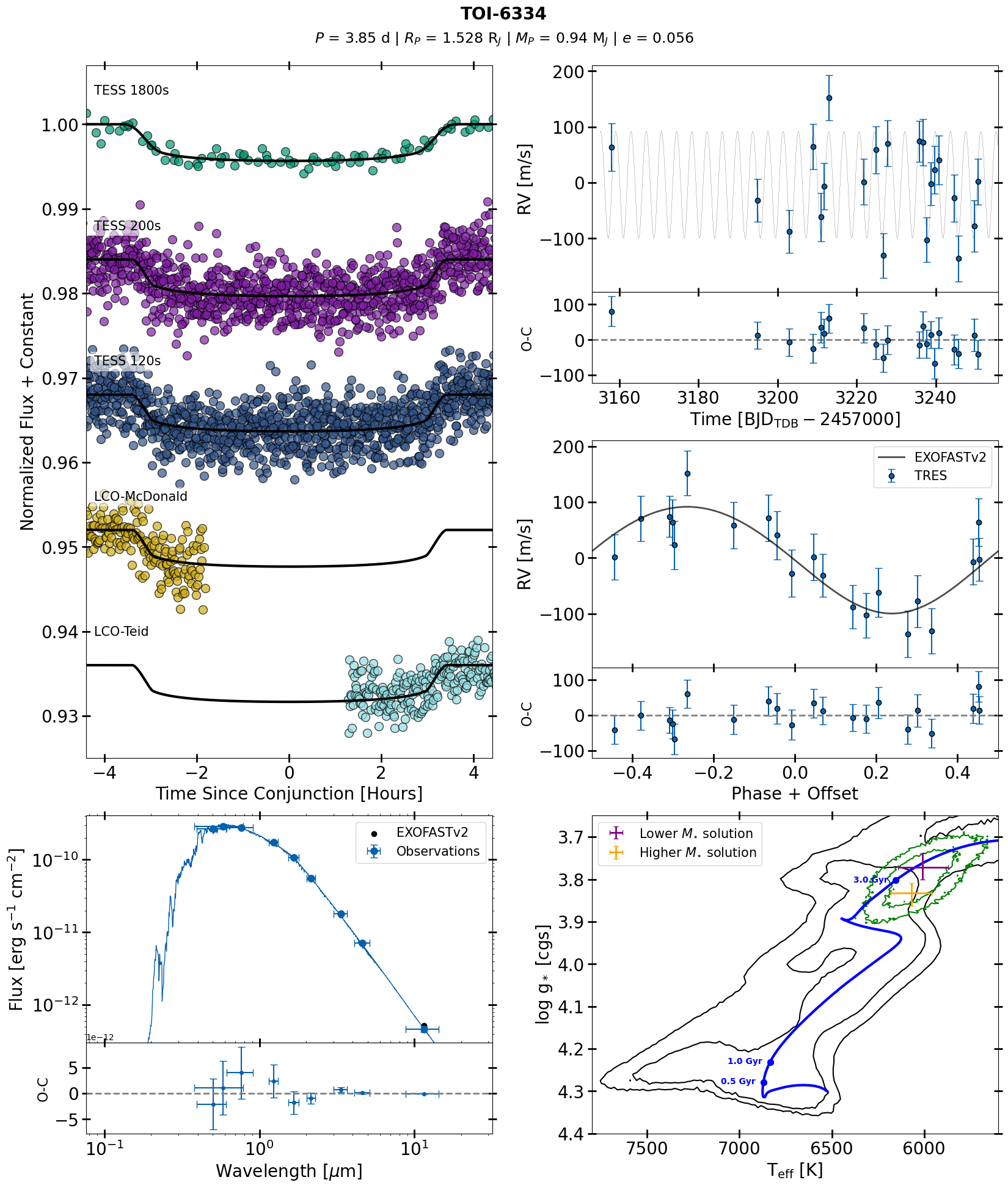}
    \caption{Same as Figure~\ref{fig:toi3041}, but for TOI-6334. In the evolutionary plot in the \textbf{bottom right}, the purple cross represents the median and 68\% confidence interval of the lower-mass subgiant solution, while the orange cross represents that of the higher-mass dwarf solution. Both  of these are listed in Table \ref{tab:bimodal}.}
    \label{fig:toi6334}
\end{figure*}

\begin{figure*}
    \centering    
    \includegraphics[width=\textwidth,height=\textheight,keepaspectratio]{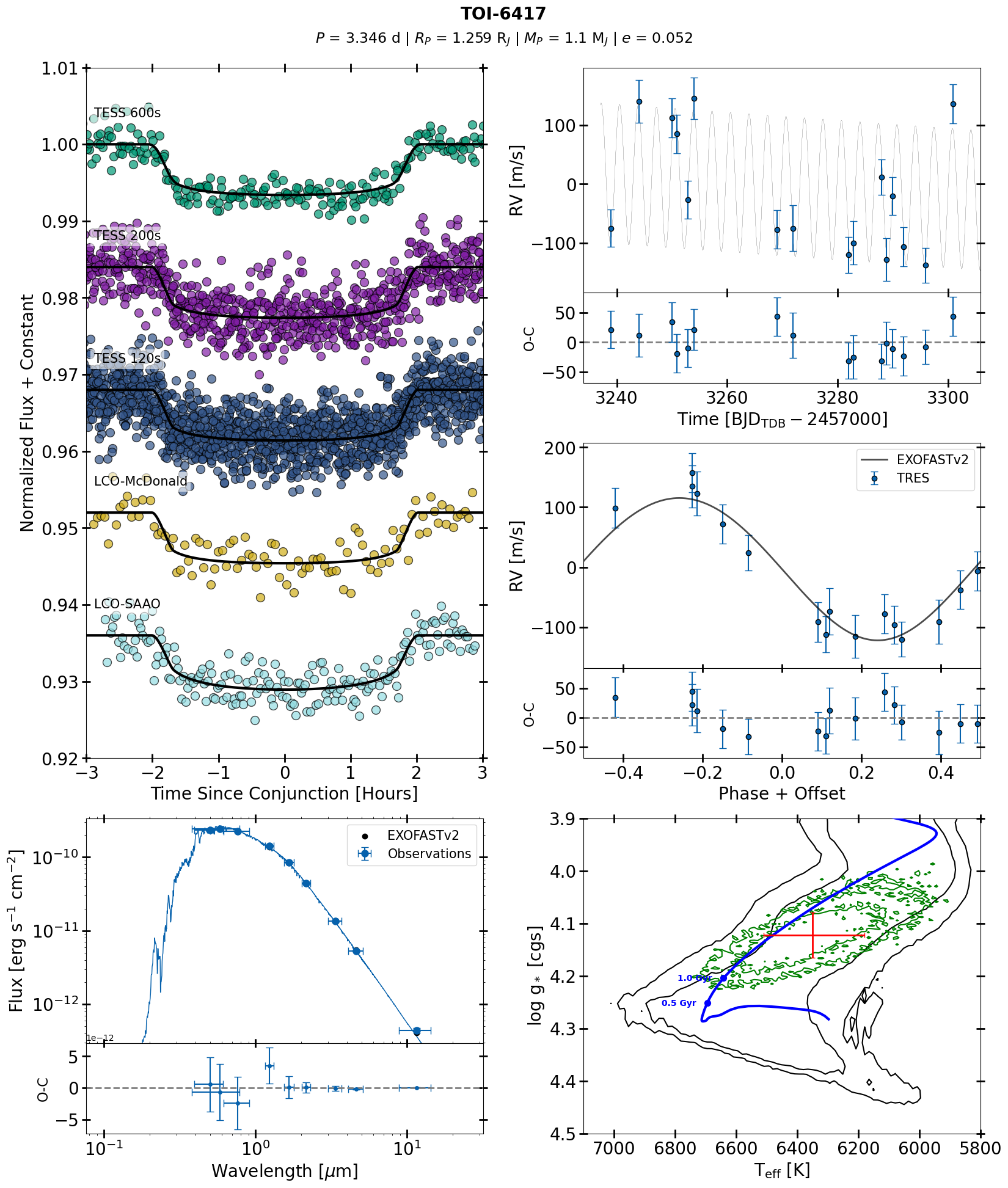}
    \caption{Same as Figure~\ref{fig:toi3041}, but for TOI-6417.}
    \label{fig:toi6417}
\end{figure*}

\begin{figure*}
    \centering    
    \includegraphics[width=\textwidth,height=\textheight,keepaspectratio]{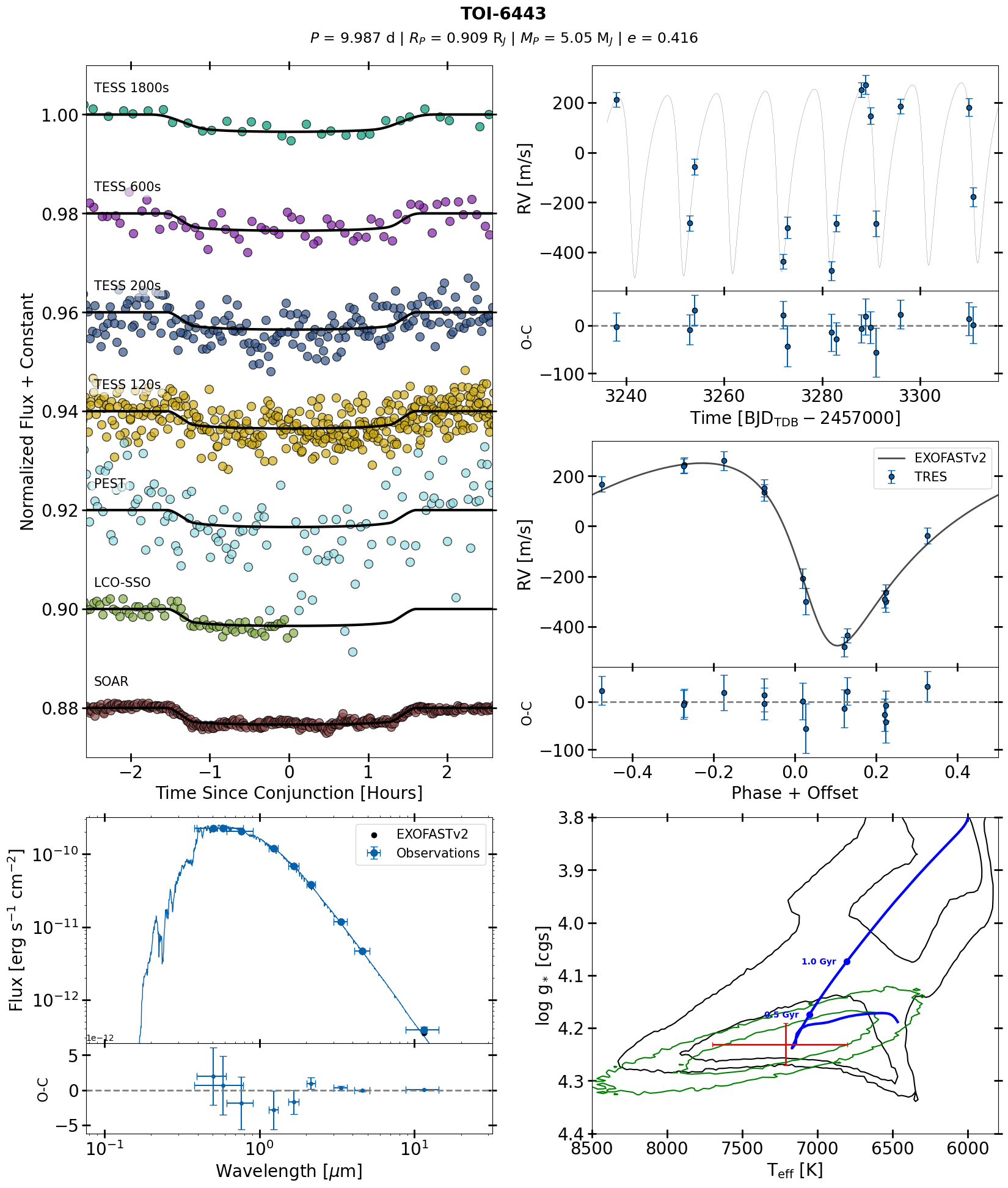}
    \caption{Same as Figure~\ref{fig:toi3041}, but for TOI-6443.}
    \label{fig:toi6443}
\end{figure*}

\begin{figure*}
    \centering    
    \includegraphics[width=\textwidth,height=\textheight,keepaspectratio]{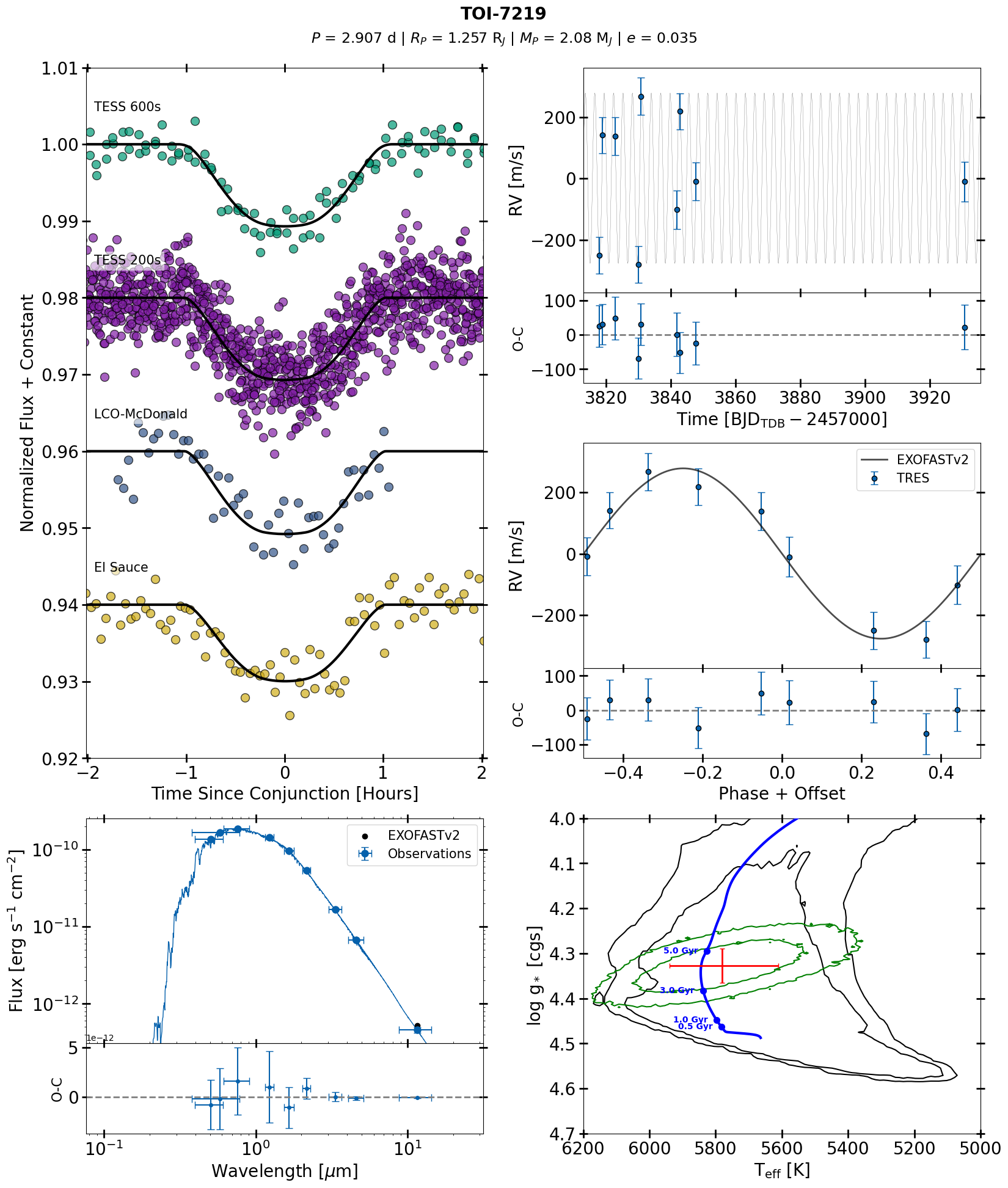}
    \caption{Same as Figure~\ref{fig:toi3041}, but for TOI-7219.}
    \label{fig:toi7219}
\end{figure*}

\begin{figure*}
    \centering    
    \includegraphics[width=\textwidth,height=\textheight,keepaspectratio]{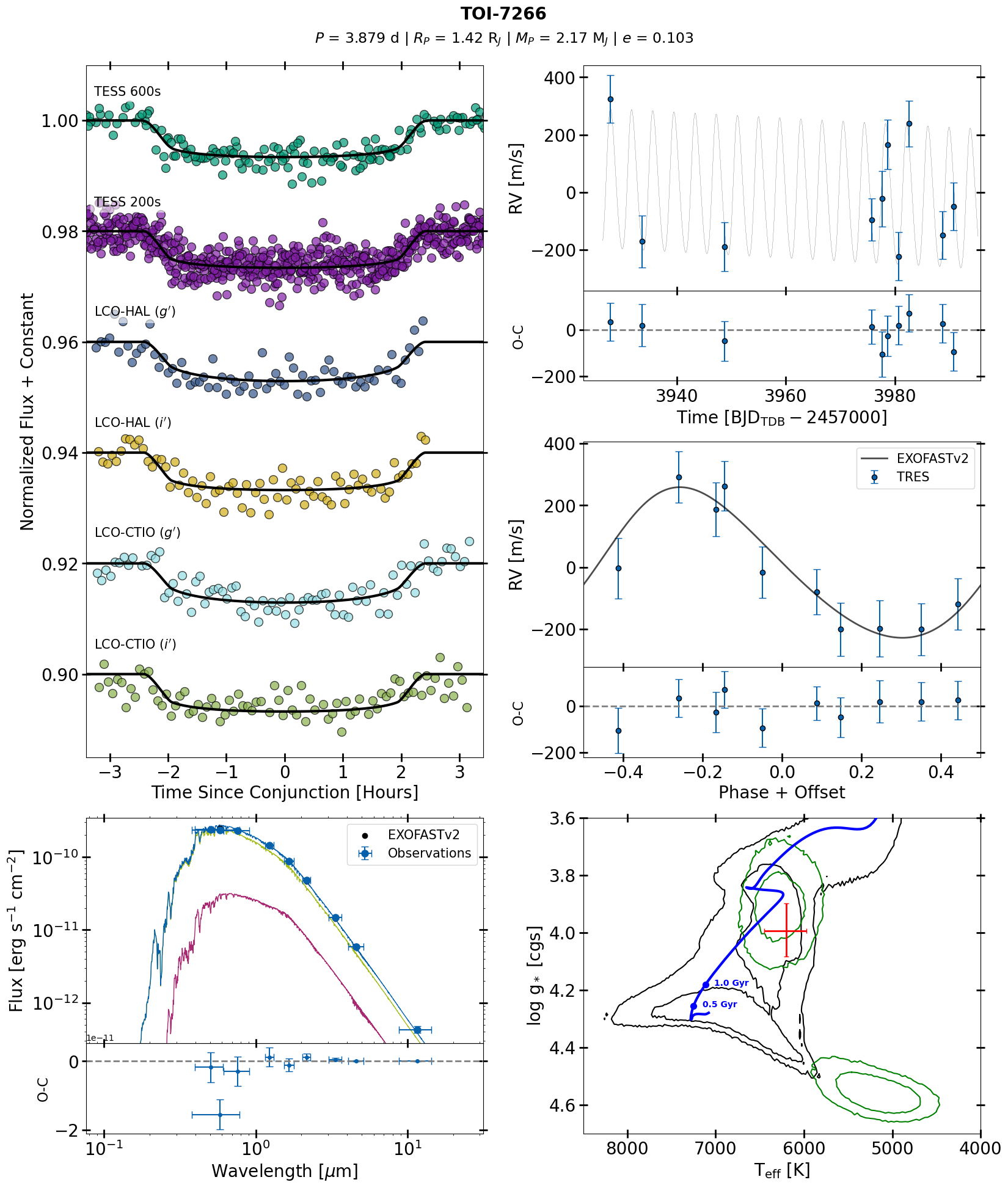}
    \caption{Same as Figure~\ref{fig:toi3041}, but for TOI-7266.In the SED plot in the \textbf{bottom left}, model atmospheres are shown for TOI-7266\,A (green), TOI-7266\,B (magenta), as well as the summed atmosphere (blue). All \gaia, 2MASS, and WISE observations were blended.}
    \label{fig:toi7266}
\end{figure*}

\begin{figure*}
    \centering    
    \includegraphics[width=\textwidth,height=\textheight,keepaspectratio]{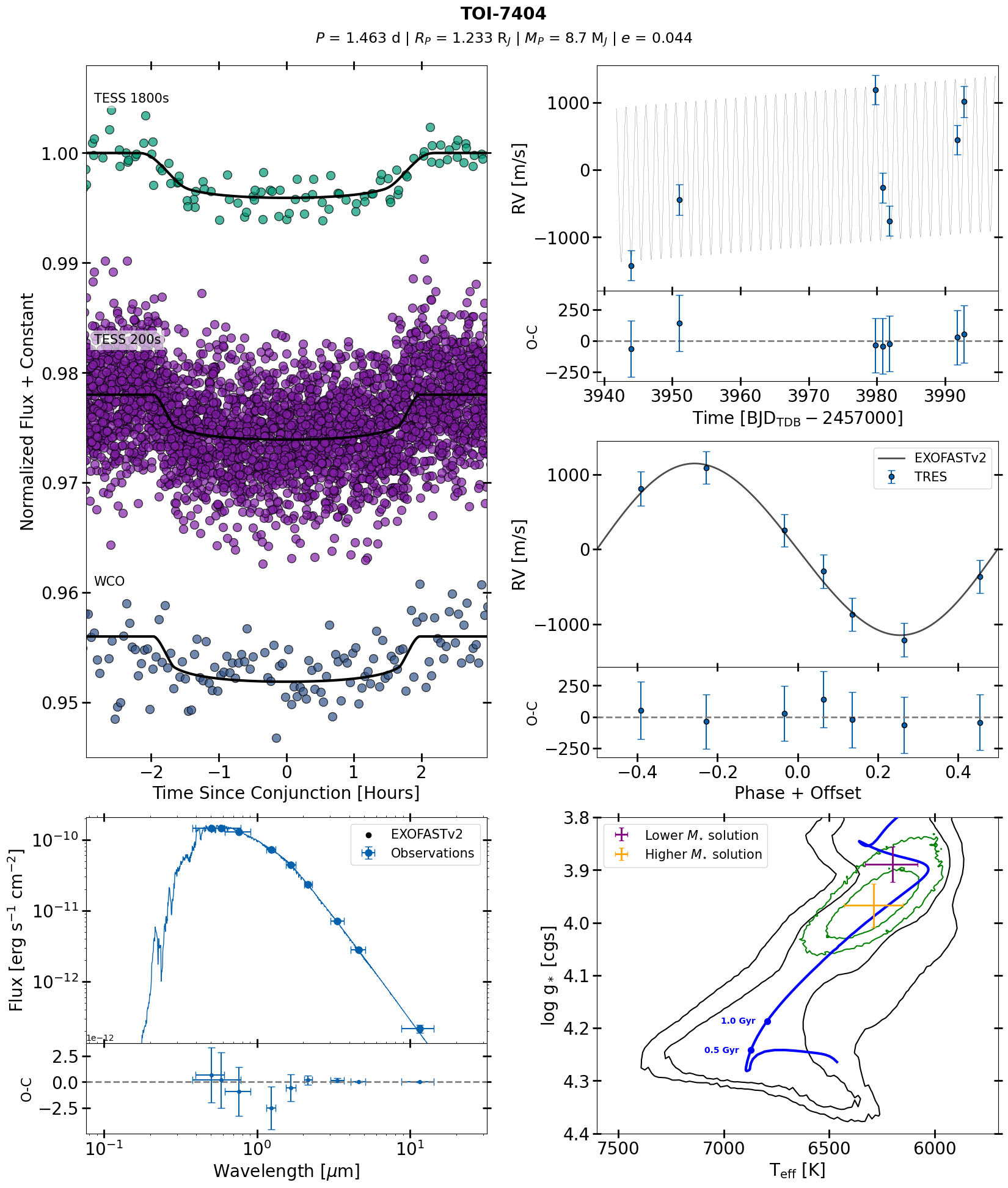}
    \caption{Same as Figure~\ref{fig:toi3041}, but for TOI-7404. In the evolutionary plot in the \textbf{bottom right}, the purple cross represents the median and 68\% confidence interval of the lower-mass subgiant solution, while the orange cross represents that of the higher-mass dwarf solution. Both  of these are listed in Table \ref{tab:bimodal}.}
    \label{fig:toi7404}
\end{figure*}

\begin{figure*}
    \centering    
    \includegraphics[width=\textwidth,height=\textheight,keepaspectratio]{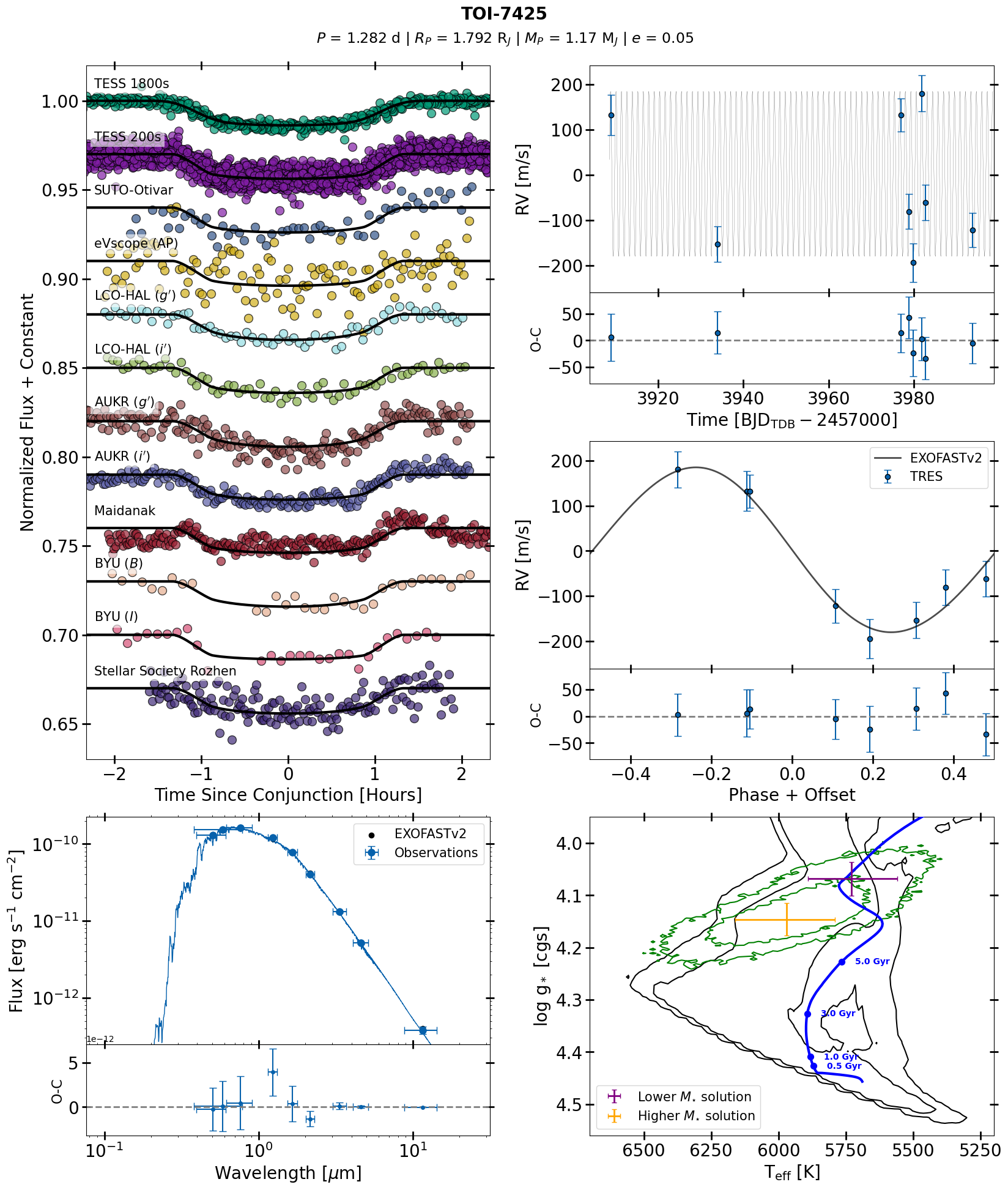}
    \caption{Same as Figure~\ref{fig:toi3041}, but for TOI-7425. In the evolutionary plot in the \textbf{bottom right}, the purple cross represents the median and 68\% confidence interval of the lower-mass subgiant solution, while the orange cross represents that of the higher-mass dwarf solution. Both  of these are listed in Table \ref{tab:bimodal}.}
    \label{fig:toi7425}
\end{figure*}

\begin{figure*}
    \centering    
    \includegraphics[width=\textwidth,height=\textheight,keepaspectratio]{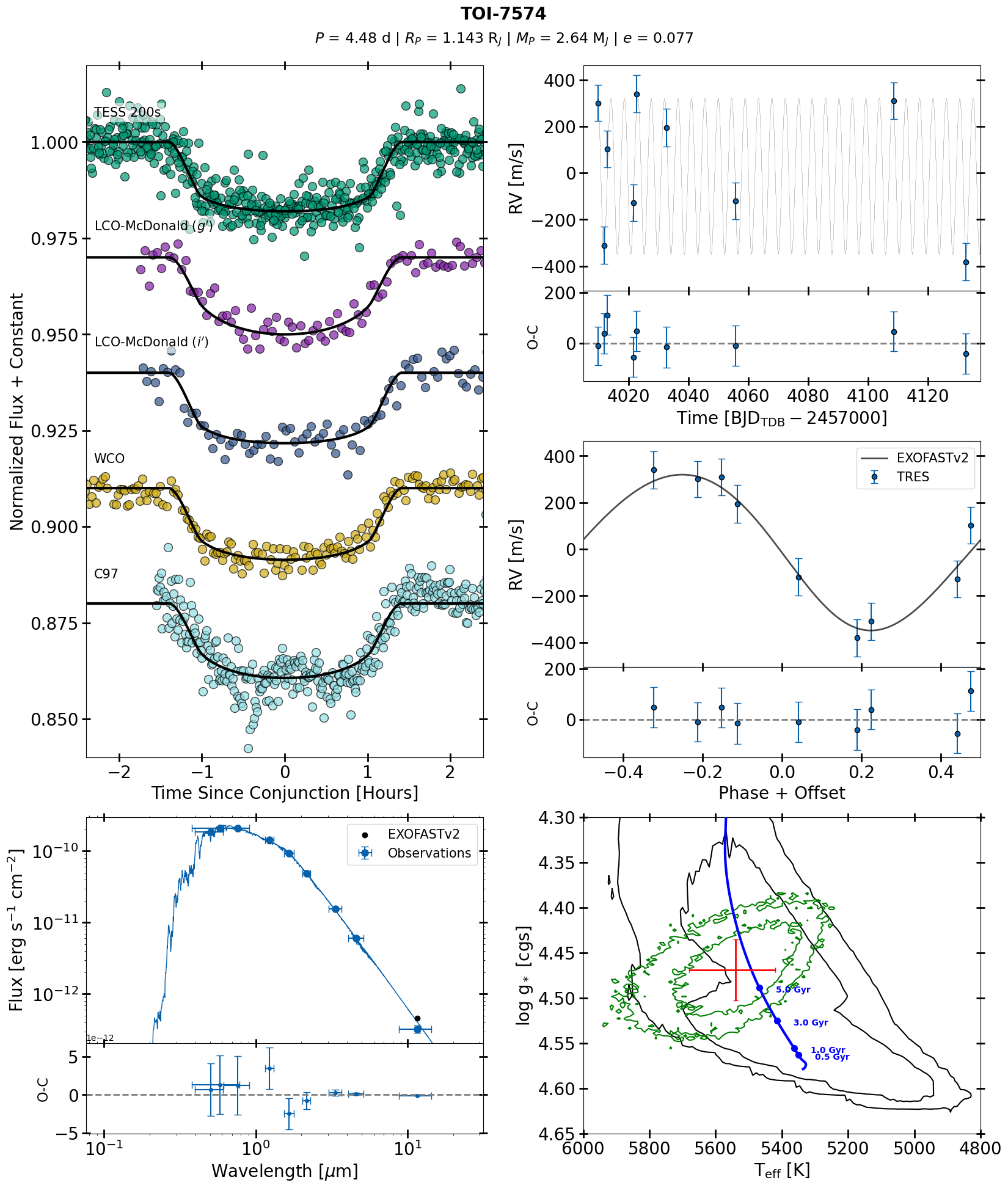}
    \caption{Same as Figure~\ref{fig:toi3041}, but for TOI-7574.}
    \label{fig:toi7574}
\end{figure*}



\clearpage

\begingroup
\footnotesize\itshape
\setlength{\parindent}{0pt}
\setlength{\parskip}{1.5pt plus 1pt}
\newcommand{\affil}[2]{\par\hangindent=1em\hangafter=1 $^{#1}$#2}

\affil{1}{Center for Data Intensive and Time Domain Astronomy, Department of Physics and Astronomy, Michigan State University, East Lansing, MI 48824, USA}
\affil{2}{Center for Astrophysics | Harvard \& Smithsonian, 60 Garden St, Cambridge, MA 02138, USA}
\affil{3}{Department of Astronomy, Indiana University, 727 East 3rd Street, Bloomington, IN 47405-7105, USA}
\affil{4}{Facultad de Ingenier\'ia y Ciencias, Universidad Adolfo Ib\'{a}\~{n}ez, Av. Diagonal las Torres 2640, 7941169 Pe\~{n}alol\'{e}n, Santiago, Chile}
\affil{5}{Al-Fulaij Observatory, Oman}
\affil{6}{University of Wisconsin - Madison}
\affil{7}{Hamilton College, Clinton, NY 13323, USA}
\affil{8}{Brown University, Providence, RI 02912, USA}
\affil{9}{Institute of Advanced Physical Studies, 111 Tsarigradsko Shose Blvd., Sofia, Bulgaria}
\affil{10}{Department of Astronomy, Faculty of Physics, Sofia University, 5 James Bourchier Blvd., Sofia, Bulgaria}
\affil{11}{Physics and Engineering Department, Austin College, Sherman, TX 75090, USA}
\affil{12}{Lomonosov Moscow State University, Universitetskiy prospekt, 13, Moscow, Russia, 119192}
\affil{13}{Instituto de Astrof\'isica de Canarias (IAC), Calle V\'ia L\'actea s/n, 38200, La Laguna, Tenerife, Spain}
\affil{14}{Astrobiology Research Unit, Universit\'e de Li\`ege, 19C All\'ee du 6 Ao\^ut, 4000 Li\`ege, Belgium}
\affil{15}{Department of Earth, Atmospheric and Planetary Science, Massachusetts Institute of Technology, 77 Massachusetts Avenue, Cambridge, MA 02139, USA}
\affil{16}{Department of Astronomy and Space Sciences, Faculty of Science, Ankara University, Tandogan 06100, Ankara, T\"{u}rkiye}
\affil{17}{Ankara University, Astronomy and Space Sciences Research and Application Center (Kreiken Observatory), Incek Blvd., TR-06837, Ahlatl{\i}bel, Ankara, T\"urkiye}
\affil{18}{Silesian University of Technology, Akademicka 16, Gliwice, Poland}
\affil{19}{Acton Sky Portal (Acton, MA USA)}
\affil{20}{SETI Institute \& Unistellar Citizen Scientist}
\affil{21}{Stellar Society Bulgaria}
\affil{22}{Department of Physics and Astronomy, University of Lethbridge, Lethbridge, Alberta, T1K 3M4, Canada}
\affil{23}{Dipartimento di Fisica "E.R. Caianiello", Universit\`{a} di Salerno, Via Giovanni Paolo II 132, Fisciano, I-84084, Italy}
\affil{24}{Istituto Nazionale di Fisica Nucleare, Sezione di Napoli, Via Cintia, Napoli, I-80126, Italy}
\affil{25}{Astrophysics Group, Keele University, Staffordshire ST5 5BG, UK}
\affil{26}{NASA Exoplanet Science Institute - Caltech/IPAC, Pasadena, CA 91125, USA}
\affil{27}{Komaba Institute for Science, The University of Tokyo, 3-8-1 Komaba, Meguro, Tokyo 153-8902, Japan}
\affil{28}{SETI Institute, 339 Bernardo Ave, Suite 200, Mountain View, CA 94043}
\affil{29}{SkyMapper, 2121 Harrison Street, Suite C, San Francisco, CA 94110}
\affil{30}{Department of Astronomy, University of California, Berkeley, CA 94720, USA}
\affil{31}{El Sauce Observatory, Coquimbo Province, Chile}
\affil{32}{NASA Exoplanet Science Institute - Caltech/IPAC, Pasadena, CA 91125, USA | Five College Astronomy Department, Smith College, Northampton, MA 01063, USA}
\affil{33}{Department of Astronomy, The Ohio State University, 140 West 18th Avenue, Columbus, OH 43210, USA}
\affil{34}{Center for Cosmology and Astroparticle Physics, The Ohio State University, 191 W. Woodruff Avenue, Columbus, OH 43210, USA}
\affil{35}{Instituto de Astrof\'\i sica de Canarias (IAC), 38205 La Laguna, Tenerife, Spain}
\affil{36}{Departamento de Astrof\'\i sica, Universidad de La Laguna (ULL), 38206, La Laguna, Tenerife, Spain}
\affil{37}{Grand Pra Observatory, 1984 Les Hauderes, Switzerland}
\affil{38}{Observatori de Ca l'Ou, Sant Mart\'{i} Sesgueioles, GEECAT, Barcelona, Spain}
\affil{39}{Max-Planck-Institut f\"{u}r Astronomie, K\"{o}nigstuhl 17, D-69117 Heidelberg, Germany}
\affil{40}{Brigham Young University}
\affil{41}{SUPA Physics and Astronomy, University of St\,Andrews, Fife, KY16\,9SS Scotland, UK}
\affil{42}{Graduate School of Social Data Science, Hitotsubashi University, 2-1 Naka, Kunitachi, Tokyo 186-8601, Japan}
\affil{43}{NASA Ames Research Center, Moffett Field, CA 94035, USA}
\affil{44}{Departamento de Astronom\'{i}a, Universidad de Chile, Casilla 36-D, Santiago, Chile}
\affil{45}{Department of Astronomy, The University of Texas at Austin, Austin, TX 78712, USA}
\affil{46}{Department of Physics, Lomonosov Moscow State University, Moscow 119991, Russia}
\affil{47}{Hamilton College, 198 College Hill Rd, Clinton, NY 13323, USA}
\affil{48}{Instituto de Astrof\'{i}sica de Andaluc\'{i}a, IAA-CSIC, Glorieta de la Astronom\'{i}a s/n, 18008 Granada, Spain}
\affil{49}{Villa 39 Observatory, Landers, CA, USA}
\affil{50}{Department of Physics and Astronomy, Wellesley College, Wellesley, MA 02481, USA}
\affil{51}{Waffelow Creek Observatory}
\affil{52}{Astrobiology Center, 2-21-1 Osawa, Mitaka, Tokyo 181-8588, Japan}
\affil{53}{National Astronomical Observatory of Japan, 2-21-1 Osawa, Mitaka, Tokyo 181-8588, Japan}
\affil{54}{Departamento de Astronom\'{\i}a y Astrof\'{\i}sica, Universidad de Valencia, E-46100 Burjassot, Valencia, Spain}
\affil{55}{Observatorio Astron\'omico, Universidad de Valencia, E-46980 Paterna, Valencia, Spain}
\affil{56}{Facility for Rare Isotope Beams, Michigan State University, East Lansing, MI 48824, United States of America}
\affil{57}{Odessa I.I.Mechnikov National University, Odessa, Ukraine}
\affil{58}{Brierfield Observatory Bowral N.S.W. Australia}
\affil{59}{AAVSO - Frustaglia Observatory, Spain}
\affil{60}{Unistellar, 19 Rue Vacon, 13001 Marseille, France}
\affil{61}{Department of Physics and Astronomy, University of Notre Dame, 225 Nieuwland Science Hall, Notre Dame, IN 46556, USA}
\affil{62}{Milwaukee Astronomical Society}
\affil{63}{Department of Physics and Kavli Institute for Astrophysics and Space Research, Massachusetts Institute of Technology, Cambridge, MA 02139, USA}
\affil{64}{Kotizarovci Observatory, Sarsoni 90, 51216 Viskovo, Croatia}
\affil{65}{Hazelwood Observatory}
\affil{66}{Department of Physics and Astronomy, West Virginia University, Morgantown, WV 26506-6315, USA}
\affil{67}{Center for Gravitational Waves and Cosmology, West Virginia University, Chestnut Ridge Research Building, Morgantown, WV, USA}
\affil{68}{Sternberg Astronomical Institute Lomonosov Moscow State University, Universitetskii prospekt, 13, Moscow 119991, Russia}
\affil{69}{Department of Astronomy, McPherson Laboratory, The Ohio State University, 140 W 18th Ave, Columbus, Ohio 43210, USA}
\affil{70}{Perth Exoplanet Survey Telescope, Perth, Western Australia, Australia}
\affil{71}{Department of Natural and Environmental Sciences, Western Colorado University, Gunnison, CO}
\affil{72}{U.S. Air Force Academy, Colorado Springs, CO, USA}
\affil{73}{Department of Multi-Disciplinary Sciences, Graduate School of Arts and Sciences, The University of Tokyo, 3-8-1 Komaba, Meguro, Tokyo 153-8902, Japan}
\affil{74}{Department of Physics and Astronomy, Union College, Schenectady, NY, USA}
\affil{75}{Department of Physics, Engineering and Astronomy, Stephen F. Austin State University, 1936 North St, Nacogdoches, TX 75962, USA}
\par
\endgroup
\normalfont\normalsize


\bsp	
\label{lastpage}
\end{document}